\documentclass[a4paper,12pt]{article}
\usepackage[utf8]{inputenc}
\usepackage{cancel}
\usepackage{ulem}
\usepackage{amsfonts}
\usepackage{amssymb}
\usepackage{graphicx}
\usepackage{amsmath}
\usepackage{enumerate}
\usepackage{mathtools}
\usepackage{subfloat}
\usepackage{color}
\usepackage{float}
\usepackage{here}
\usepackage{cite}
\usepackage{mathrsfs}
\usepackage{float,epsfig}
\usepackage{dcolumn}

\usepackage{tabularx, ragged2e, booktabs, caption}
\usepackage{blindtext}
\usepackage{caption} 
\usepackage{subcaption}
\usepackage{tablefootnote}
\usepackage{booktabs} 
\usepackage[export]{adjustbox}
\usepackage{authblk}

\usepackage{ulem}

\usepackage{graphicx}
\usepackage{bm}
\usepackage{amsmath,amssymb,amsthm}
\usepackage[colorlinks=true,linkcolor=blue,citecolor=red]{hyperref}
\newcommand{\be}{\begin{equation}}
\newcommand{\ee}{\end{equation}}
\newcommand{\bea}{\setlength\arraycolsep{2pt} \begin{eqnarray}}
\newcommand{\eea}{\end{eqnarray}}

\def\0{{\sst{(0)}}}
\def\1{{\sst{(1)}}}
\def\2{{\sst{(2)}}}
\def\3{{\sst{(3)}}}
\def\4{{\sst{(4)}}}
\def\5{{\sst{(5)}}}
\def\6{{\sst{(6)}}}
\def\7{{\sst{(7)}}}
\def\8{{\sst{(8)}}}
\def\sst#1{{\scriptscriptstyle #1}}

\usepackage{array,multirow,makecell}
\setcellgapes{1pt}
\makegapedcells
\newcolumntype{R}[1]{>{\raggedleft\arraybackslash }b{#1}}
\newcolumntype{L}[1]{>{\raggedright\arraybackslash }b{#1}}
\newcolumntype{C}[1]{>{\centering\arraybackslash }b{#1}}

\usepackage{tikz,xcolor,hyperref}

\definecolor{lime}{HTML}{A6CE39}
\DeclareRobustCommand{\orcidicon}{%
	\begin{tikzpicture}
	\draw[lime, fill=lime] (0,0) 
	circle [radius=0.16] 
	node[white] {{\fontfamily{qag}\selectfont \tiny ID}};
	\draw[white, fill=white] (-0.0625,0.095) 
	circle [radius=0.007];
	\end{tikzpicture}
	\hspace{-2mm}
}
\foreach \x in {A, ..., Z}{%
	\expandafter\xdef\csname orcid\x\endcsname{\noexpand\href{https://orcid.org/\csname orcidauthor\x\endcsname}{\noexpand\orcidicon}}
}

\title{\bf Topology of Born-Infeld-AdS Black Hole
Phase Transitions: Bulk and CFT Sides}

\author{
 Md Sabir Ali\orcidS{}$^{1,2,3,4}$\footnote{alimd.sabir3@gmail.com},
 Hasan  El Moumni\orcidH{}$^5$\thanks{h.elmoumni@uiz.ac.ma (Corresponding author)} ,   Jamal  Khalloufi\orcidJ{}$^{5}$\footnote{jamalkhalloufi@gmail.com },
 Karima  Masmar\orcidK{}$^{6}$\footnote{karima.masmar@gmail.com }\footnote{Authors are in alphabetical order.}
\\
{
\small $^{1}$ Department of Physics, Mahishadal Raj College (affiliated to Vidyasagar University), Purba Medinipur, West Bengal, 721628, India\\
\small $^{2}$ Key Laboratory of Quantum Theory and Applications of MoE, Lanzhou Center for Theoretical Physics, Lanzhou University, Lanzhou 730000, China\\

\small $^{3}$ Key Laboratory of Theoretical Physics of Gansu Province, Institute of Theoretical Physics $\&$ Research Center of Gravitation, Lanzhou University, Lanzhou 730000, China \vspace{0.1cm}\\

\small $^{4}$ School of Physical Science and Technology, Lanzhou University, Lanzhou 730000, China \vspace{0.1cm}

\small $^{5}$ LPTHE, Physics Department, Faculty of Sciences, Ibnou Zohr University, B.P 8106, Agadir, Morocco. \\
\small $^{6}$ Laboratory of High Energy Physics and Condensed Matter, Faculty of
Sciences Ain Chock, B.P 5366, HASSAN II University, Casablanca, Morocco. 
}
}
\vspace{-3.6em}

\date{\today}
\begin{document} 
\maketitle
\begin{abstract}
The thermodynamic criticality of the AdS black holes serves as an important structure during the thermal phase transition. This paper discusses about the critical points and their topology during thermal phase transitions of the Born-Infeld AdS black holes. We make such investigations using two different topological approaches, namely, using Duan's topological current $\phi$-mapping theory, and the off-shell free energy. Within Duan's formalism, we observe that for a given value of the Born-Infeld parameter $b$, there exists an associated electric charge parameter $Q$, which is highly sensitive to the topological phase transitions. This way we examine the connections of the first-order phase transition and the topological nature of the critical points. We find that the topological nature has a possible breakdown in certain parametric ranges. In effect, we determine the unconventional and the conventional phase critical points as the creation (topologically vortex) and annihilation (topologically anti-vortex) points (pairs). As the second approach, we call the off-shell free energy to determine the topological classes: of which one corresponds to the AdS-Schwarzschild black hole phases, while the other provides a possible topological phase transition. Here we also reveal a novel phase transition between two unstable phases, namely, the unstable small black hole and the intermediate black holes. For certain parametric values of the Born-Infeld parameter and the pressure, we also study the different topological descriptions that inevitably correspond to the AdS-Reissner-Nordstr$\Ddot{o}$m black hole phases. As a consistency check of the critical points during the topological phase transitions, we study the vortex/anti-vortex annihilation thermodynamics from local as well as global thermodynamic viewpoints.     
 At the end, a special emphasis has been put on phase transition from the conformal field theory perspective revealing that the Born-Infeld-AdS black hole and its dual CFT share identical topology. Consequently, any topological transition that occurs in the bulk corresponds to a parallel transition in the dual CFT.
{\noindent}

\end{abstract}

\tableofcontents


\section{Introduction}

Born-Infeld (BI) electrodynamics is a prominent example of Nonlinear Electrodynamics (NLED). It introduces modifications to the conventional laws of electrodynamics in order to achieve a finite value for the self-energy of a point charge. In other words, it can be seen as
 a natural generalization of the Maxwell theory to probe the short-distance behavior of the electromagnetic fields \cite{Born:1934gh}. Therefore, the effects of the nonlinearity of the Born-Infeld electrodynamics are very useful for studying the very strong field regime of gravity, such as black holes, neutron stars\cite{Prasetyo:2017hrb,Prasetyo:2021kfx}, or any other compact objects\cite{Pereira:2018mnn,Kim:2022xum,Jaramillo:2023lgk} that have the effects of strong gravitational force.  In addition to the relevance of the Born-Infeld electrodynamics in the astrophysics, it has major impacts on string theory, black hole physics, cosmology, and quantum field theory. Indeed, the Born-Infeld action is a part of the effective action for the world-volume dynamics of D-branes in string theory \cite{Fradkin:1985qd,Tseytlin:1997csa,Gibbons:2001gy}, and in the context of cosmology, it provides alternatives to standard cosmological models \cite{Gibbons:2001sx,Dyadichev:2001su,Bouhmadi-Lopez:2013lha,Jana:2016uvq,Albarran:2018mpg,Jayawiguna_2018,Kibaroglu:2022olh}. Born-Infeld-type actions have also connections to certain aspects of quantum field theory and noncommutative geometry \cite{Hofman:1998iy,Hofman:1998iv,Serie:2003nf}.
 
The thermodynamics of black holes in particular, or any thermal systems in general, and their critical points analysis are of great importance in the general relativistic theory of gravity. Especially, the Born-Infeld black hole solutions in the presence of a cosmological constant \cite{Cai:2004eh,Dey:2004yt} have some rich phase transitions behaviour. Furthermore, the thermodynamic properties and phase structure of BI black holes in various physical fields have also been the subject of intense investigation \cite{Gunasekaran:2012dq,Bai:2022vmx,He:2022opa,Tao:2017fsy,Meng:2018vtl,Bahrami-Asl:2018xnd,Wen:2022hkv}. Moreover, such 
 black holes in AdS spacetime in four dimensions show an unusual behavior called the reentrant phase transitions that were uncommon from the viewpoint of black hole thermodynamics. Such behavior is reflected in a thermal system comprising of nicotine/water mixture \cite{NARAYANAN1994135}. Few of the phenomena arising from such a strong field regime have been tested so far\cite{Banerjee:2012zm,Mo:2014qsa,Johnson:2015fva,Zhang:2017lhl,Hendi:2018hdo,Simovic:2019zgb,Kumar:2019zbp,Chen:2019bwt,NaveenaKumara:2020biu,Dehghani:2020jcw,Jing:2020sdf,Kusuma:2021yfs,Kumar:2022fyq}. Among them include the gravitational waves emerging from the black hole mergers \cite{LIGOScientific:2016aoc}, the gravitational lensing and shadows of the black holes \cite{EventHorizonTelescope:2019dse, EventHorizonTelescope:2019ggy, EventHorizonTelescope:2022wkp}, and few more to cite. But the black hole thermodynamics remains to be tested though significant progress has been made on the thermodynamics of black holes. In this sense, recently, an attempt has been made to test Hawking's area theorem using the gravitational wave (GW) data \cite{Isi:2020tac, Prasad:2020xgr, Gerosa:2022fbk}. Therewith, a wealth of research has spurred numerous authors to investigate the near-horizon geometry of black holes through the analysis of gravitational wave (GW) data, revealing a remarkable alignment with observational data. This convergence between theory and observation is reinforced by a range of indirect indicators supporting the idea that black holes adhere to the principles of thermodynamic systems. Notably, a recent analysis using LIGO/Virgo's GW data \cite{Carullo:2021yxh} has provided insights into the Bekenstein-Hod universal bound, offering compelling evidence that the rate of information emission from black holes aligns with theoretical expectations. Therefore, it is universally accepted that the black hole should behave as a thermal system when quantum field effects on the horizon geometry are taken into consideration. There are other compelling corroborations, such as gauge/gravity duality \cite{Maldacena:1997re, Witten:1998qj}, the van der Waals-like phase transitions \cite{Kubiznak:2012wp}, the reentrant phase transitions behavior \cite{Altamirano:2013ane, Altamirano:2013uqa, Altamirano:2014tva, Kubiznak:2015bya,Hendi:2015soe, Frassino:2014pha, Wei:2014hba, Hennigar:2015esa, Sherkatghanad:2014hda, Hennigar:2015wxa, Xu:2019yub, Zou:2016sab, Zou:2013owa, Chabab:2020xwr,Chabab:2019kfs,NaveenaKumara:2020biu,Ali:2023iuz}, the superfluid analogs of the liquid He$^4$ \cite{Hennigar:2016xwd, Xu:2018fag, Hennigar:2016ekz,Hendi:2015hoa,Hendi:2016yof} etc., for the AdS black hole systems which opened the door to unravel the thermodynamics of black holes. \\

The thermal phase transition is an elegant phenomenon for a black hole system, particularly near critical points. Such points are utilized to study the behavior of various thermodynamic quantities. The thermal transition appears in different forms for various black hole systems in the asymptotically AdS spacetime. The presence of critical points provides us with the second-order phase transition that occurs locally in a physical system. This way, the critical exponents are determined near critical points using the mean-field theory. Most of the AdS black hole systems outshine the same critical exponents near the critical points. Therefore, the critical exponents are universal and generic features of the physical systems when analyzed from the viewpoint of the mean-field theory. An exception to this case was found for the AdS black holes in Lovelock gravity where the physical systems comprise the isolated critical points and thereby, violating the mean field exponents \cite{Frassino:2016vww, Wu:2022plw}. Additionally, the thermality of some AdS black holes shows multiple critical points, such as the reentrant phase transitions, the triple point phase structure \cite{Wei:2014hba, Zhang:2020obn, Cui:2021qpu, Frassino:2014pha, Hennigar:2016ekz, Hull:2021bry,Astefanesei:2021vcp}, and as an extreme case, the infinite critical points for a superfluid black hole. 

 Understanding the holographic interpretation of the extended thermodynamics entails a certain degree of difficulty \cite{Kastor:2014dra,Karch:2015rpa,Chabab:2015ytz,Johnson:2014yja,Belhaj:2015uwa,Dolan:2014cja,Zhang:2014uoa,McCarthy:2017amh}. The AdS/CFT correspondence facilitates the depiction of the extended thermodynamics within the bulk as a representation
of the corresponding CFT at finite temperature. Specifically, the parameters characterizing the black holes, such as mass, entropy, and the Hawking temperature, can be directly correlated to the energy, entropy, and temperature of thermal states in the dual CFT \cite{Hawking:1982dh,Witten:1998zw}. However, a parallel correlation between the bulk pressure and the pressure of the dual field theory is absent. Furthermore, the thermodynamic volume of black holes does not align with the CFT volume \cite{Johnson:2014yja}. In contrast, the Smarr relation linking bulk quantities should correspond to the Euler relation for CFT quantities, with the latter devoid of dimension-dependent factors.

 These challenges can be mitigated by expanding the parameter space of bulk thermodynamics. It has been proposed that modifying the bulk pressure $P$ is tantamount to adjusting the central charge $C$ or the number of colors $N$ in the dual gauge theory, as elucidated through a holographic dictionary \cite{Kastor:2009wy,Kastor:2014dra,Karch:2015rpa,Johnson:2014yja,Dolan:2014cja}:
\begin{equation}
C=k \frac{l^{D-2}}{16 \pi G}, \label{eq:central}
\end{equation}
where $l$ denotes the AdS radius, $k$ is a numerical factor determined by specific features of the holographic system \cite{Karch:2015rpa}. If the variation of Newton’s constant $G$ is taken into account, the central charge $C$ can be adjusted to maintain the integrity of the field theory.
 These observations for various AdS black hole systems help us to understand some of the intriguing aspects of black hole thermodynamics near the critical points,  and to get a better understanding of the AdS/CFT interpretation, which gives us significant clues that thermodynamics is a strong root to quantum gravity.\\

In many branches of modern theoretical physics, the study of the topological phenomena of any physical system has been an active area of research ranging from particles to condensed matter physics. Such ideas were developed and used for many years. In gravitational systems, they are very recent and still in their initial stage. In 1980, Duan Yishi developed a phenomenological model to study the topological charge and current of a physical system \cite{Duan:1984ws}. Such an idea is an elegant prescription to investigate the kinetics of various topological defects, such as collisions, coalesces, the splitting of various defects, and the production/annihilation of vortex/antivortex pairs. In gravitational systems, this prescription has been adopted recently to study the formation of the light rings, the thermodynamic phase structure of the black holes \cite{PhysRevD.105.104003, PhysRevD.107.046013, Wei:2022dzw}. From thermodynamic perspectives, the topological charge and the current are studied using Duan's prescription for (non)-charged and rotating black holes, the black holes in Gauss-Bonnet gravity and many more \cite{Bai:2022klw,Wu:2022whe, Liu:2022aqt, Fan:2022bsq, Wu:2023sue, Chatzifotis:2023ioc, Wu:2023xpq, Li:2023ppc, Yerra:2023hui, Du:2023nkr, Alipour:2023uzo, Zhang:2023tlq, TranNHung:2023nmb, Gogoi:2023qku, Wu:2023fcw}. The light rings structure and the geodesics (in)stability have been investigated for a few types of black hole systems \cite{Ye:2023gmk,Cunha:2017qtt, Wei:2022mzv, Delgado:2021jxd, Wei:2020rbh}, kinematic of particles \cite{Yin:2023pao} and CFT thermodynamics \cite{Zhang:2023uay}. Furthermore, the winding numbers and the topological numbers are also obtained by applying the residue theorem to a characteristic complex function constructed out of the off-shell Gibbs free energy \cite{ Fang:2022rsb}. The residue method can easily recover the results that are obtained using the topological current method. Besides, a new perception is lent from the complex analysis and which is based on the correspondence between the winding numbers and the Riemann surface foliation has been constructed in \cite{Xu:2023vyj}.

Motivated by all these findings, we aim in our current paper to find the topology of the Born-Infeld AdS black holes using two approaches: (1) using Duan's topological current $\phi$-mapping method, and (2) by constructing the off-shell Gibbs free energy. We discuss the paper as follows. In Section~\ref{Duan's appraoch}, we extensively discuss Duan's topological current method to study the thermodynamic critical points and their connection to the topological phase transitions. The calculation of the winding numbers and also the unconventional, conventional phase transitions, and the emergence of possible topological classes are vividly described and identified in this section. On the other hand, The analysis regarding the off-shell free energy is part of Section~\ref{off-shell_FE}. The detailed prescriptions to the stable/unstable phase generations and annihilation which in topologically equivalent descriptions correspond, respectively, to the vortex/anti-vortex creation and the annihilation points is a part of the discussion of this section. In Section~\ref{vortex-antivortex}, we briefly present the vortex/anti-vortex annihilation in a thermodynamics context. 
 Then, in Section~\ref{adscft}, we employ the holographic thermodynamics framework to examine the potential existence of such phase transitions, both within the bulk and the corresponding dual CFT.
Finally, in Section~\ref{summary} we summarize our results and conclude the paper.

\section{ Anti-de Sitter black hole solution in the  Born-Infeld electrodynamics framework}

The foundation of our exploration lies in the realm of four-dimensional Einstein gravity, within which the  action of gravity coupled to the Born-Infeld nonlinear electrodynamics is obtained in \cite{Dey:2004yt} as 
\begin{equation}\label{1x}
	\mathcal{S} = \dfrac{1}{16 \pi} \int d^4x\sqrt{-g}\left[  R-2 \Lambda + 4 b^2 \left( 1 - \sqrt{1+\dfrac{F_{\mu \nu} F^{\mu \nu}}{2 b^2}}\right)  \right]  ,
\end{equation}
where $R$ is the Ricci scalar curvature, $\Lambda$ denotes the cosmological constant, and which is related to the Anti-de-Sitter radius as $\Lambda = -3/l^2$. The constant $b$ stands for the Born-Infeld parameter with the dimension of mass that relates to the string tension $\alpha'$ as $b = 1 / \left( 2 \pi \alpha ' \right) $ \cite{Gibbons:2001gy}. Moreover, the electromagnetic tensor field $F_{\mu \nu}$ is obtained from the vector potential $A_{\mu}$ as 
 $F_{\mu \nu} = \partial_{\mu} A_{\nu}-\partial_{\nu} A_{\mu}$. 
 
 The $ansatz$ associated with the statically and spherically symmetric black hole solution is 
\begin{equation}\label{2x}
		ds^2  = - f(r) dt^2 + \dfrac{dr^2}{f(r)}+ r^2 d\Omega^2,
\end{equation}
with, $d\Omega$ is the metric on a unit $2-$sphere and the blackening function $f(r)$ is found to be \cite{Dey:2004yt,Fernando:2003tz,Cai:2004eh}
\begin{equation}\label{3x}	
	f(r) = 1 + \dfrac{r^2}{l^2}-\dfrac{m}{r}+\dfrac{2 b^2 r^2}{3}\left( 1 - \sqrt{1+\dfrac{16 \pi^2 Q^2}{b^2 r^2}}\right) +\dfrac{64 \pi^2 Q^2}{3 r^2} {}_2\mathcal{F}_1\left[ \dfrac{1}{4} , \dfrac{1}{2}, \dfrac{5}{4}, - \dfrac{16 \pi^2 Q^2}{b^2 r^2} \right], 
\end{equation}
in which ${}_2\mathcal{F}_1\left[ a,b,c,d \right]$ is the hypergeometric function, the black hole mass $M$ can be expressed by the help of the integration constant $m$ as $M = m/8\pi$, the electric charge of the black hole per unit volume $\omega=4\pi$ is denoted by $Q$. The non-zero component of the gauge potential can be expressed as follows
\begin{equation}\label{4x}	
	A_t(r) = - \dfrac{4 \pi Q}{r} {}_2\mathcal{F}_1\left[ \dfrac{1}{4} , \dfrac{1}{2}, \dfrac{5}{4}, - \dfrac{16 \pi^2 Q^2}{b^2 r^2} \right]. 
\end{equation}
The charged-Anti-de Sitter black hole solution \cite{Kubiznak:2012wp, Dehyadegari:2016nkd} can easily be found by considering the limit $b \to \infty$ in the both blackening and the gauge potential functions. The formula of Hawking temperature of Born-Infeld AdS black hole is 
\begin{equation}\label{16}	
	T = \dfrac{1}{4 \pi} \left. \dfrac{\partial f(r)}{\partial r}\right| _{r=r_h} = \dfrac{1}{4 \pi r_h} + \dfrac{3 r_h}{4 \pi l^2} + \dfrac{b^2 r_h}{2\pi} \left( 1 - \sqrt{1+\dfrac{16 \pi^2 Q^2}{b^2 r^2}}\right). 
\end{equation}
Herein $r_h$ denotes the event horizon radius which is obtained as the largest positive real solution of the equation $f(r)=0$. The electric potential at infinity with respect to the event horizon reads as 
\begin{equation}\label{6x}	
	\wp = \dfrac{\partial M}{\partial Q} = \dfrac{4 \pi Q}{r_h} {}_2\mathcal{F}_1\left[ \dfrac{1}{4} , \dfrac{1}{2}, \dfrac{5}{4}, - \dfrac{16 \pi^2 Q^2}{b^2 r_h^2} \right]. 
\end{equation}
To formulate the first law of thermodynamics linked to the Born-Infeld-AdS black hole, it is essential to revisit its fundamental components. These include entropy denoted as $S$, pressure represented by $P$, and the corresponding volume, designated as $V$ \cite{Gunasekaran:2012dq}.
\begin{equation}\label{7x}	
\begin{split}
	S &=\int \dfrac{1}{T} \dfrac{\partial M}{\partial r_h} d r_h = \dfrac{r_h^2}{4}, \quad P = - \dfrac{\Lambda}{8 \pi}, \quad V = \dfrac{\partial M}{\partial P} = \dfrac{r_h^3}{4}, \\
	\mathcal{B}& = \dfrac{\partial M}{\partial b} =  \dfrac{b r_h^3}{6 \pi} \left( 1 - \sqrt{1+\dfrac{16 \pi^2 Q^2}{b^2 r^2}}\right) + \dfrac{4 \pi Q^2}{3 b r_h}  {}_2\mathcal{F}_1\left[ \dfrac{1}{4} , \dfrac{1}{2}, \dfrac{5}{4}, - \dfrac{16 \pi^2 Q^2}{b^2 r_h^2} \right],
\end{split}
\end{equation}
the additional quantity $\mathcal{B}$ is conjugate to the parameter $b$ that  and which is interpreted as the Born-Infeld vacuum polarization \cite{Gunasekaran:2012dq}, to that the first law and the Smarr formula take the forms
\begin{equation}\label{8x}	
	\begin{split}
		dM &= T dS + \wp dQ + V dP + \mathcal{B}db,\\
		M& = 2 T S + \wp Q - 2 P V - \mathcal{B} b.
	\end{split}
\end{equation}
It's crucial to note that the thermodynamic quantities $M$, $Q$, $S$, and $V$ are expressed per unit volume $\omega$.
 The study of Born-Infeld-AdS black hole thermodynamics in the extended phase space is significant due to its relevance to fundamental questions in gravitational physics, gauge-gravity duality, and the exploration of novel phenomena associated with black hole solutions in the presence of nonlinear electrodynamics. An extension to AdS backgrounds leads us to explore the unexpected reentrant phase structure and the nature of the critical points.

\section{First approach: Duan's topological current $\phi$-mapping theory}\label{Duan's appraoch}

The temperature of the Born-Infeld-AdS black hole can be expressed as a function of the entropy $S$, pressure
$P$, electric charge $Q$ and the Born-Infeld parameter $b$
\begin{equation}\label{1}
T = T(S,P,Q,b).
\end{equation}
Following \cite{PhysRevD.105.104003} and \cite{PhysRevD.107.046013}, one can reduce the degree of dependency associated with temperature by requiring  the following constraint
\begin{equation}\label{2}
	\left. \dfrac{\partial T}{\partial S}\right)_{P,Q,b}  = 0.
\end{equation}
Then we introduce a new thermodynamic function
\begin{equation}\label{3}
	\Phi = \dfrac{1}{\sin(\theta)} \tilde{T}(S,Q,b),
\end{equation}
where $\tilde{T}$ is the black hole temperature obtained via \eqref{2} upon eliminating pressure $P$, and $\frac{1}{\sin(\theta)}$ denotes an auxiliary factor recalled for simplification reasons in our study of the topology of black hole thermodynamics.

Hence, we construct a vector field $\phi = (\phi^S,\phi^{\theta})$, such that
\begin{equation}\label{4}
	\phi^S = \left. \dfrac{\partial \Phi}{\partial S}\right)_{\theta,Q,b}, \quad \quad \quad \phi^{\theta} = \left. \dfrac{\partial \Phi}{\partial \theta}\right)_{S,Q,b}.
\end{equation}
The direction of the newly introduced vector $\phi$ is perpendicular to the horizontal lines at $\theta = 0$ and $\theta = \pi$, which can be treated as two boundaries in the parameter space. Moreover, the zero point of $\phi$ is always located at $\theta = \pi/2$. The critical point is exactly spotted at the zero point of $\phi$. Indeed, the  remaining condition to determine the critical point is given by \footnote{In fact, there are two conditions to determine the critical point:\begin{equation*}
\left. \dfrac{\partial T}{\partial S}\right)_{P,Q,b}  = 0 \quad \text{and} \quad \left. \dfrac{\partial^2 T}{\partial S^2}\right)_{P,Q,b}  = 0,
\end{equation*} 
but the first one is the same as reducing the dependency order  Eq.\eqref{2}.}
\begin{equation}\label{5}
	\left. \dfrac{\partial^2 \tilde{T}}{\partial S^2}\right)_{Q,b}  = 0.
\end{equation}

Using, the prescription of Duan’s $\phi$-mapping topological current theory \cite{Duan:1979ucg, Duan:1984ws}, we can build the topological current as
\begin{equation}\label{6}
	j^{\mu} = \dfrac{1}{2 \pi} \epsilon^{\mu\nu\rho}\epsilon_{ij}\dfrac{\partial n^i}{\partial x^{\nu}}\dfrac{\partial n^j}{\partial  x^{\rho}}, \quad \quad \quad \mu,\nu,\rho = 0,1,2,
\end{equation}
where $x^\nu = (t,r,\theta)$, and the normlized vector $n^i$ is defined as $n^i= \frac{\phi^i}{\left| \left| \phi\right| \right| }, \; i = (1,2)$, with $\phi^1 = \phi^S$ and $\phi^2 = \phi^{\theta}$. It is easy to show that this topological current is conserved, i.e.,
\begin{equation}\label{7}
\dfrac{\partial j^{\mu}}{\partial x^{\mu}}	 = 0.
\end{equation}
Thanks to the Jacobi tensor
\begin{equation}\label{8}
	 \epsilon^{ij}J^{\mu}\left( \frac{\phi}{x}\right)  = \epsilon^{\mu\nu\rho}\dfrac{\partial \phi^i}{\partial x^{\nu}}\dfrac{\partial \phi^j}{\partial  x^{\rho}},
\end{equation}
and the two-dimensional Laplacian Green function
\begin{equation}\label{9}
	\Delta_{\phi^i} \ln\left| \left| \phi\right| \right| = 2 \pi \delta(\phi) ,
\end{equation}
 that allow us  to  express the topological current  as
\begin{equation}\label{10}
	j^{\mu} = \delta^2(\phi)J^{\mu}\left( \frac{\phi}{x}\right) .
\end{equation}
Now, it is obvious that $j^{\mu}$ is nonzero only at the zero points of $\phi$. In addition, according to the $delta$-function theory, one can obtain the density of the topological current,
\begin{equation}\label{11}
	j^{0} = \sum_{k=1}^{N}\beta_k \eta_k \delta^2(\vec{x}-\vec{z_k}),
\end{equation}
in which $\vec{z_k}$ stands for the $k^{th}$ solution of the equation $\phi^i(x^j) = 0$, and $N$ is the number of solutions it comprises of. The positive Hopf index $\beta_k$ measures the number of the loops that $\phi^i$ makes in the vector $\phi$ space when $x^\mu$ goes around the zero point $z_k$. The Brouwer degree $\eta_k = \text{sign} \left(J^{0}(\phi/x)_{z_k} \right) = \pm 1 $. Consequently, the topological charge at a given parameter region $\Sigma$ can be calculated via
\begin{equation}\label{12}
	\mathcal{Q} = \int_{\Sigma}j^0 d^2x = \sum_{k=1}^{N}\beta_k \eta_k = \sum_{k=1}^{N} w_k,
\end{equation}
where $w_k$ is the winding number for the $k^{th}$ zero point of $\phi$.
Then an elegant correspondence can be established by associating 
 each critical point with a topological charge, such a charge equals the winding number.  Since $\eta_k$ can be positive or negative, these critical points have two different topological properties. 
 
 When we consider $\Sigma$ as the complete thermodynamic parameter space for a thermodynamic system, we are in a position to calculate its topological charge. Then, different thermodynamic systems can then be classified into various classes.

Topology tells us that if the contour encloses the critical point, it will have a nonzero topological charge; otherwise, it will be null.
To compute the topological charge, we shall build a contour that is parameterized by  $\vartheta \in \left] 0,\pi \right[  $ as 
\begin{equation}\label{13}
	r = a_r \cos(\vartheta) + r_0, \quad \quad \theta =  a_{\theta} \sin(\vartheta) + \dfrac{\pi}{2},
\end{equation}
with $a_r$ and $a_\theta$ are some aritrary constants and  $r_0$ is the $r$-coordinate of at the critical point. Then we define a new quantity measuring the deflection of the vector field along a given contour
\begin{equation}\label{14}
	\Omega(\vartheta) = \int_{0}^{\vartheta} \epsilon_{ij} n^i \dfrac{\partial n^j}{\partial \vartheta} d\vartheta.
\end{equation}
For $\vartheta = 2 \pi$, the contour surrounds the zero point and the topological charge is given by 
\begin{equation}\label{15}
	\mathcal{Q} = \dfrac{1}{2 \pi} \Omega(2 \pi).
\end{equation}
In what follows, we would like to examine  topologically the thermodynamical properties of the Born-Infeld-AdS black hole by calculating the topological charge. 
The Hawking temperature is obtained by Eq.\eqref{16}  
and using it in Eq.\eqref{2} and eliminating the pressure term, we can get from Eq.\eqref{3}, the thermodynamic function $\Phi$ 
\begin{equation}\label{17}
	\Phi = \dfrac{\csc(\theta)}{4 \pi S^{3/2}}\left( S-\dfrac{8 \pi^2 Q^2}{\sqrt{1 + \dfrac{\pi^2 Q^2}{b^2 S^2}}}\right).
\end{equation}

\begin{figure}[!ht]
		\centering 
	\begin{subfigure}[h]{0.48\textwidth}
		\centering \includegraphics[scale=0.6]{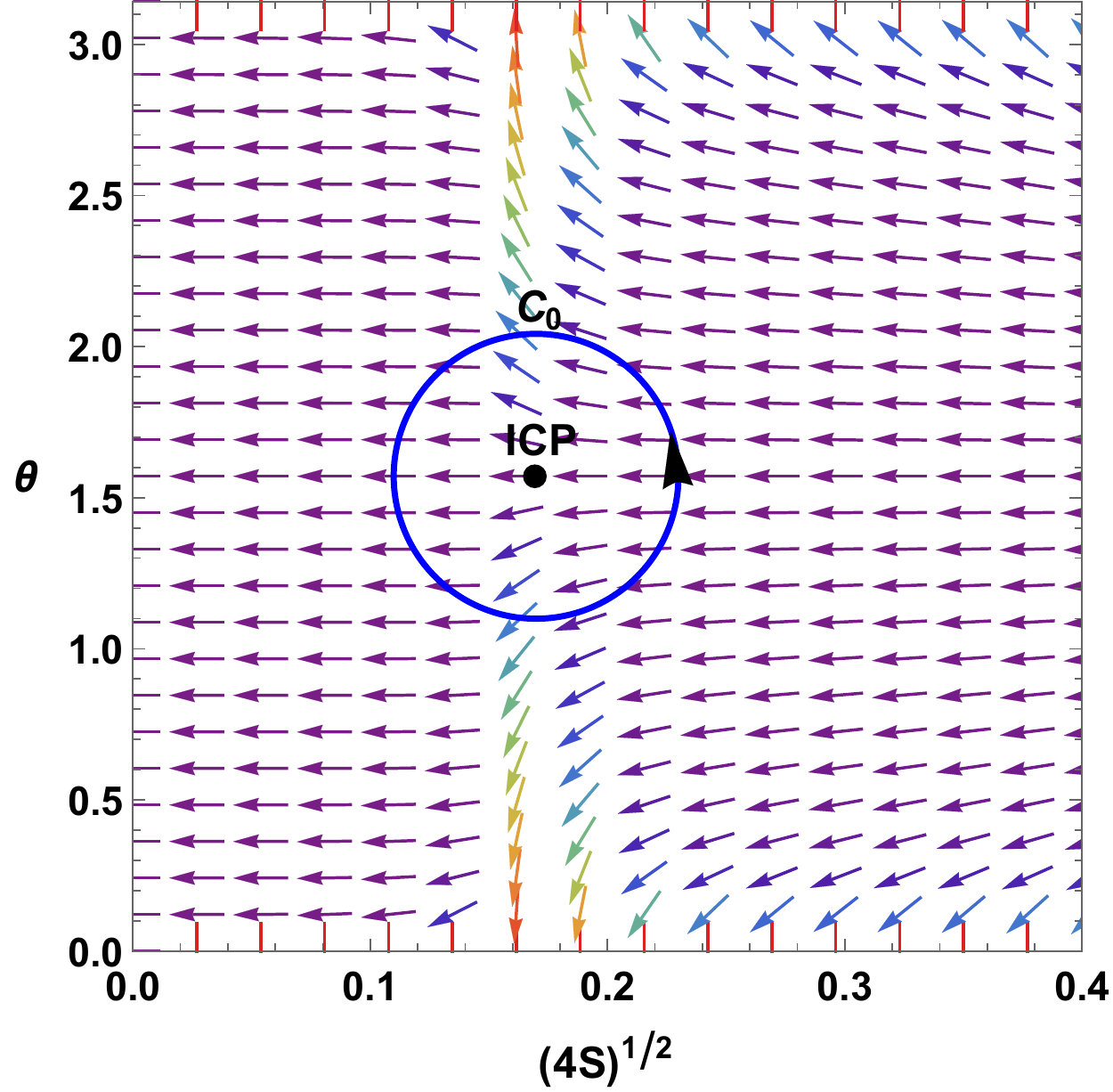}
		\caption{$Q = Q_I$}
		\label{f1_1}
	\end{subfigure}
	\hspace{1pt}	
	\begin{subfigure}[h]{0.48\textwidth}
		\centering \includegraphics[scale=0.6]{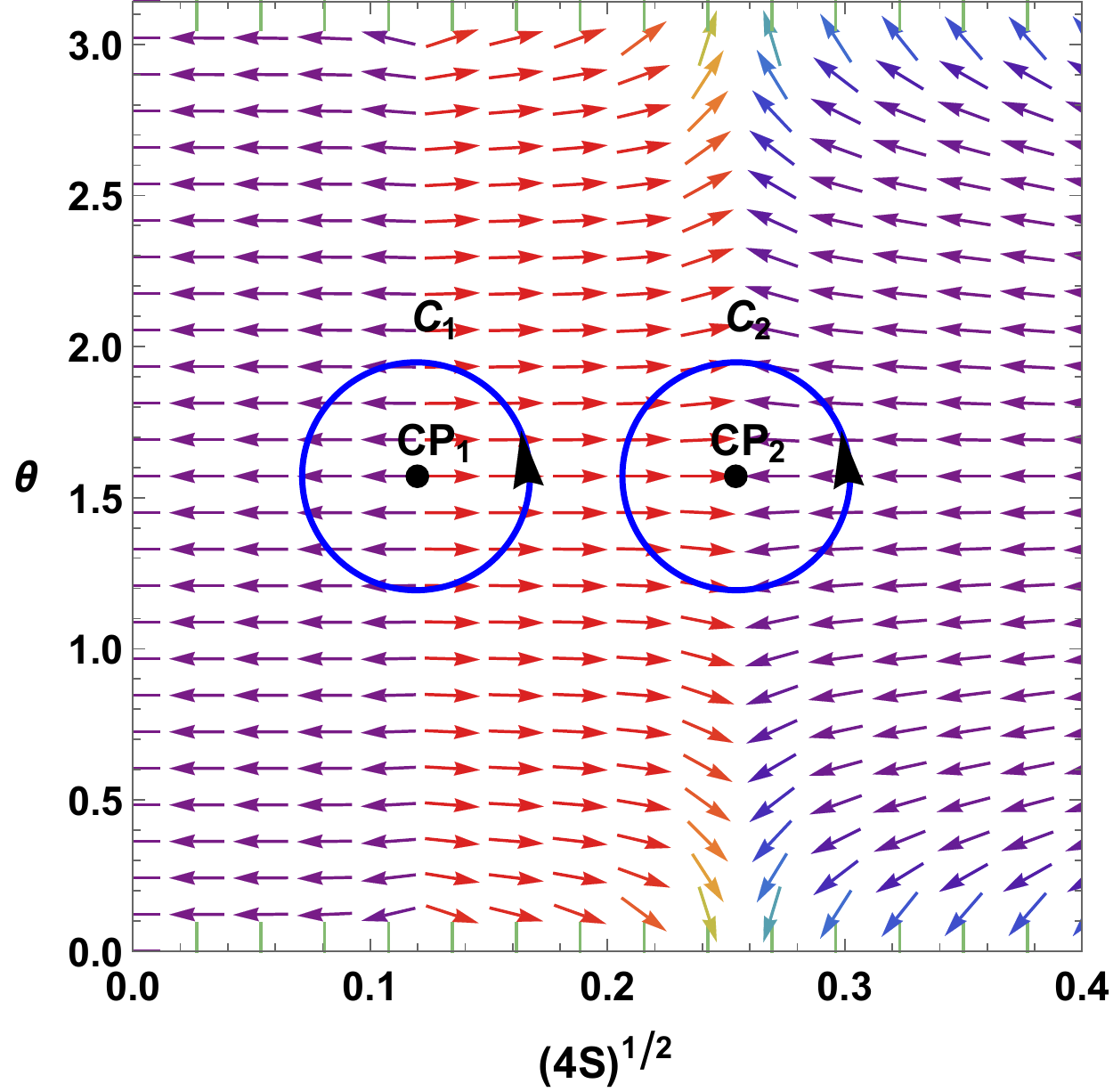}
		\caption{$Q>Q_I$}
		\label{f1_2}	
	\end{subfigure}
	\caption{\footnotesize\it Normalized vector field $n^i$ in
		the $(S,\theta)$ plane for (a) $Q=Q_I = 0.00803854$ and (b) $Q = 0.0095>Q_I$  with $b = 3.5$. }
	\label{f1}
\end{figure}

We have previously shown in \cite{Ali:2023wkq} that for a Born-Infeld-AdS black hole, the existence of  the critical point is controlled by the following constraint 
\begin{equation}\label{18}
	b \geq \sqrt{4 \pi P \left( 1 + \sqrt{2}\right)}.
\end{equation}
Let us consider the special case associated with
\begin{equation}\label{19}
	P_I = \dfrac{b^2}{4 \pi \left( 1 + \sqrt{2}\right)},
\end{equation} 
which we utilize in Eqs.~\eqref{2} and \eqref{5} to determine the corresponding electric charge $Q_I$, and entropy $S_I$ for a given $b$. 
For the Born-Infeld parameter $b = 3.5$, we find that $Q_I = 0.00803854$ and $S_I = 0.00721538$. Thus, for $Q<Q_I$, there is no critical point. In fact, the point determined by $(P_I, Q_I,S_I)$ is an isolated critical point analogous to that observed in \cite{PhysRevD.107.046013}. Moreover, we plot in Fig.\ref{f1}, the normalized vector field $n^i$ in the $(S,\theta)$ plane for $Q=Q_I$ and $Q = 0.0095>Q_I$ to illustrate such a situation,
where we observe the existence of an isolated critical point, $ICP$, and two critical points, $CP_1$ and $CP_2$ which merge together as we increase value of the the electric charge.

\begin{figure}[!ht]
	\centering 
	\begin{subfigure}[h]{0.48\textwidth}
		\centering \includegraphics[scale=0.6]{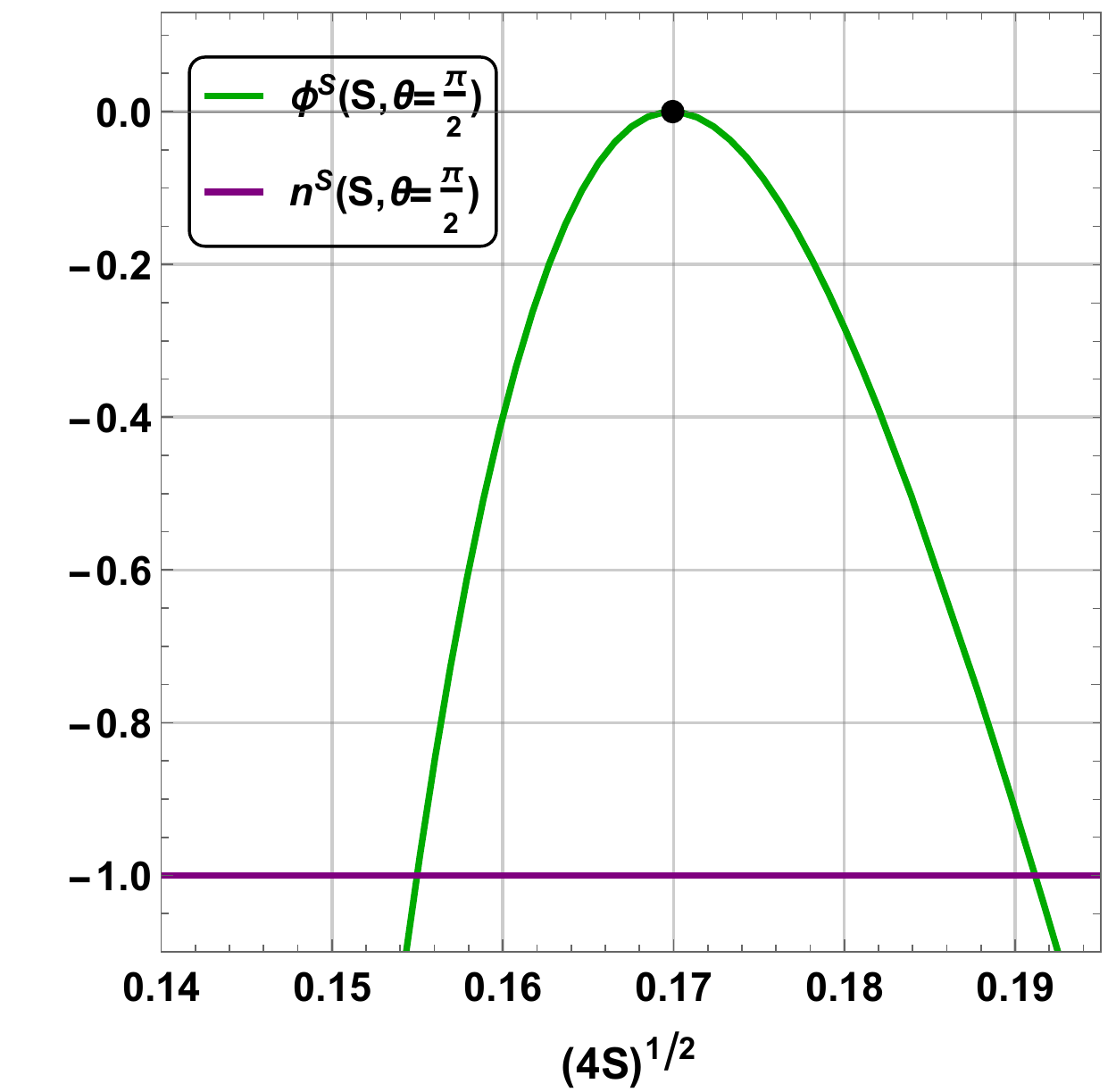}
		\caption{}
		\label{f2_1}
	\end{subfigure}
	\hspace{1pt}	
	\begin{subfigure}[h]{0.48\textwidth}
		\centering \includegraphics[scale=0.6]{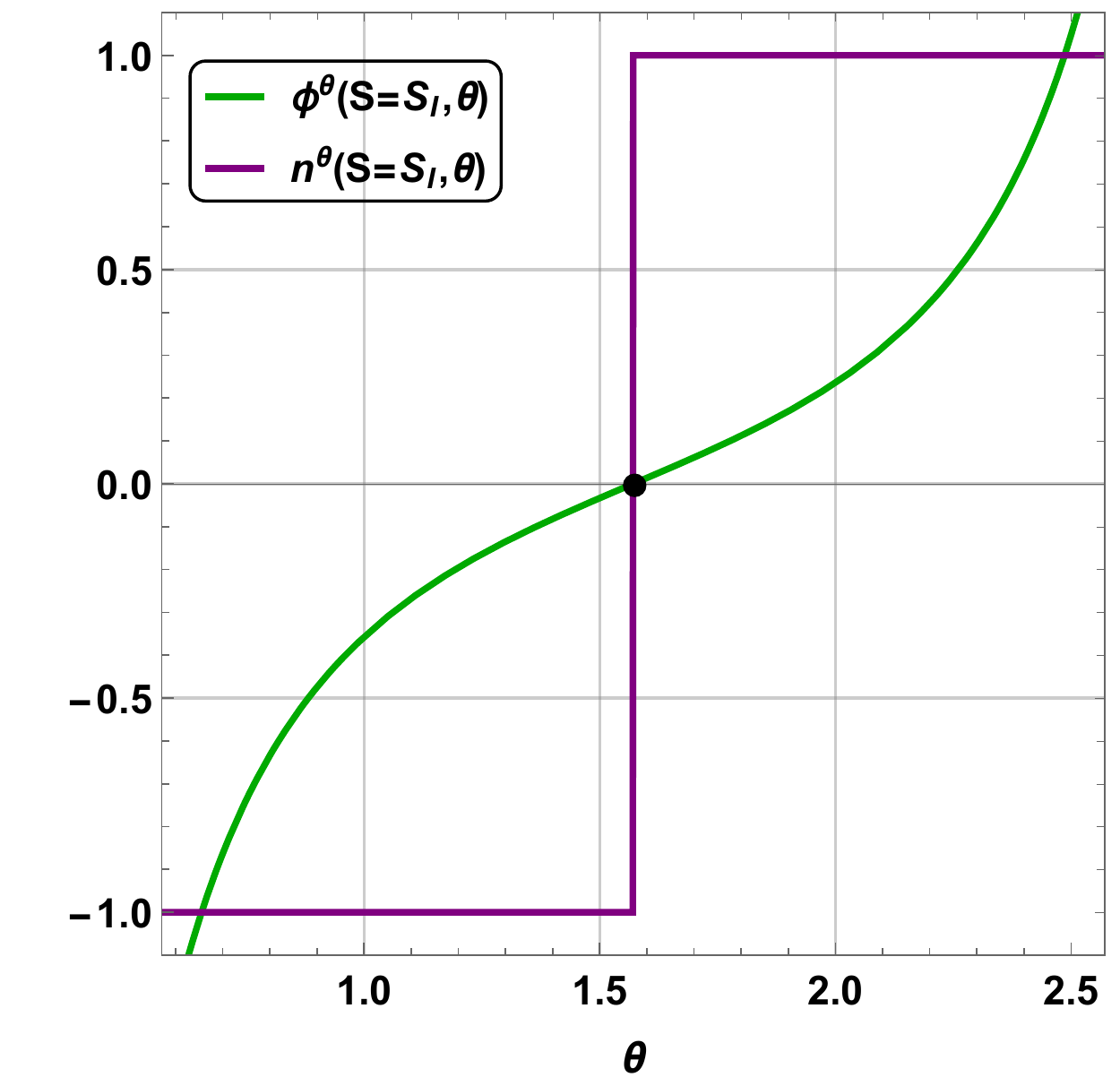}
		\caption{}
		\label{f2_2}	
	\end{subfigure}
	\caption{\footnotesize\it (a) Vector components $\phi^S$ and $n^S$ as a function of $S$ for $\theta = \dfrac{\pi}{2}$. (b) Vector components $\phi^\theta$ and $n^\theta$ as a function of $\theta$ for $S = S_I = 0.00721538$. $Q=Q_I = 0.00803854$ and $b = 3.5$. }
	\label{f2}
\end{figure}

Rigorously speaking, we suggest studying first what happens when the charge reaches the value
$Q = Q_I$ associated with the isolated critical point. According to \cite{PhysRevD.107.046013}, the black hole system does not exhibit an isolated critical point in the zero of $\phi$ and hence it does not verify Eq.\eqref{2}. Such an assertion, in fact, is incorrect. 
 Indeed, we depict in Fig.\ref{f2_1} the vector field component $\phi^S$ and the normalized vector $n^S$ as a function of the entropy $S$ with $\theta = \frac{\pi}{2}$ for $Q=Q_I$. 

\begin{figure}[!ht]
	\centering 
	\begin{subfigure}[h]{0.48\textwidth}
		\centering \includegraphics[scale=0.6]{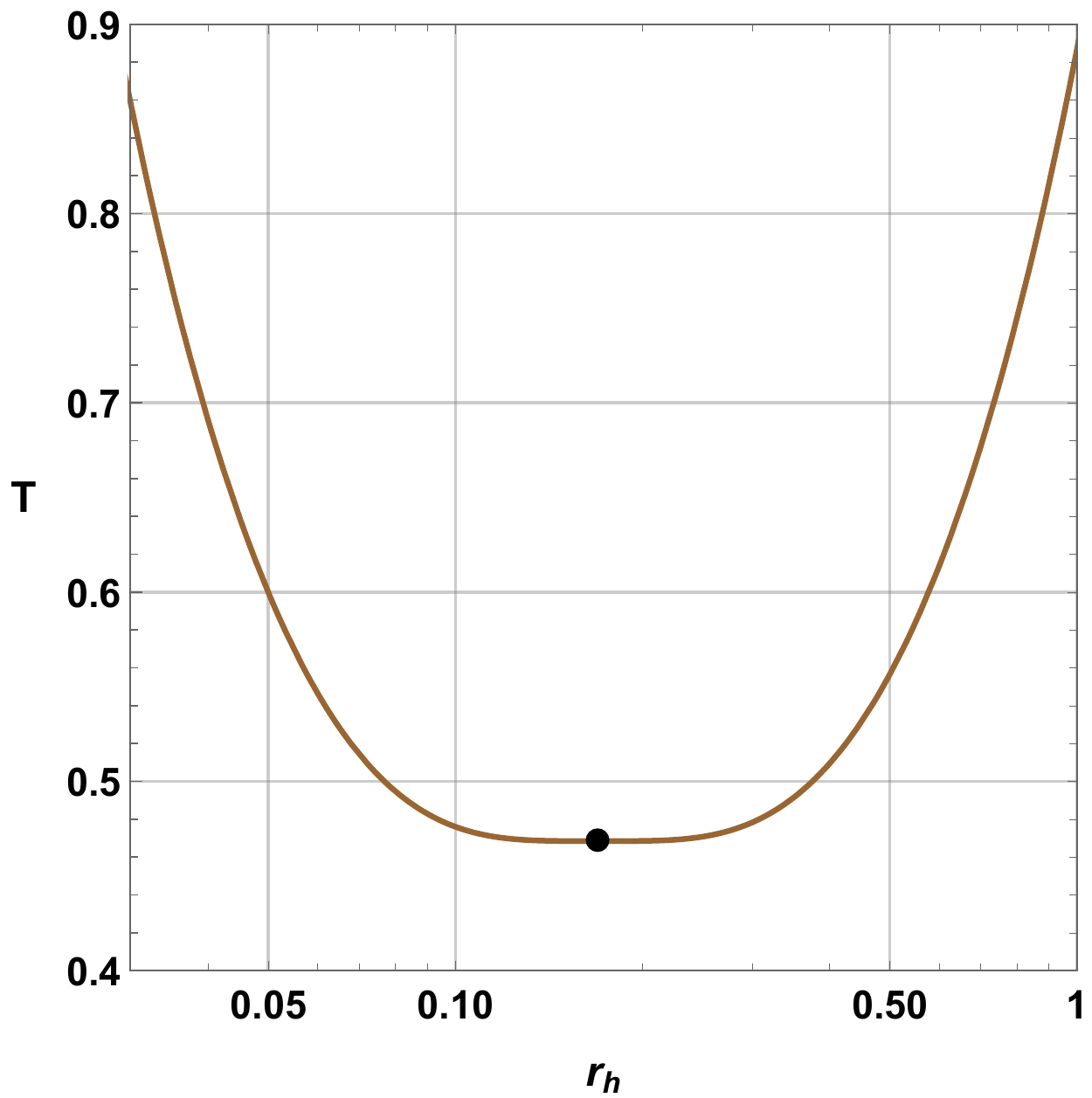}
		\caption{}
		\label{f3_1}
	\end{subfigure}
	\hspace{1pt}	
	\begin{subfigure}[h]{0.48\textwidth}
		\centering \includegraphics[scale=0.6]{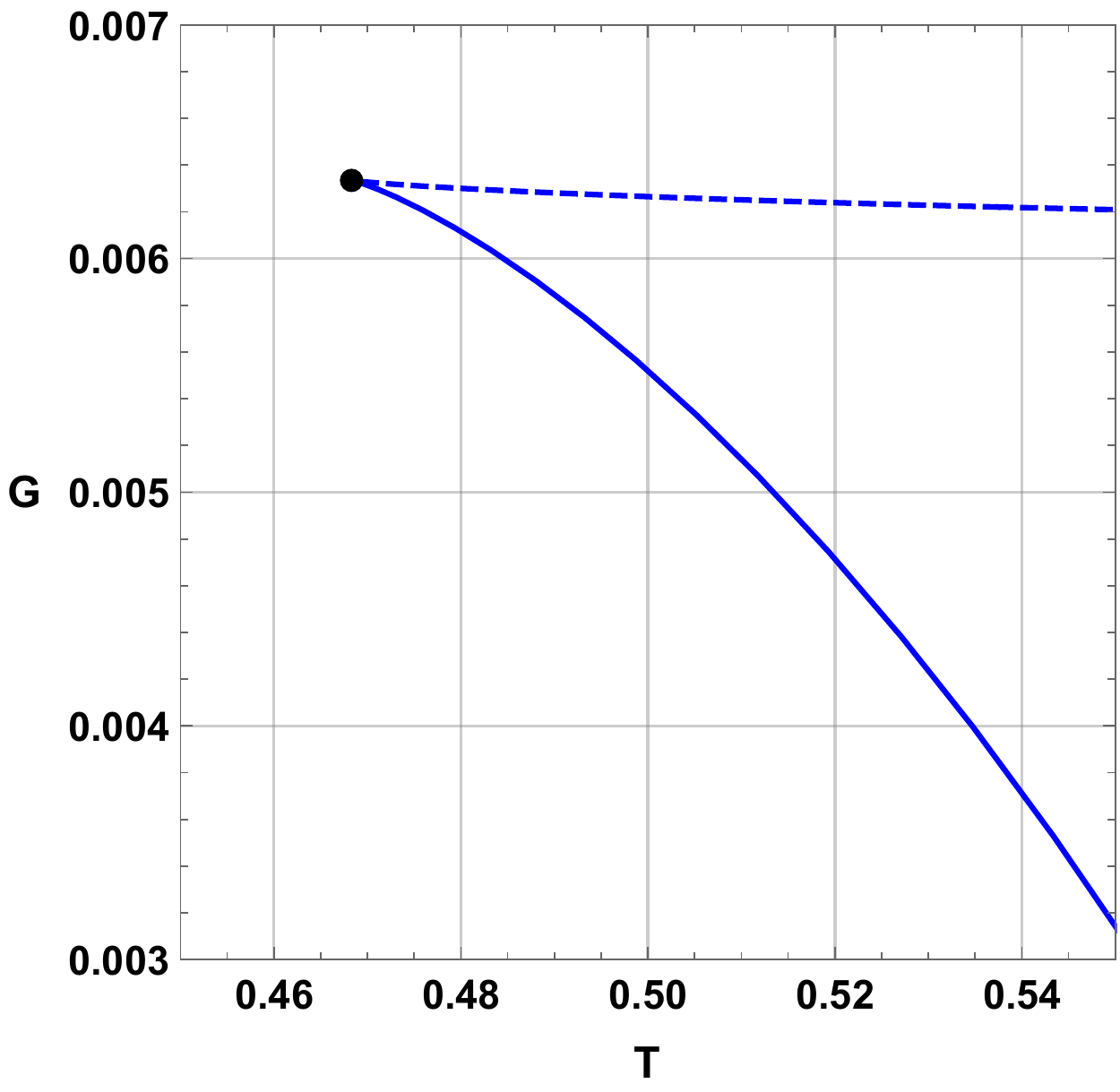}
		\caption{}
		\label{f3_2}	
	\end{subfigure}
	\caption{\footnotesize\it (a) Hawking temperature as a function of horizon radius $r_h$ and (b) and free Gibbs energy as a function of temperature with $Q=Q_I = 0.00803854$, $P = P_I = 0.403785$ and $b = 3.5$. }
	\label{f3}
\end{figure}

Obviously, the vector field component $\phi^S$ vanishes at $S=S_I$ (isolated critical point marked by the black dot) without changing the sign. It is a semi-stable point. Whereas, the normalized vector field component $n^S$ does not vanish because $\phi^S$ and $\left| \left| \phi\right| \right| $, simultaneously tend to zero, keeping the ratio $n^S = \frac{\phi^S}{\left| \left| \phi\right| \right| }$ constant and negative ($n^S (S,\frac{\pi}{2}) = -1$). Therefore, the $\phi^S$ vanishes at the isolated critical point. Moreover, we observe in Fig.\ref{f2_1} that $\phi^\theta$ vanishes too at the isolated critical point $(S = S_I, \theta = \frac{\pi}{2})$, whereas $n^\theta$ becomes discontinuous. Thus, the vector field $\phi$ vanishes at the isolated critical point and obeys Eq.\eqref{2}. In order to compute the topological charge $\mathcal{Q}(ICP)$, we consider the contour $C_0$ in Fig.\ref{f1_1}. We further use \eqref{15}, to find 
\begin{equation}\label{20}
	\mathcal{Q}(ICP) = 0,
\end{equation} 
which is in agreement with the result found in \cite{PhysRevD.107.046013}. 
For more detailed understanding of the thermodynamics associated with the isolated critical point, we depict in Fig.\ref{f3}, the variations of the Hawking temperature and Gibbs free energy. 

We observe that the thermodynamic behavior is Schwarzchild-like and the Gibbs free energy forms a cusp which is a signature of a transition between an unstable phase (small black holes) and a \textit{locally} stable one (large black holes) as in Hawking-Page phase transition.
 Moreover, the Hawking temperature admits an inflection point that corresponds to a minimum. Therefore, the isolated critical point satisfies Eq.\eqref{5}.

\begin{figure}[!ht]
	\centering 
	\begin{subfigure}[h]{0.48\textwidth}
		\centering \includegraphics[scale=0.6]{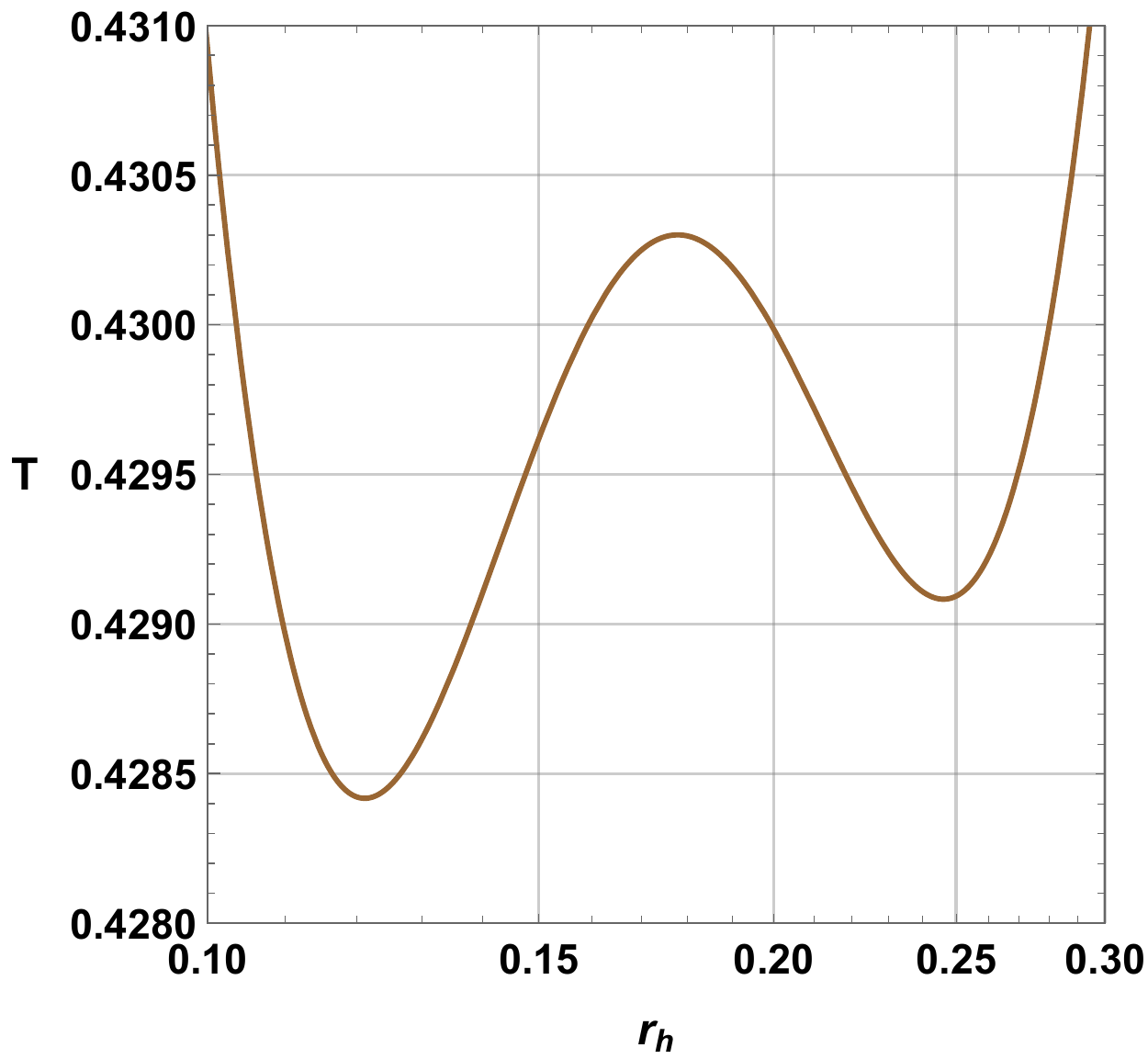}
		\caption{}
		\label{f4_1}
	\end{subfigure}
	\hspace{1pt}	
	\begin{subfigure}[h]{0.48\textwidth}
		\centering \includegraphics[scale=0.6]{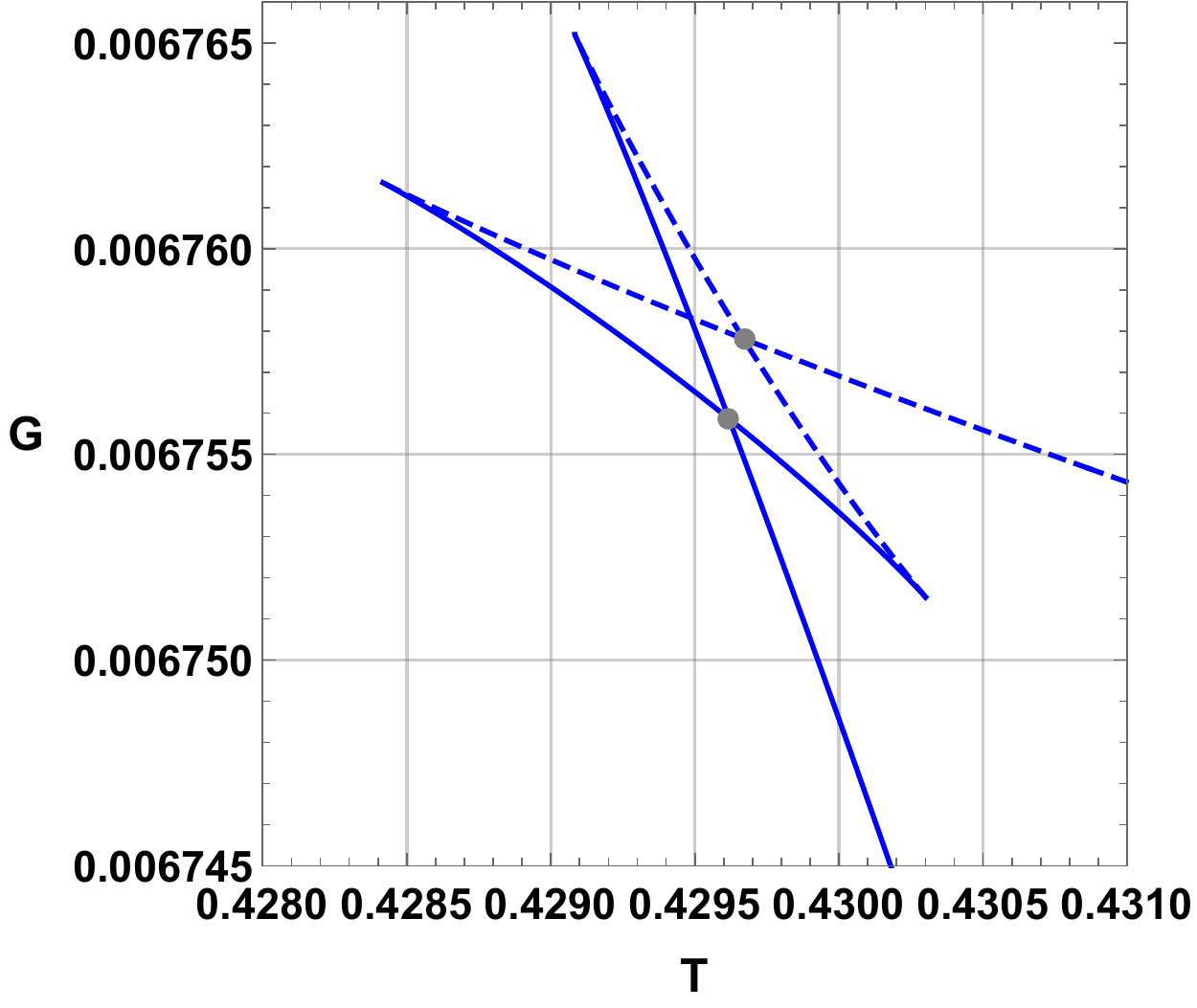}
		\caption{}
		\label{f4_2}	
	\end{subfigure}
	\caption{\footnotesize\it (a)
 Hawking temperature as a function of horizon radius $r_h$ and (b) and free Gibbs energy as a function of temperature near the isolated critical point $ICP$ with $Q= 0.0085$, $P = 0.3315735$ and $b = 3.5$. }
	\label{f4}
\end{figure}

With a view of probing the isolated critical point formation, we plot in Fig.\ref{f4} the Hawking temperature and Gibbs free energy with parameters near those of the isolated critical point.
 We notice the exhibition of two swallowtail shapes in the free energy-temperature diagram which means that there is a phase transition between two unstable phases (dashed curves), that correspond to a decreasing temperature in terms of $r_h$. This phase transition (upper gray dot in Fig.\ref{f4_2}) is unconventional because it occurs between two unstable phases (unstable small black holes and intermediate black holes). Besides, we observe a first-order phase transition between two stable phases (solid curves), which correspond to an increasing temperature in terms of $r_h$. This phase transition (lower gray dot in Fig.\ref{f4_2}) is conventional because it occurs between two \textit{locally} stable phases (small black holes and large black holes). As the electric charge gets decreased further below its critical value $Q_I$, the two critical points (conventional and unconventional) coincide and form the isolated critical point observed previously. Thus, such an isolated critical point is an annihilation process between a conventional critical point and an unconventional one.

Next, we consider the second situation as in Fig.\ref{f1_2}, where we observe the formation of two critical points when the electric charge $Q>Q_I$.  Topologically speaking, this is a vortex/anti-vortex pair creation. Using Eq.\eqref{15} and the two contours $C_1$ and $C_2$, the topological charges of these two points are : 
\begin{equation}\label{21}
	\mathcal{Q}(CP_1) = +1, \quad \quad \mathcal{Q}(CP_2) = -1
\end{equation} 
This result is in good agreement with \cite{PhysRevD.105.104003} and \cite{PhysRevD.107.046013}, where the positive charge characterizes the unconventional critical point and the negative one describes the conventional critical point. Moreover, the total topological charge of the whole system is as follows
\begin{equation}\label{22}
	\mathcal{Q} = \mathcal{Q}(CP_1) + \mathcal{Q}(CP_2) = 0,
\end{equation} 
which is the same as \eqref{20} when $Q=Q_I$. That is to say, the total topological charge is preserved, and no topological phase transition occurs. Thus, the two black holes ( $Q=Q_I$ and $Q = 0.0095>Q_I$) belong to the same topological class.

\begin{figure}[!ht]
	\centering 
	\begin{subfigure}[h]{0.48\textwidth}
		\centering \includegraphics[scale=0.6]{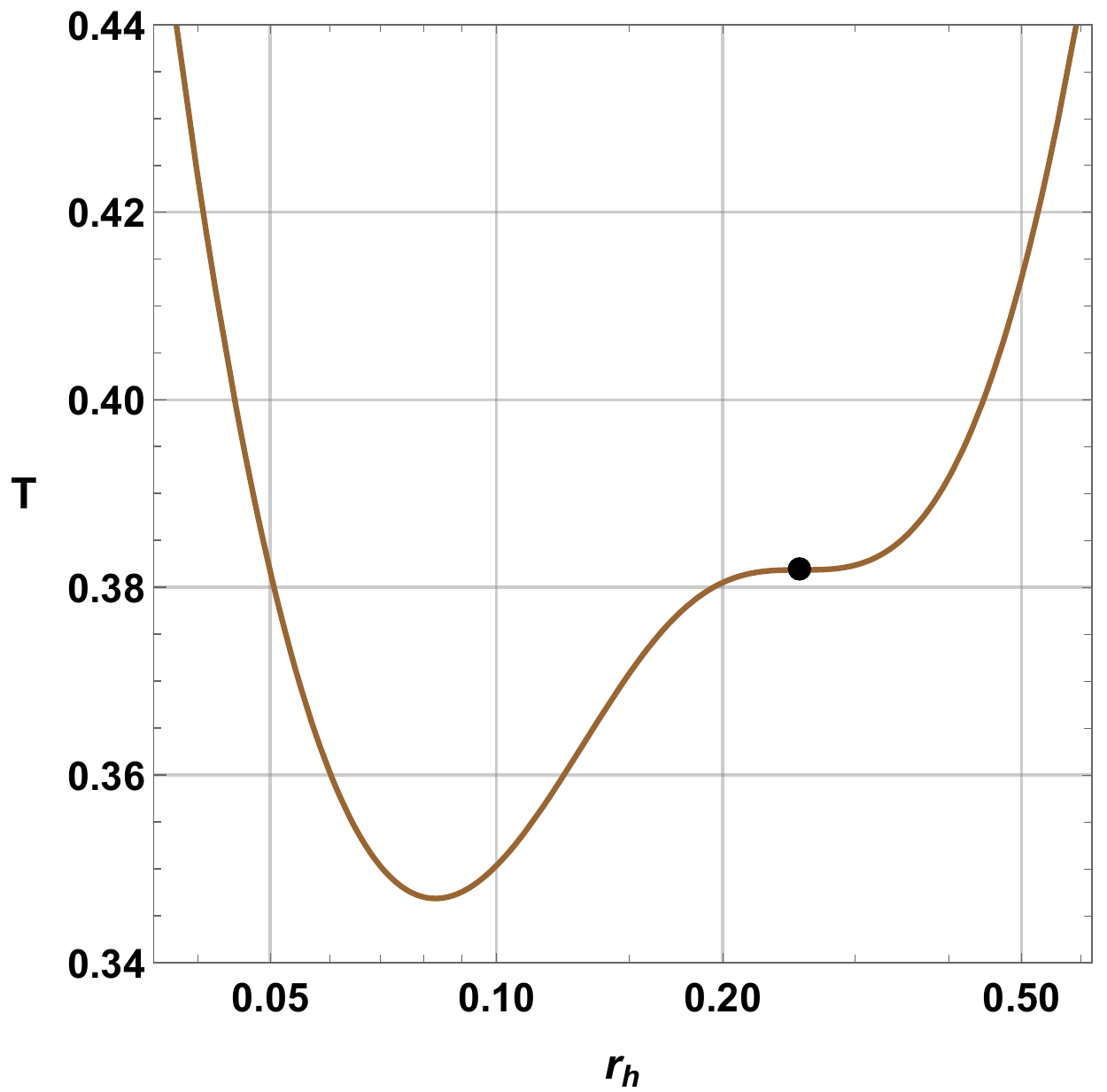}
		\caption{}
		\label{f5_1}
	\end{subfigure}
	\hspace{1pt}	
	\begin{subfigure}[h]{0.48\textwidth}
		\centering \includegraphics[scale=0.6]{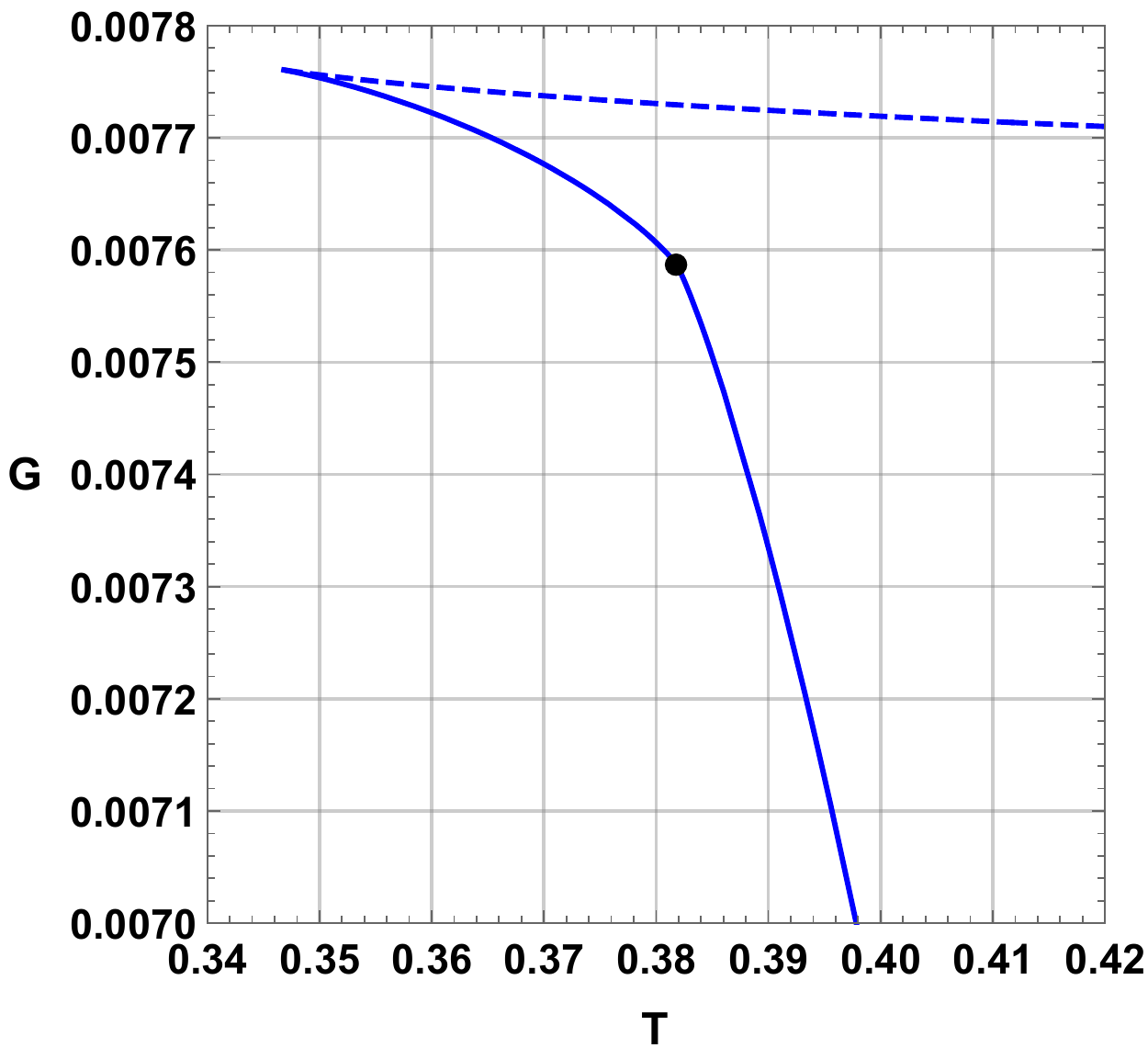}
		\caption{}
		\label{f5_2}	
	\end{subfigure}
	\caption{\footnotesize\it (a) Hawking temperature as a function of horizon radius $r_h$ and (b) and free Gibbs energy as a function of temperature with $Q= 0.0095$, $P =P_{CP_2} = 0.26282$ and $b = 3.5$. }
	\label{f5}
\end{figure}

Herein, we will focus on the thermodynamics of these two critical points. In fact, we plot in Fig.\ref{f5} the Hawking temperature and Gibbs free energy corresponding to conventional critical point $CP_2$.

 We observe that there exists a second phase transition between two \textit{locally} stable phases (small and large black holes) and we see an inflection point in the right branch of temperature where it is an increasing function in terms of horizon radius (the critical point, $CP_2$, is represented by the black dot).

\begin{figure}[!ht]
	\centering 
	\begin{subfigure}[h]{0.48\textwidth}
		\centering \includegraphics[scale=0.6]{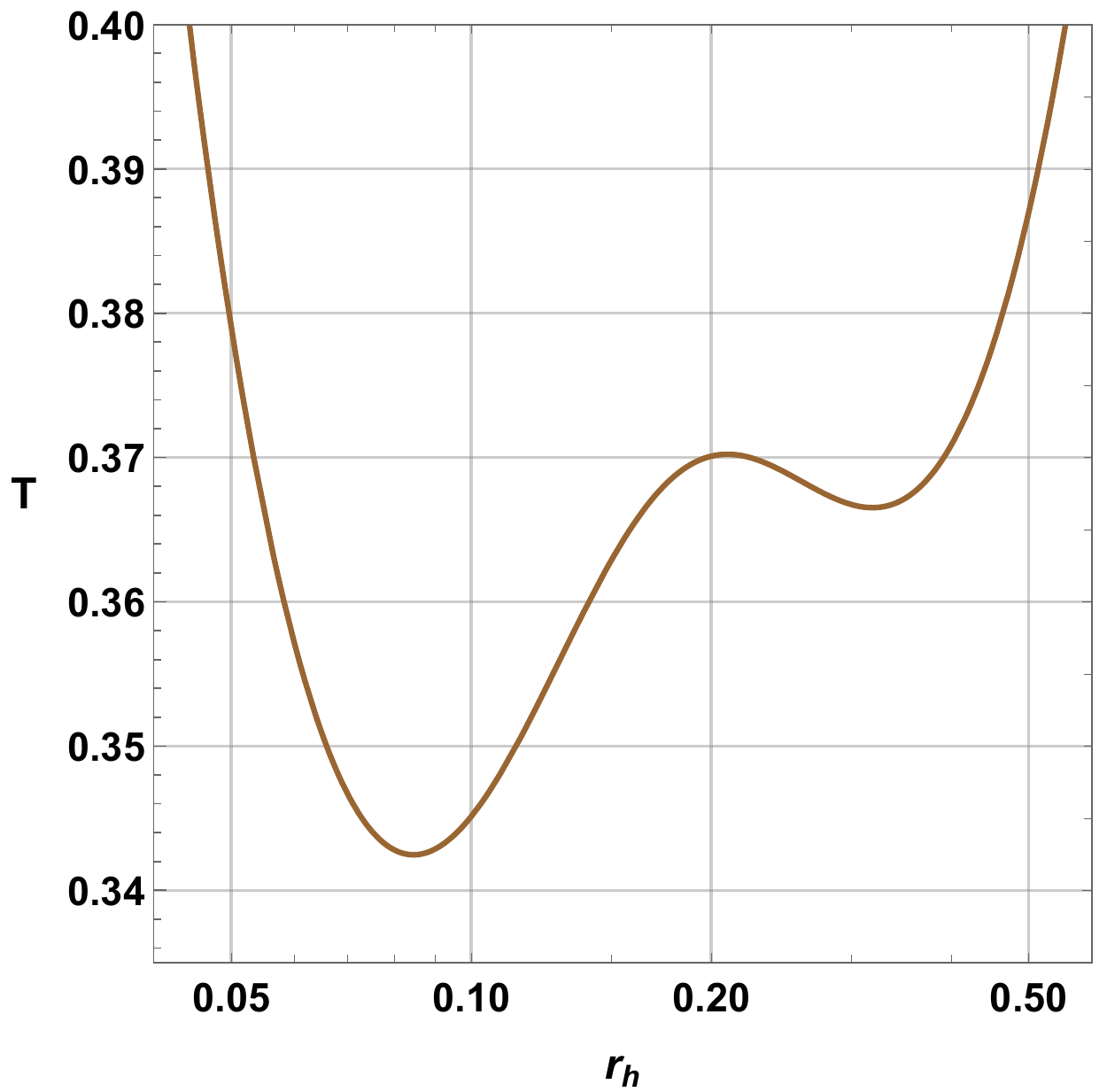}
		\caption{}
		\label{f6_1}
	\end{subfigure}
	\hspace{1pt}	
	\begin{subfigure}[h]{0.48\textwidth}
		\centering \includegraphics[scale=0.6]{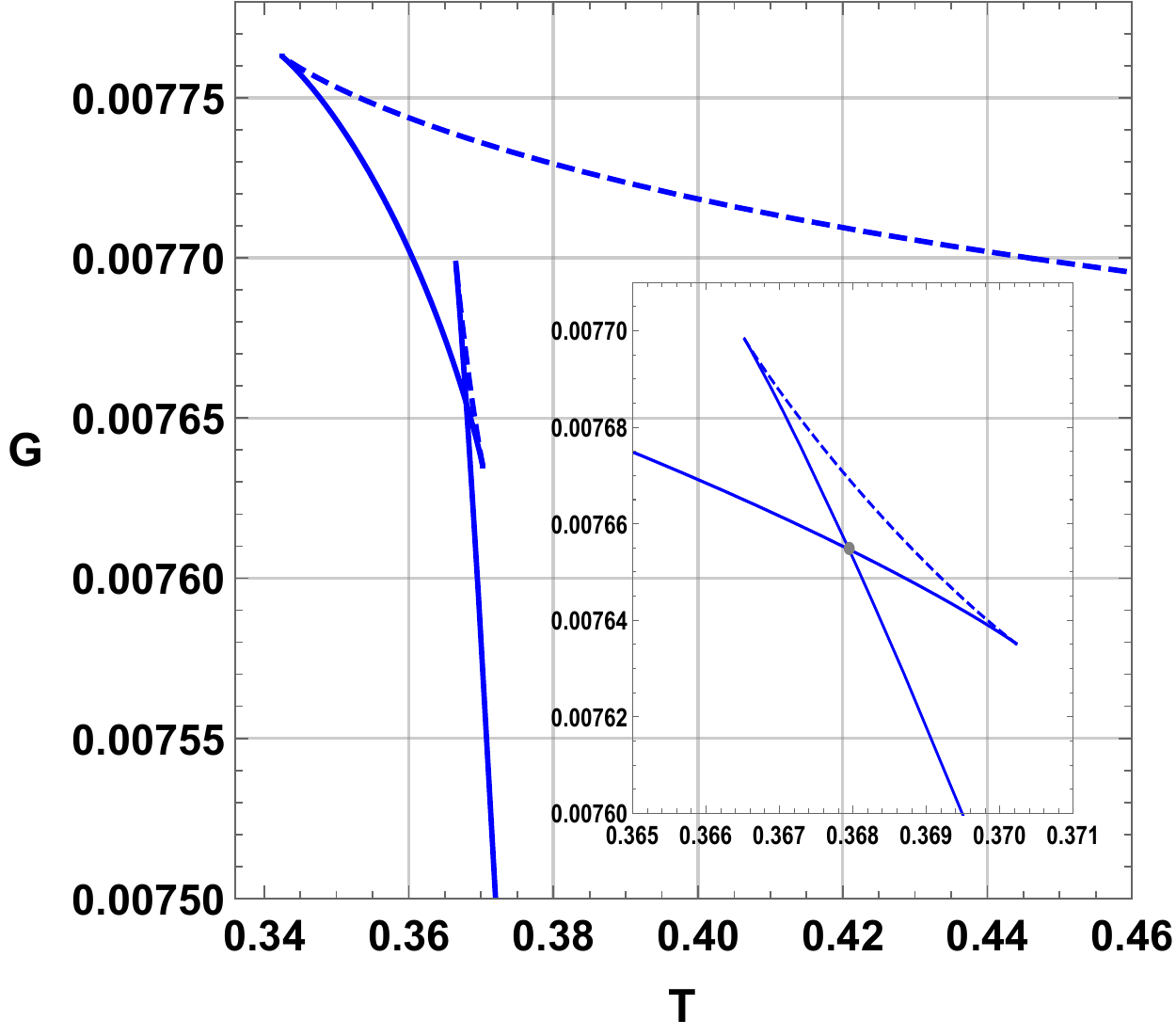}
		\caption{}
		\label{f6_2}	
	\end{subfigure}
	\caption{\footnotesize\it (a) Hawking temperature as a function of horizon radius $r_h$ and (b) and free Gibbs energy as a function of temperature near the conventional critical point $CP_2$ with $Q= 0.0095$, $P = 0.2367915$ and $b = 3.5$. }
	\label{f6}
\end{figure}

In Fig.\ref{f6}, we plot the Hawking temperature and Gibbs free energy with the same electric charge and pressure near to that of the conventional critical point $CP_2$.
 This figure reveals that a first-order phase transition happens between two \textit{locally} stable phases analogous to that observed in Reissner–Nordström-AdS black hole. Nevertheless, we observe a cusp in the free energy-temperature diagram as in AdS-Schwarzchild black hole case, where the temperature essentially tends to $+\infty$ when $r_h\to 0$.

\begin{figure}[!ht]
	\centering 
	\begin{subfigure}[h]{0.48\textwidth}
		\centering \includegraphics[scale=0.6]{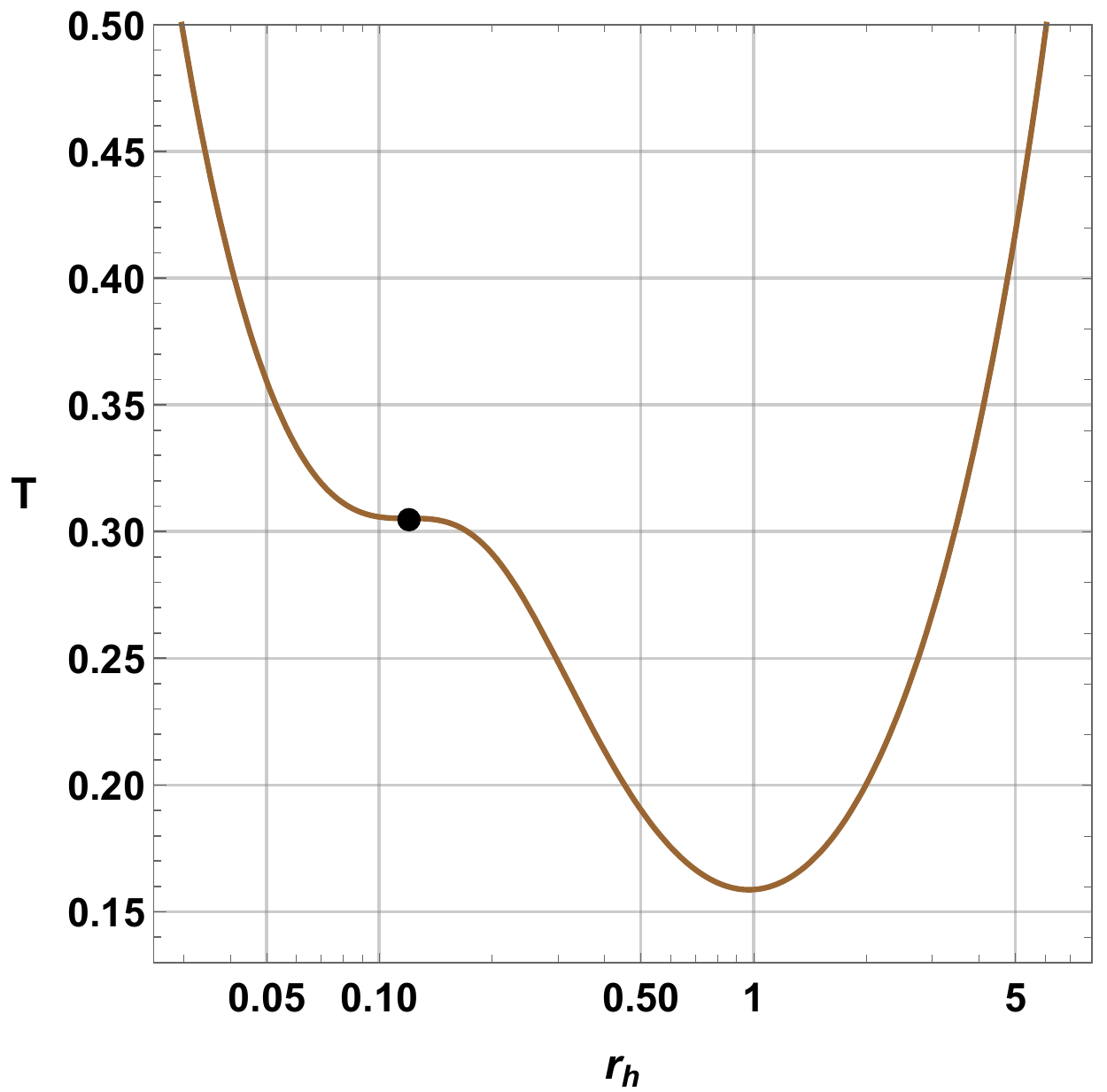}
		\caption{}
		\label{f7_1}
	\end{subfigure}
	\hspace{1pt}	
	\begin{subfigure}[h]{0.48\textwidth}
		\centering \includegraphics[scale=0.6]{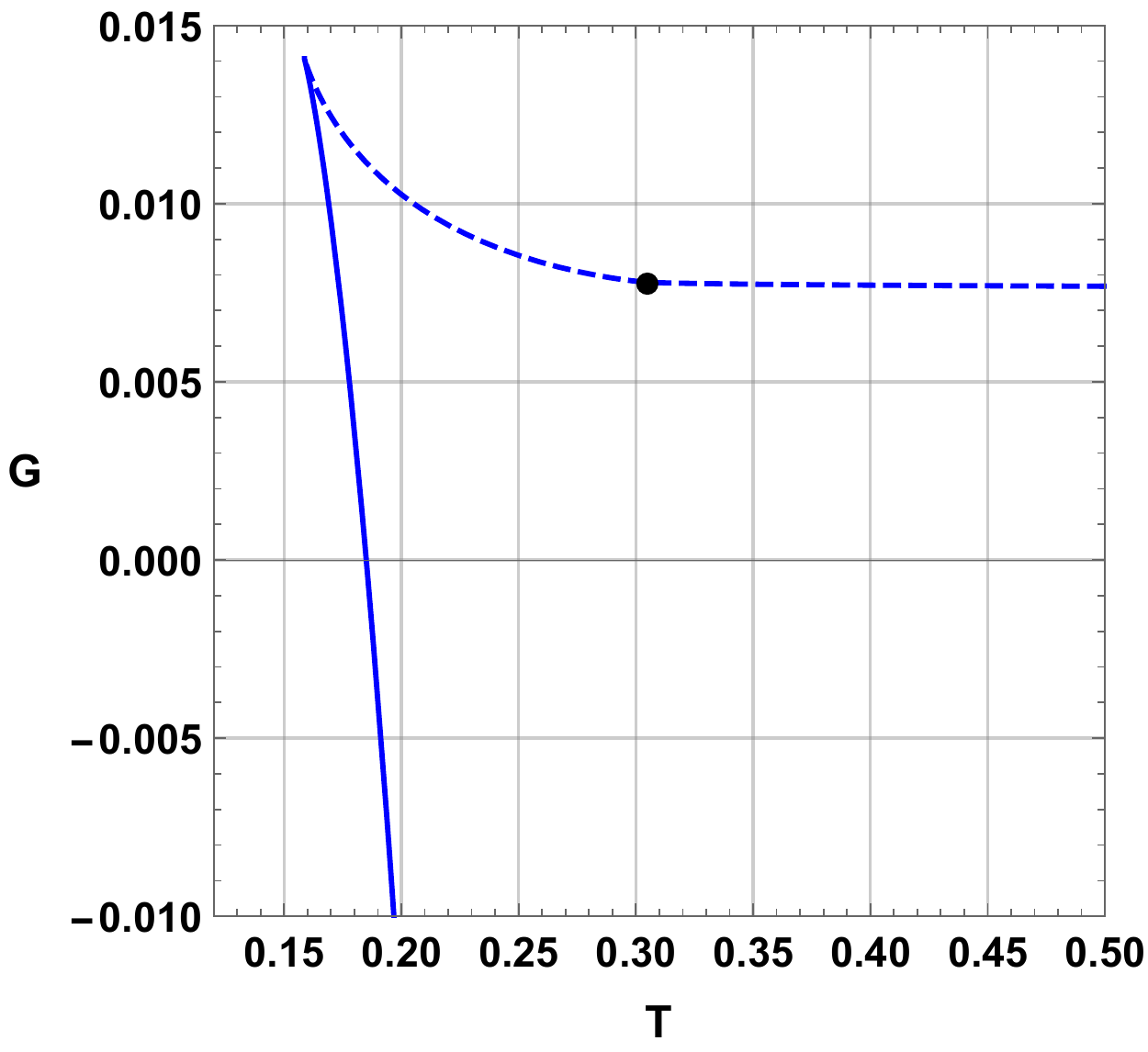}
		\caption{}
		\label{f7_2}	
	\end{subfigure}
	\caption{\footnotesize\it (a) Hawking temperature as a function of horizon radius $r_h$ and (b) and free Gibbs energy as a function of temperature with $Q= 0.0095$, $P =P_{CP_1} = 0.0401615$ and $b = 3.5$. }
	\label{f7}
\end{figure}

Concerning the unconventional critical point, $CP_1$,  its Hawking temperature, and the Gibbs free energy are depicted in  Fig.\ref{f7}.

The figure unveils the existence of an inflection point in the left branch of temperature where it is a decreasing function in terms of horizon radius (the critical point, $CP_1$ is represented by the black dot). This inflection point is a signature of the existence of a second-order phase transition, whereas we see in the free energy-temperature diagram that this transition occurs between two unstable phases and the system is equivalent to AdS-Schwarzchild black hole.

\begin{figure}[!ht]
	\centering 
	\begin{subfigure}[h]{0.48\textwidth}
		\centering \includegraphics[scale=0.6]{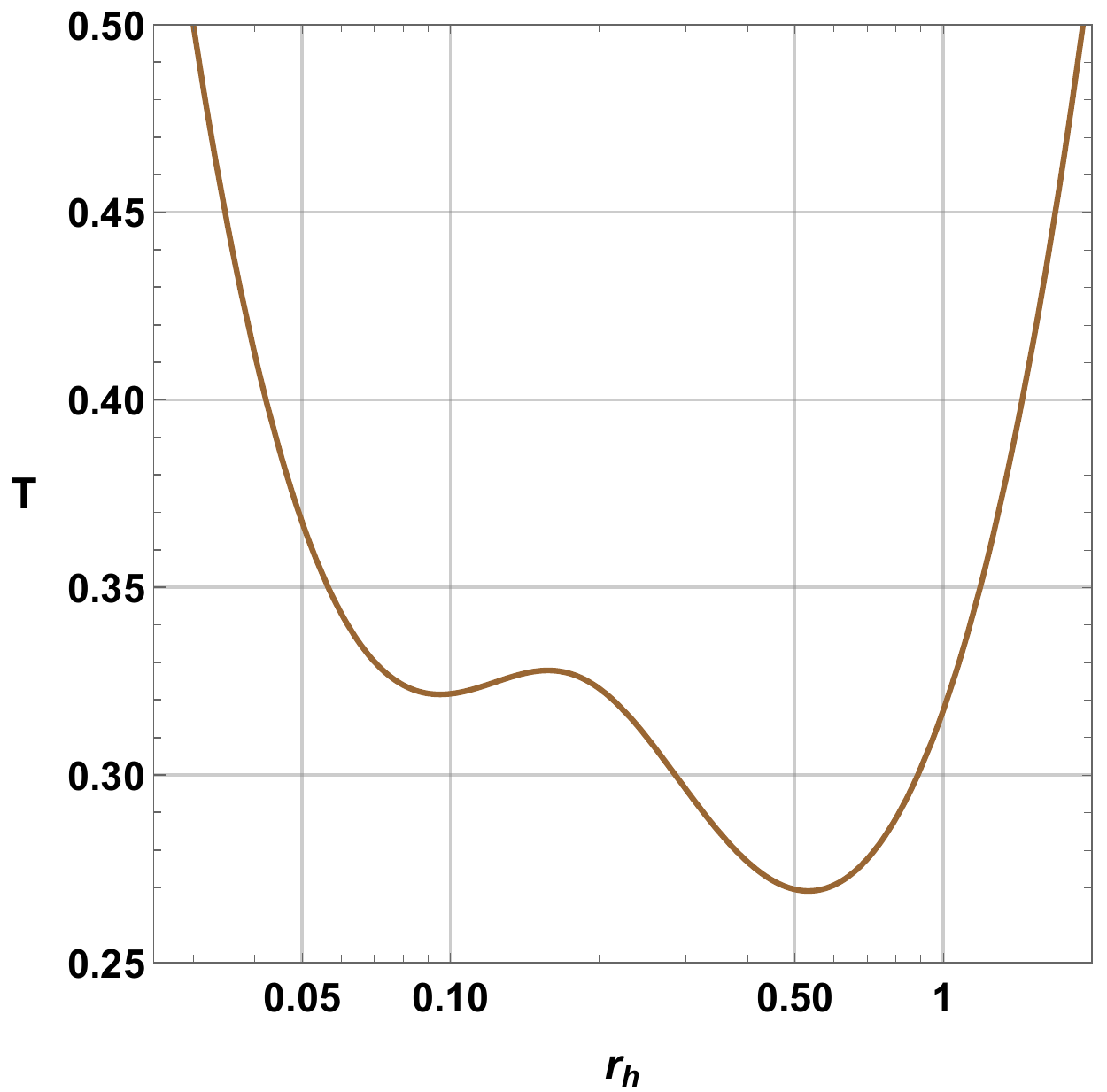}
		\caption{}
		\label{f8_1}
	\end{subfigure}
	\hspace{1pt}	
	\begin{subfigure}[h]{0.48\textwidth}
		\centering \includegraphics[scale=0.6]{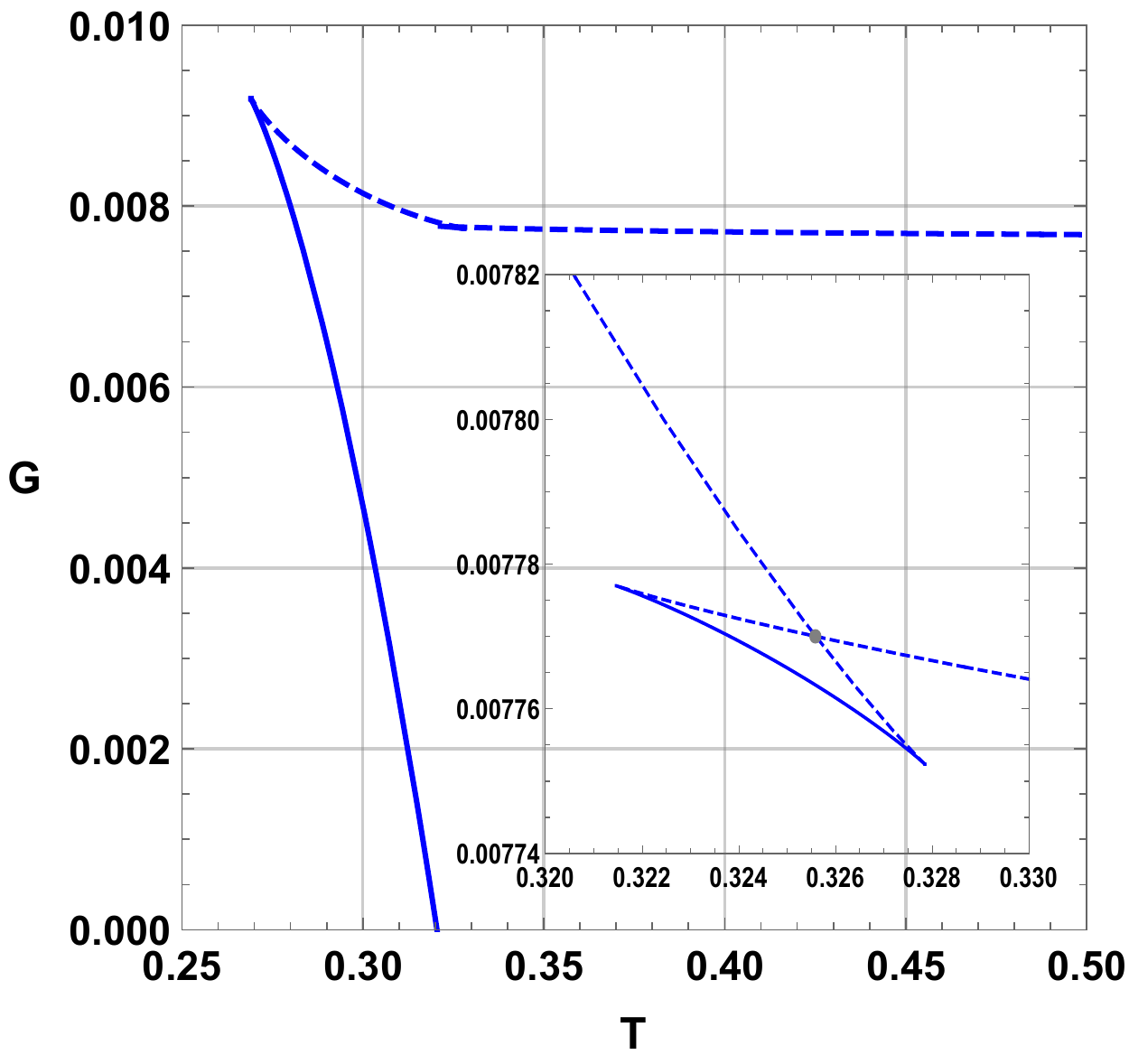}
		\caption{}
		\label{f8_2}	
	\end{subfigure}
	\caption{\footnotesize\it (a) Hawking temperature as a function of horizon radius $r_h$ and (b) and free Gibbs energy as a function of temperature near the unconventional critical point $CP_1$ with $Q= 0.0095$, $P = 0.119366$ and $b = 3.5$. }
	\label{f8}
\end{figure}

As in the previous case, in Fig.\ref{f8}, we will envisage the situation near the unconventional critical point $CP_1$, and then we address a portrait of Hawking temperature and Gibbs free energy with the same electric charge and the pressure in the vicinity of $CP_1$. 
The system exhibits an unconventional phase transition between two unstable phases that occurs in the left branch of the temperature-horizon radius diagram. This phase transition is unconventional (unstable) because the intermediate state is \textit{locally} stable and it takes place in the upper branch of the free energy-temperature diagram.

\begin{figure}[!ht]
	\centering 
 \includegraphics[scale=0.6]{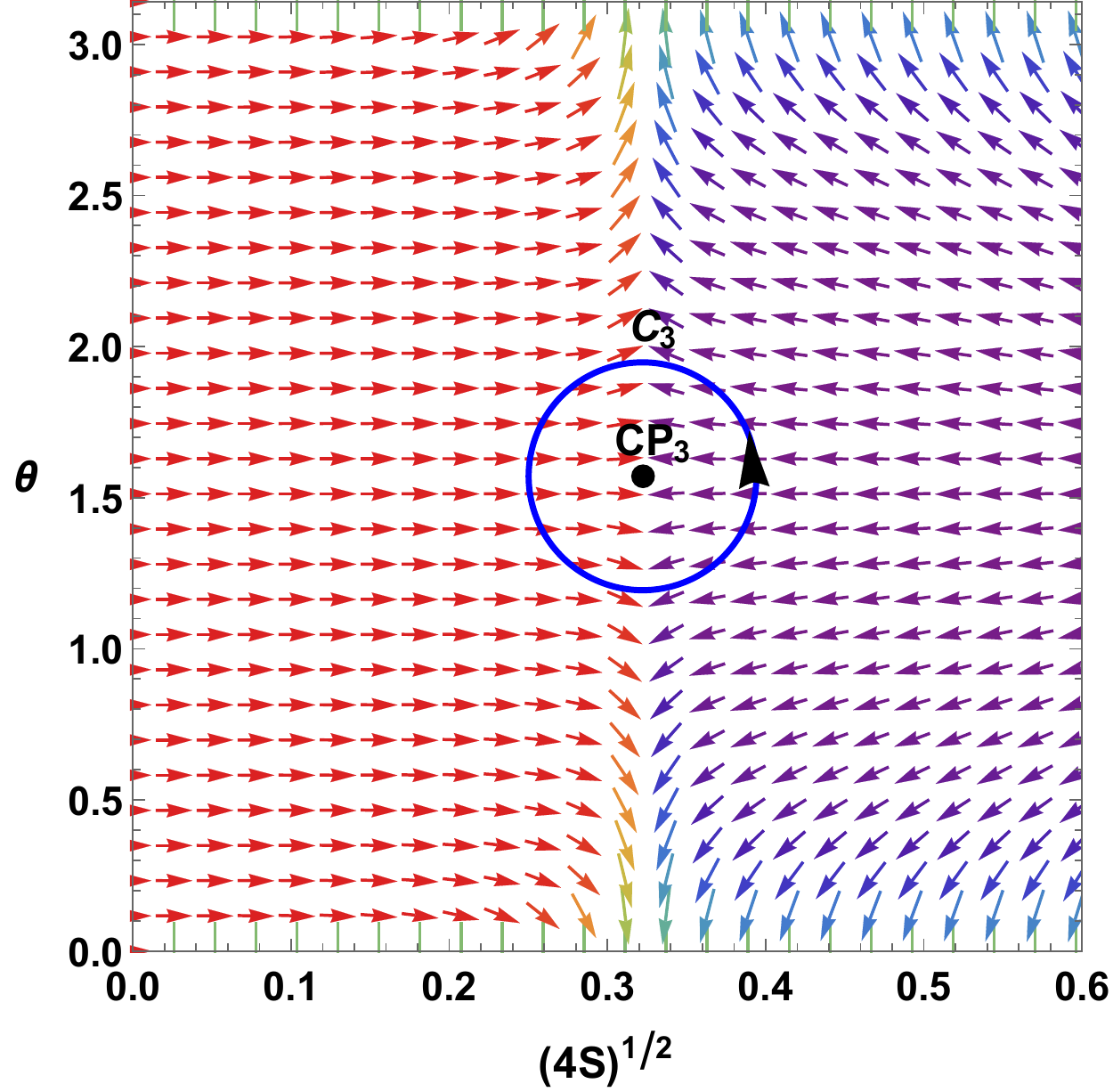}
	\caption{\footnotesize\it Normalized vector field $n^i$ in the $(S,\theta)$ plane for (a) $Q=Q_m = 0.0113682$  with $b = 3.5$. }
	\label{f9}
\end{figure}

In an earlier work \cite{Ali:2023wkq}, we have shown that when the electric charge is greater than some critical value $Q_m$, the Born-Infeld-AdS black hole becomes equivalent to  AdS-Reissner-Nordström one. The topological critical electric charge $Q_m$ is given by : 
\begin{equation}\label{23}
	Q_m = \dfrac{1}{8 \pi b},
\end{equation} 
and its associated topological illustration is depicted in Fig.\ref{f9}, where the portrait of the normalized vector field $n^i$ in the $(S,\theta)$ plane for $Q=Q_m$ is constructed.
 One can remark that there exists only one conventional stable critical point, $CP_3$, whereas the unconventional critical point has disappeared. The topological charge for this critical point is : 
\begin{equation}\label{24}
	\mathcal{Q}(CP_3) = -1,
\end{equation} 
which also is obviously the total topological charge of the system (because we have just one critical point). Thus we are in the presence of a \textit{topological phase transition}. Indeed, for $Q_I<Q<Q_m$, the total topological charge $\mathcal{Q} = 0$, but when $Q\geq Q_m$, the total topological charge $\mathcal{Q} = -1$. Thus the two black holes ($Q_I<Q<Q_m$ and $Q\geq Q_m$) belong to different topological classes.

This topological transition is analog to the Berezinskii–Kosterlitz–Thouless transition (BKT transition) observed in the two-dimensional XY model \cite{J.M.Kosterlitz_1973}. Here, the critical charge $Q_m$ plays the role of the critical temperature $T_{\text{BKT}}$. Indeed, for $Q<Q_m$, the vortex and the anti-vortex ($CP_2$ and $CP_1$) form a pair, they are a bound state of vortex–anti-vortex pair, while when $Q\geq Q_m$ a free vortex becomes consistent. Consequently, the total topological charge is altered, and therefore the system belongs to another topological class.

\begin{figure}[!ht]
\centering 
\begin{subfigure}[h]{0.48\textwidth}
	\centering \includegraphics[scale=0.6]{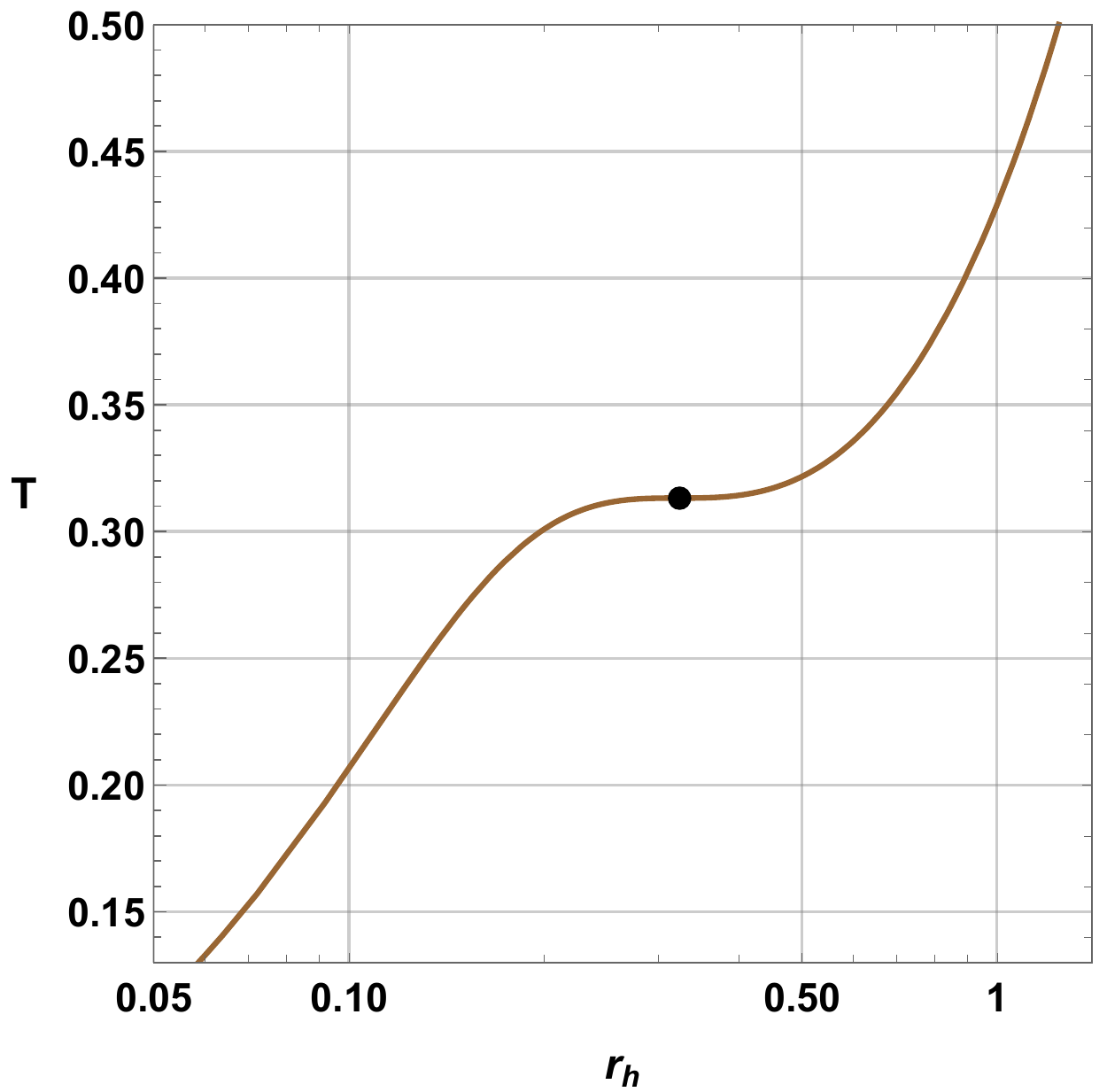}
	\caption{}
	\label{f10_1}
\end{subfigure}
\hspace{1pt}	
\begin{subfigure}[h]{0.48\textwidth}
	\centering \includegraphics[scale=0.6]{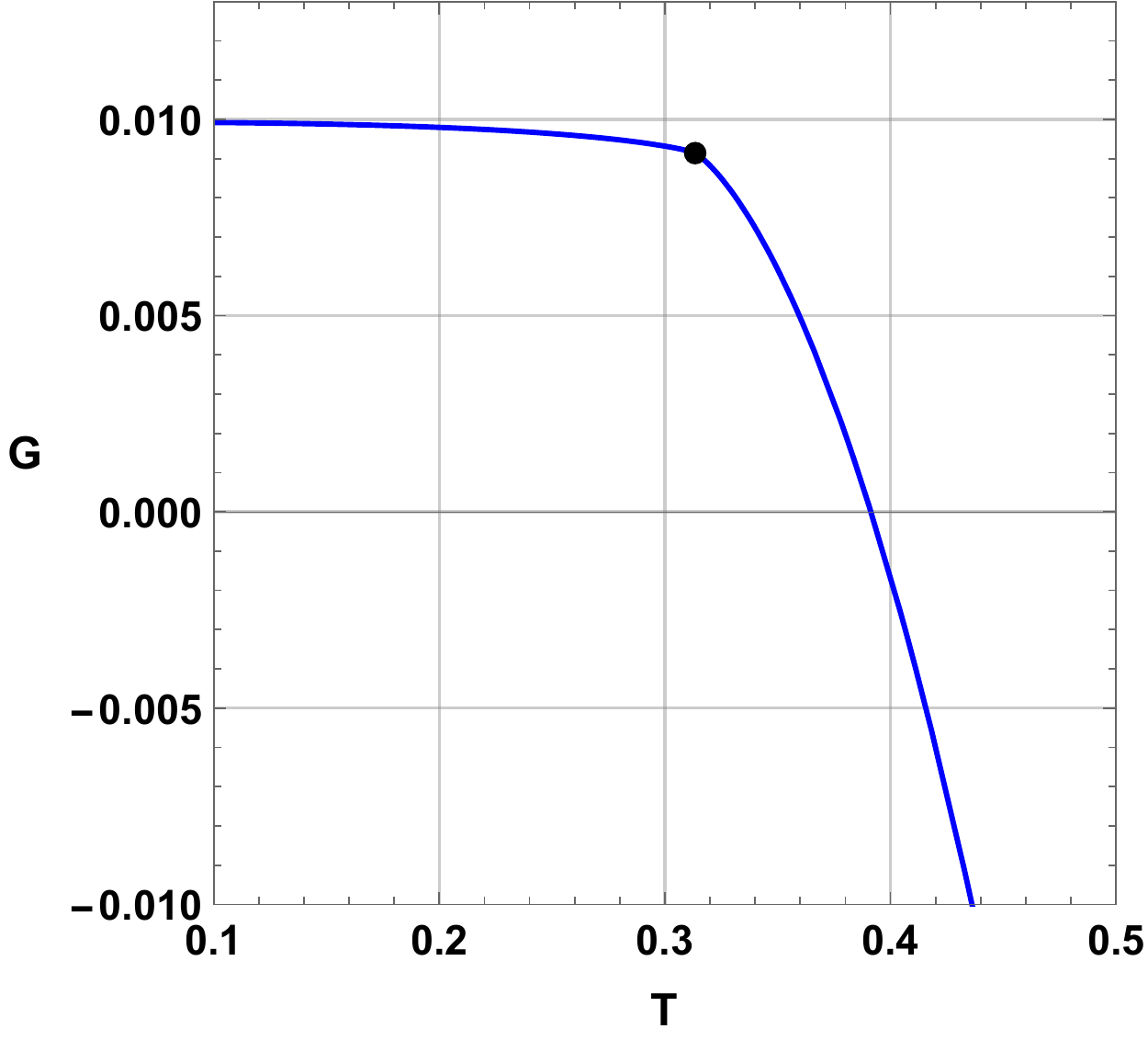}
	\caption{}
	\label{f10_2}	
\end{subfigure}
\caption{\footnotesize\it (a) Hawking temperature as a function of horizon radius $r_h$ and (b) and free Gibbs energy as a function of temperature with $Q=Q_m = 0.0113682$, $P =P_{CP_3} = 0.175429$ and $b = 3.5$. }
\label{f10}
\end{figure}

To establish the electric charge dependency of this topological transition governed only by the black hole electric charge, we plot now in Fig.\ref{f10} the Hawking temperature and Gibbs free energy corresponding to conventional critical point $CP_3$. 
 We observe that a second kind of phase transition takes place between two stable phases (small and large black holes) and a unique inflection point is present in the right branch of temperature. Moreover, the temperature is a monotonic increasing function in terms of horizon $r_h$  as in AdS-Reissner–Nordström black hole. Besides, the free energy-temperature diagram reveals that there is no unstable phase, and no cusp shape is observed under the charge $Q<Q_m$.

\begin{figure}[!ht]
	\centering 
	\begin{subfigure}[h]{0.48\textwidth}
		\centering \includegraphics[scale=0.6]{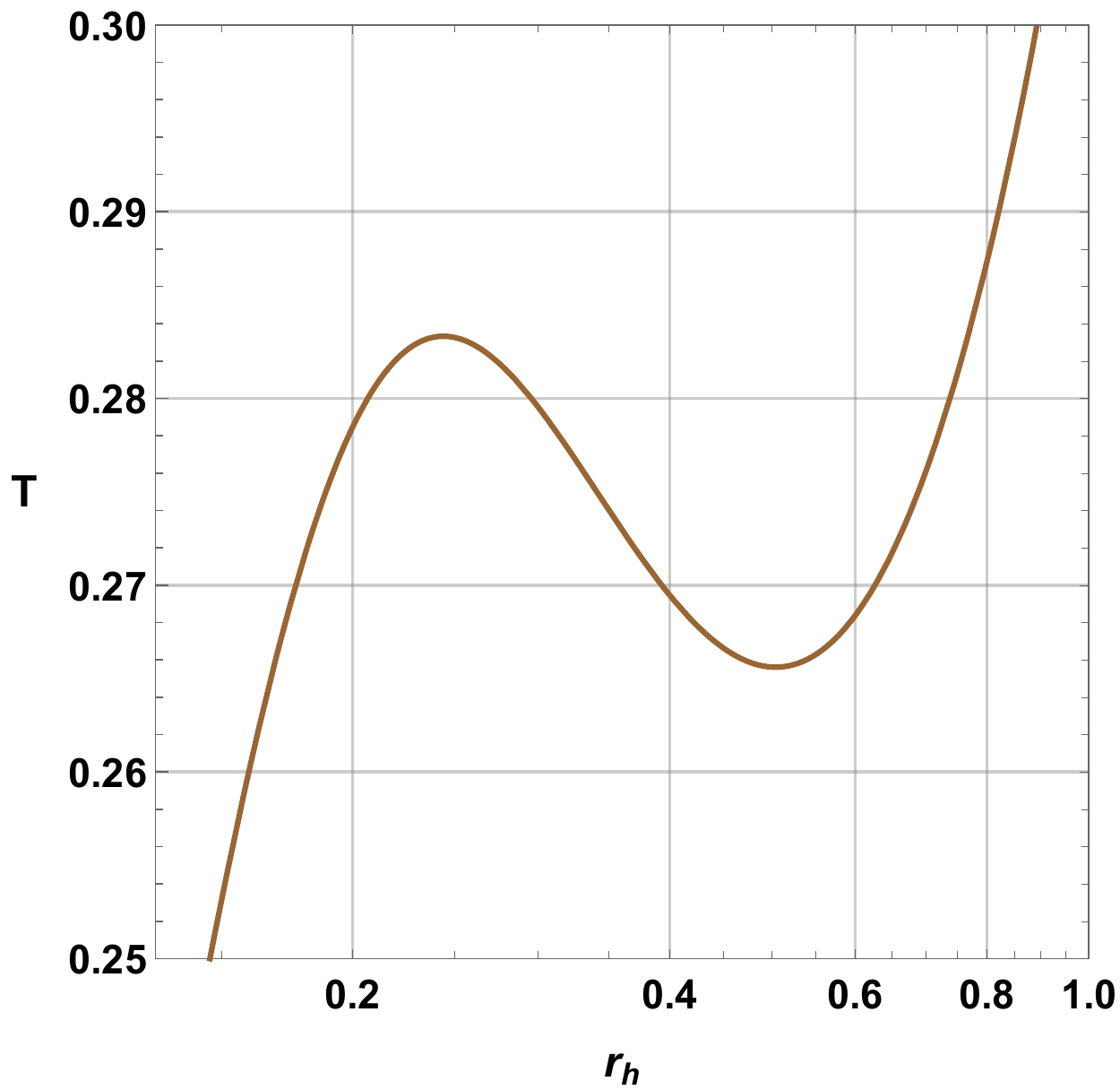}
		\caption{}
		\label{f11_1}
	\end{subfigure}
	\hspace{1pt}	
	\begin{subfigure}[h]{0.48\textwidth}
		\centering \includegraphics[scale=0.6]{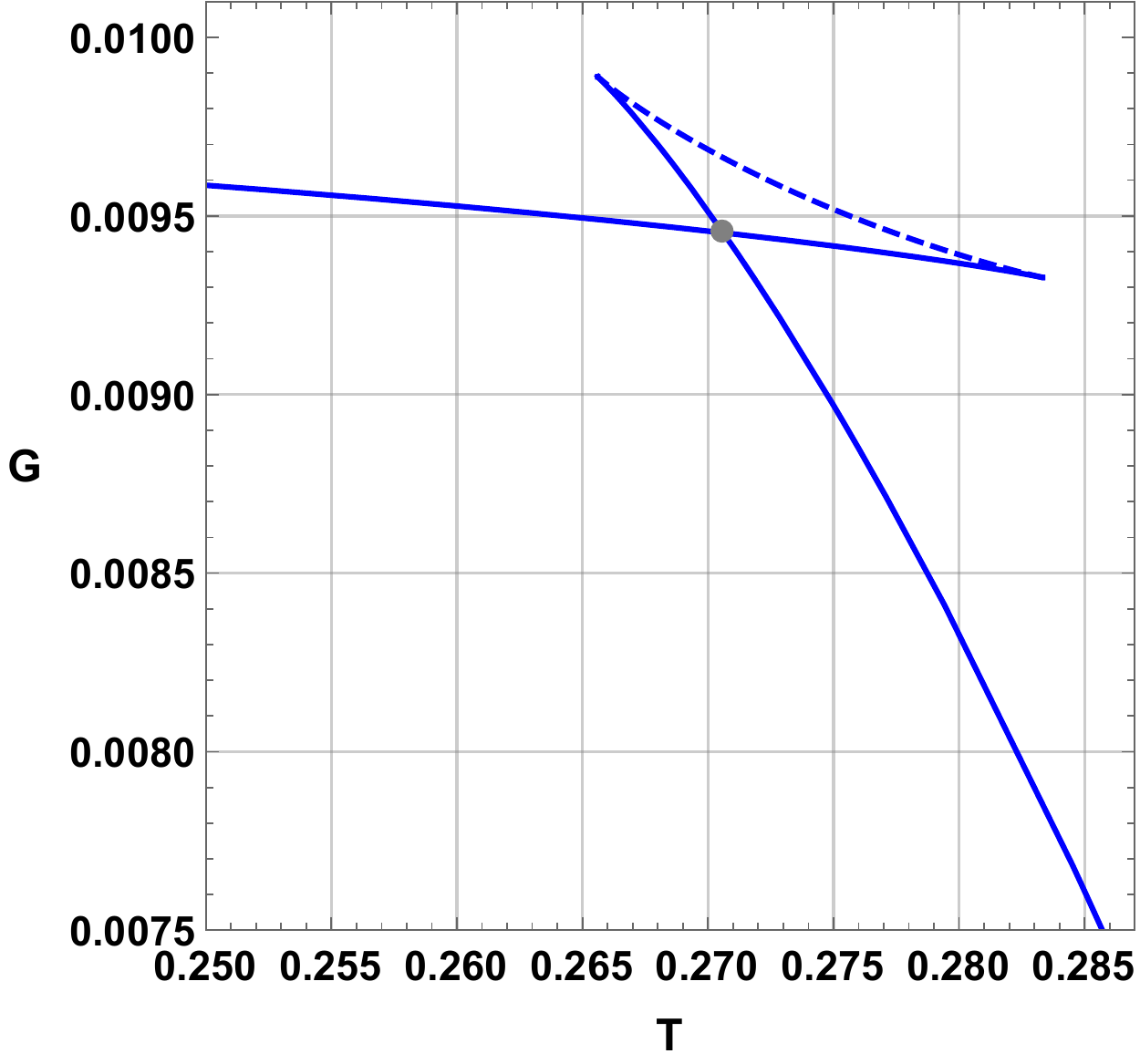}
		\caption{}
		\label{f11_2}	
	\end{subfigure}
	\caption{\footnotesize\it (a) Hawking temperature as a function of horizon radius $r_h$ and (b) and free Gibbs energy as a function of temperature near the conventional critical point $CP_3$ with $Q=Q_m = 0.0113682$, $P =0.119366$ and $b = 3.5$. }
	\label{f11}
\end{figure}

Afterward, we move to the vicinity of the conventional critical point $CP_3$ and plot the thermal and free energy variation 
in Fig.\ref{f11}  
As in AdS-Reissner–Nordström black hole case,  a first-order phase transition between two stable phases is clearly observed. While no cusp in the free energy-temperature diagram appears for $Q<Q_m$ as in the previous situation. Moreover, the temperature vanishes when $r_h\to 0$ which is an extreme case. All these remarks consolidate that the system is perfectly equivalent to AdS-Reissner–Nordström black hole.

In Table.\ref{Table1}, we summarize some thermodynamic differences between the two topological phases, 
\begin{table}[!ht]
\begin{center}
\begin{tabular}{ |C{5.3cm}||C{5.7cm}|C{4.2cm}|  }
	\hline
	\centering Electric charge $Q$ &    $Q_I<Q<Q_m$ &    $Q\geq Q_m$  \\
	\hline
	Topological charge $\mathcal{Q}$   & 0    & -1\\
	\hline
	Critical points&   2  & 1  \\
	\hline
	Vortex/anti-vortex &   Bound vortex/anti-vortex pair  & Free vortex  \\
	\hline
	Extremal black hole  & No & Yes\\
	\hline
	Hawking-Page-like transition &   Yes  & No\\
	\hline
	Hawking Temperature &  $\lim_{r \to 0} T(r) = +\infty  $  & $\exists r_e \in \mathbb{R}^+\quad T(r_e) = 0 $   \\
	\hline
	Gibbs free energy&$\left.  \dfrac{\partial }{\partial r}  G(T,r)\right| _{r=0}>0$ & $\left.  \dfrac{\partial }{\partial r}  G(T,r)\right| _{r=0}\leq 0$\\
	\hline
\end{tabular}
\end{center}
	\caption{\footnotesize\it Thermodynamic characteristics of the two topological phases.}
\label{Table1}
\end{table}
particularly the existence of Hawking-Page-like transition for $Q<Q_m$. Indeed, the first derivative of Gibbs free energy at $r_h = 0$ is given by 
\begin{equation}\label{25}
	\left.  \dfrac{\partial }{\partial r}  G(T,r)\right| _{r=0} = \frac{1}{8 \pi }-b Q, 
\end{equation} 
which is negative when $Q<Q_m$ and then there is always a Hawking-Page-like transition between thermal radiations and small/large black holes: For $Q<Q_t = 0.0103638$\footnote{With $b=3.5$ and $P=0.119366$.}, there is a Hawking-Page-like transition between thermal radiations and large black holes, for  $Q>Q_t$, there is a Hawking-Page-like transition between thermal radiations and small black holes, and for $Q= Q_t$ there exits a triple point where the three phases coexist. At the topological transition, the first derivative of Gibbs free energy in \eqref{25} vanishes. Moreover, for $Q<Q_m$, there is no extremal black hole and the system is Schwarzchild-like, whereas, for $Q\geq Q_m$, the black hole admits an extremal state where the Hawking temperature vanishes and the extremal radius $r_e$ is given by 
\begin{equation}\label{26}
r_e = \sqrt{\frac{-2 b^2 l^4+2 \sqrt{b^4 l^8+48 \pi ^2 b^2 l^4 Q^2 \left(4 b^2 l^2+3\right)}-3 l^2}{12 b^2 l^2+9}},
\end{equation}
which is real and positive for $Q\geq Q_m$ and imaginary otherwise.

Having established the ability of Duan's topological current $\phi$-mapping formalism to be an efficient topological tool to prob the Born-Infeld-AdS black hole phase structure and the confirmation of the existence of an isolated critical, we turn our attention in the next section to a second topological approach based on the off-shell free energy.

\section{Second approach: the off-shell free energy}\label{off-shell_FE}

Inspired by \cite{Wei:2022dzw}, in this section, we use another topological approach that allows us to investigate the locally thermodynamic stable and unstable black hole solutions. The off-shell-free energy is obtained to be 
\begin{equation}\label{27}
	G = M-TS,
\end{equation}
where $M$ is the black hole mass, $S$ denotes its entropy and $T$ is a parameter having the dimension of temperature which we can vary freely. When $T$ takes the value of the Hawking temperature, the off-shell free energy coincides with Gibbs free energy describing the black hole configurations. Following \cite{Wei:2022dzw}, we define a new vector filed $\phi = (\phi^{r_h},\phi^{\theta})$ as 
\begin{equation}\label{28}
	\phi = \left(- \dfrac{\partial G}{\partial r_h}, -\cot(\theta) \csc(\theta)\right) ,
\end{equation}
and then we recall again Duan’s $\phi$-mapping topological current theory introduced in the first approach to study the topology of such a black hole solution by the use of this new vector field.

 In our previous work \cite{Ali:2023wkq}, we have demonstrated that depending upon the parameter $b$, the critical behavior of Born-Infeld-AdS black hole could emerge in S-type or RN-type black hole \footnote{S-type and RN-type stand for Schwarzchild-type and Reissner–Nordström-type respectively.} \cite{Dehyadegari:2017hvd}. 
Writing these conditions in terms of pressure $P$ for a fixed $b$ is an easy task.  The condition to have a critical behavior within the S-type configuration reads :
\begin{equation}\label{29}
	\dfrac{\left( \sqrt{6\sqrt{3}-9}-1\right) b^2}{4 \pi}<P<\dfrac{b^2}{4\pi \left( 1+\sqrt{2}\right) }.
\end{equation} 
While if the RN-type configuration is considered, one can find such conditions as :
\begin{equation}\label{30}
	P<\dfrac{\left( \sqrt{6\sqrt{3}-9}-1\right) b^2}{4 \pi}.
\end{equation}
Thus, no critical behavior is observed otherwise, as we have seen in the previous section that when $P=P_I$ 
corresponds to the isolated critical point.

\subsection{S-type black hole}

\begin{figure}[!ht]
	\centering 
	\begin{subfigure}[h]{0.48\textwidth}
		\centering \includegraphics[scale=0.6]{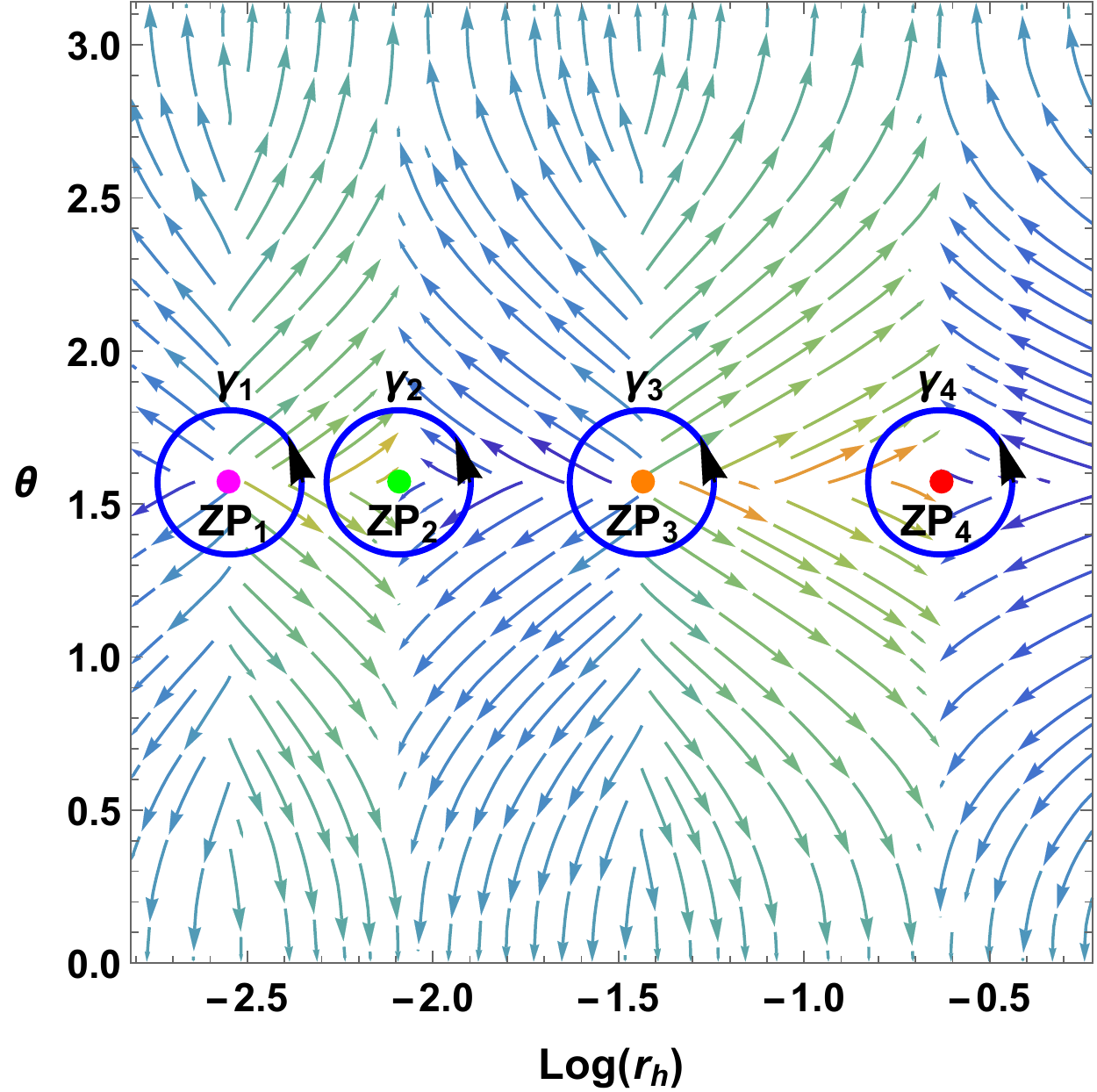}
		\caption{}
		\label{f12_1}
	\end{subfigure}
	\hspace{1pt}	
	\begin{subfigure}[h]{0.48\textwidth}
		\centering \includegraphics[scale=0.6]{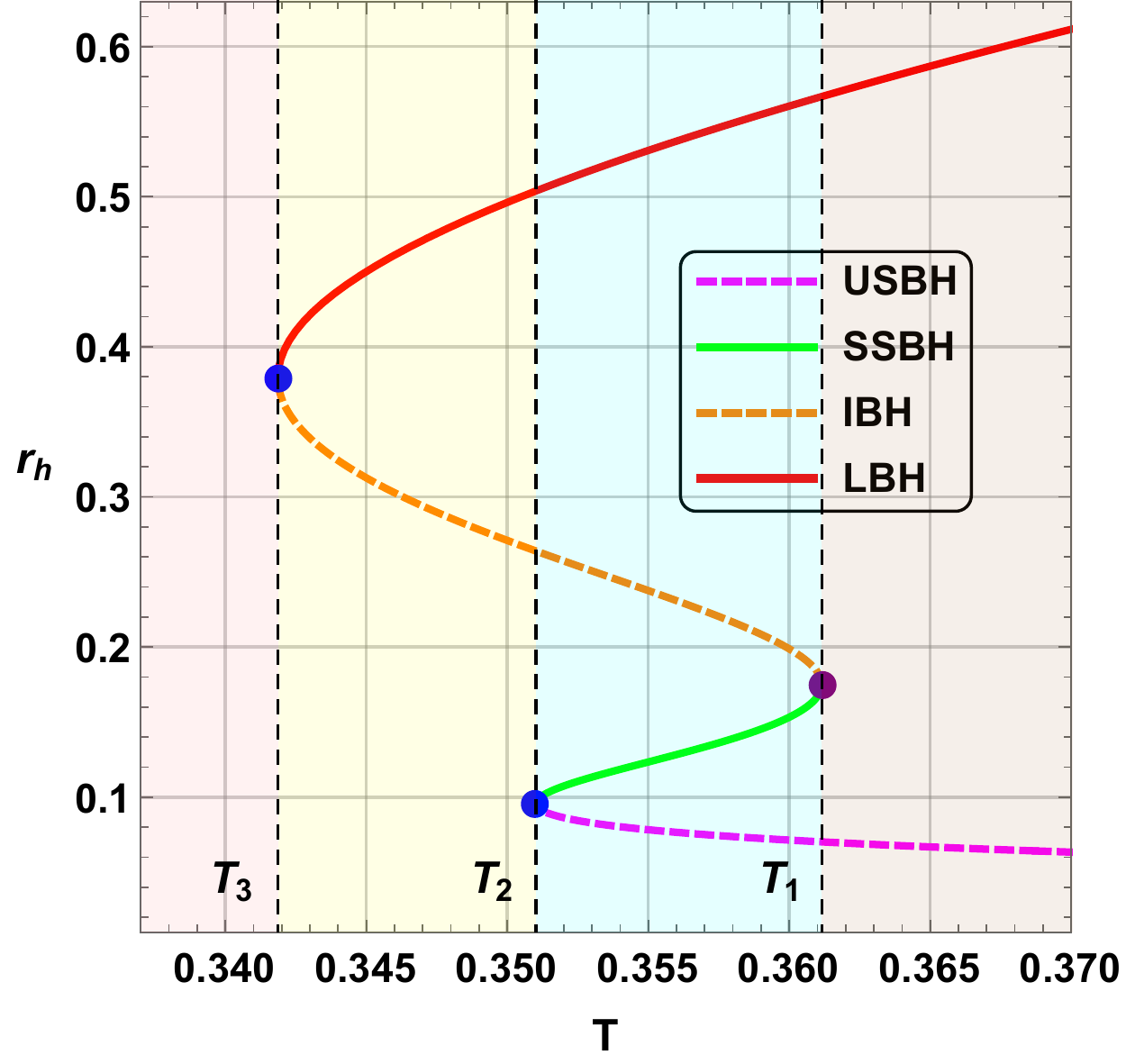}
		\caption{}
		\label{f12_2}	
	\end{subfigure}
	\caption{\footnotesize\it (a) Normalized vector field $n^i$ in
		the $(r_h,\theta)$ plane for $T = 0.335$ and (b) zero points of $\phi^{r_h}$ shown in $(r_h,T)$ plane with $P= 0.2$, $Q= 0.0093$ and $b = 3.5$. }
	\label{f12}
\end{figure}

We envisage the situation when critical behavior takes place in the S-type black hole. For $b=3.5$, we set $P = 0.2$ which respects the condition in Eq.\eqref{29}. Then, we plot in Fig.\ref{f12_1} the normalized vector field $n^i$, such that $\phi^i/\left| \left| \phi\right| \right| $, in the $(r_h,\theta)$ plane for $Q= 0.0093$  and $T = 0.335$, 
we observe that we have four zeros. Using Eq.\eqref{15}, and the four contours $\gamma_{i=1\cdots4}$, the topological charges of these four zeros are : 
\begin{equation}\label{31}
	\mathcal{Q}(ZP_1) = +1,  \quad \mathcal{Q}(ZP_2) = -1, \quad  \mathcal{Q}(ZP_3) = +1, \quad  \mathcal{Q}(ZP_4) = -1.
\end{equation}
These four zeros correspond to the four solutions associated with Born-Infeld-AdS black hole in this situation  \cite{Ali:2023wkq,Wei:2022dzw}. Indeed, the first zero point $ZP_1$ (magenta dot) corresponds to an unstable small black hole (USBH), the second zero point $ZP_2$ (green dot) is associated with a locally stable small black hole (SSBH), the third zero point $ZP_3$ (orange dot) is linked to an intermediate black hole, and the fourth zero one $ZP_4$ (red dot) corresponds to a large black hole (LBH). Thus, the stable phases ($ZP_2$ and $ZP_4$) have a negative topological charge ($-1$) and the unstable phases  ($ZP_1$ and $ZP_3$) have a positive topological charge ($+1$). The total topological charge of the system in this case is null :
\begin{equation}\label{32}
	\mathcal{Q} = \mathcal{Q}(ZP_1) + \mathcal{Q}(ZP_2)+ \mathcal{Q}(ZP_3)+\mathcal{Q}(ZP_4) = 0.
\end{equation}
Therefore, the system is topologically equivalent to AdS-Schwarzchild black hole \cite{Yerra:2022coh}. Next, we display in Fig.\ref{f12_2}, the zero points of $\phi^{r_h}$ in $(r_h,T)$ plane for  $Q= 0.0093$. From this panel, we observe that there are four black hole branches: unstable small black hole (magenta dashed line), stable small black hole (green solid line), intermediate black hole (orange dashed line), and large black hole (red solid line). Moreover, we observe a vortex/anti-vortex creation point (purple dot) and  two vortex/anti-vortex annihilation points (blue points). For $T>T_1$ (brown region), there is only a pair of vortex/anti-vortex corresponding to LBH/USBH phases and the total topological charge in this region is zero \footnote{The topological charge of a vortex (stable phase) is $-1$ and that of an anti-vortex (unstable phase) is $+1$.}. For $T = T_1$, we have a vortex/anti-vortex creation point (purple dot), that is to say, we have a generation of the stable small black hole phase (vortex) and the unstable intermediate black hole phase (anti-vortex). For $T_2<T<T_1$ (cyan region), there are two pairs of vortex/anti-vortex corresponding to LBH/USBH and SSBH/IBH phases, besides, the total topological charge in this region is also zero.  For $T=T_2$, we have a  vortex/anti-vortex annihilation point (blue dot), indicating that we are in the presence of annihilation of the stable small black hole phase (vortex) and the unstable small black hole phase (anti-vortex). Moreover, in the $T_3<T<T_2$ domain (yellow region), there is only a pair of vortex/anti-vortex corresponding to LBH/IBH  phases and the total topological charge in this region is zero. Lastly, for $T=T_3$, we have a second vortex/anti-vortex annihilation point (blue dot), that is to say, we have an annihilation of the stable large black hole phase (vortex) and the unstable intermediate black hole phase (anti-vortex).  For $T<T_3$ (pink region), there is no black hole solution (no vortexes) and then the total topological charge is zero. Therefore, the total topological charge is always zero and is independent of the temperature indicating the system belongs to the same topological class of AdS-Schwarzchild black hole.

\begin{figure}[!ht]
	\centering 
	\begin{subfigure}[h]{0.48\textwidth}
		\centering \includegraphics[scale=0.6]{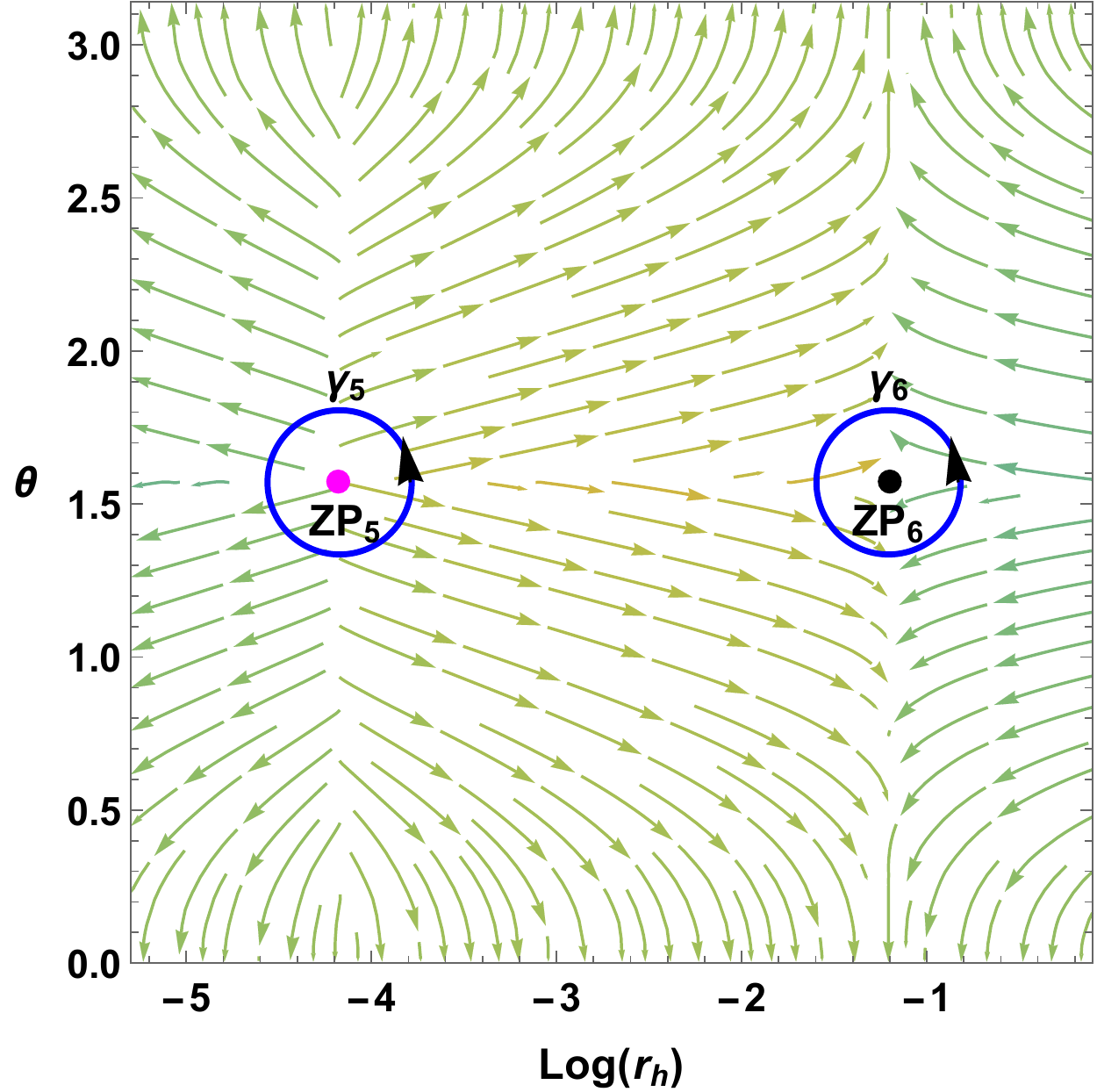}
		\caption{}
		\label{f13_1}
	\end{subfigure}
	\hspace{1pt}	
	\begin{subfigure}[h]{0.48\textwidth}
		\centering \includegraphics[scale=0.6]{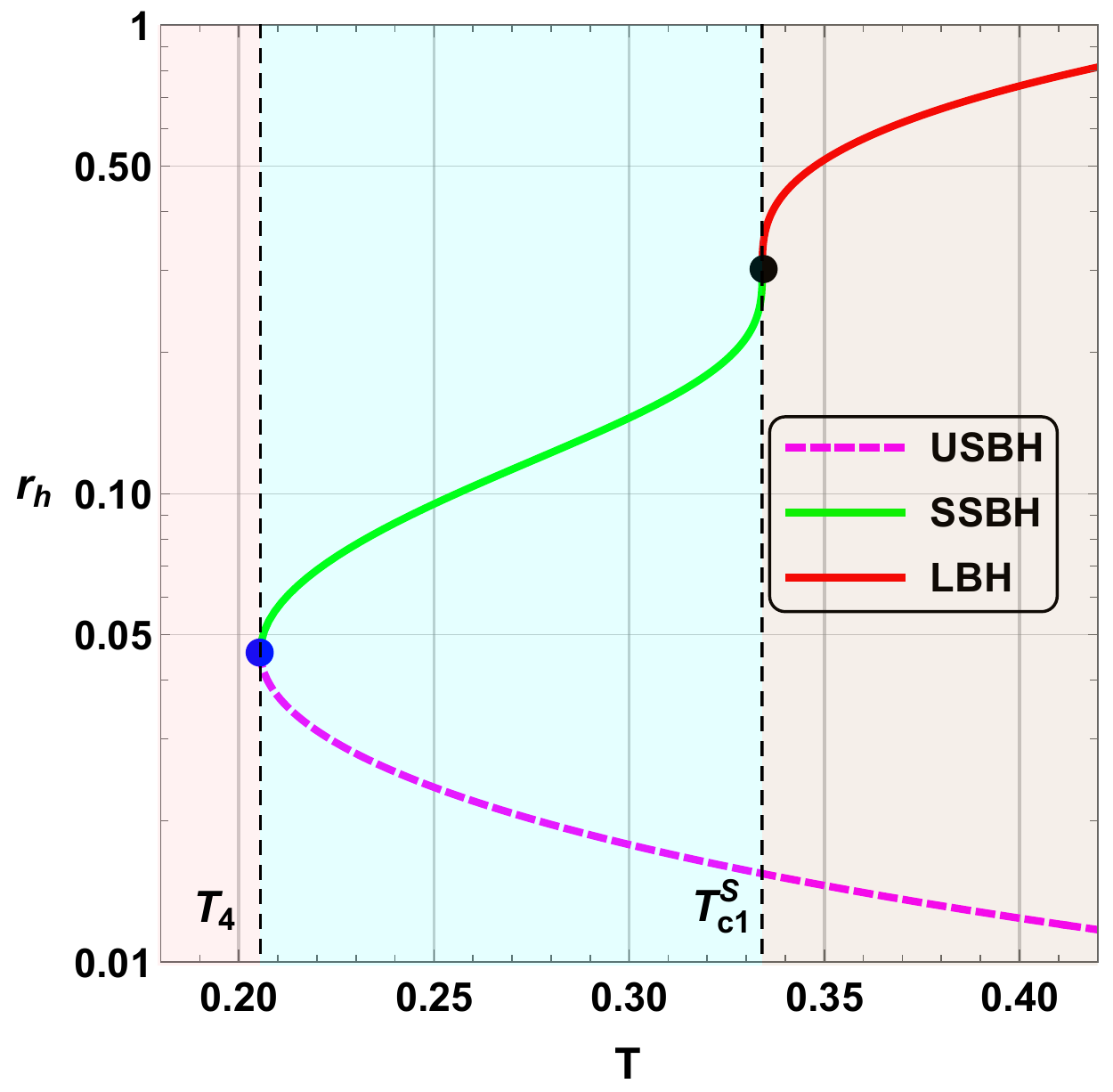}
		\caption{}
		\label{f13_2}	
	\end{subfigure}
	\caption{\footnotesize\it (a) Normalized vector field $n^i$ in
		the $(r_h,\theta)$ plane for $T = T_{c1}^{S}=  0.334094$ and (b) zero points of $\phi^{r_h}$ shown in $(r_h,T)$ plane with $P= 0.2$, $Q=Q_{c1}^{S}= 0.0107116$ and $b = 3.5$. }
	\label{f13}
\end{figure}

Let us now consider the conventional critical point situation where a second-order phase transition between two stable phases SSBH and LBH takes place.  We illustrate in Fig.\ref{f13_1}, the normalized vector field $n^i$, in the $(r_h,\theta)$ plane for $Q=Q_{c1}^{S}= 0.0107116$  and $T = T_{c1}^{S}=  0.334094$.
One can notice that two zero points appear in such cases. Using Eq.\eqref{15}, and the two contours $\gamma_5$ and $\gamma_6$, the topological charges of these two zero points are : 
\begin{equation}\label{33}
	\mathcal{Q}(ZP_5) = +1,  \quad \mathcal{Q}(ZP_6) = -1.
\end{equation}
 The first zero point denoted $ZP_5$ (magenta dot) corresponds to an unstable small black hole (USBH) whereas the second zero point $ZP_6$ (black dot) corresponds to the conventional critical point. Indeed, $ZP_6$ can be regarded as a degenerate zero point that corresponds to a superposition of two stable phases (two vortexes: SBH and LBH phases), and an unstable phase (an anti-vortex: IBH phase) which explain the negative topological charge. In this case, the system presents a vanishing topological charge:
 \begin{equation}\label{34}
 	\mathcal{Q} =  \mathcal{Q}(ZP_5) + \mathcal{Q}(ZP_6) = 0.
 \end{equation}
In Fig. \ref{f13_2}, we illustrate the zero points of $\phi^{r_h}$ in $(r_h,T)$ plane for $Q= Q=Q_{c1}^{S}= 0.0107116$. Within such a panel, we observe that there are three black hole branches: USBH (magenta dashed line), SSBH (green solid line), and LBH (red solid line), and in fact, the IBH phase is no longer available. Moreover, we remark a critical point (black dot) and just one vortex/anti-vortex annihilation point (blue dot) and there is no other vortex/anti-vortex annihilation point. Thus the critical point can be regarded as a superposition of vortex/anti-vortex creation and annihilation points. For $T>T_{c1}^{S}$ (brown region), there is just a pair of vortex/anti-vortex corresponding to LBH/USBH phase reminiscences as observed in the previous situation, Fig.\ref{f12_2} and a null total topological charge is computed in this region. For $T=T_{c1}^{S}$, we notice the existence of a critical point that corresponds to a superposition of vortex/anti-vortex creation point and vortex/anti-vortex annihilation point. Indeed, at the critical point, we have a vortex/anti-vortex creation point associated with SSBH/IBH phase generation and a vortex/anti-vortex annihilation point that corresponds to LBH/IBH phase annihilation. That is to say, the LBH phase (vortex) is annihilated with the generated IBH phase (anti-vortex) and only the generated SSBH phase (vortex) will remain. In the situation where $T_4<T<T_{c1}^{S}$ (cyan region), we observe a pair of vortex/anti-vortex corresponding to SSBH/USBH  phases, then the total topological charge in this region is zero. Arriving at $T=T_4$,  a vortex/anti-vortex annihilation point (blue dot) is observed, in other words, we have an annihilation of SSBH and USBH phases. When $T<T_4$ (pink region), there is no black hole solution (no vortexes) and then the total topological charge is zero. Therefore, the total topological charge is always zero independent of temperature.

\begin{figure}[!ht]
	\centering 
	\begin{subfigure}[h]{0.48\textwidth}
		\centering \includegraphics[scale=0.6]{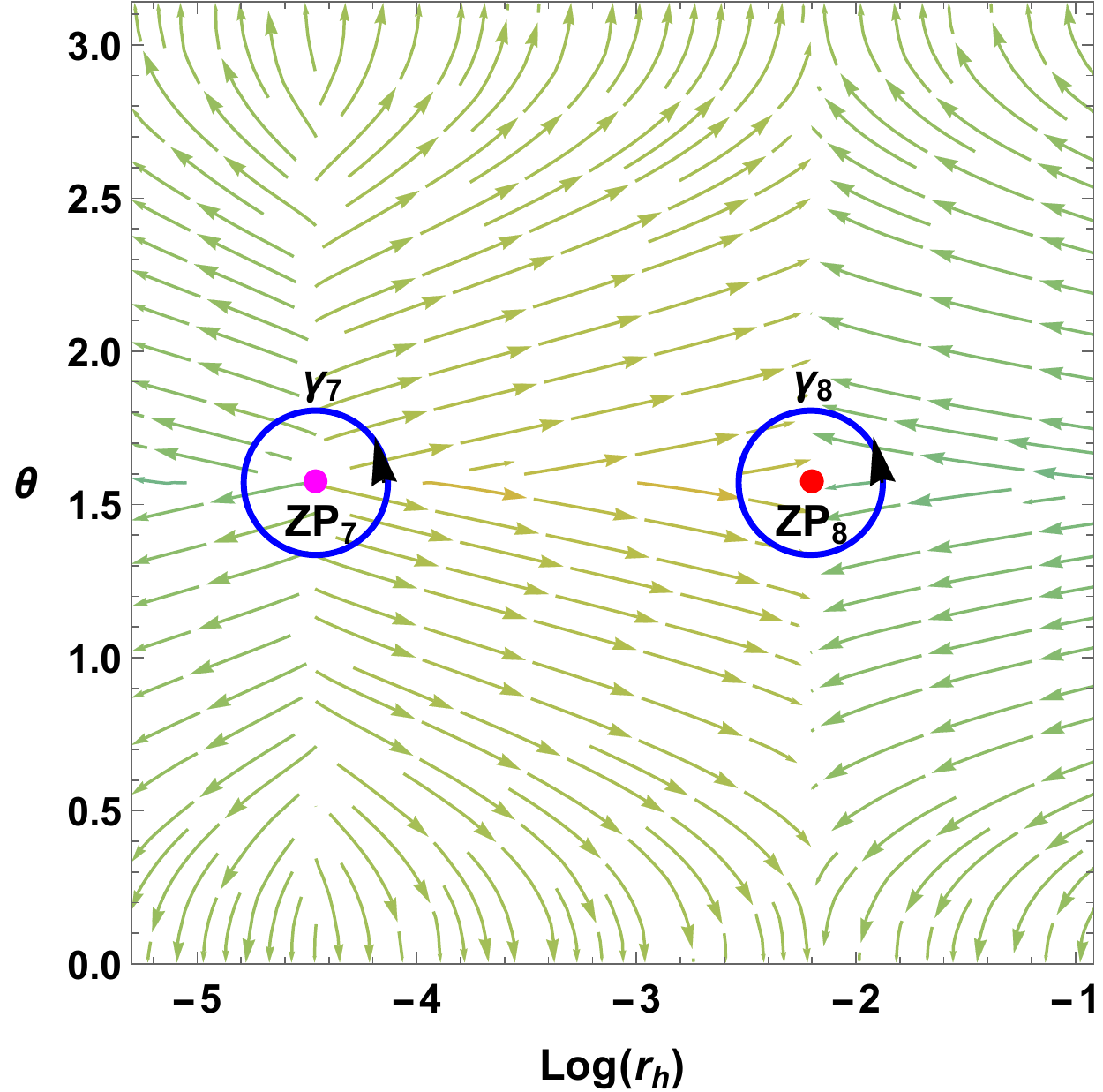}
		\caption{}
		\label{f14_1}
	\end{subfigure}
	\hspace{1pt}	
	\begin{subfigure}[h]{0.48\textwidth}
		\centering \includegraphics[scale=0.6]{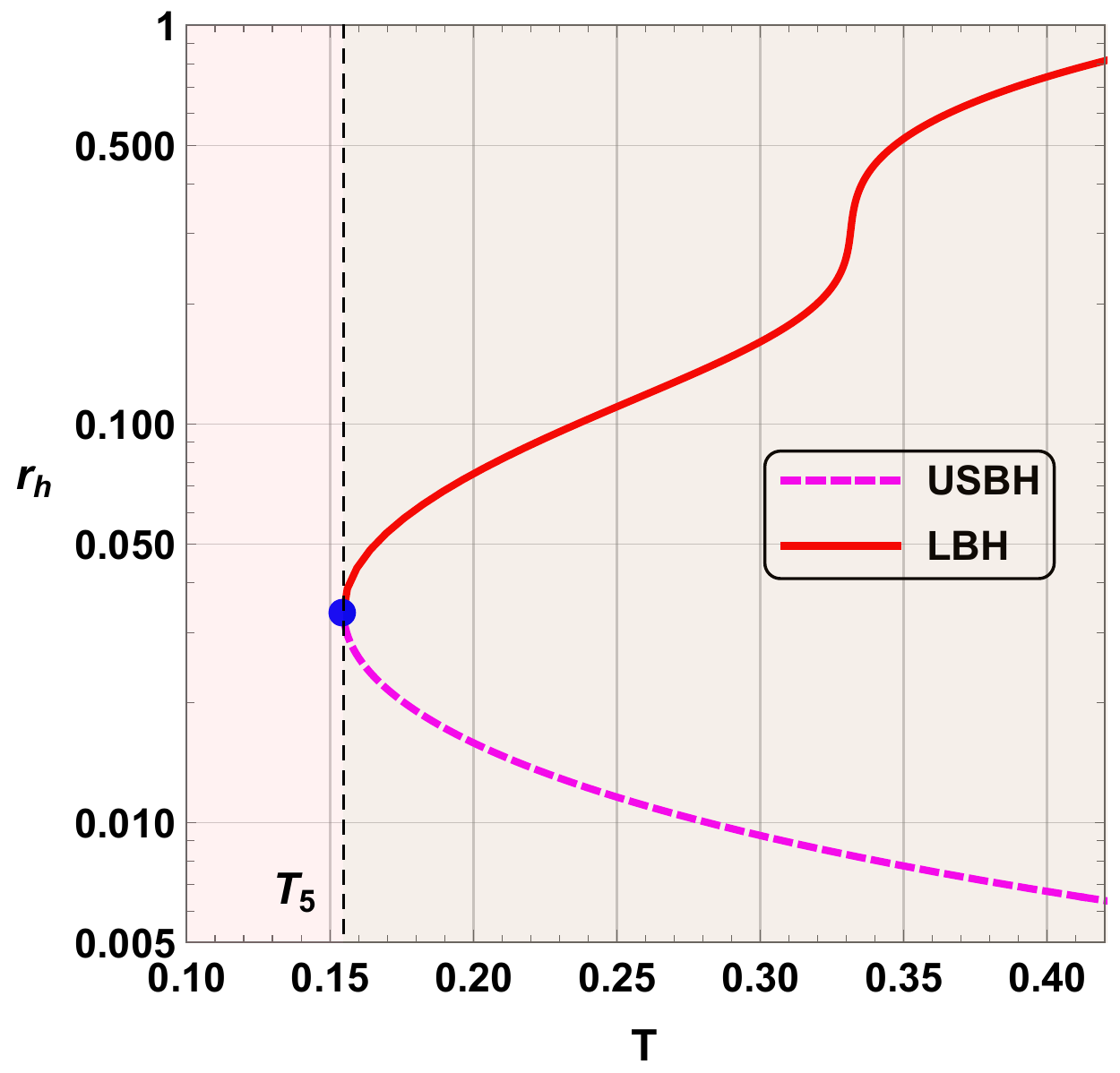}
		\caption{}
		\label{f14_2}	
	\end{subfigure}
	\caption{\footnotesize\it (a) Normalized vector field $n^i$ in
		the $(r_h,\theta)$ plane for $T =  0.25$ and (b) zero points of $\phi^{r_h}$ shown in $(r_h,T)$ plane with $P= 0.2$, $Q= 0.011 < Q_m$ and $b = 3.5$. }
	\label{f14}
\end{figure}

We regard now the case when the electrical charge gets increased but still below the topological critical charge $Q_m$. We plot in Fig.\ref{f14_1} the normalized vector field $n^i$, in the $(r_h,\theta)$ plane for $Q= 0.011$ ($Q_{c1}^{S}<Q<Q_m$)  and $T =  0.25$.
The figure unveils two zero points. Using Eq.\eqref{15}, and the two contours $\gamma_7$ and $\gamma_8$, the topological charges of these two zero points are found to be  
\begin{equation}\label{35}
	\mathcal{Q}(ZP_7) = +1,  \quad \mathcal{Q}(ZP_8) = -1.
\end{equation}
The first zero point $ZP_7$ (magenta dot) corresponds to the USBH while the second zero point $ZP_8$ (red dot) is associated with the LBH. The total topological charge of the system in this case is zero: 
\begin{equation}\label{36}
	\mathcal{Q} =  \mathcal{Q}(ZP_7) + \mathcal{Q}(ZP_8) = 0,
\end{equation}

 The zero points of $\phi^{r_h}$ in $(r_h,T)$ plane for  $Q= 0.011$ ($Q_{c1}^{S}<Q<Q_m$) are displayed in  Fig.\ref{f14_2}.  A simple examination reveals that the system owns two black hole branches: USBH (magenta dashed line) and LBH (red solid line). The SSBH phase has disappeared because the creation point that generates the SSBH phase is no longer present. Moreover, one can remark just only the vortex/anti-vortex annihilation point (blue dot) and there is no vortex/anti-vortex creation point, like in the AdS-Schwarzchild case. For the region $T>T_5$ (brown region), there is a pair of vortex/anti-vortex corresponding to LBH/USBH phases and the total topological charge in this region is zero. For $T=T_5$, we have a vortex/anti-vortex annihilation point (blue dot), that is to say, we have an annihilation of LBH and USBH phases. In the $T<T_5$ domain (pink region), there is no black hole solution (no vortexes) and then the total topological charge remains null.

\begin{figure}[!ht]
	\centering 
	\begin{subfigure}[h]{0.48\textwidth}
		\centering \includegraphics[scale=0.6]{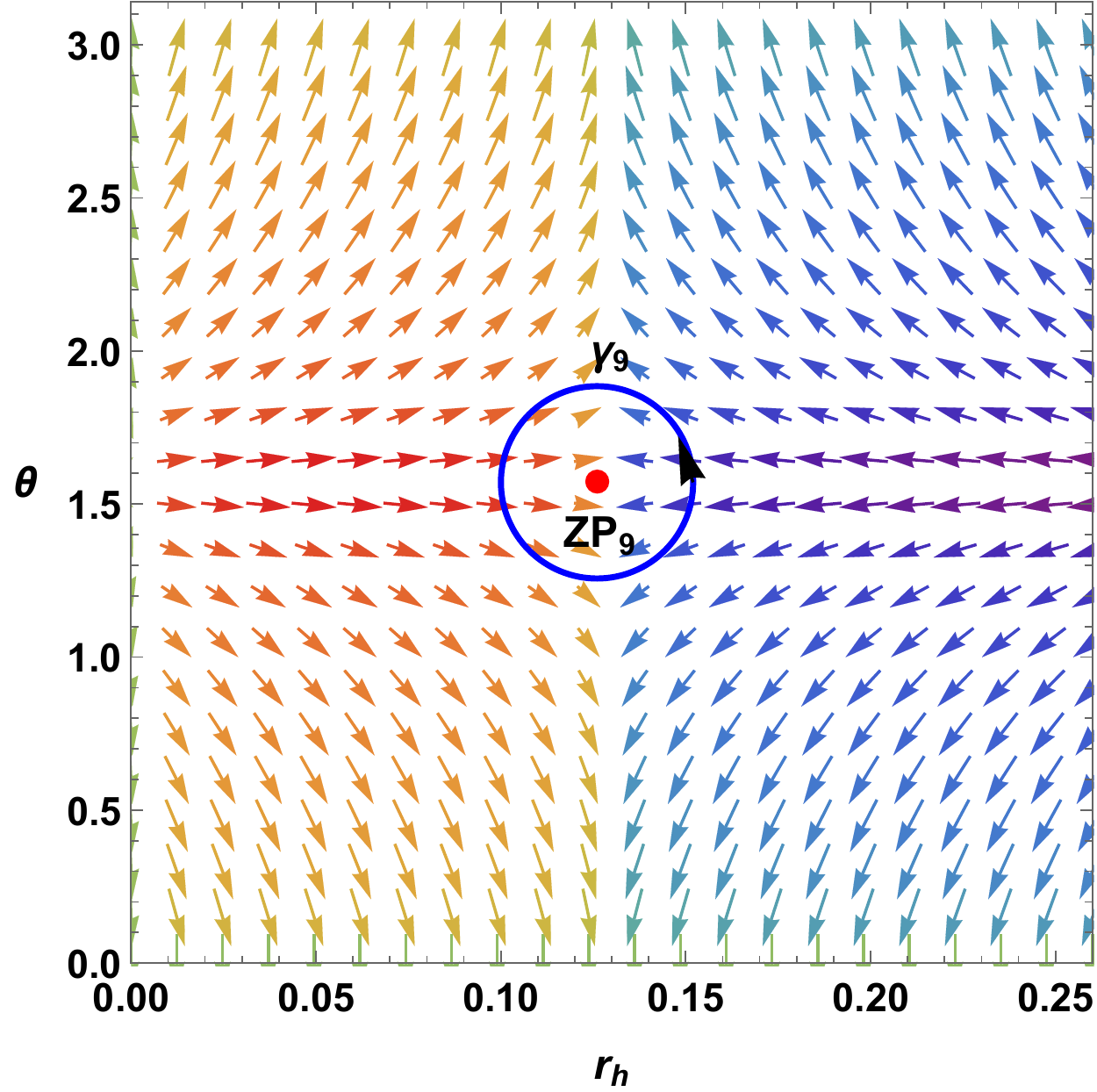}
		\caption{}
		\label{f15_1}
	\end{subfigure}
	\hspace{1pt}	
	\begin{subfigure}[h]{0.48\textwidth}
		\centering \includegraphics[scale=0.6]{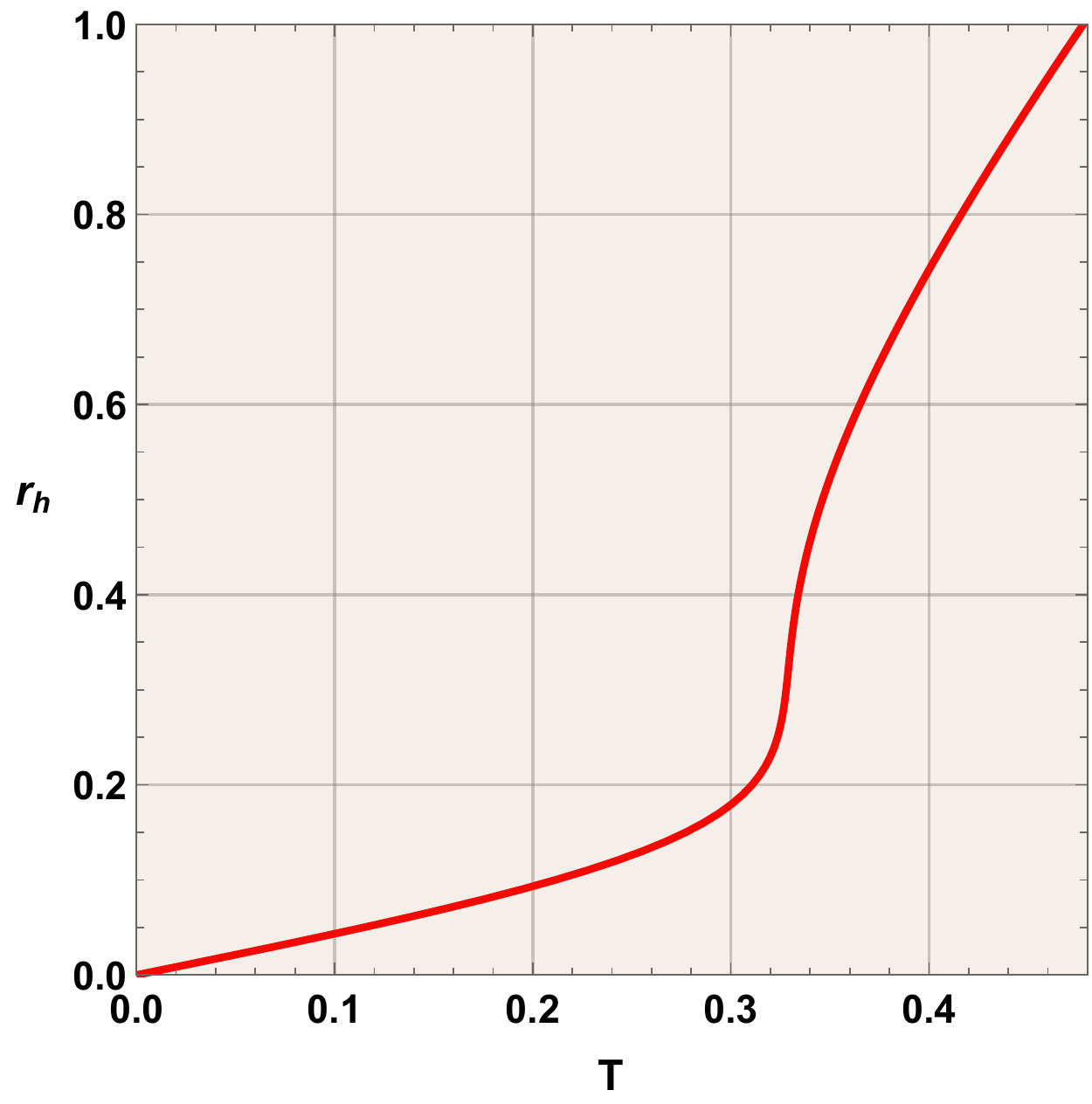}
		\caption{}
		\label{f15_2}	
	\end{subfigure}
	\caption{\footnotesize\it (a) Normalized vector field $n^i$ in
		the $(r_h,\theta)$ plane for $T =  0.25$ and (b) zero points of $\phi^{r_h}$ shown in $(r_h,T)$ plane with $P= 0.2$, $Q= Q_m = 0.0113682$ and $b = 3.5$. }
	\label{f15}
\end{figure}

Henceforth, electric charge is equal to the topological critical charge $Q_m$ and we depict in Fig.\ref{f15_1} the normalized vector field $n^i$, in the $(r_h,\theta)$ plane for $Q= Q_m = 0.0113682$   and $T =  0.25$, and in Fig.\ref{f15_2} the zero points of $\phi^{r_h}$ in $(r_h,T)$ plane for   $Q= Q_m = 0.0113682$
Only one zero point is present. Using Eq.\eqref{15}, and the  contour $\gamma_9$, the topological charge of this zero point is : 
\begin{equation}\label{37}
	\mathcal{Q}(ZP_9) = -1.
\end{equation}
This zero point $ZP_9$ (green dot) corresponds to LBH and the total topological charge of the system in this case is $-1$ : 
\begin{equation}\label{38}
	\mathcal{Q} =  \mathcal{Q}(ZP_9) = -1,
\end{equation}
which is in total agreement, as shown in the figure at the right panel, where only one black hole branch corresponding to LBH exists. The USBH phase has disappeared. Moreover, we do not observe any  vortex/anti-vortex creation or annihilation point. Therefore, there is a topological transition, and the total topology changes. Thus, the system belongs to a new topological class of the  Reissner-Nordström black hole \cite{Wei:2022dzw}.

\begin{figure}[!ht]
	\centering 
	\begin{subfigure}[h]{0.48\textwidth}
		\centering \includegraphics[scale=0.6]{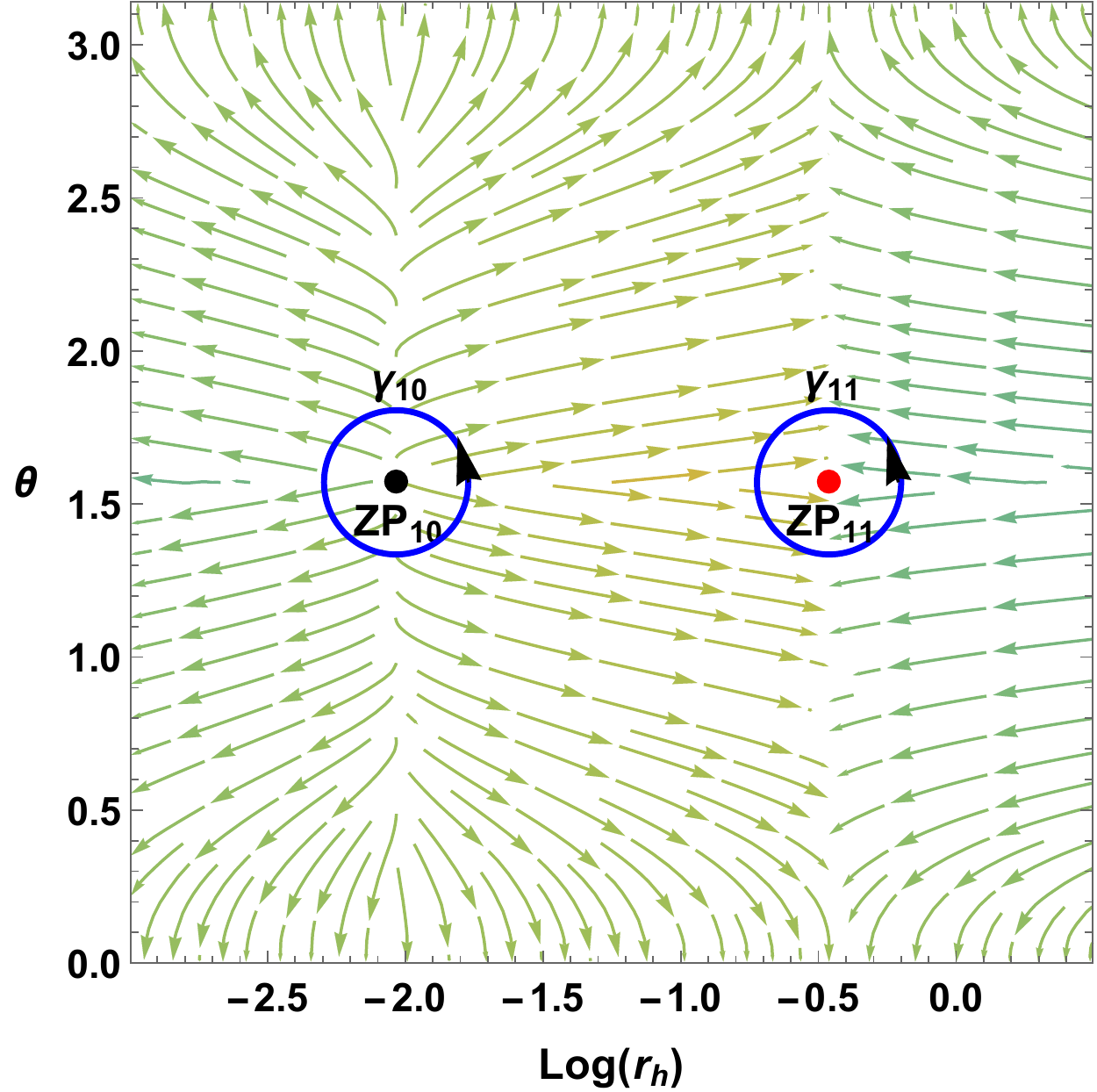}
		\caption{}
		\label{f16_1}
	\end{subfigure}
	\hspace{1pt}	
	\begin{subfigure}[h]{0.48\textwidth}
		\centering \includegraphics[scale=0.6]{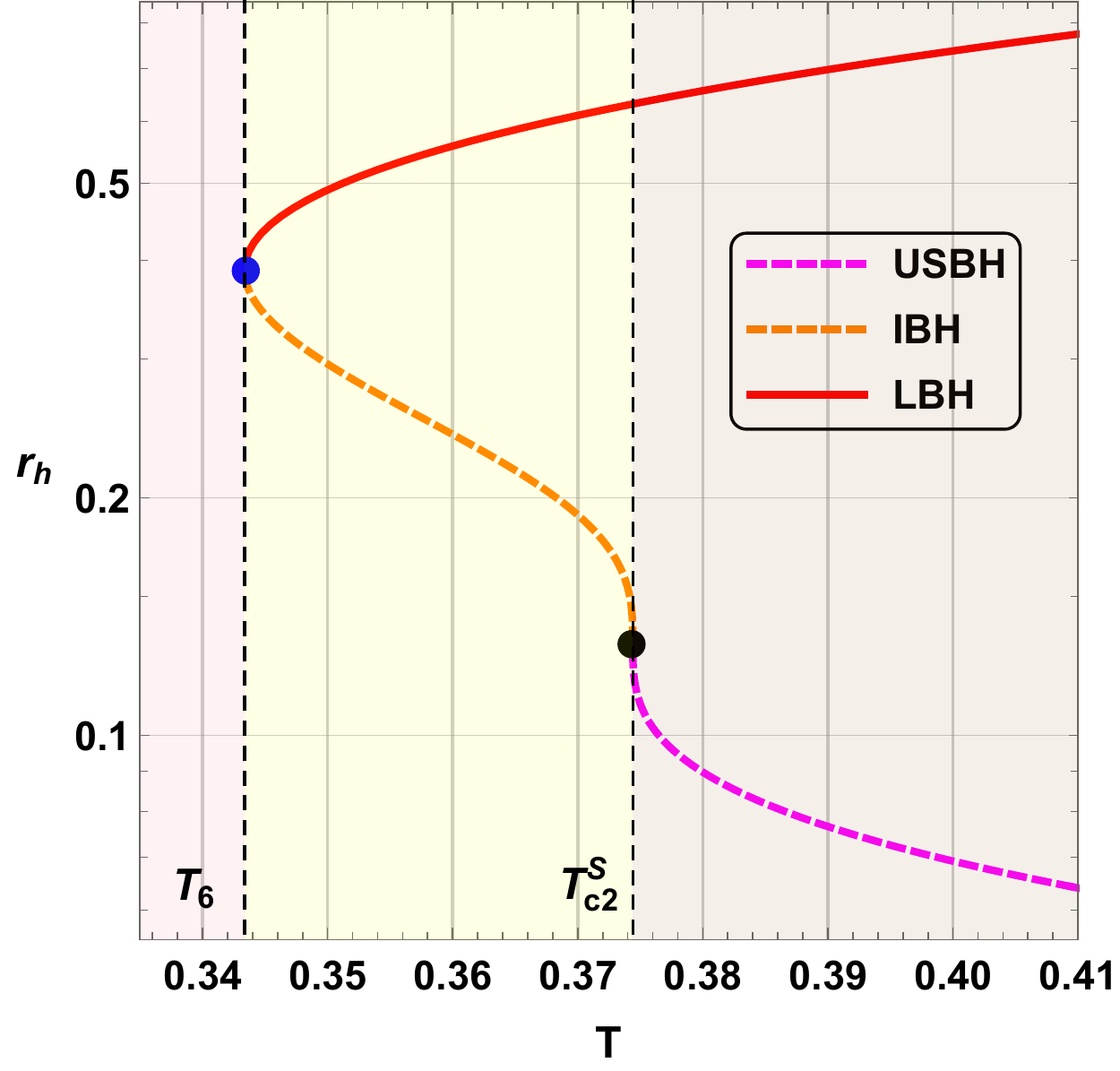}
		\caption{}
		\label{f16_2}	
	\end{subfigure}
	\caption{\footnotesize\it (a) Normalized vector field $n^i$ in
		the $(r_h,\theta)$ plane for $T = T_{c2}^{S}=  0.374398$ and (b) zero points of $\phi^{r_h}$ shown in $(r_h,T)$ plane with $P= 0.2$, $Q=Q_{c2}^{S}= 0.0089223$ and $b = 3.5$. }
	\label{f16}
\end{figure}

Let us now consider the unconventional critical point situation where an unconventional transition between two unstable phases USBH and IBH occurs.  Again, we plot in Fig.\ref{f16_1} the normalized vector field $n^i$, in the $(r_h,\theta)$ plane for $Q=Q_{c2}^{S}= 0.0089223$  and $T = T_{c2}^{S}=  0.374398$, 
the black hole system exhibits two zero points associated with the  $\gamma_{10}$ and $\gamma_{11}$ contours, and the topological charges are estimated to be
\begin{equation}\label{39}
	\mathcal{Q}(ZP_{10}) = +1,  \quad \mathcal{Q}(ZP_{11}) = -1.
\end{equation}
The second zero point $ZP_{11}$ (red dot) corresponds to LBH whereas the first zero point $ZP_{10}$ (black dot) denotes the unconventional critical point. Indeed, $ZP_{10}$ is a degenerate zero point corresponding to a superposition of two unstable phases (two anti-vortexes: USBH and IBH phases), and a stable phase (a vortex: SBH phase) which explain its positive topological charge, while the total topological charge of the system in this case is zero: 
\begin{equation}\label{40}
	\mathcal{Q} =  \mathcal{Q}(ZP_{10}) + \mathcal{Q}(ZP_{11})=0.
\end{equation}
We display in Fig.\ref{f16_2} the zero points of $\phi^{r_h}$ in $(r_h,T)$ plane for $Q= Q=Q_{c2}^{S}= 0.0089223$.  It is to be mentioned that in such a case, there are the existence of three black hole branches: USBH (magenta dashed line), IBH (orange dashed line) and LBH (red solid line). The SSBH phase has disappeared. Moreover, such a panel reveals a critical point (black dot) and just one vortex/anti-vortex annihilation point (blue dot), besides there is no vortex/anti-vortex creation point. Thus the critical point corresponds to a superposition of vortex/anti-vortex creation and annihilation points. For $T>T_{c2}^{S}$ (brown region), there is just a pair of vortex/anti-vortex corresponding to LBH/USBH phases and the total topological charge in this region is also zero. While for $T=T_{c2}^{S}$, we are in the presence of a critical point associated with a superposition of a vortex/anti-vortex creation point and a vortex/anti-vortex annihilation point. Indeed, at the critical points, we have a vortex/anti-vortex creation point that corresponds to SSBH/IBH phases generation and that the vortex/anti-vortex annihilation point corresponds to SSBH/USBH phases annihilation. That is to say, the USBH phase (anti-vortex) is annihilated with the generated SSBH phase (vortex), and only the generated IBH phase (anti-vortex) remains unaffected. In the thermal interval of $T_6<T<T_{c2}^{S}$ (yellow region), there is also a pair of vortex/anti-vortex corresponding to LBH/IBH  phases, thereby resulting in the total topological charge in this region as zero. In the the domain of  $T<T_6$ (pink region), there is no black hole solution (no vortexes) and consequently, the total topological charge is also null.

\begin{figure}[!ht]
	\centering 
	\begin{subfigure}[h]{0.48\textwidth}
		\centering \includegraphics[scale=0.6]{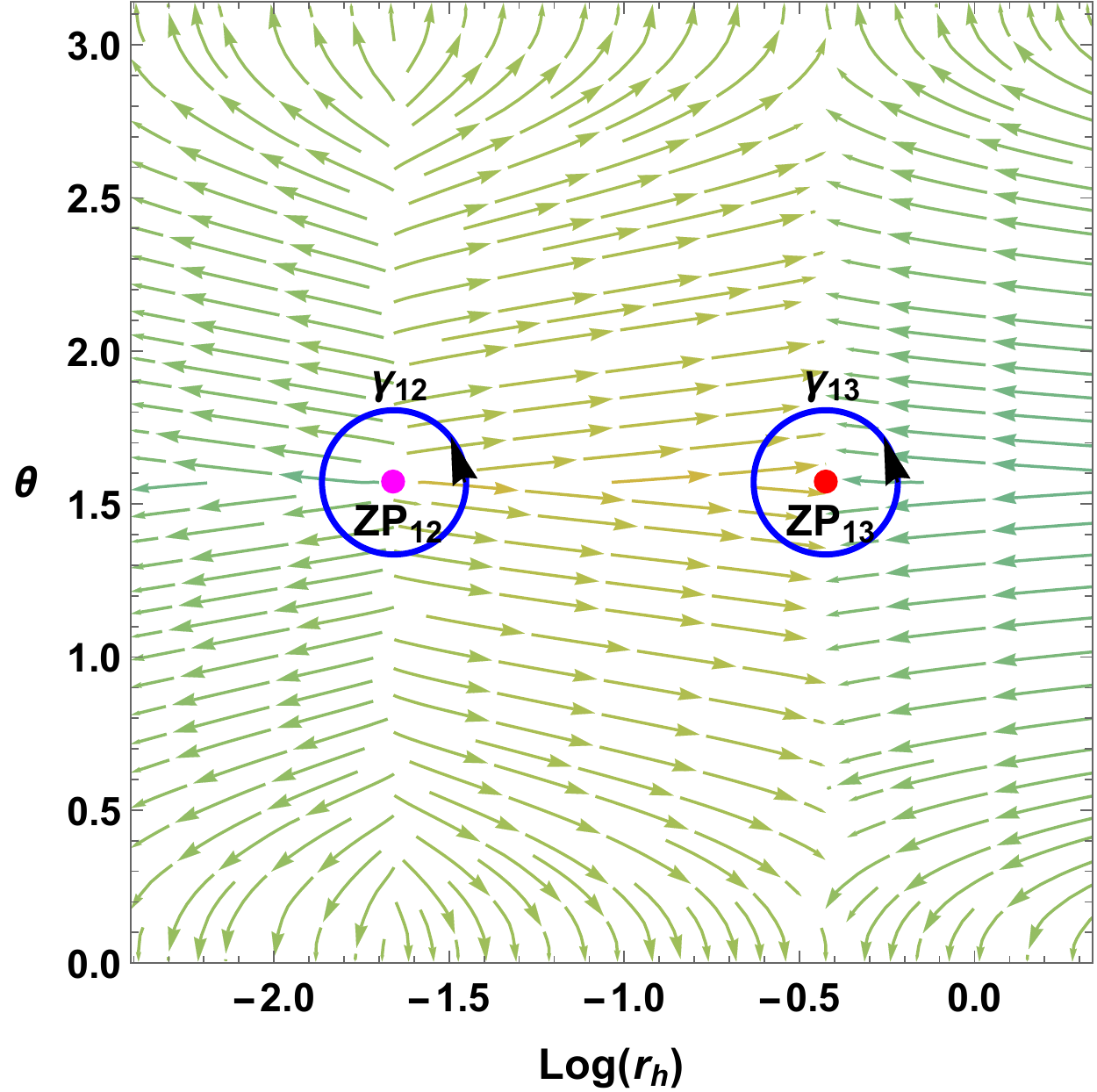}
		\caption{}
		\label{f17_1}
	\end{subfigure}
	\hspace{1pt}	
	\begin{subfigure}[h]{0.48\textwidth}
		\centering \includegraphics[scale=0.6]{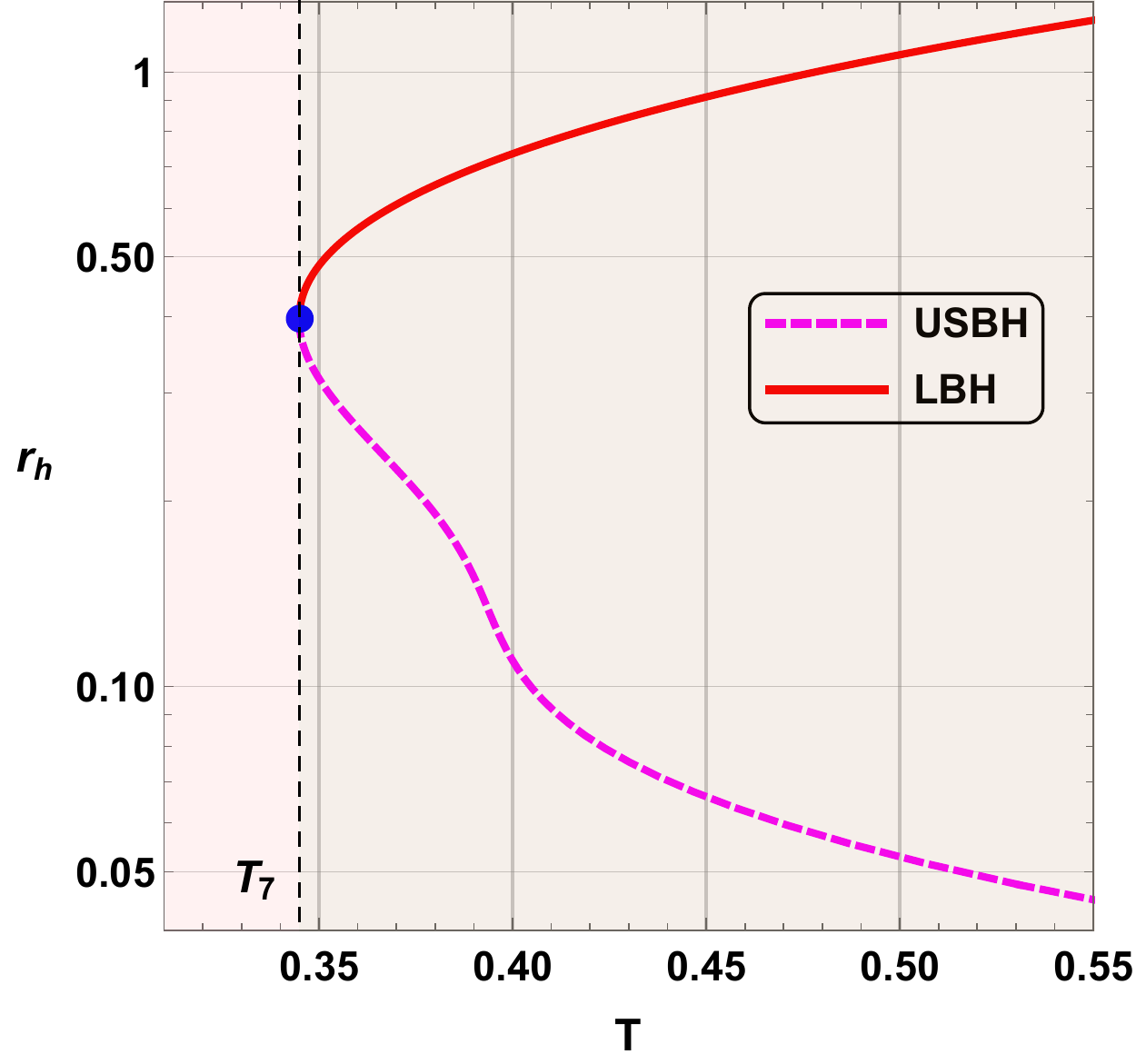}
		\caption{}
		\label{f17_2}	
	\end{subfigure}
	\caption{\footnotesize\it (a) Normalized vector field $n^i$ in
		the $(r_h,\theta)$ plane for $T =  0.38$ and (b) zero points of $\phi^{r_h}$ shown in $(r_h,T)$ plane with $P= 0.2$, $Q=0.0085$ and $b = 3.5$. }
	\label{f17}
\end{figure}

Moving to the case below the unconventional critical charge $Q_{c2}^S$, we depict the same previous quantities as in Fig.\ref{f17_1} and Fig.\ref{f17_2} for the particular values of charge $Q= 0.0085$  and  temperature $T = 0.38$.
Two zero points are present in this case. By recalling the relation, Eq.\eqref{15}, and the contours $\gamma_{12}$ and $\gamma_{13}$, we evaluate the topological charges of such zero point to be  
\begin{equation}\label{41}
	\mathcal{Q}(ZP_{12}) = +1,  \quad \mathcal{Q}(ZP_{13}) = -1.
\end{equation}
The first zero point $ZP_{12}$ (magenta dot) corresponds to USBH and the second zero point $ZP_{13}$ (red dot) is linked to LBH. The total topological charge of the system in this case is zero : 
\begin{equation}\label{42}
	\mathcal{Q} =  \mathcal{Q}(ZP_{12}) + \mathcal{Q}(ZP_{13}) = 0,
\end{equation}
The right panel unveils the existence of  two black hole branches: USBH (magenta dashed line) and LBH (red solid line). The IBH phase is no longer present. Moreover, we notice just one  vortex/anti-vortex annihilation point (blue dot) and there is no vortex/anti-vortex creation point, like in the AdS-Schwarzchild black hole system. In the domain of $T>T_7$ (brown region), there is a pair of vortex/anti-vortex associated with LBH/USBH phases and then the total topological charge is null. For $T=T_7$,  a vortex/anti-vortex annihilation point (blue dot) persists, traducing an annihilation of LBH and USBH phases. In the interval $T<T_7$ (pink region), there is no black hole solution (no vortexes) and the total topological charge in this region is also zero.

Finally, we summarize in Tab.\ref{Table2} the topological characteristics of S-type Born-Infeld-AdS black hole as the function of electrical charge showing a topological transition between AdS-Schwarzchild and  AdS-Reissner-Nordström topological classes.
\begin{table}[!ht]
	\small
	\begin{center}
		\begin{tabular}{|c||c|c|c|C{2cm}|}
			\hline
			\multirow{2}{*}{ Electric charge $Q$}   &   \multicolumn{3}{c}{ $Q<Q_m$} \vline  &       \multirow{2}{*}{ $Q\geq Q_m$}  \\
			\cline{2-4}
			& $Q<Q_{c2}^{S}$ & $Q_{c2}^{S}<Q<Q_{c1}^{S}$ & $Q>Q_{c1}^{S}$ &  \\	
			\hline
			Vortex/anti-vortex creation points &\multirow{2}{*}{ $0$} & \multirow{2}{*}{ $1$} & \multirow{2}{*}{$0$} & \multirow{2}{*}{$0$ }\\
			(Stable/unstable phases generation) & &  &  & \\
			\hline
			Vortex/anti-vortex annihilation points &\multirow{2}{*}{ $1$} & \multirow{2}{*}{ $2$} & \multirow{2}{*}{$1$} & \multirow{2}{*}{$0$ }\\
			(Stable/unstable phases annihilation) & &  &  & \\
			\hline
			Vortexes (stable phases) & $1$ & $2$ & $1$ & $1$\\
			\hline
			Anti-vortexes (unstable phases) & $1$ & $2$ & $1$ & $0$ \\
			\hline
			Topological charge $\mathcal{Q}$   &   \multicolumn{3}{c}{ $0$} \vline  &       $-1$\\
			\hline
			\multirow{3}{*}{Topological class}  &   \multicolumn{3}{c}{\multirow{3}{*}{ AdS-Schwarzchild}} \vline   &   AdS-  \\
			&  \multicolumn{3}{c}{} \vline & Reissner- \\
			&  \multicolumn{3}{c}{} \vline  & Nordström \\
			\hline
				
		\end{tabular}
	\end{center}
	\caption{\footnotesize\it Topological characteristics of S-type Born-Infeld-AdS black hole.}
	\label{Table2}
\end{table}

\subsection{RN-type black hole}
Having discussed in precise details all the possible topological phase transitions in the so-called S-type black hole, we turn our attention in this section to the situation where the critical behavior takes place in RN-type black hole. 
 
\begin{figure}[!ht]
	\centering 
	\begin{subfigure}[h]{0.48\textwidth}
		\centering \includegraphics[scale=0.6]{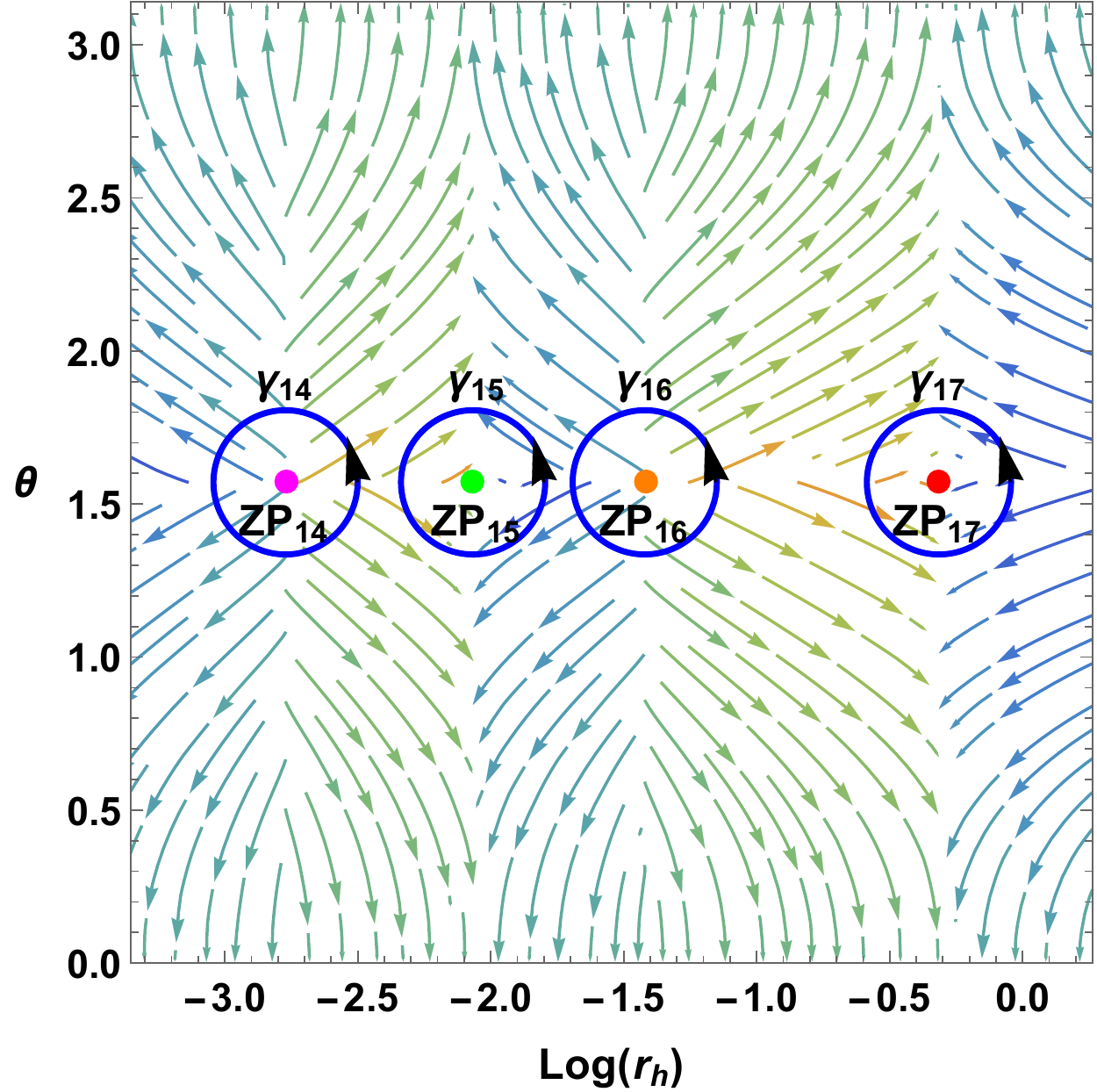}
		\caption{}
		\label{f18_1}
	\end{subfigure}
	\hspace{1pt}	
	\begin{subfigure}[h]{0.48\textwidth}
		\centering \includegraphics[scale=0.6]{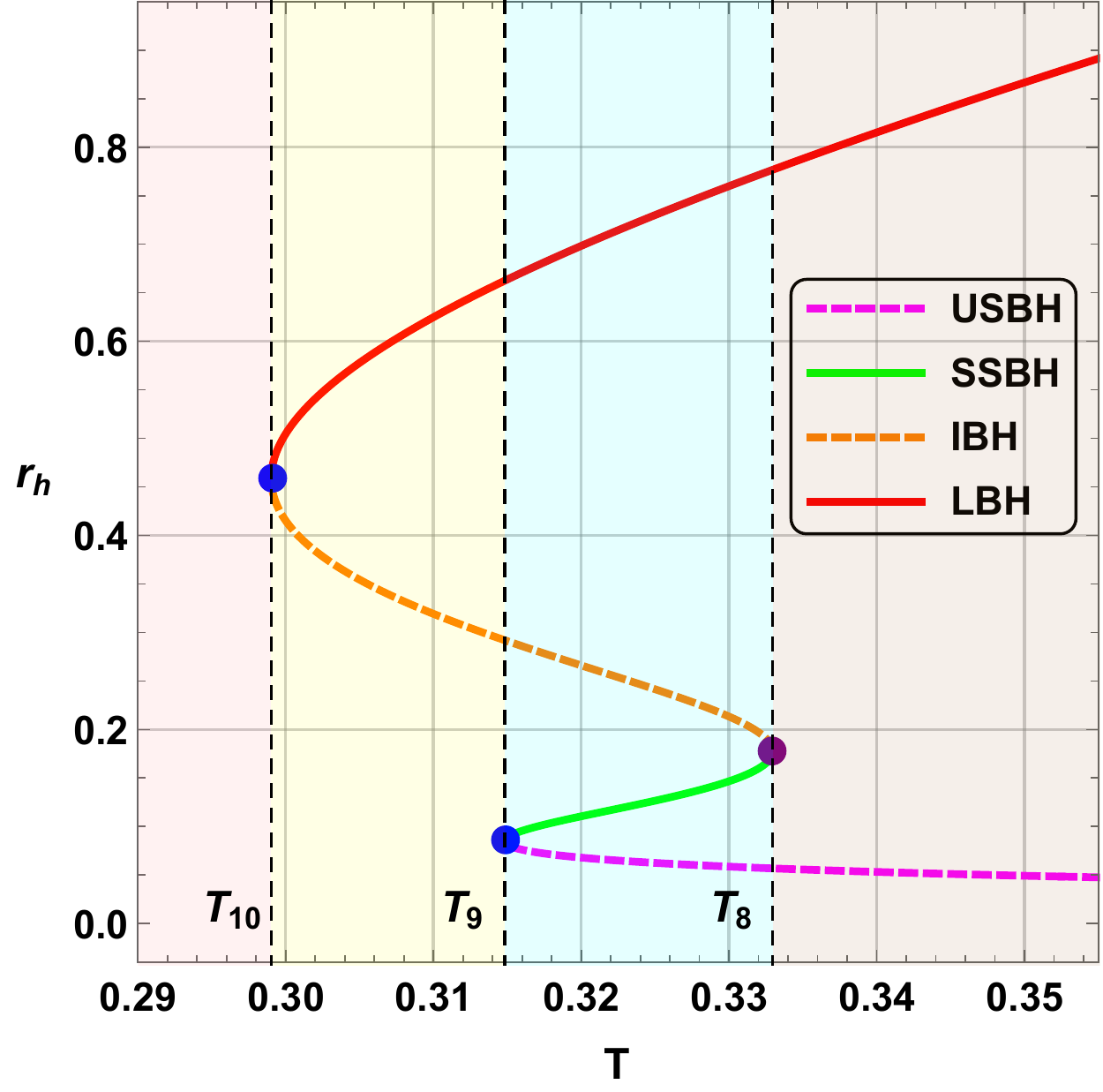}
		\caption{}
		\label{f18_2}	
	\end{subfigure}
	\caption{\footnotesize\it (a) Normalized vector field $n^i$ in
		the $(r_h,\theta)$ plane for $T = 0.325$ and (b) zero points of $\phi^{r_h}$ shown in $(r_h,T)$ plane with $P= 0.15$, $Q= 0.00966$ and $b = 3.5$. }
	\label{f18}
\end{figure}

The starting point is to consider the Born-Infeld parameter $b=3.5$ and we set $P = 0.15$ which respects the condition \eqref{30}. We plot in Fig.\ref{f18_1} and Fig.\ref{f18_2} the normalized vector field $n^i$, in the $(r_h,\theta)$ plane and the zero points of $\phi^{r_h}$ in $(r_h,T)$ diagram for $Q= 0.00966$  and $T = 0.325$, 
Herein, we observe the four zero points and we denote their contours according to Eq.\eqref{15} as $\gamma_{14}$, $\gamma_{15}$, $\gamma_{16}$ and $\gamma_{17}$, respectively. The topological charges of these four zero points are : 
\begin{equation}\label{43}
	\mathcal{Q}(ZP_{14}) = +1,  \quad \mathcal{Q}(ZP_{15}) = -1, \quad  \mathcal{Q}(ZP_{16}) = +1, \quad  \mathcal{Q}(ZP_{17}) = -1.
\end{equation}
These four zero points correspond to the four solutions of the Born-Infeld-AdS black hole in this situation \cite{Ali:2023wkq,Wei:2022dzw}. Indeed, the first zero point $ZP_{14}$ (magenta dot) corresponds to an unstable small black hole (USBH), the second zero point $ZP_{15}$ (green dot) is associated with a stable small black hole (SSBH), the third zero point $ZP_{16}$ (orange dot) fits with an intermediate black hole, and the fourth zero point $ZP_{17}$ (red dot) corresponds to a large black hole (LBH). Thus, the stable phases ($ZP_{15}$ and $ZP_{17}$) have a negative topological charge ($-1$) and the unstable phases  ($ZP_{14}$ and $ZP_{16}$) have a positive topological charge ($+1$). Summarizing the total
 topological charge of the system leads to a vanishing result: 
\begin{equation}\label{44}
	\mathcal{Q} = \mathcal{Q}(ZP_{14}) + \mathcal{Q}(ZP_{15})+ \mathcal{Q}(ZP_{16})+\mathcal{Q}(ZP_{17}) = 0.
\end{equation}
Therefore, the system is topologically equivalent to AdS-Schwarzchild black hole, and is in good agreement with the finding of \cite{Yerra:2022coh}. From Fig.\ref{f18_2}, we observe that there are four black hole branches: USBH (magenta dashed line), SSBH (green solid line), IBH (orange dashed line), and LBH (red solid line). Moreover, we detect a vortex/anti-vortex creation point (purple dot) and  two vortex/anti-vortex annihilation points (blue points). For $T>T_8$ (brown region), there is only a pair of vortex/anti-vortex corresponding to LBH/USBH phases and the total topological in this region is zero. When $T = T_8$,  a vortex/anti-vortex creation point (purple dot) exists which is equivalent to the generation of the SSBH phase (vortex) and the IBH phase (anti-vortex). In the interval $T_9<T<T_8$ (cyan region), there are both pairs of vortex/anti-vortex corresponding to LBH/USBH and SSBH/IBH phases, then the total topological charge in this region is also null.  For $T=T_9$, we observe a vortex/anti-vortex annihilation point (blue dot), in fact, an annihilation of the SSBH phase (vortex) and the USBH phase (anti-vortex). In the domain $T_{10}<T<T_9$ (yellow region), there is only a pair of vortex/anti-vortex corresponding to LBH/IBH  phases and the total topological charge in this region is zero. Reaching the temperature $T=T_{10}$, the system exhibits a second vortex/anti-vortex annihilation point (blue dot), namely, we found an annihilation of the SSBH (vortex) and the IBH phase (anti-vortex).  When $T<T_{10}$ (pink region), there is no black hole solution (no vortexes), and then the total topological charge is zero. Therefore, one can deduce that
the total topological charge is always vanishing independently of temperature and the system belongs to the same topological class of AdS-Schwarzchild black hole.

\begin{figure}[!ht]
	\centering 
	\begin{subfigure}[h]{0.48\textwidth}
		\centering \includegraphics[scale=0.6]{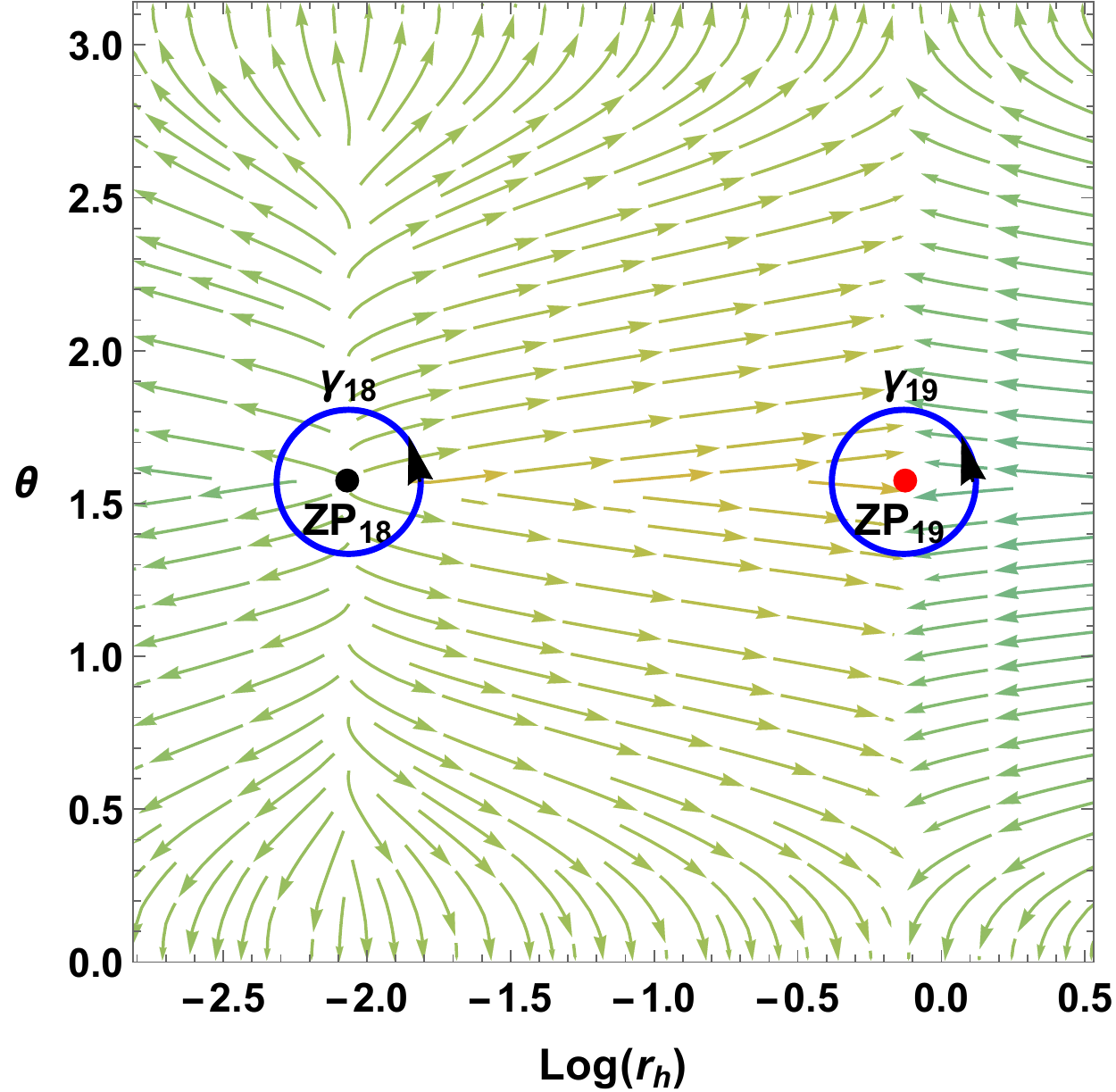}
		\caption{}
		\label{f19_1}
	\end{subfigure}
	\hspace{1pt}	
	\begin{subfigure}[h]{0.48\textwidth}
		\centering \includegraphics[scale=0.6]{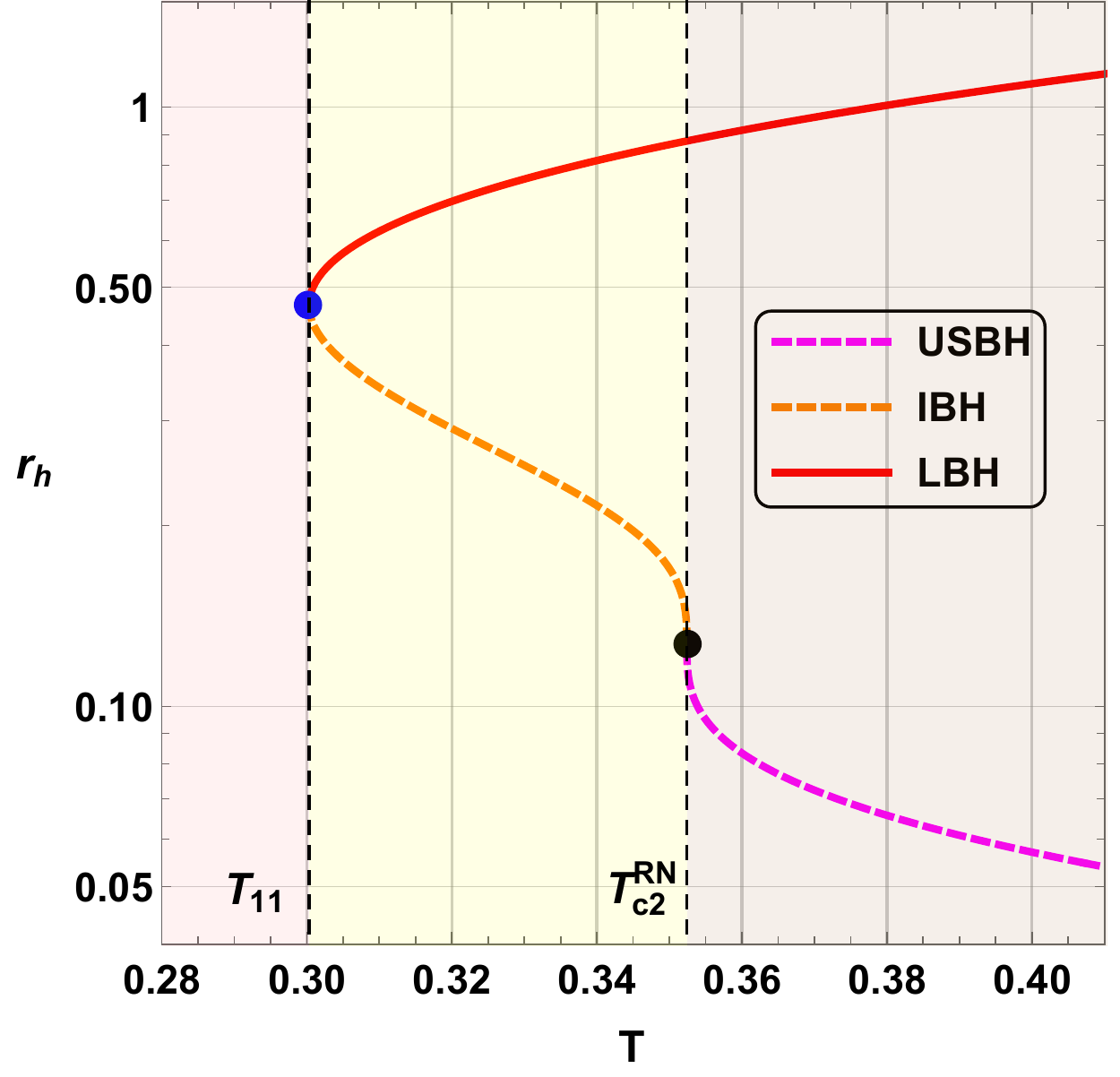}
		\caption{}
		\label{f19_2}	
	\end{subfigure}
	\caption{\footnotesize\it (a) Normalized vector field $n^i$ in
		the $(r_h,\theta)$ plane for $T = T_{c2}^{RN}=  0.352409$ and (b) zero points of $\phi^{r_h}$ shown in $(r_h,T)$ plane with $P= 0.15$, $Q=Q_{c2}^{RN}= 0.0091107$ and $b = 3.5$. }
	\label{f19}
\end{figure}

The next relevant consideration is the unconventional critical point situation where the system shows an unconventional transition between two unstable phases of USBH and IBH.  In Fig.\ref{f19_1} and Fig.\ref{f19_2}, we plot the normalized vector field $n^i$ and the zero points of $\phi^{r_h}$ in the $(r_h,\theta)$ and  $(r_h,T)$ planes respectively for $Q=Q_{c2}^{RN}= 0.0091107$  and $T = T_{c2}^{RN}=  0.352409$, 

Two zero points associated with $\gamma_{18}$ and $\gamma_{19}$ contours are noticed in this case and their topological charges are : 
\begin{equation}\label{45}
	\mathcal{Q}(ZP_{18}) = +1,  \quad \mathcal{Q}(ZP_{19}) = -1.
\end{equation}
The second zero point $ZP_{19}$ (red dot) corresponds to LBH whereas the first zero point $ZP_{18}$ (black dot) corresponds to the unconventional critical point. Indeed, $ZP_{18}$ is a degenerate zero point associated with a superposition of two unstable phases (two anti-vortexes: USBH and IBH phases), and a stable phase (a vortex: SBH phase) and which explains its positive topological charge. Thus, the total topological charge of the system is still null: 
\begin{equation}\label{46}
	\mathcal{Q} =  \mathcal{Q}(ZP_{18}) + \mathcal{Q}(ZP_{19}) = 0.
\end{equation}
Moreover, we observe that there are three black hole branches: USBH (magenta dashed line), IBH (orange dashed line), and LBH (red solid line). The SSBH phase has disappeared from the phase structure. In addition, we notice a critical point (black dot) and just one  vortex/anti-vortex annihilation point (blue dot) and there is no vortex/anti-vortex creation point. Thus the critical point corresponds to a superposition of vortex/anti-vortex creation and annihilation points. For $T>T_{c2}^{RN}$ (brown region), there is just a pair of vortex/anti-vortex corresponding to LBH/USBH phases and the total topological charge in this region is also zero. While, when $T=T_{c2}^{RN}$, a critical point corresponding to a superposition of a vortex/anti-vortex creation point and a vortex/anti-vortex annihilation point rises in the phase portrait. Indeed, the creation point corresponds to SSBH/IBH phase generation, and  annihilation one relates to SSBH/USBH phase annihilation. 
In thermal interval $T_{11}<T<T_{c2}^{RN}$ (yellow region), there is also a pair of vortex/anti-vortex corresponding to LBH/IBH  phases, and hence the total topological charge in this region is zero. For $T=T_{11}$, we have a  vortex/anti-vortex annihilation point (blue dot), that is to say, we have an annihilation of the LBH (vortex) and the IBH phase (anti-vortex).  For $T<T_{11}$ (pink region), there is no black hole solution (no vortexes) and therefore, the total topological charge is also zero.

\begin{figure}[!ht]
	\centering 
	\begin{subfigure}[h]{0.48\textwidth}
		\centering \includegraphics[scale=0.6]{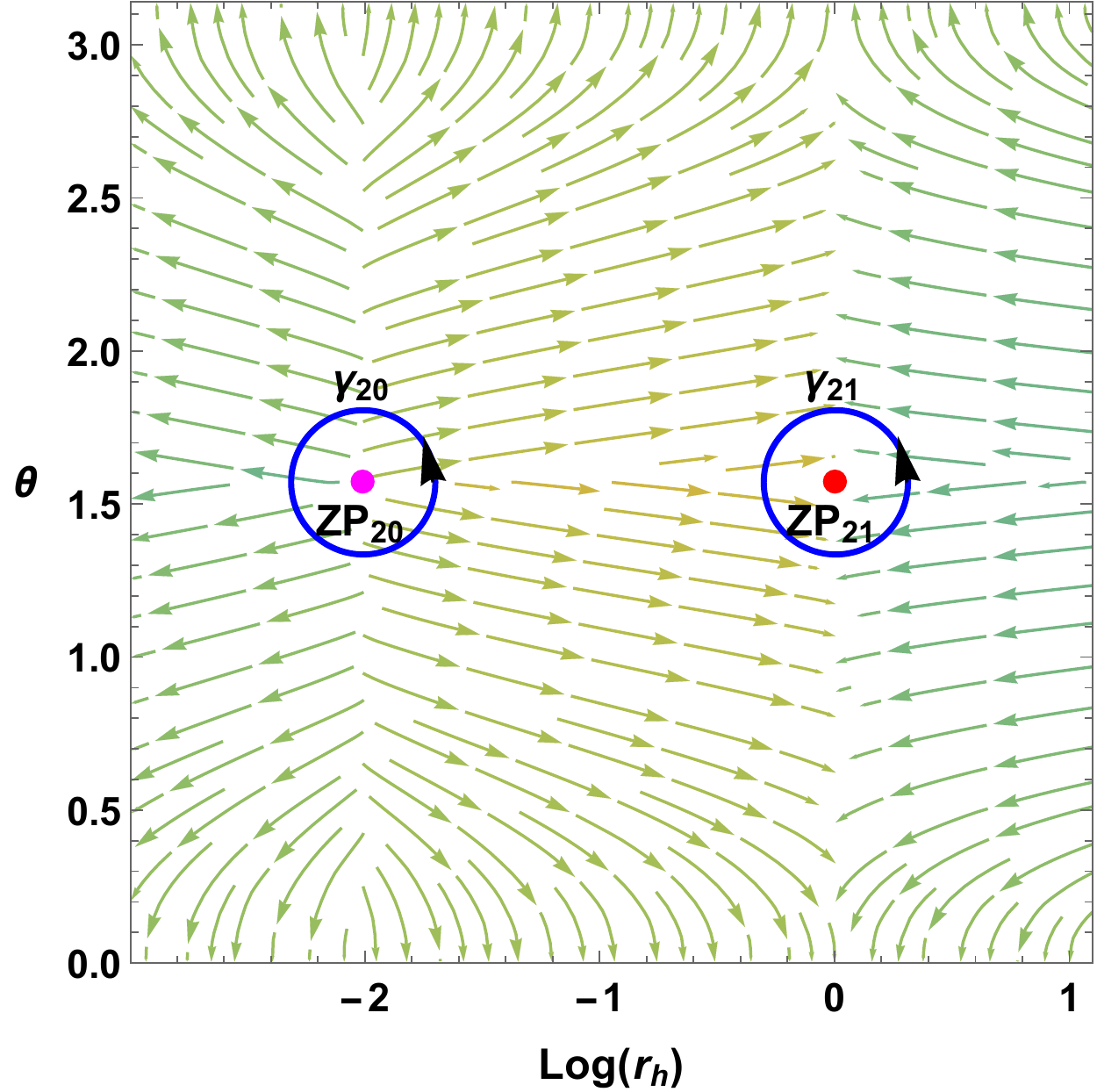}
		\caption{}
		\label{f20_1}
	\end{subfigure}
	\hspace{1pt}	
	\begin{subfigure}[h]{0.48\textwidth}
		\centering \includegraphics[scale=0.6]{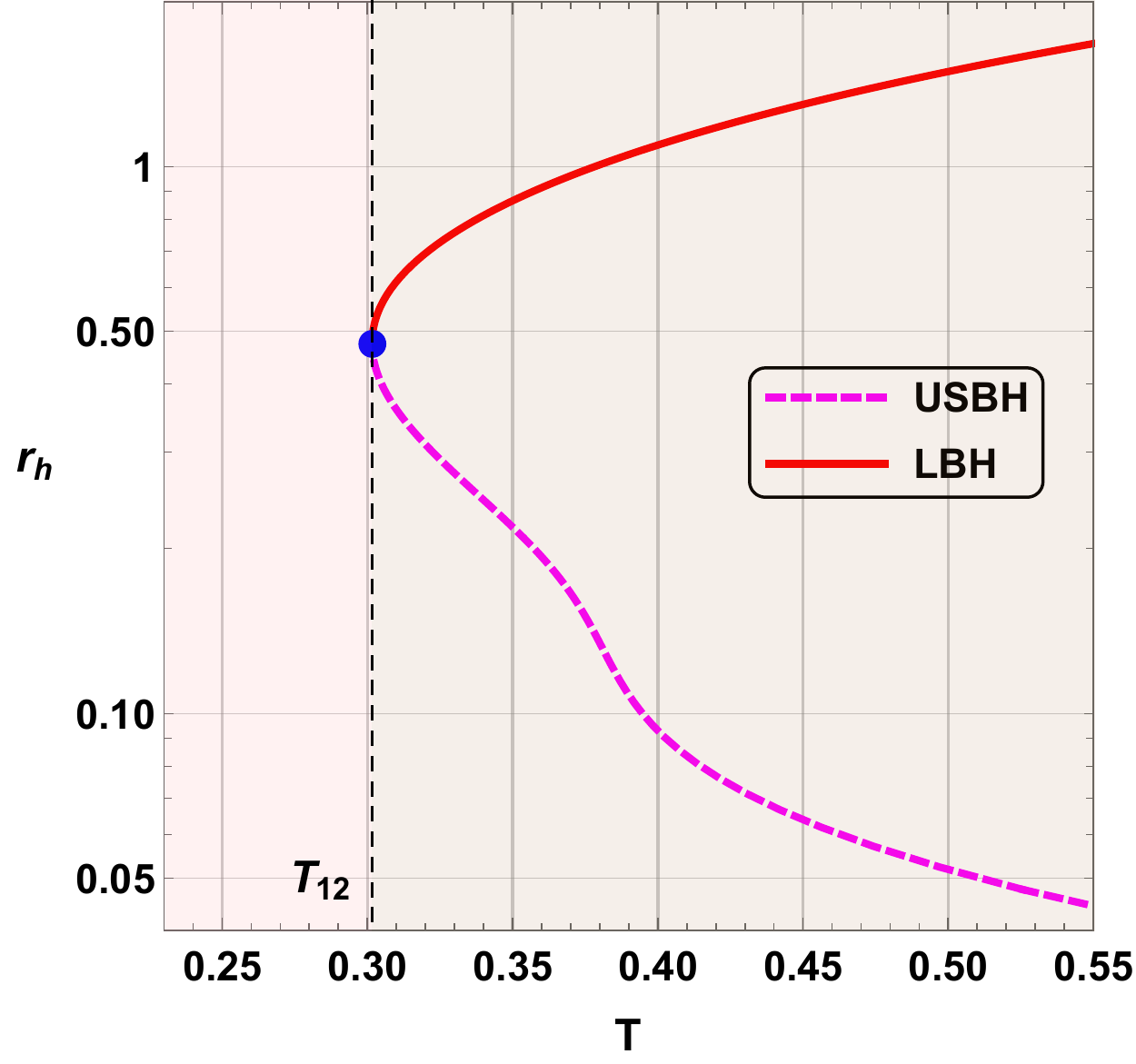}
		\caption{}
		\label{f20_2}	
	\end{subfigure}
	\caption{\footnotesize\it (a) Normalized vector field $n^i$ in
		the $(r_h,\theta)$ plane for $T =  0.38$ and (b) zero points of $\phi^{r_h}$ shown in $(r_h,T)$ plane with $P= 0.15$, $Q=0.0085$ and $b = 3.5$. }
	\label{f20}
\end{figure}

Hereafter, we consider the situation when the electric charge is below the unconventional critical charge $Q_{c2}^{RN}$. We depict in Fig.\ref{f20_1} the normalized vector field $n^i$, in the $(r_h,\theta)$ plane and in Fig.\ref{f20_2} the zero points of $\phi^{r_h}$ in $(r_h,T)$ plane for $Q= 0.0085$  and $T =  0.38$, 
The situation where two zero points persist is observed, and in which  the two contours $\gamma_{20}$ and $\gamma_{21}$ give the topological charges 
\begin{equation}\label{47}
	\mathcal{Q}(ZP_{20}) = +1,  \quad \mathcal{Q}(ZP_{21}) = -1.
\end{equation}
The first zero point $ZP_{20}$ (magenta dot) corresponds to USBH and the second zero point $ZP_{21}$ (red dot) to LBH. We also observe that there are two black hole branches: USBH (magenta dashed line) and LBH (red solid line). The IBH phase has disappeared. Moreover, we observe just one vortex/anti-vortex annihilation point (blue dot) and there is no vortex/anti-vortex creation point, like in the AdS-Schwarzchild case. For $T>T_{12}$ (brown region), there is a pair of vortex/anti-vortex corresponding to LBH/USBH phases and then the total topological charge is zero. For $T=T_{12}$, we have a vortex/anti-vortex annihilation point (blue dot) i.e. annihilation of LBH and USBH phases. In the case of $T<T_{12}$ (pink region), no black hole solution (no vortexes) persists and the total topological charge in this region is also null.

\begin{figure}[!ht]
	\centering 
	\begin{subfigure}[h]{0.48\textwidth}
		\centering \includegraphics[scale=0.6]{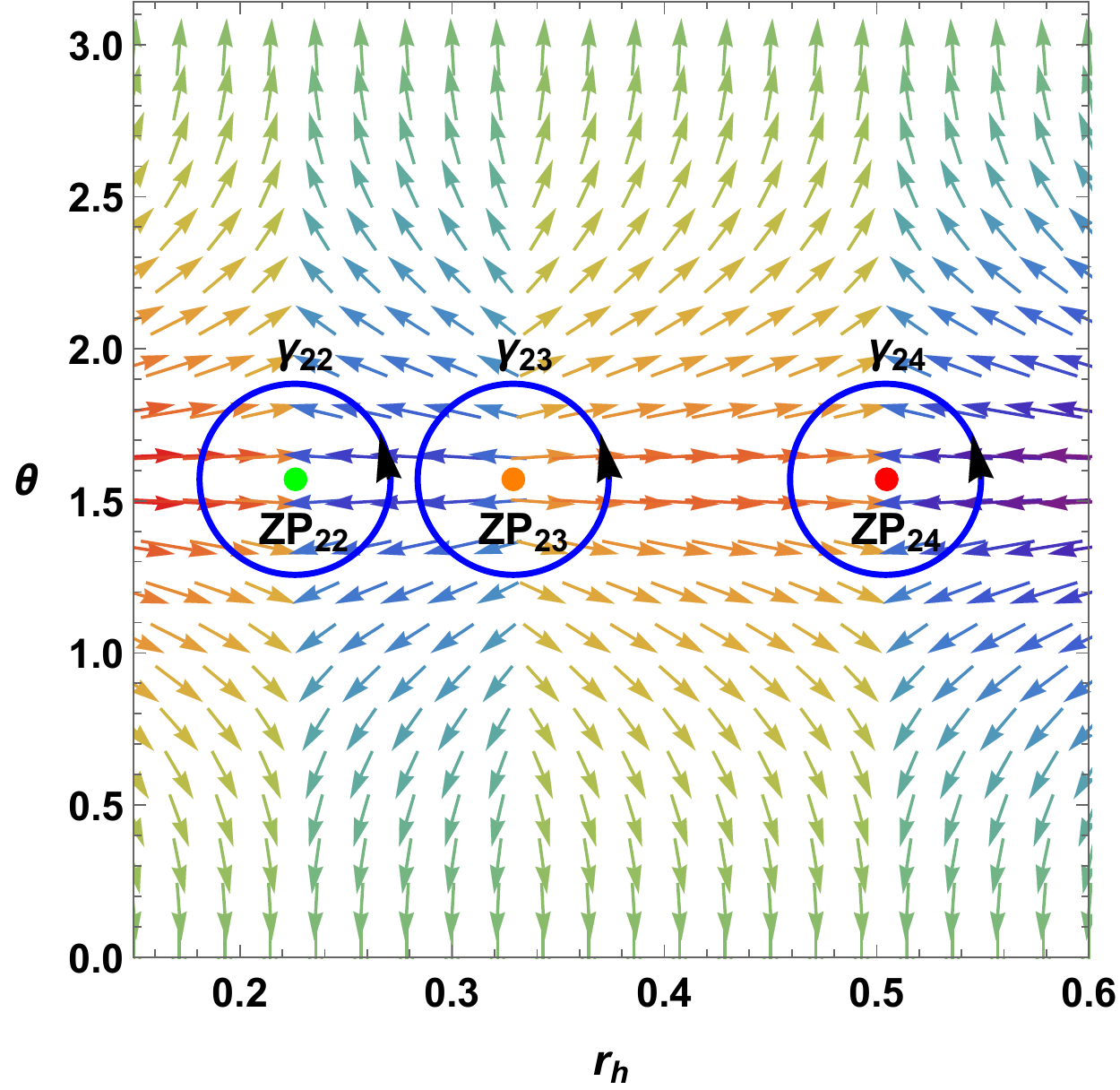}
		\caption{}
		\label{f21_1}
	\end{subfigure}
	\hspace{1pt}	
	\begin{subfigure}[h]{0.48\textwidth}
		\centering \includegraphics[scale=0.6]{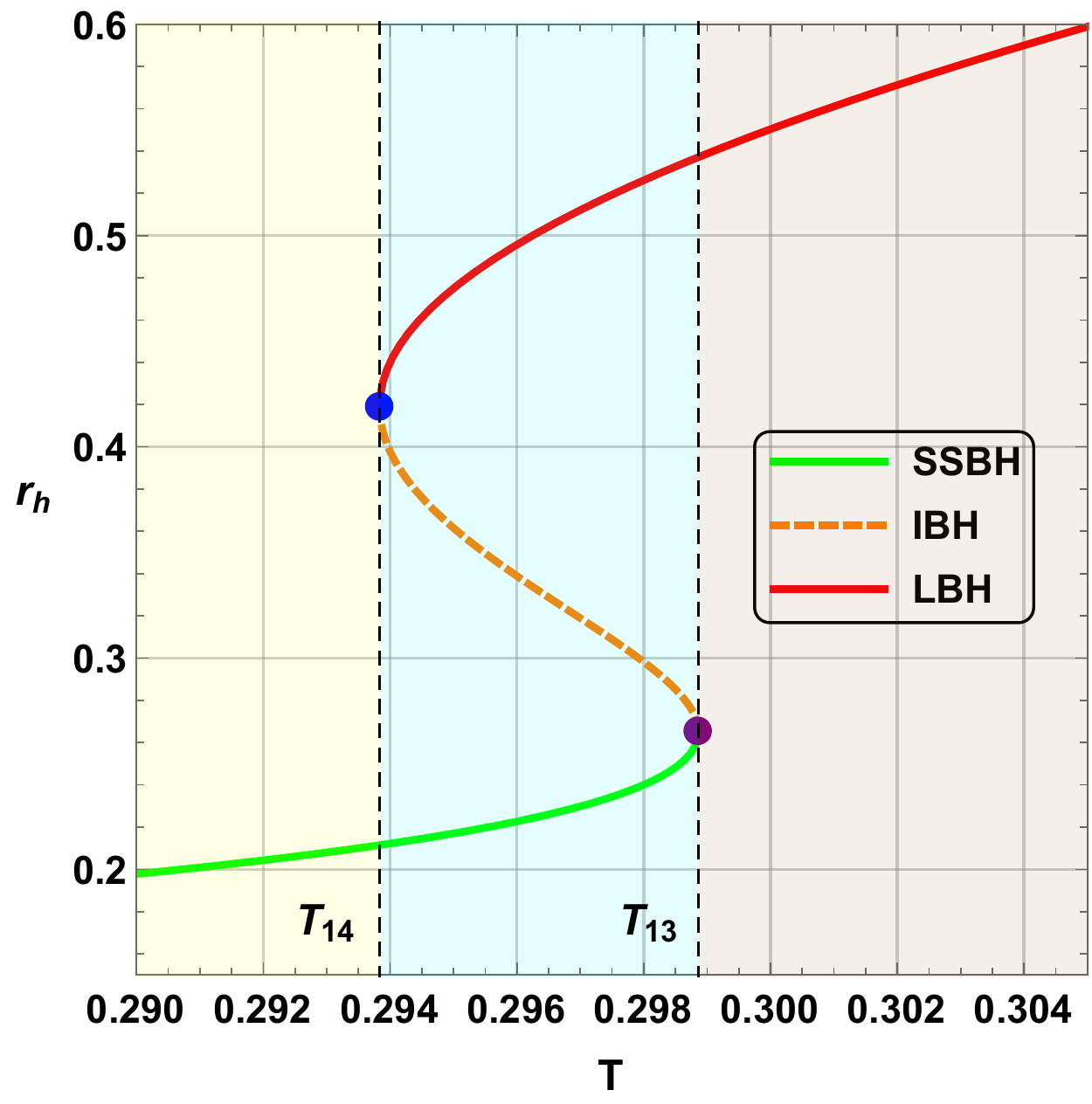}
		\caption{}
		\label{f21_2}	
	\end{subfigure}
	\caption{\footnotesize\it (a) Normalized vector field $n^i$ in
		the $(r_h,\theta)$ plane for $T =  0.2965$ and (b) zero points of $\phi^{r_h}$ shown in $(r_h,T)$ plane with $P= 0.15$, $Q= Q_m = 0.0113682$ and $b = 3.5$. }
	\label{f21}
\end{figure}

Now we consider in Fig.\ref{f21_1} and Fig.\ref{f21_2}, the situation when the electric charge is equal to the topological critical charge $Q_m$ which is below the conventional critical point $Q_{c1}^{RN}$, i.e. the  $Q= Q_m = 0.0113682$   and $T =  0.2965$
The figure unveils the presence of three zero points associated with  the  contour $\gamma_{22}$, $\gamma_{23}$ and $\gamma_{24}$. The topological charges of these zero points are as follows: 
\begin{equation}\label{49}
	\mathcal{Q}(ZP_{22}) = -1, \quad \mathcal{Q}(ZP_{23}) = +1, \quad \mathcal{Q}(ZP_{24}) = -1.
\end{equation}
The first zero point $ZP_{22}$ (green dot) corresponds to SSBH, the second zero point $ZP_{23}$ (orange dot)  to IBH, and the third zero point $ZP{24}$ (red dot) to LBH, and consequently, the total topological charge of the system in such a case is 
\begin{equation}\label{50}
	\mathcal{Q} =  \mathcal{Q}(ZP_{22}) + \mathcal{Q}(ZP_{23}) + \mathcal{Q}(ZP_{24}) =  -1.
\end{equation}
Otherwise, we observe a  vortex/anti-vortex creation point (purple dot) and a vortex/anti-vortex annihilation point (blue dot). For $T> T_{13}$ (brown region), there is only one free vortex that corresponds to the LBH phase and then the total topological charge is equal to $-1$. For $T = T_{13} $, we notice a vortex/anti-vortex creation point (purple dot) interpreted as the generation of the SSBH (vortex) and the IBH (anti-vortex). Considering $T_{14}<T< T_{13}$ (cyan region), we found two vortexes associated with SSBH and LBH phases and one anti-vortex corresponding to the IBH phase, and therefore, the total topological charge is also equal to $-1$. For $T = T_{14} $, we have a vortex/anti-vortex annihilation point (blue dot), that is to say, we have an annihilation of the LBH (vortex) and the IBH (anti-vortex). Within $T<T_{14}$ (yellow region), there is only one free vortex that corresponds to the SSBH phase and then the total topological charge is equal to $-1$. Therefore, there is a topological transition and the total topology of the black hole changes. Thus, the system belongs now to the topological class of The  Reissner-Nordström black hole \cite{Wei:2022dzw}.

\begin{figure}[!ht]
	\centering 
	\begin{subfigure}[h]{0.48\textwidth}
		\centering \includegraphics[scale=0.6]{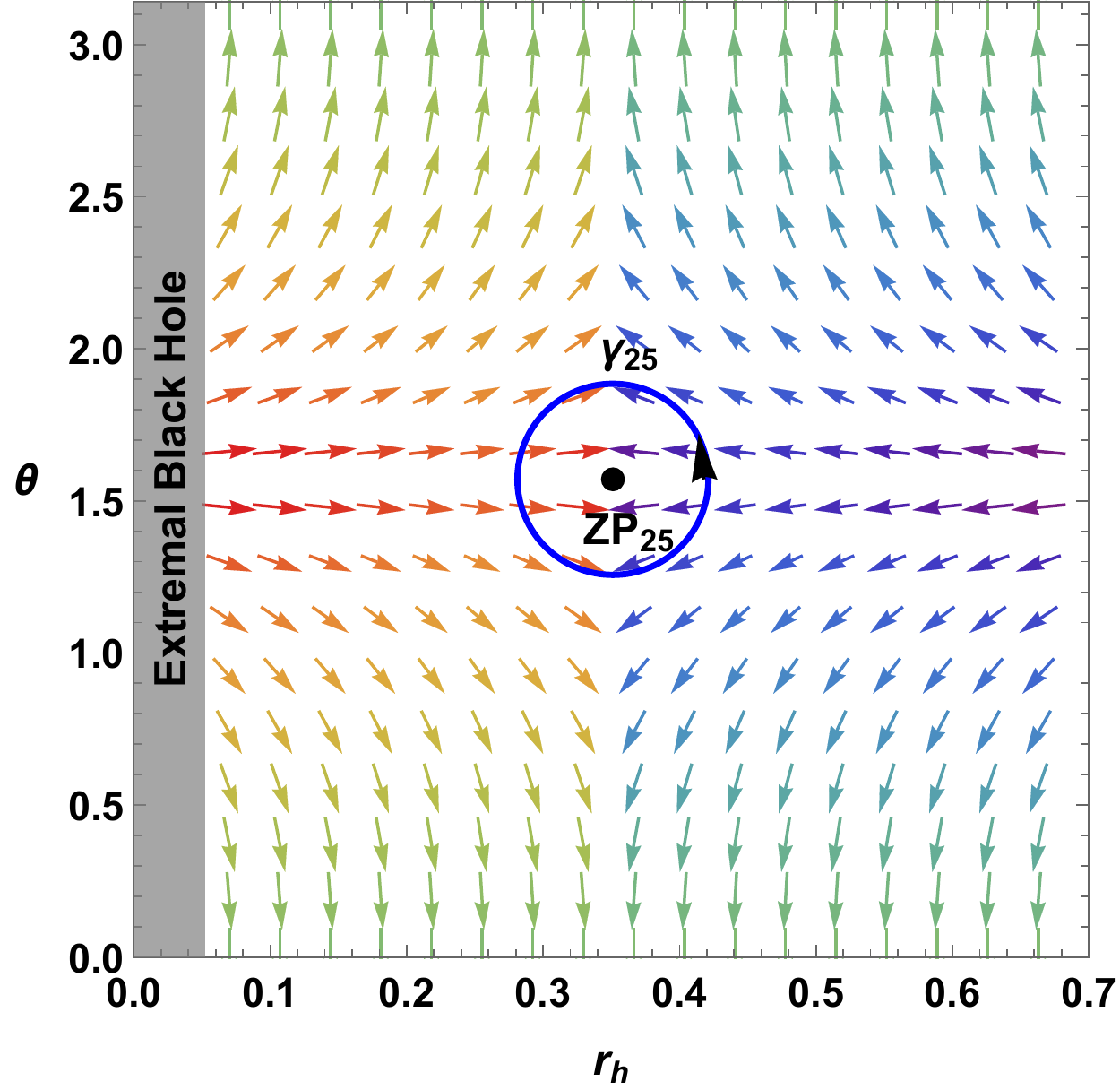}
		\caption{}
		\label{f22_1}
	\end{subfigure}
	\hspace{1pt}	
	\begin{subfigure}[h]{0.48\textwidth}
		\centering \includegraphics[scale=0.6]{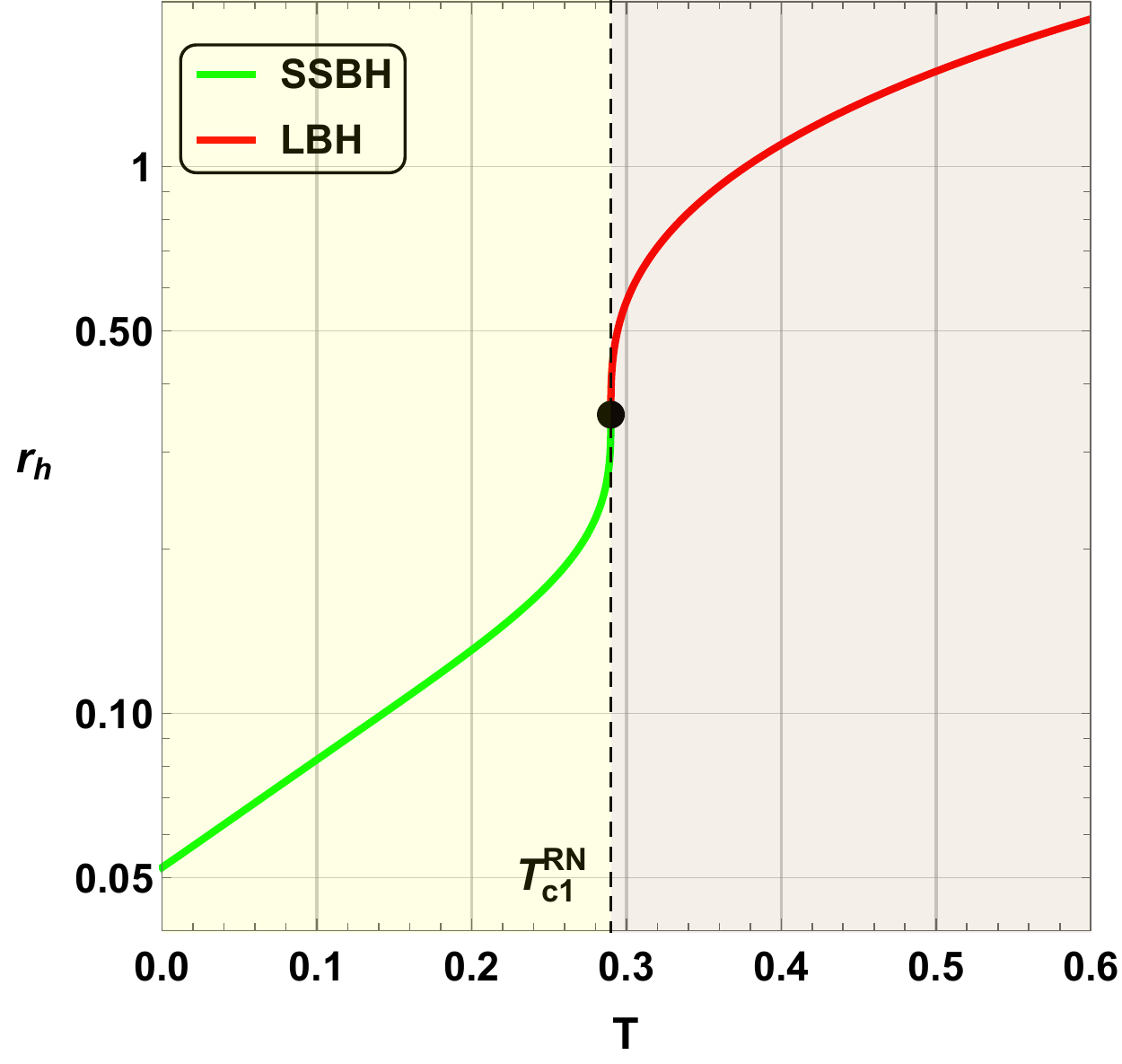}
		\caption{}
		\label{f22_2}	
	\end{subfigure}
	\caption{\footnotesize\it (a) Normalized vector field $n^i$ in
		the $(r_h,\theta)$ plane for $T = T_{c1}^{RN}=  0.289918$ and (b) zero points of $\phi^{r_h}$ shown in $(r_h,T)$ plane with $P= 0.15$, $Q=Q_{c1}^{RN}=0.0122199$ and $b = 3.5$. }
	\label{f22}
\end{figure}

Further, let us now focus on the conventional critical point situation where we have a second-order phase transition between two stable phases SSBH and LBH. Such a situation is illustrated in Fig.\ref{f22_1} and Fig.\ref{f22_2}, for the specific values of charge $Q=Q_{c1}^{RN}= 0.0122199$  and temperature $T = T_{c1}^{RN}=  0.289918$.
 A solitary zero point $ZP_{25}$ (black dot) is observed in the normalized vector field behavior and the topological charge of this zero point is evaluated to be : 
\begin{equation}\label{51}
	\mathcal{Q}(ZP_{25}) = -1.
\end{equation}
 Indeed, $ZP_{25}$ is a degenerate zero point that corresponds to a superposition of two stable phases (two vortexes: SBH and LBH phases), and an unstable phase (an anti-vortex: IBH phase) which explain the negative topological charge. Moreover, we observe a forbidden region (gray region) corresponding to an extremal black hole solution with an extremal horizon $r_e =  0.052$. The total topological charge of the system in this case is also equal to $-1$ : 
\begin{equation}\label{52}
	\mathcal{Q} =   \mathcal{Q}(ZP_{25}) = -1.
\end{equation}
From the right panel, we remark that there are no vortex/anti-vortex creation or annihilation points. Thus the critical point is nothing more than a superposition of vortex/anti-vortex creation and annihilation points. For $T>T_{c1}^{RN}$ (brown region), there is only one free vortex that corresponds to the LBH phase. For $T=T_{c1}^{RN}$, we have a vortex/anti-vortex creation point that corresponds to SSBH/IBH phases generation and a vortex/anti-vortex annihilation point that corresponds to LBH/IBH phases annihilation. That is to say, the LBH phase (vortex) is annihilated with the generated IBH phase (anti-vortex) and only the generated SSBH phase (vortex) will remain. Under the critical temperature $T<T_{c1}^{RN}$ (yellow region), there is only one free vortex that corresponds to the SSBH phase and then the total topological charge is equal to $-1$.  Therefore, The total topological charge is always equal to $-1$ independent of the temperature.

\begin{figure}[!ht]
	\centering 
	\begin{subfigure}[h]{0.48\textwidth}
		\centering \includegraphics[scale=0.6]{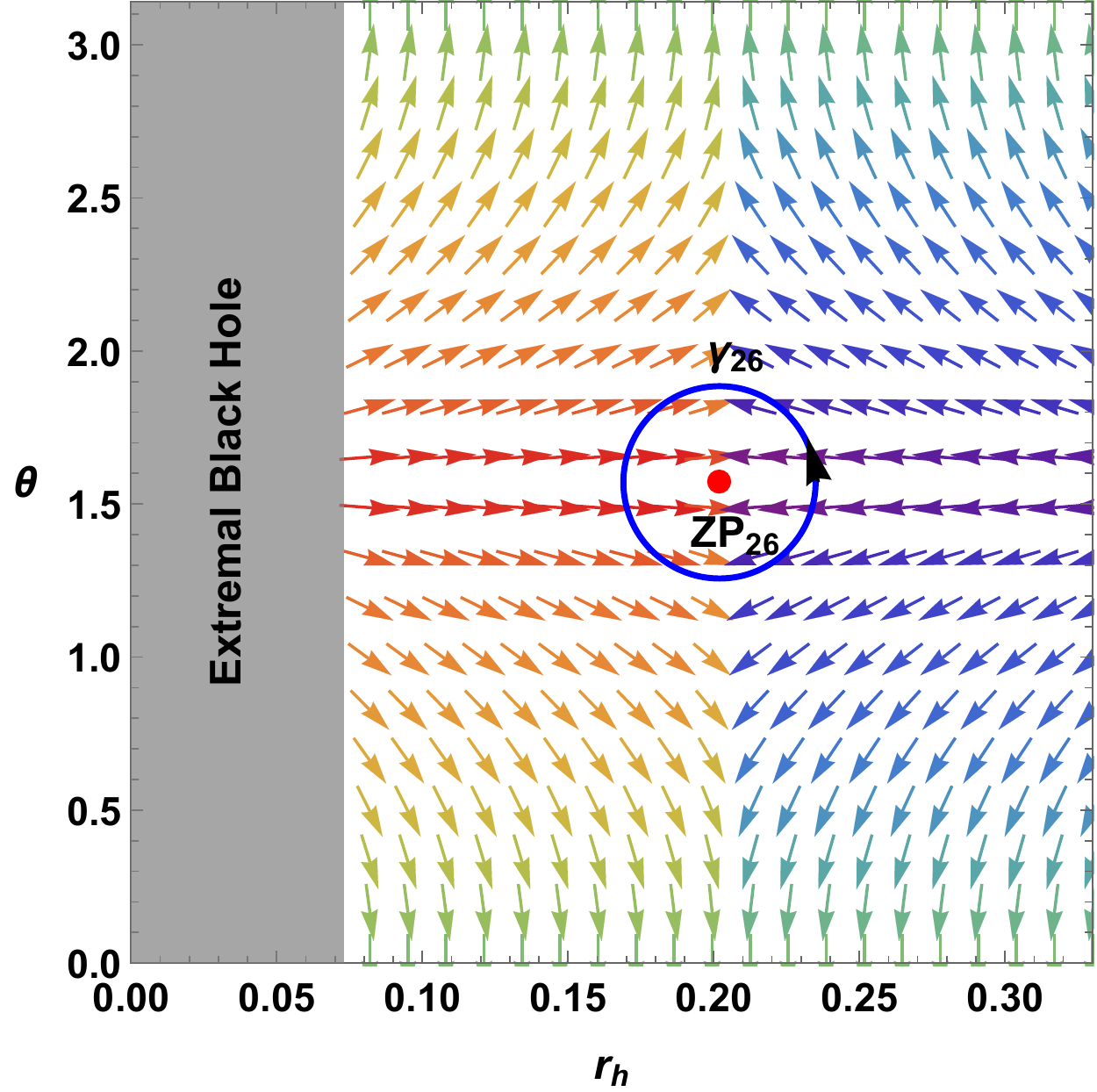}
		\caption{}
		\label{f23_1}
	\end{subfigure}
	\hspace{1pt}	
	\begin{subfigure}[h]{0.48\textwidth}
		\centering \includegraphics[scale=0.6]{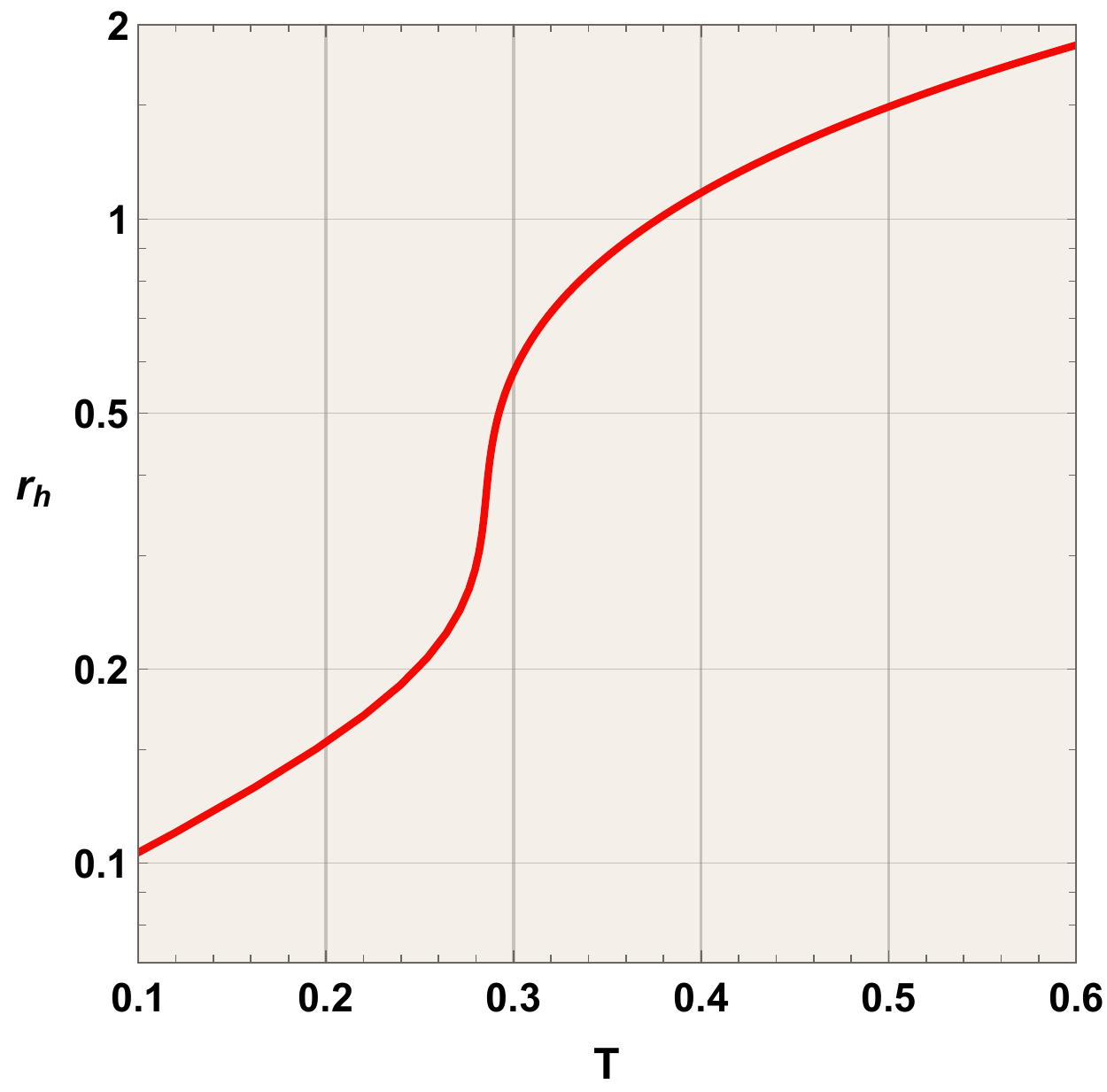}
		\caption{}
		\label{f23_2}	
	\end{subfigure}
	\caption{\footnotesize\it (a) Normalized vector field $n^i$ in
		the $(r_h,\theta)$ plane for $T =  0.25$ and (b) zero points of $\phi^{r_h}$ shown in $(r_h,T)$ plane with $P= 0.15$, $Q= 0.013 >Q_{c1}^{RN}$ and $b = 3.5$. }
	\label{f23}
\end{figure}

We consider now the situation when the electric charge gets increased further. We illustrate again in Fig.\ref{f23_1} and Fig.\ref{f23_2}  the normalized vector field $n^i$, in the $(r_h,\theta)$ plane  and the zero points of $\phi^{r_h}$ in $(r_h,T)$ plane for
for $Q= 0.013$ ($Q>Q_{c1}^{RN}$)  and $T =  0.25$, 
Just one zero point $ZP_{26}$ (red dot) associated with  LBH is shown. Using \eqref{15} and the contour $\gamma_{26}$  the topological charge of this zero point is obtained to be : 
\begin{equation}\label{53}
	\mathcal{Q}(ZP_{26}) = -1.
\end{equation}
 Additionally, we notice a forbidden region (gray region) corresponding to an extremal black hole with an extremal horizon at $r_e = 0.073$. The total topological charge of the system in this case is equal to $-1$ : 
\begin{equation}\label{54}
	\mathcal{Q} =  \mathcal{Q}(ZP_{26}) = -1.
\end{equation}
 Fig.\ref{f23_2} reveals that there is only one black hole branch that corresponds LBH (red solid line). The SSBH phase has disappeared because the creation point that generates the SSBH phase has disappeared. Moreover, we do not observe any vortex/anti-vortex creation or annihilation point.  Therefore, The total topological charge remains $-1$.\\
 
Finally, we list in Table.\ref{Table3} all the topological characteristics of RN-type Born-Infeld-AdS black hole as the function of electrical charge showing a topological transition between AdS-Schwarzchild and  AdS-Reissner-Nordström topological classes.
\begin{table}
	\begin{center}
		\begin{tabular}{|c||c|c|c|c|}
			\hline
		\multirow{2}{*}{ Electric charge $Q$}   &   \multicolumn{2}{c}{ $Q<Q_m$} \vline  &      \multicolumn{2}{c}{ $Q\geq Q_m$}  \vline \\
			\cline{2-5}
		 & $Q<Q_{c2}^{RN}$ & $Q>Q_{c2}^{RN}$ & $Q<Q_{c1}^{RN}$ & $Q>Q_{c1}^{RN}$ \\	
		 \hline
		 Vortex/anti-vortex creation points &\multirow{2}{*}{ $0$} & \multirow{2}{*}{ $1$} & \multirow{2}{*}{$1$} & \multirow{2}{*}{$0$ }\\
		 (Stable/unstable phases generation) & &  &  & \\
		 \hline
		 Vortex/anti-vortex annihilation points &\multirow{2}{*}{ $1$} & \multirow{2}{*}{ $2$} & \multirow{2}{*}{$1$} & \multirow{2}{*}{$0$ }\\
		 (Stable/unstable phases annihilation) & &  &  & \\
		 \hline
		 Vortexes (stable phases) & $1$ & $2$ & $2$ & $1$\\
		 \hline
		 Anti-vortexes (unstable phases) & $1$ & $2$ & $1$ & $0$ \\
		 \hline
		 	Topological charge $\mathcal{Q}$   &   \multicolumn{2}{c}{ $0$} \vline  &      \multicolumn{2}{c}{ $-1$}  \vline \\
		 \hline
		 Topological class  &   \multicolumn{2}{c}{ AdS-Schwarzchild} \vline  &      \multicolumn{2}{c}{ AdS-Reissner-Nordström }  \vline \\
		 \hline
		\end{tabular}
	\end{center}
	\caption{\footnotesize\it Topological characteristics of RN-type Born-Infeld-AdS black hole.}
	\label{Table3}
\end{table}


\section{ Vortex/anti-vortex annihilation thermodynamics}
\label{vortex-antivortex}

In this section, we extend our discussion to probe the vortex/anti-vortex annihilation and creation thermodynamics. 

\begin{figure}[!ht]
	\centering 
	\begin{subfigure}[h]{0.48\textwidth}
		\centering \includegraphics[scale=0.6]{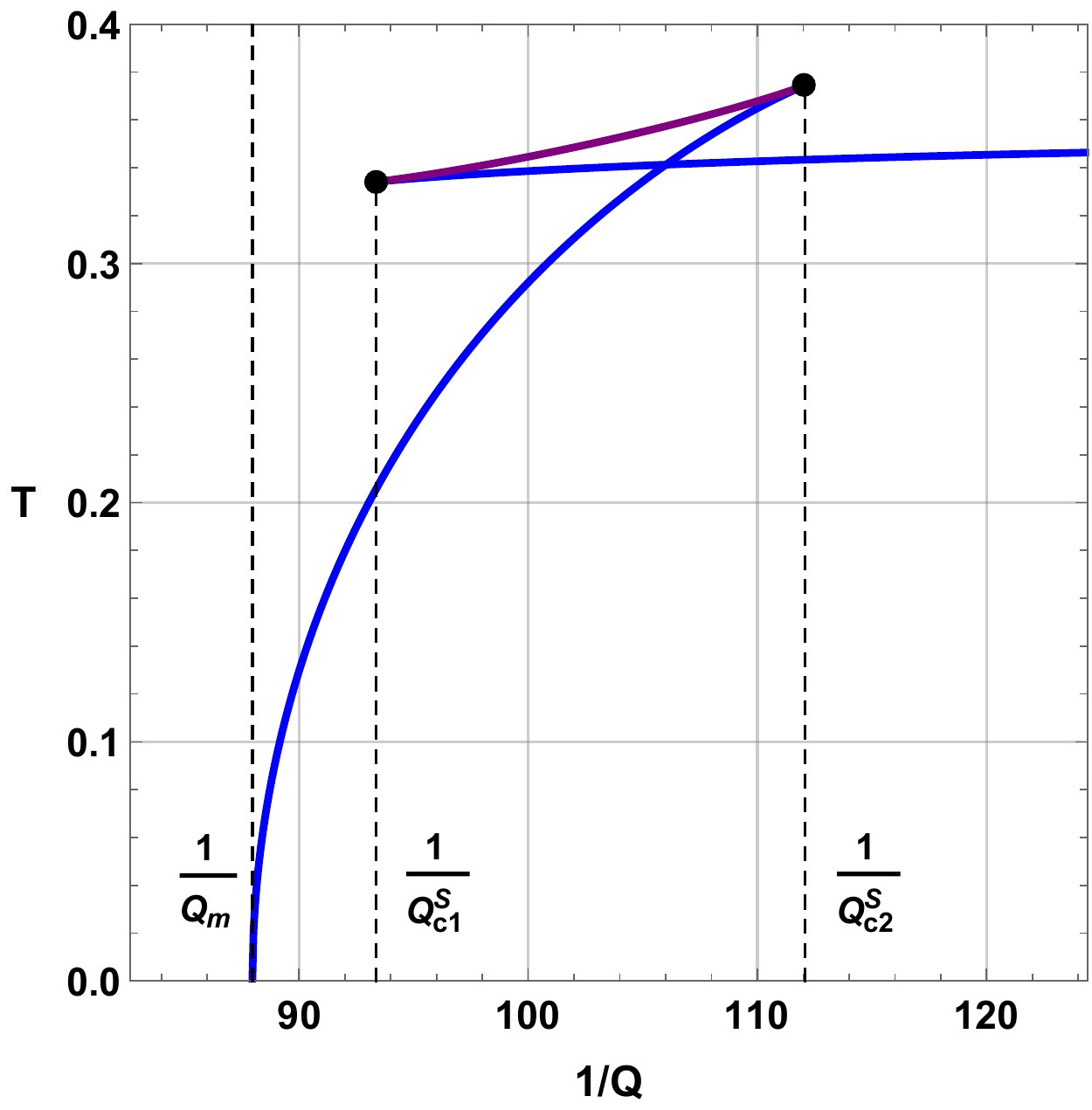}
		\caption{}
		\label{f24_1}
	\end{subfigure}
	\hspace{1pt}	
	\begin{subfigure}[h]{0.48\textwidth}
		\centering \includegraphics[scale=0.6]{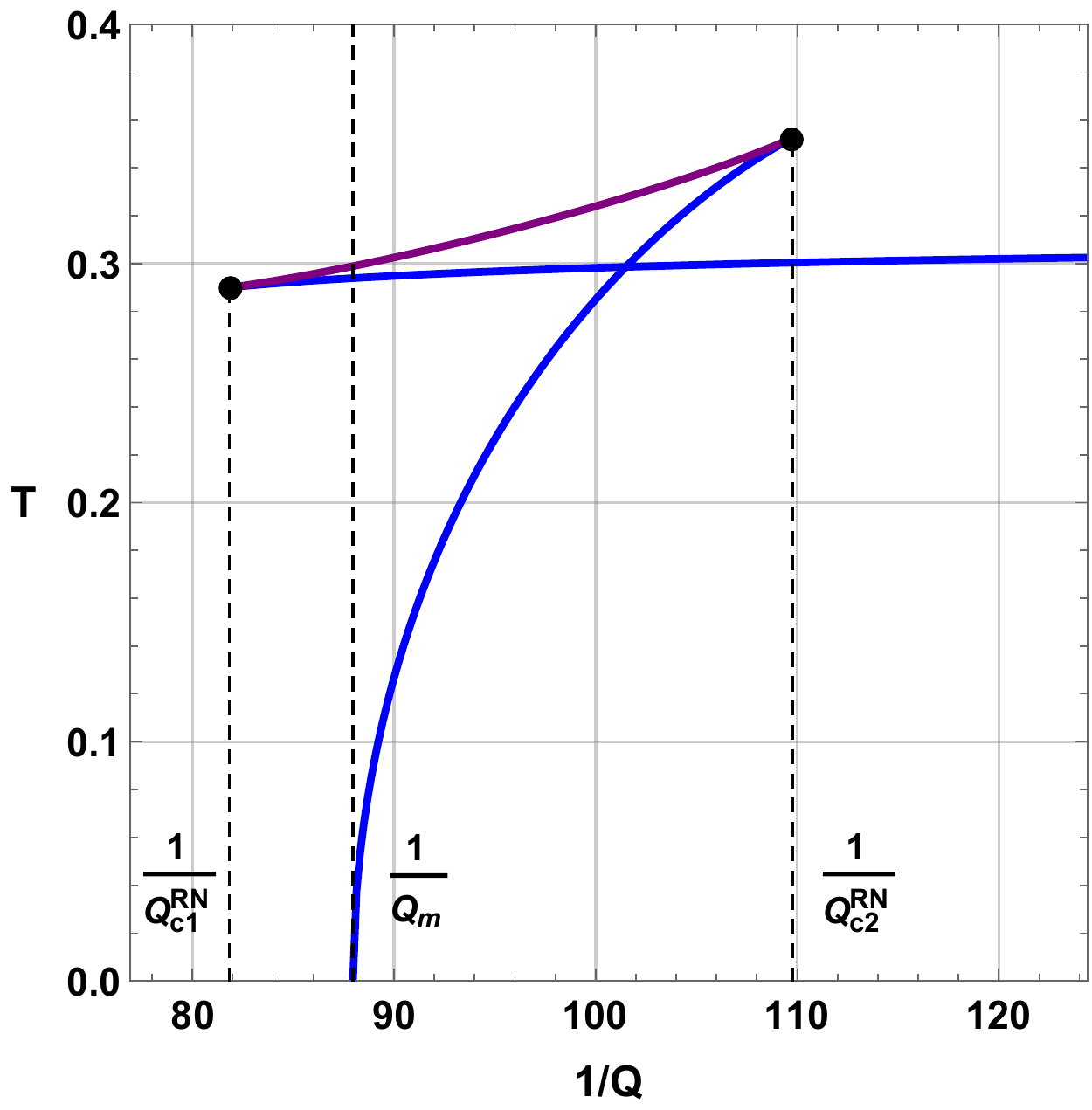}
		\caption{}
		\label{f24_2}	
	\end{subfigure}
	\caption{\footnotesize\it Vortex/anti-vortex creation and annihilation points in $(\frac{1}{Q},T)$ plane for (a) S-type and (b) RN-type black holes with $P = 0.15$ and $P = 0.2$ respectively and $b= 3.5$.}
	\label{f24}
\end{figure}

\begin{figure}[!ht]
	\centering 
	\begin{subfigure}[h]{0.48\textwidth}
		\centering \includegraphics[scale=0.6]{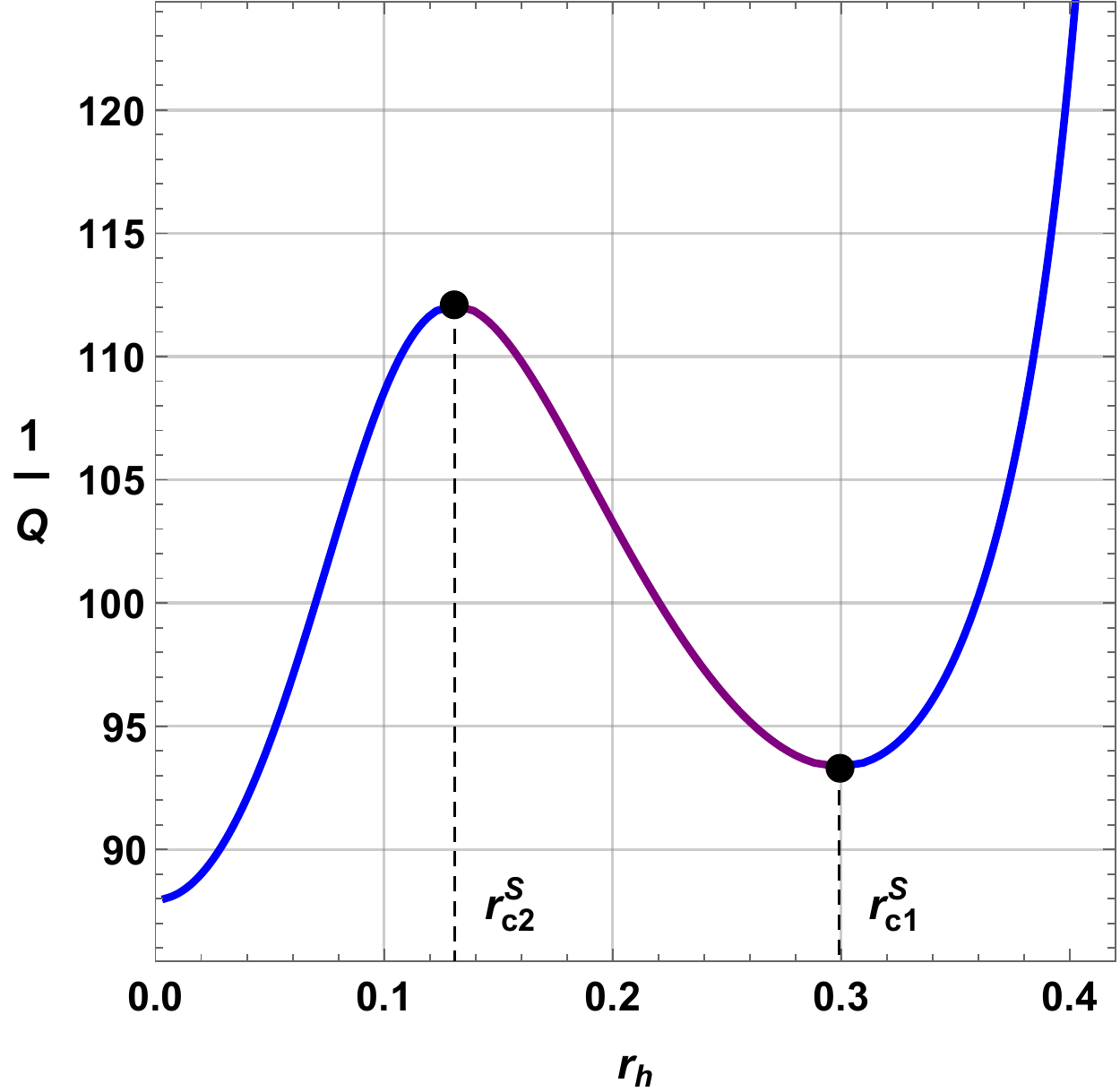}
		\caption{}
		\label{f25_1}
	\end{subfigure}
	\hspace{1pt}	
	\begin{subfigure}[h]{0.48\textwidth}
		\centering \includegraphics[scale=0.6]{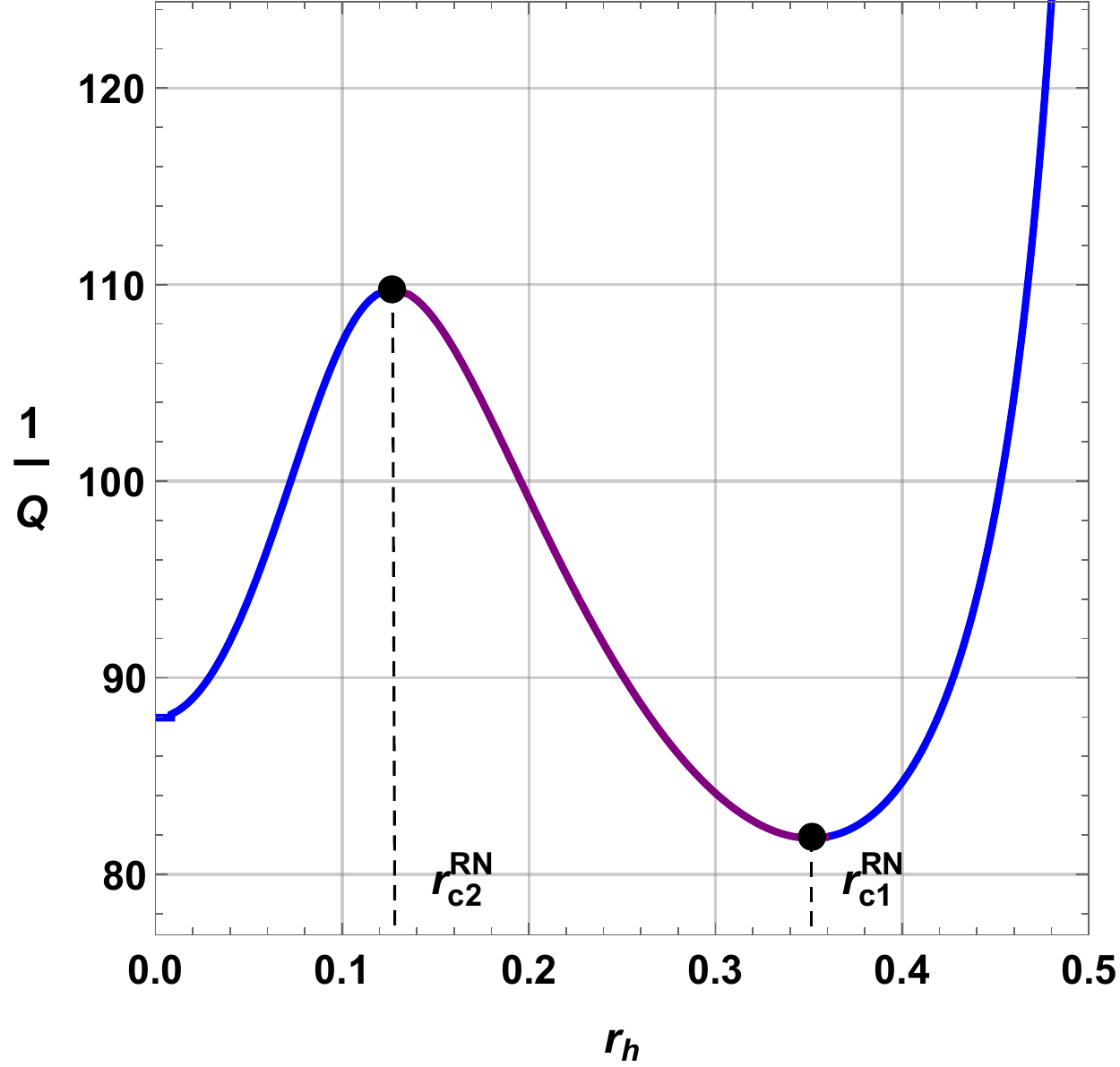}
		\caption{}
		\label{f25_2}	
	\end{subfigure}
	\caption{\footnotesize\it Vortex/anti-vortex creation and annihilation points in $(r_h,\frac{1}{Q})$ plane for (a) S-type and (b) RN-type black holes with $P = 0.15$ and $P = 0.2$ respectively and $b= 3.5$.}
	\label{f25}
\end{figure}

To this end, we plot in Fig.\ref{f24} the vortex/anti-vortex creation and annihilation points in $(\frac{1}{Q},T)$ plane, in which the blue and purple lines representing the annihilation and creation points, respectively. 
One can observe that the curve has the same shape as the free energy in terms of temperature in the first-order phase transition scenario which is characterized by a swallowtail shape. Black dots stand for conventional and unconventional critical points discussed previously. Therefore, we can interpret the temperature as the annihilation/creation free energy and the electric charge (precisely the inverse $1/Q$) as the annihilation/creation temperature. Indeed, one notices from the $(r_h,\frac{1}{Q})$ plane depicted in Fig.\ref{f25}, 
that the curve has the same shape as the temperature in terms of horizon radius in the first-order phase transition scenario. Thus, the temperature $T$ plays the role of the free energy, and the inverse of electric charge $\frac{1}{Q}$ can be considered as  temperature. With this {\bf analogy} at hand, we can define three thermodynamical phases: LBH and SSBH annihilation phases which are both locally stable (blue lines in Fig.\ref{f24} and  Fig.\ref{f25} ) and SSBH creation phase which is an unstable phase (purple lines in Fig.\ref{f24} and  Fig.\ref{f25}). Then, we shall study these phases and the transition between them in local and global thermodynamical views for S-type and RN-type black holes.

\subsection{Local thermodynamical view}

In this subsection, we suggest examining the thermodynamics of the vortex/anti-vortex creation and annihilation taking into account {\tt only the local stability} of different phases. Howsoever, we are interested in phases that minimize locally the temperature (which plays the role of free energy) without caring about the global stability of the black hole.

\begin{figure}[!ht]
	\centering 
	\begin{subfigure}[h]{0.48\textwidth}
		\centering \includegraphics[scale=0.6]{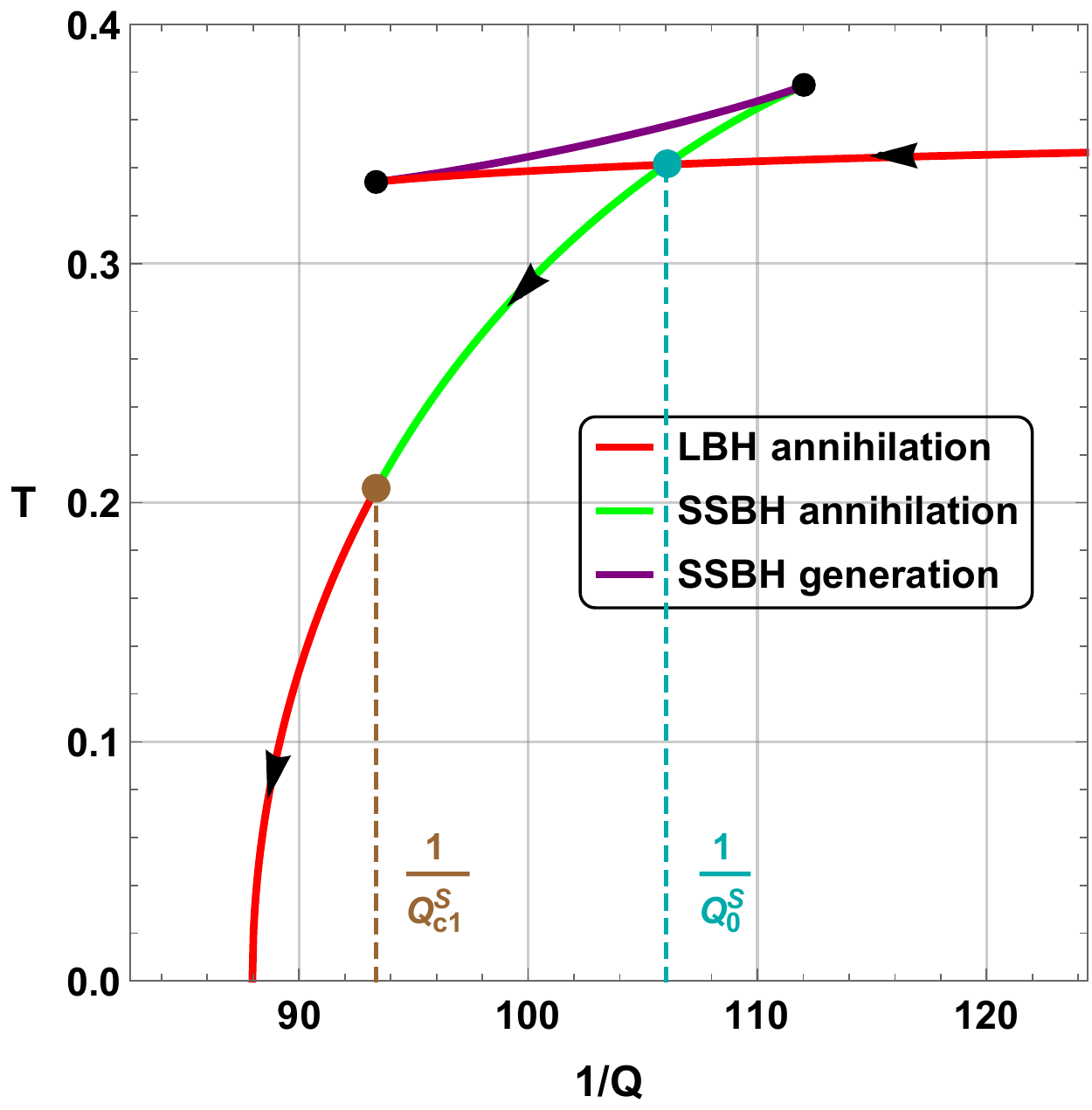}
		\caption{}
		\label{f26_1}
	\end{subfigure}
	\hspace{1pt}	
	\begin{subfigure}[h]{0.48\textwidth}
		\centering \includegraphics[scale=0.6]{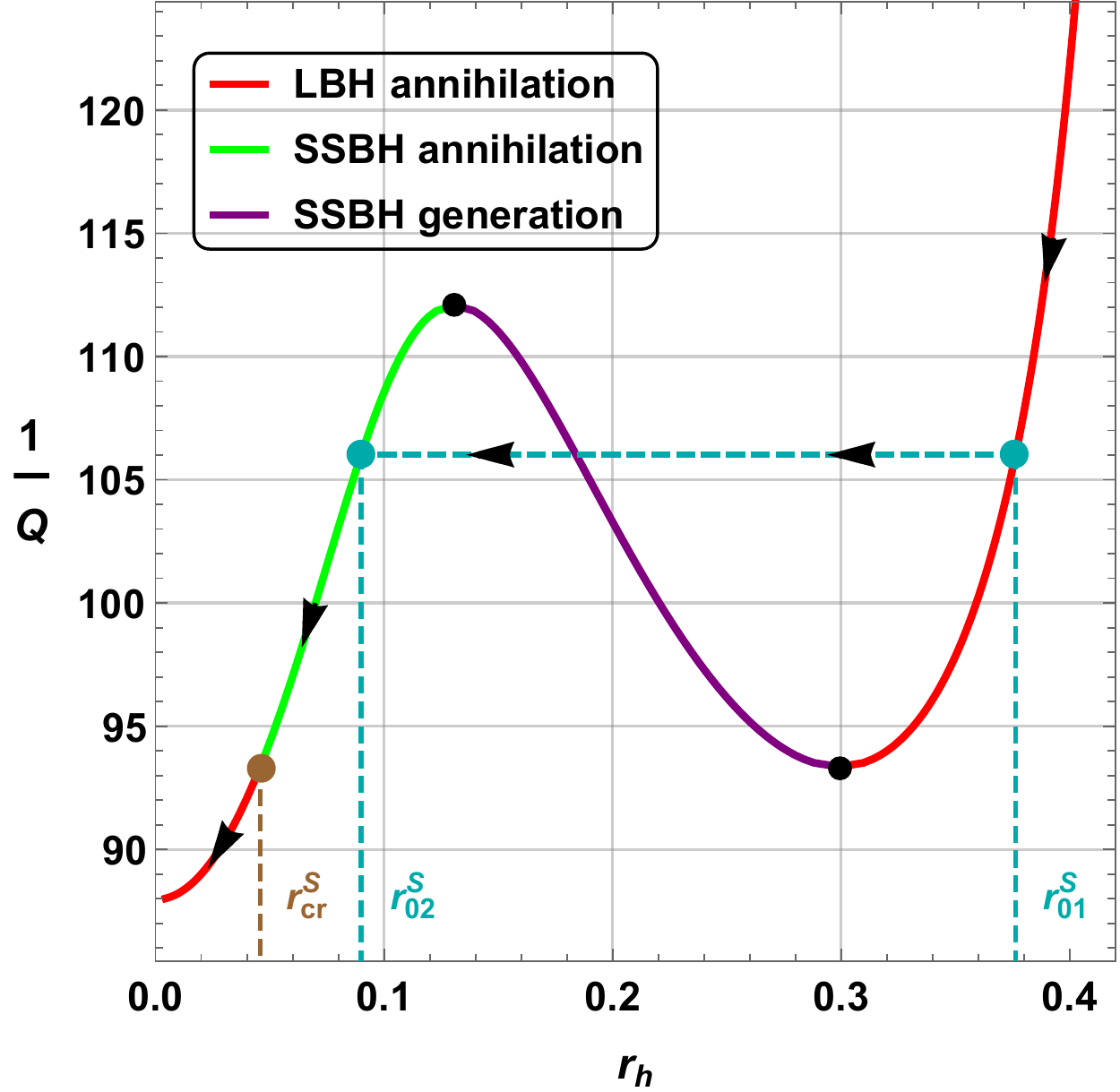}
		\caption{}
		\label{f26_2}	
	\end{subfigure}
	\caption{\footnotesize\it Stable/unstable phases generation and annihilation points in (a) $(\frac{1}{Q}, T)$ and (b) $(r_h,\frac{1}{Q})$ planes for  S-type black hole with $P = 0.15$ and $b= 3.5$.}
	\label{f26}
\end{figure}

Let us begin with a S-type black hole, we display in Fig.\ref{f26} the stable/unstable phases generation and annihilation points in  $(\frac{1}{Q}, T)$ and $(r_h,\frac{1}{Q})$ planes for  S-type black hole.
Where, red, green, and purple lines stand for large and stable small black hole annihilation phases and small black hole generation phase respectively. We observe that for large $\frac{1}{Q}$ (small electric charge $Q$), we have LBH annihilation which is the stable phase in this region as $\frac{1}{Q}$ (which plays the role of temperature) is an increasing function in terms of $r_h$.  For $\frac{1}{Q} = \frac{1}{Q_0^{S}}$, we have the intersection of LBH and SSBH annihilation phases which is a signature of the first phase transition. Indeed, at $\frac{1}{Q} = \frac{1}{Q_0^{S}}$, there is a first-order phase transition between LBH and SSBH annihilation phases, where SSBH generation phase (purple line) plays the role of the unstable phase as we can see in Fig.\ref{f26_2} that $\frac{1}{Q}$ is a decreasing function  in terms of $r_h$. Therefore, the LBH annihilation point disappears at $r_{01}^S$ and the SSBH annihilation appears at $r_{02}^S$ as it is indicated in Fig.\ref{f26_2} where the darker cyan line represents the first order phase transition path. For $\frac{1}{Q_{c1}^S}<\frac{1}{Q} < \frac{1}{Q_0^{S}}$, The SSBH annihilation phase is the preferred (stable) state. For $\frac{1}{Q} = \frac{1}{Q_{c1}^S}$, we have a crossover from SSBH to LBH annihilation phases, then the SSBH annihilation point transforms to LBH annihilation one as we can remark when we compare Fig.\ref{f13_2} with Fig.\ref{f14_2}. Indeed, from such a comparison, we remark that the annihilation point (blue dot) is an SSBH annihilation in Fig.\ref{f13_2} and it becomes an LBH annihilation in Fig.\ref{f14_2}. This transition is called a {\bf crossover} because it is a smooth transition and there is any discontinuity behavior in the free energy function or in its derivatives. Indeed, we observe that there is any anomaly (discontinuity) in the behavior of temperature $T$ in terms of the inverse of electric charge at $\frac{1}{Q} = \frac{1}{Q_{c1}^S}$ and in the behavior of $\frac{1}{Q}$ in terms of $r_h$ at $r_h = r_{cr}$. For $\frac{1}{Q_{m}}<\frac{1}{Q} < \frac{1}{Q_{c1}^{S}}$, the stable phase is the LBH annihilation phase again, but for $\frac{1}{Q} \leq \frac{1}{Q_{m}}$ the annihilation points disappear. Thus, at the topological transition, the annihilation phases disappear completely in the S-type black hole.

\begin{figure}[!ht]
	\centering 
	\begin{subfigure}[h]{0.48\textwidth}
		\centering \includegraphics[scale=0.6]{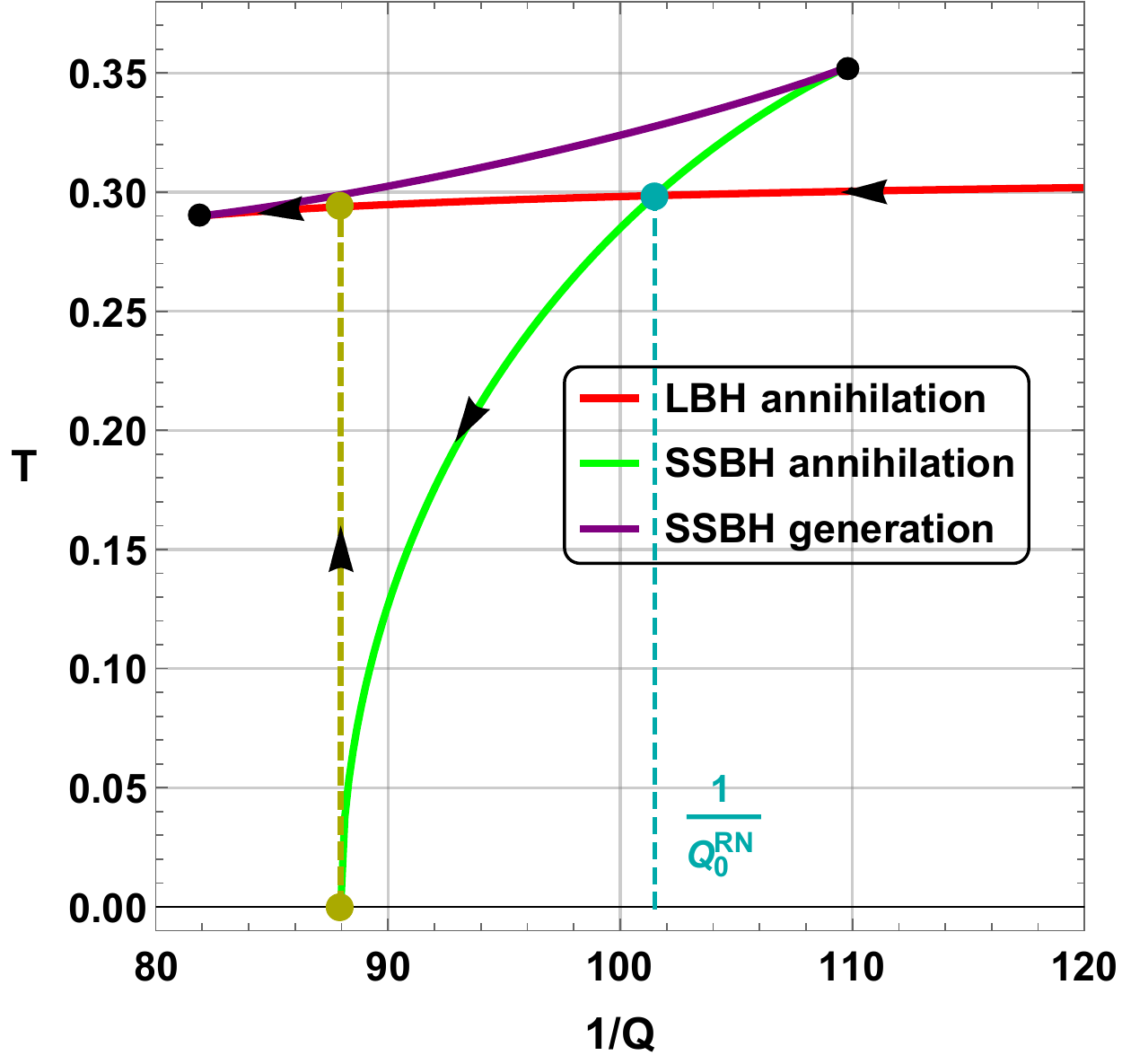}
		\caption{}
		\label{f27_1}
	\end{subfigure}
	\hspace{1pt}	
	\begin{subfigure}[h]{0.48\textwidth}
		\centering \includegraphics[scale=0.6]{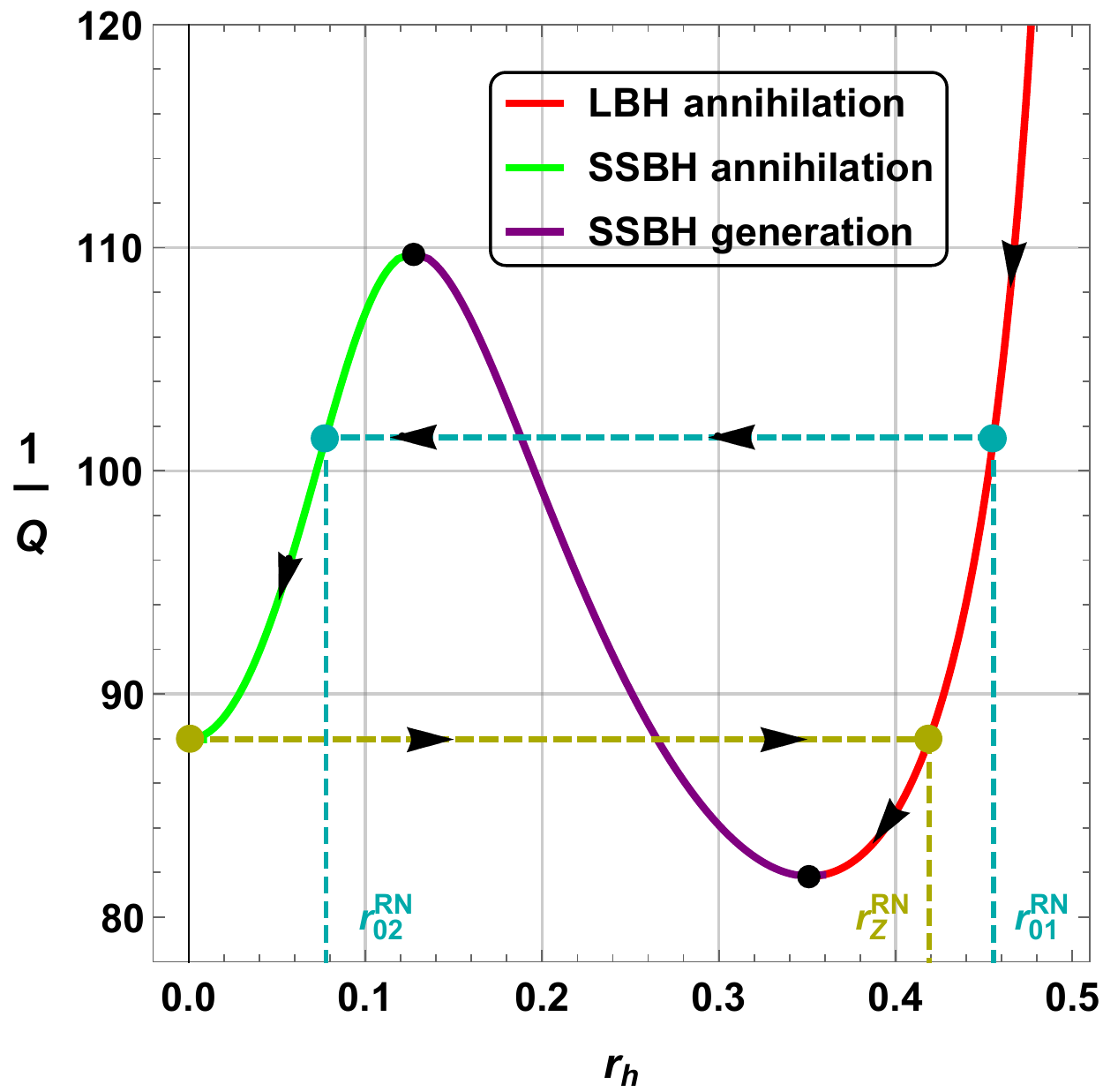}
		\caption{}
		\label{f27_2}	
	\end{subfigure}
	\caption{\footnotesize\it Stable/unstable phases generation and annihilation points in (a) $(\frac{1}{Q}, T)$ and (b) $(r_h,\frac{1}{Q})$ planes for  RN-type black hole with $P = 0.20$ and $b= 3.5$.}
	\label{f27}
\end{figure}

Concerning the RN-type black hole, we display in Fig.\ref{f27} the stable/unstable phases generation and annihilation points in  $(\frac{1}{Q}, T)$ and $(r_h,\frac{1}{Q})$ planes. 
In which, 
red, green, and purple lines stand for large and stable small black hole annihilation phases and small black hole generation phase respectively. Herein, one observes that for large $\frac{1}{Q}$ (small electric charge $Q$), we have LBH annihilation which is the stable phase in this region as $\frac{1}{Q}$ is an increasing function in terms of $r_h$.  For $\frac{1}{Q} = \frac{1}{Q_0^{RN}}$, we have the intersection of LBH and SSBH annihilation phases which is a signature of the first phase transition. Indeed, at $\frac{1}{Q} = \frac{1}{Q_0^{RN}}$, there is a first-order phase transition between LBH and SSBH annihilation phases, where SSBH generation phase (purple line) plays the role of the unstable phase as one can see in Fig.\ref{f27_2} 
that $\frac{1}{Q}$ is a decreasing function  in terms of $r_h$. Therefore, the LBH annihilation point disappears at $r_{01}^{RN}$ and the SSBH annihilation appears at $r_{02}^{RN}$, the darker cyan line represents the first order phase transition path. For $\frac{1}{Q_m}<\frac{1}{Q} < \frac{1}{Q_0^{RN}}$, The SSBH annihilation phase is the preferred (stable) state.

 Reaching the point $\frac{1}{Q} = \frac{1}{Q_m}$, we discover  a zeroth-order phase transition from SSBH to LBH annihilation phases with a jump in temperature (which plays the role of free energy) which indicated by the yellow line. Therefore, the system displays a reentrant phase transition as we can see clearly in Fig.\ref{f27_2} and the preferred state becomes the LBH annihilation phase again.
  
 Further, for $\frac{1}{Q_{c1}^{RN}}<\frac{1}{Q} < \frac{1}{Q_{m}}$, the stable phase is the LBH annihilation phase again, but for $\frac{1}{Q} \leq \frac{1}{Q_{c1}^{RN}}$ the annihilation points disappear. Thus, contrary to the S-type black hole, the RN-type black displays an annihilation process even at the topological transition. This annihilation process is related to the first phase transition between large and small black holes, and the annihilation points disappear completely after the second phase transition.

\begin{figure}[!ht]
	\centering 
	\begin{subfigure}[h]{0.48\textwidth}
		\centering \includegraphics[scale=0.6]{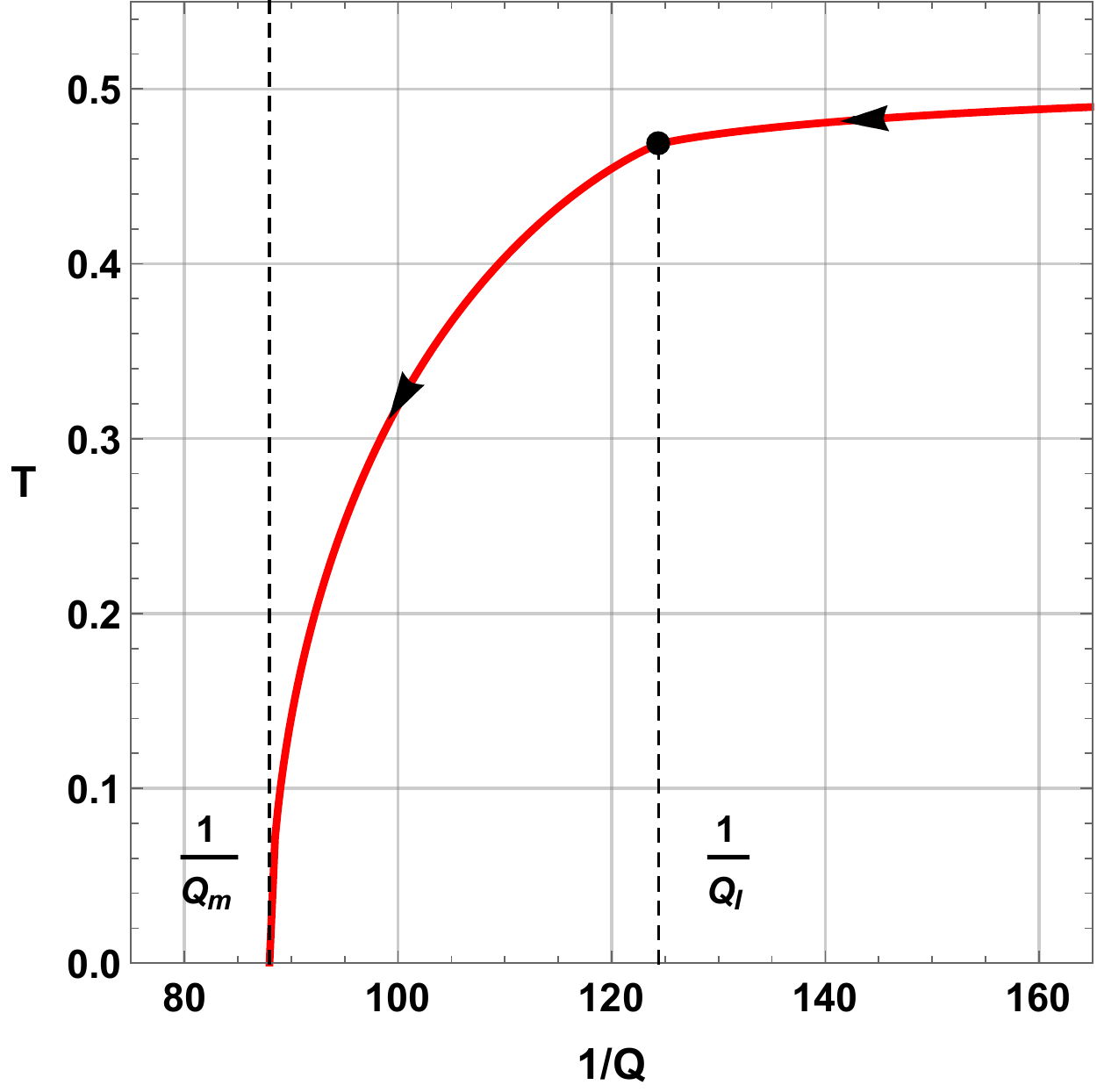}
		\caption{}
		\label{f28_1}
	\end{subfigure}
	\hspace{1pt}	
	\begin{subfigure}[h]{0.48\textwidth}
		\centering \includegraphics[scale=0.6]{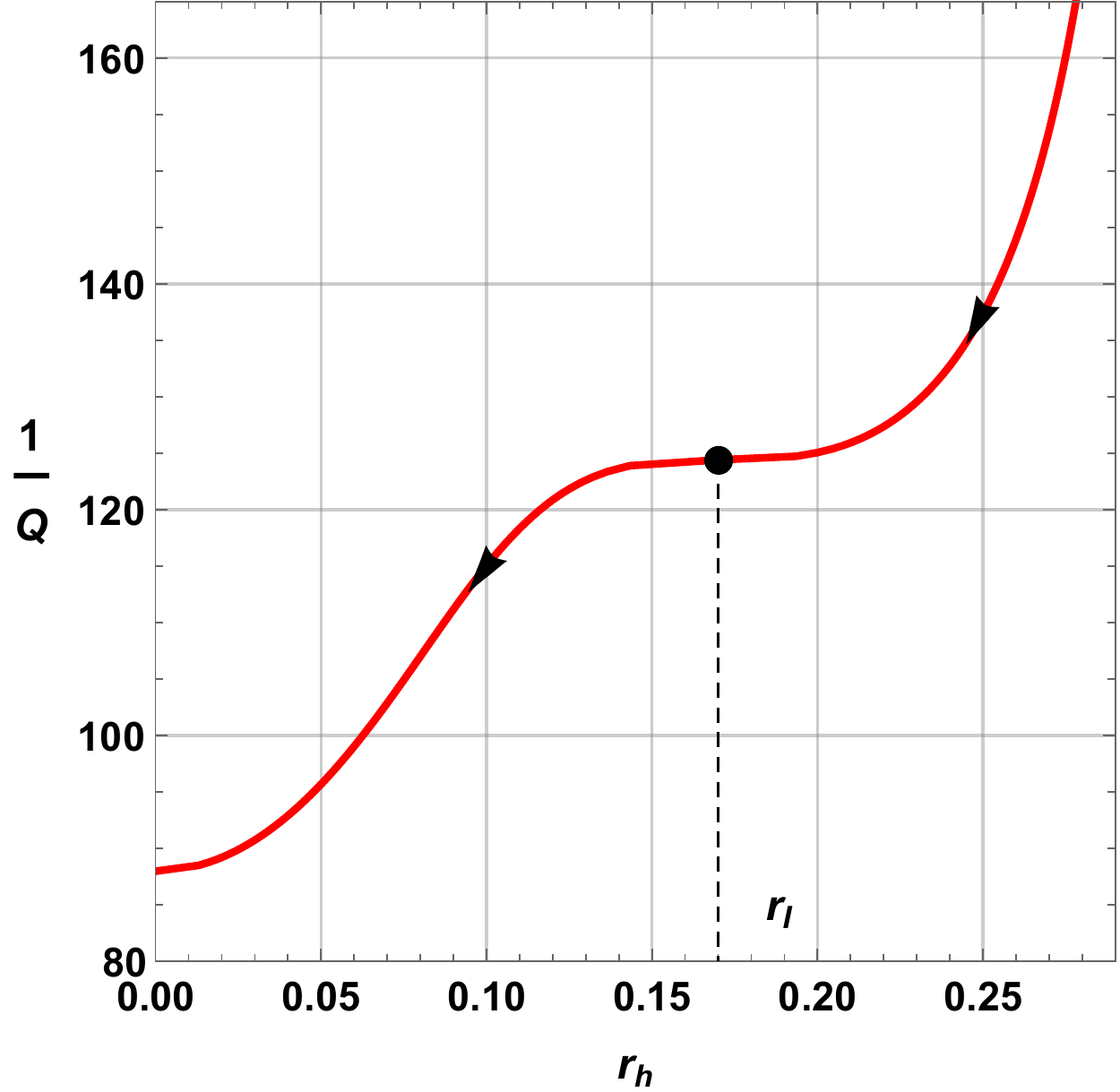}
		\caption{}
		\label{f28_2}	
	\end{subfigure}
	\caption{\footnotesize\it Stable/unstable phases  annihilation points in (a) $(\frac{1}{Q}, T)$ and (b) $(r_h,\frac{1}{Q})$ planes for the critical situation with $P = P_I = 0.403785 $ and $b= 3.5$.}
	\label{f28}
\end{figure}

This phase transition that exists in RN-type black holes as well as in S-Type black holes could lead to an eventual second-order phase transition. The critical point corresponds to the isolated critical point discussed in the first section. Indeed, we display in Fig.\ref{f28} the stable/unstable phases annihilation points in  $(\frac{1}{Q}, T)$ and $(r_h,\frac{1}{Q})$ planes for the critical pressure $P= P_I = 0.403785 $, 
one observes that the critical point corresponds to the isolated point $(Q_I,r_I)$ with $r_I = 0.169886$. Moreover, this critical behavior is special and we can not qualify it as a second-order phase transition, because the annihilation phase does not change its nature before and after the critical point and always just one phase which corresponds to LBH annihilation persists.

\subsection{Global thermodynamical view}

In this subsection, we suggest studying the annihilation process from a global thermodynamical viewpoint and taking into account the {\tt global stability} of different black hole phases. That is to say, before studying an annihilation phase, we should look if the black hole phase is globally stable or not using the free energy landscape that we have previously developed in \cite{Ali:2023wkq}. For instance, before studying the LBH annihilation phase, we should determine if the large black hole phase is globally stable or not. Indeed, if $T<T_{HP}$ (Hawking-Page temperature), the large black hole phase is not any more stable, and its annihilation has taken place at $T=T_{HP}$ or at first order phase transition temperature. 

\begin{figure}[!ht]
	\centering 
	\begin{subfigure}[h]{0.48\textwidth}
		\centering \includegraphics[scale=0.6]{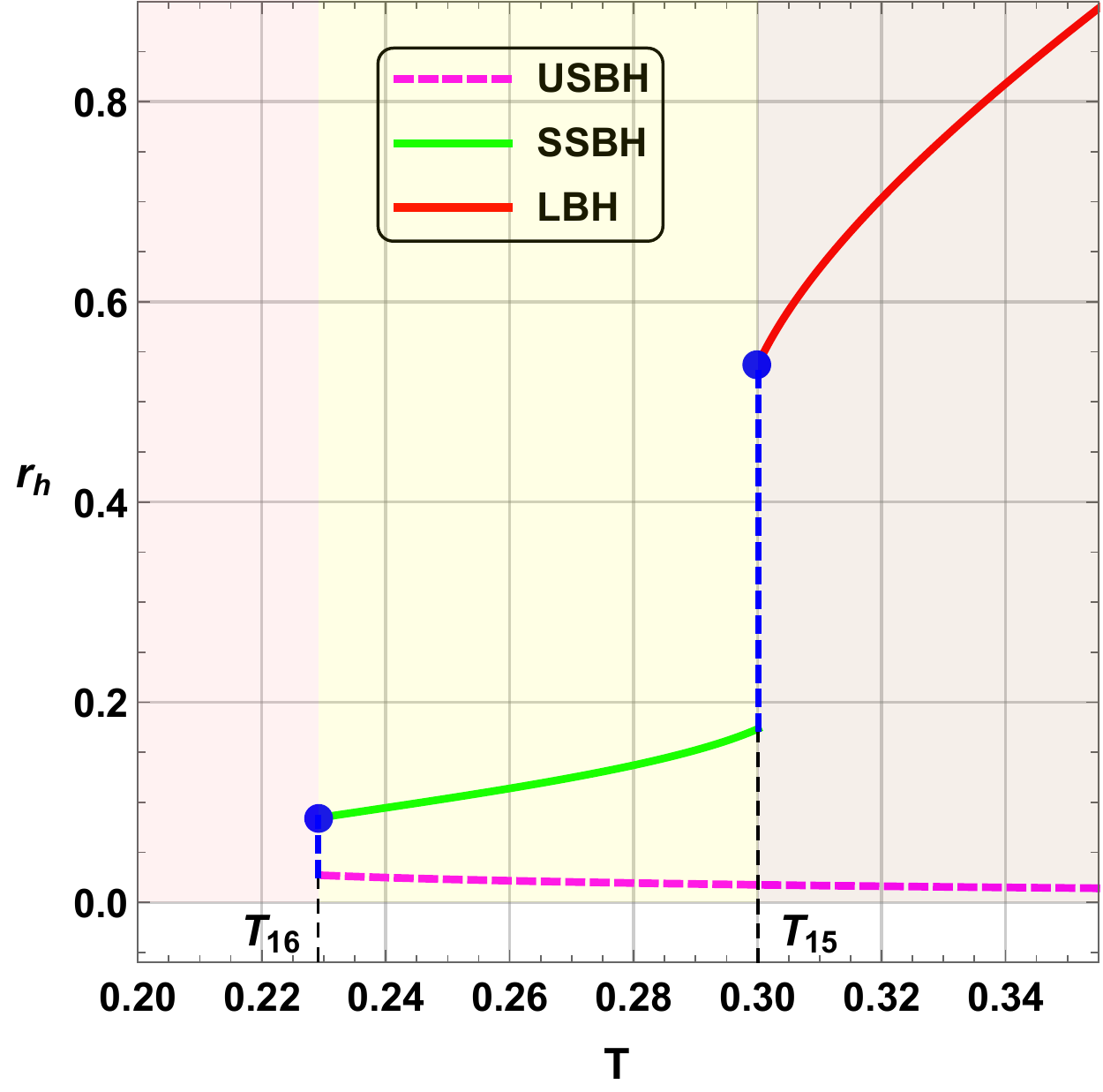}
		\caption{}
		\label{f29_1}
	\end{subfigure}
	\hspace{1pt}	
	\begin{subfigure}[h]{0.48\textwidth}
		\centering \includegraphics[scale=0.6]{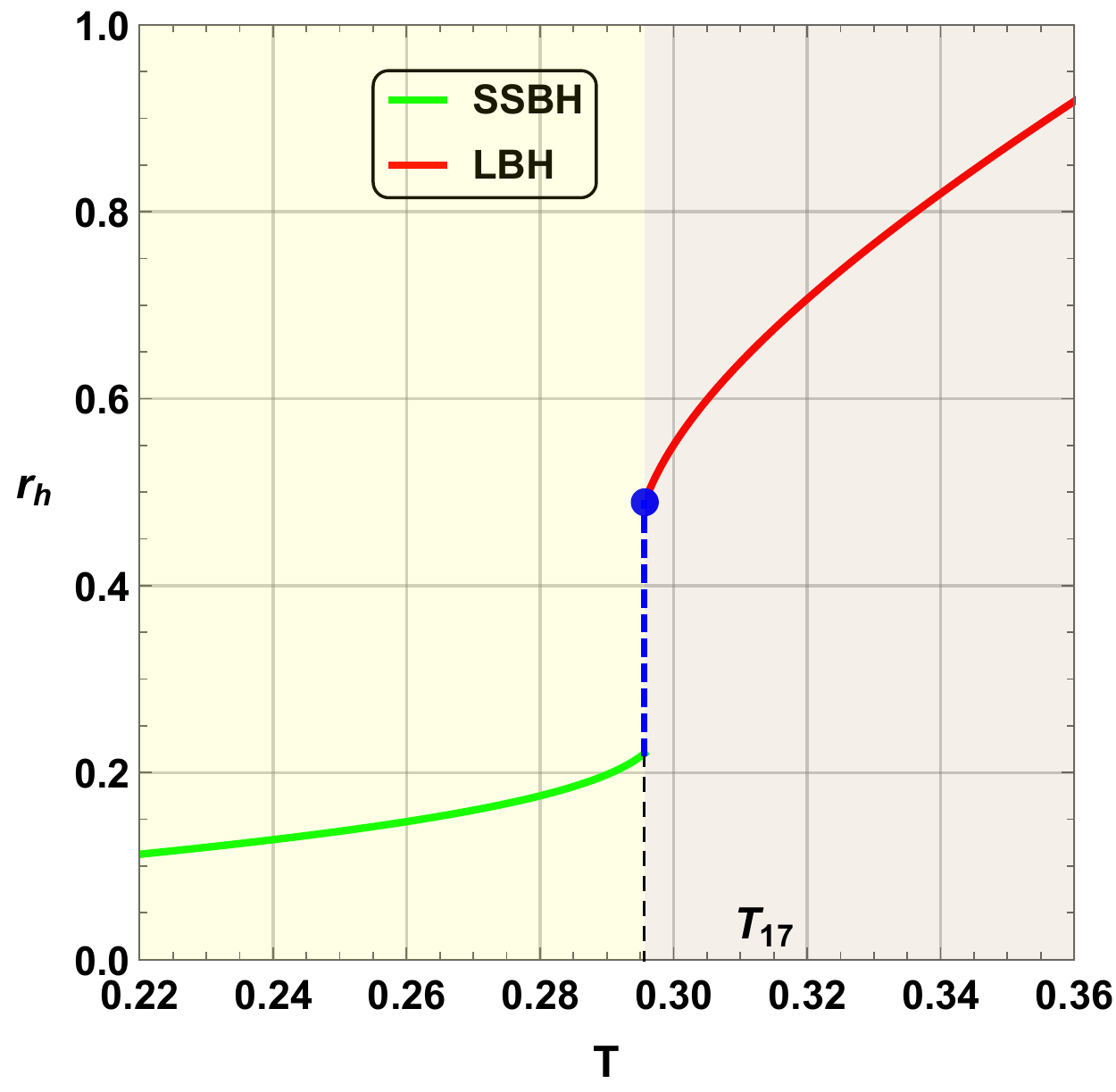}
		\caption{}
		\label{f29_2}	
	\end{subfigure}
	\caption{\footnotesize\it Zero points of $\phi^{r_h}$ shown in $(r_h,T)$ plane for (a) $Q=0.0107135 0.013<Q_m$ and (b) $Q = Q_m$ with $P= 0.15$ and $b = 3.5$. }
	\label{f29}
\end{figure}

First, we should verify that this global view conserves the topology of the black hole. We take for instance the RN-type black hole case and we use the formalism that we have developed in \cite{Ali:2023wkq} to study the global behavior of the system. We plot in Fig.\ref{f29} the zero point of $\phi^{r_h}$ in $(r_h,T)$ plane for $Q=0.0107135 0.013<Q_m$ and $Q = Q_m$ taking into account the global stability of SSBH and LBH by using the energy landscape formalism. 

Such Fig.\ref{f29_1} reveals that for $T>T_{15}$ (brown region), there is a pair of vortex/anti-vortex corresponding to LBH/USBH phases, and then the total topological charge in this region is zero. 
For $T = T_{15}$, we notice the annihilation of the LBH phase (blue dot) and the generation of the SSBH phase which corresponds to the usual first-order phase transition between LBH and SSBH \footnote{We choose the annihilation point to be the point where the annihilation (first-order phase transition) begins.}. For $T_{16}<T<T_{15}$ (yellow region), there is also a pair of vortex/anti-vortex corresponding to SSBH/IBH  phases, then the total topological charge in this region is null. In the situation where $T = T_{16}$, the system shows the annihilation of SSBH and USBH phases (blue dot) which corresponds to a Hawking-Page-like transition \footnote{The Hawking-Page transition in this situation occurs between small black holes and thermal radiations phases and not between large black holes and thermal radiations.  } \cite{Ali:2023wkq}. For $T<T_{16}$ (pink region), there is no black hole solution (no vortexes) and then the total topological charge is also zero. Therefore the total topological charge of the system is the same as in the local view. In Fig.\ref{f29_2}, the anti-vortex corresponding to the USBH phase has disappeared, and for $T>T_{17}$ (brown region) and there is only one vortex corresponding to LBH phase and the topological charge is then equal to $-1$.  Within $T = T_{17}$, we observe the annihilation of the LBH phase (blue dot) and the generation of the SSBH phase which corresponds to the usual first-order phase transition between LBH and SSBH. For $T<T_{17}$ (yellow region), there is also one vortex corresponding to the SSBH  phase, then the total topological charge in this region is also equal to $-1$. Thus, the topological transition exists also in the global view.

\begin{figure}[!ht]
	\centering 
	\begin{subfigure}[h]{0.48\textwidth}
		\centering \includegraphics[scale=0.6]{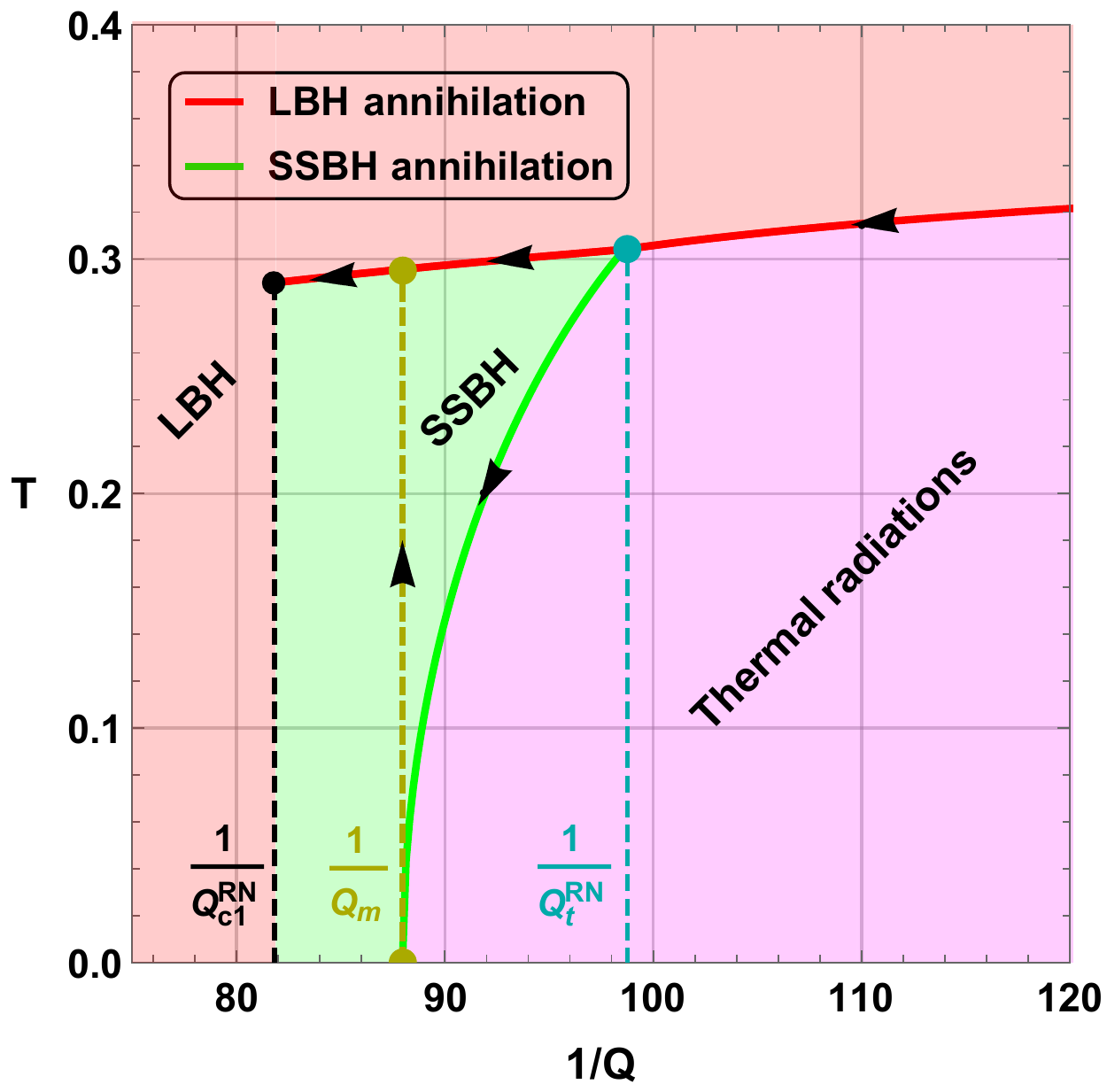}
		\caption{}
		\label{f30_1}
	\end{subfigure}
	\hspace{1pt}	
	\begin{subfigure}[h]{0.48\textwidth}
		\centering \includegraphics[scale=0.6]{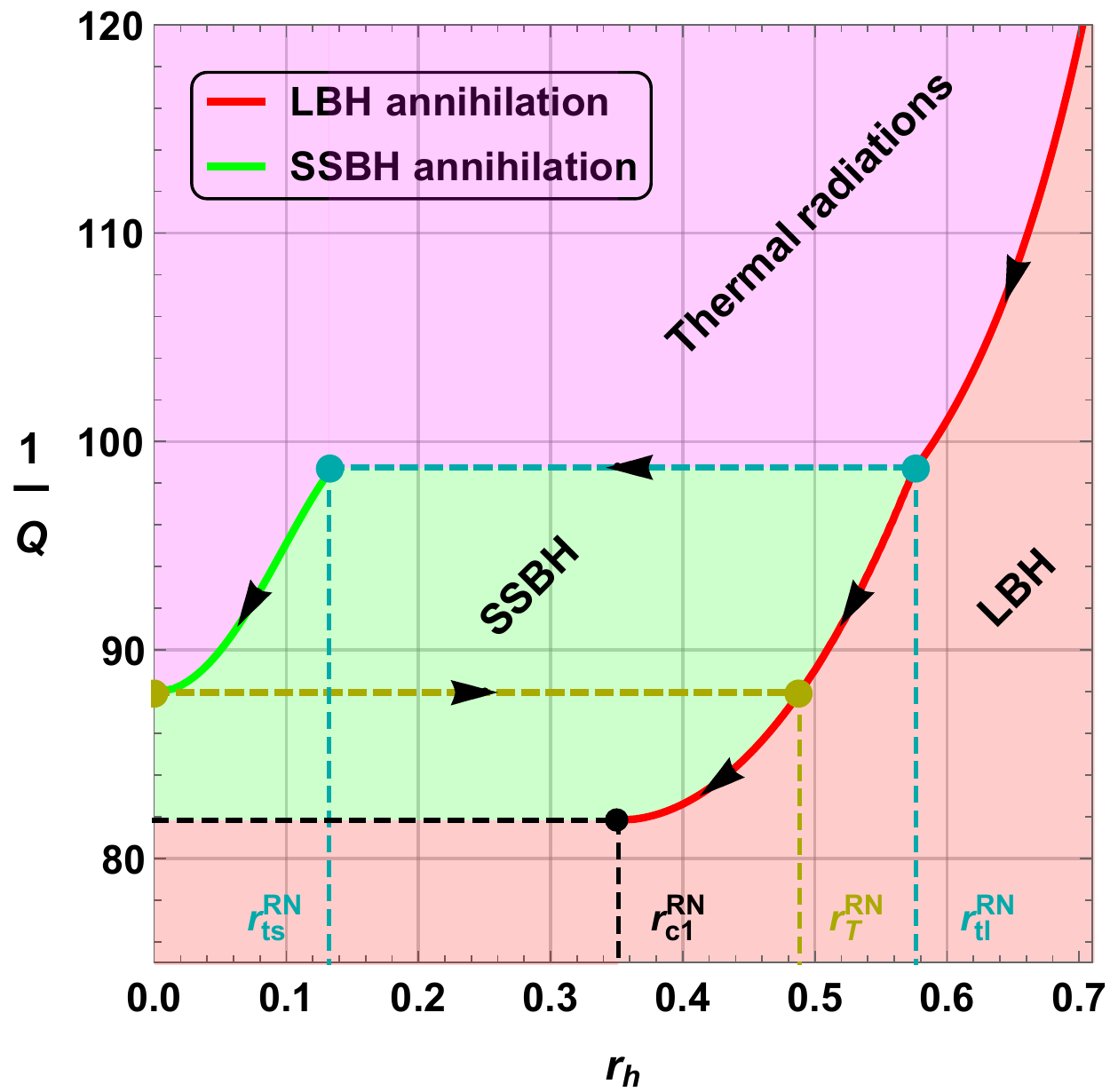}
		\caption{}
		\label{f30_2}	
	\end{subfigure}
	\caption{\footnotesize\it Stable phases annihilation points in (a) $(\frac{1}{Q}, T)$ and (b) $(r_h,\frac{1}{Q})$ planes for  RN-type black hole in global themodynamical view with $P = 0.15$ and $b= 3.5$.}
	\label{f30}
\end{figure}

We display in Fig.\ref{f30} the stable phases  annihilation points in  $(\frac{1}{Q}, T)$ and $(r_h,\frac{1}{Q})$ planes for RN-type black hole. 
The red and green lines stand for large and stable small black hole annihilation phases and stable small black hole generation phases respectively which are globally stable. In the phase portrait, there are three global stable phases corresponding to LBH (red region), SSBH (green region), and thermal radiations (magenta region) phases \cite{Ali:2023wkq}.  For large $\frac{1}{Q}$ (small electric charge $Q$), we have only the LBH annihilation phase (red curve) which represents the annihilation of LBH to thermal radiations under a hawking-Page process. With $\frac{1}{Q} = \frac{1}{Q_t^{RN}}$, we have the intersection of three black hole phases (LBH, SSBH, and thermal radiations) which corresponds to the triple point (darker cyan point) \cite{Ali:2023wkq} with $Q_t^{RN} = 0.0101266$. At this point, there is a phase zero-order transition between LBH and SSBH annihilation phases represented by the darker cyan line in Fig.\ref{f30_2} and the SSBH annihilation appears at $r_{ts}^{RN}$. It is a particular first-order phase transition because the LBH still exists at $r_{tl}^{RN}$. For $\frac{1}{Q_m}$, the SSBH annihilation phase disappears which corresponds to the topological transition and only the LBH annihilation will remain. Last, $\frac{1}{Q_m}<\frac{1}{Q} < \frac{1}{Q_{c1}^{RN}}$ case, the system is analogous to AdS-Reissner-Nordström black hole and at $\frac{1}{Q} = \frac{1}{Q_{c1}^{RN}} $, and a critical behavior is exhibited corresponding to the usual second-order phase transition between LBH and SSBH phase, then the LBH annihilation phase disappears.

\begin{figure}[!ht]
	\centering 
	\begin{subfigure}[h]{0.48\textwidth}
		\centering \includegraphics[scale=0.6]{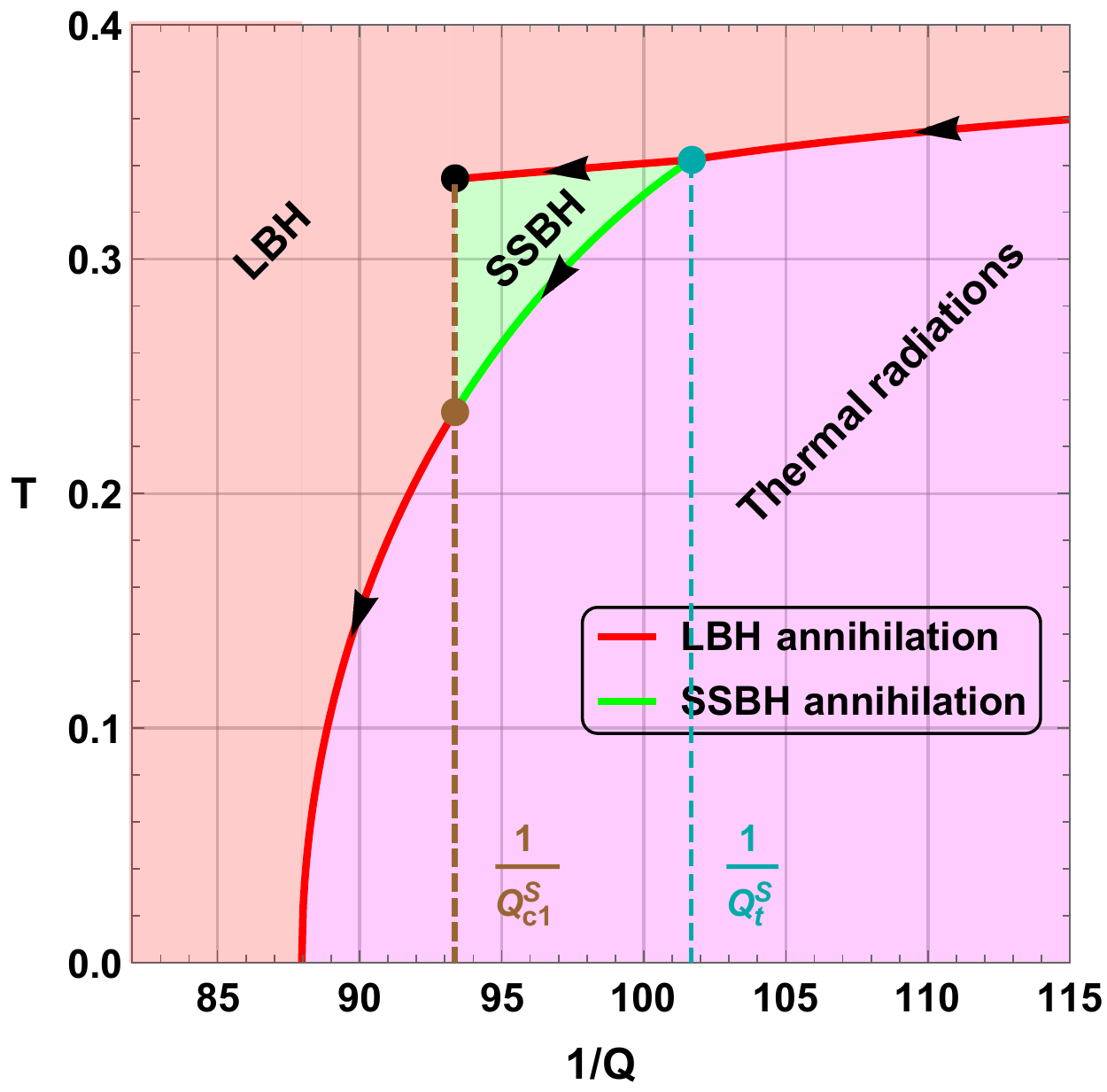}
		\caption{}
		\label{f31_1}
	\end{subfigure}
	\hspace{1pt}	
	\begin{subfigure}[h]{0.48\textwidth}
		\centering \includegraphics[scale=0.6]{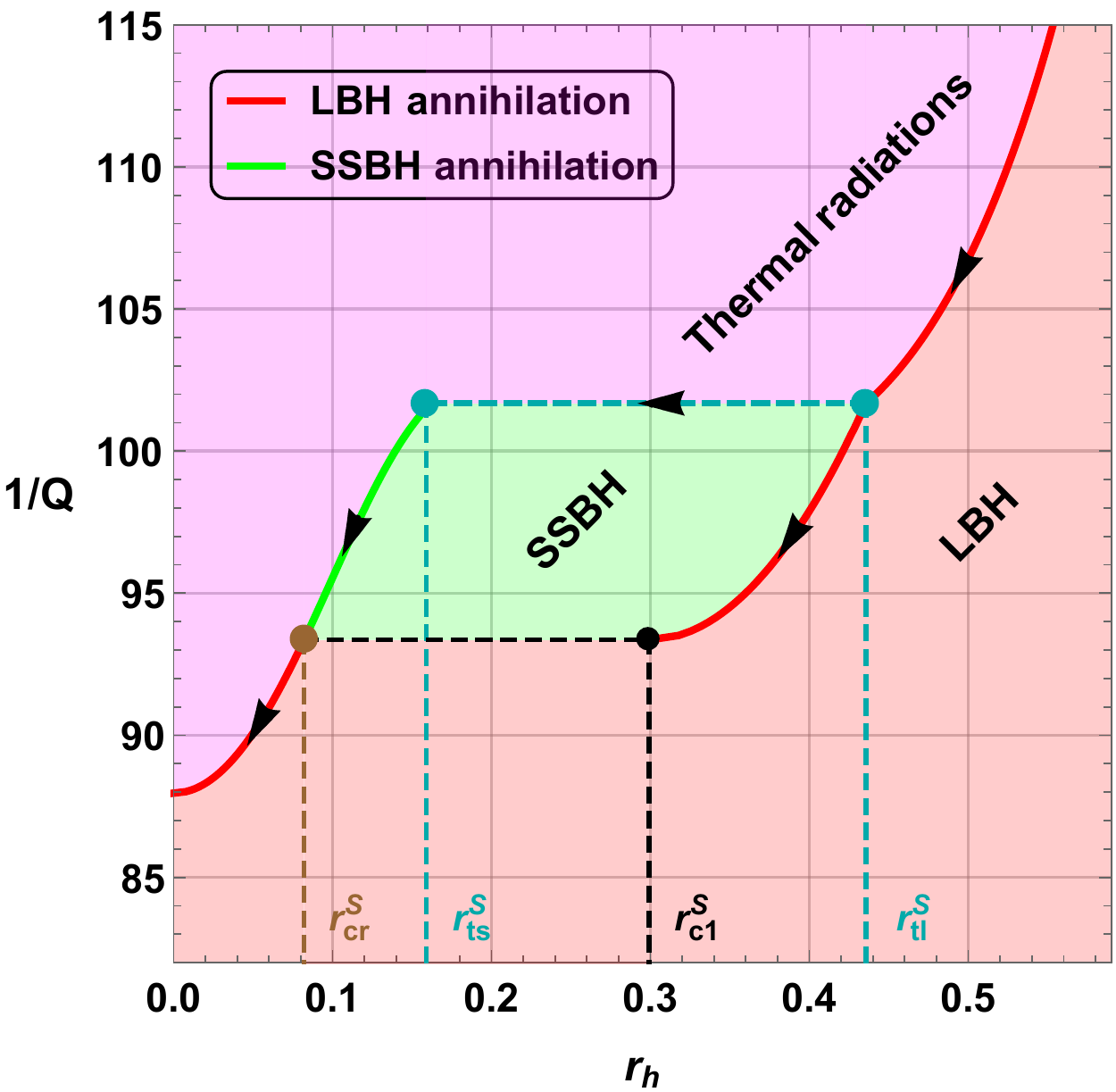}
		\caption{}
		\label{f31_2}	
	\end{subfigure}
	\caption{\footnotesize\it Stable phases annihilation points in (a) $(\frac{1}{Q}, T)$ and (b) $(r_h,\frac{1}{Q})$ planes for  S-type black hole in global themodynamical view with $P = 0.20$ and $b= 3.5$.}
	\label{f31}
\end{figure}

We return now to the S-type black hole configuration, and we display in Fig.\ref{f31} the stable phases annihilation points in  $(\frac{1}{Q}, T)$ and $(r_h,\frac{1}{Q})$ planes for S-type black hole. 
Red and green lines are associated with large and stable small black hole annihilation phases and stable small black hole generation phase respectively which are globally stable. We observe that there are three global stable phases corresponding to LBH (red region), SSBH (green region), and thermal radiations (magenta region) phases \cite{Ali:2023wkq}. Serval remarks can be elaborated, first,  for large $\frac{1}{Q}$ (small electric charge $Q$), we have only the LBH annihilation phase (red curve) which represents the annihilation of LBH to thermal radiations under a hawking-Page process. Second, for $\frac{1}{Q} = \frac{1}{Q_t^{S}}$, we have the intersection of three black hole phases (LBH, SSBH, and thermal radiations) which denote nothing than the triple point (darker cyan point)  with $Q_t^{S} = 0.0101266$. At this point, there is a phase order transition between LBH and SSBH annihilation phases represented by the darker cyan line in Fig.\ref{f31_2} and the SSBH annihilation appears at $r_{ts}^{S}$. It is a particular first-order phase transition because the LBH still exists at $r_{tl}^{S}$. For $\frac{1}{Q} = \frac{1}{Q_{c1}^{S}}$, a critical behavior occurs where the LBH annihilation phase disappears and we observe a crossover (brown dot) between SSBH and LBH annihilation phases. Last, For $\frac{1}{Q}=\frac{1}{Q_m}$, there is a topological transition characterized by a disappearance of LBH annihilation.

\begin{figure}[!ht]
	\centering 
	\begin{subfigure}[h]{0.48\textwidth}
		\centering \includegraphics[scale=0.6]{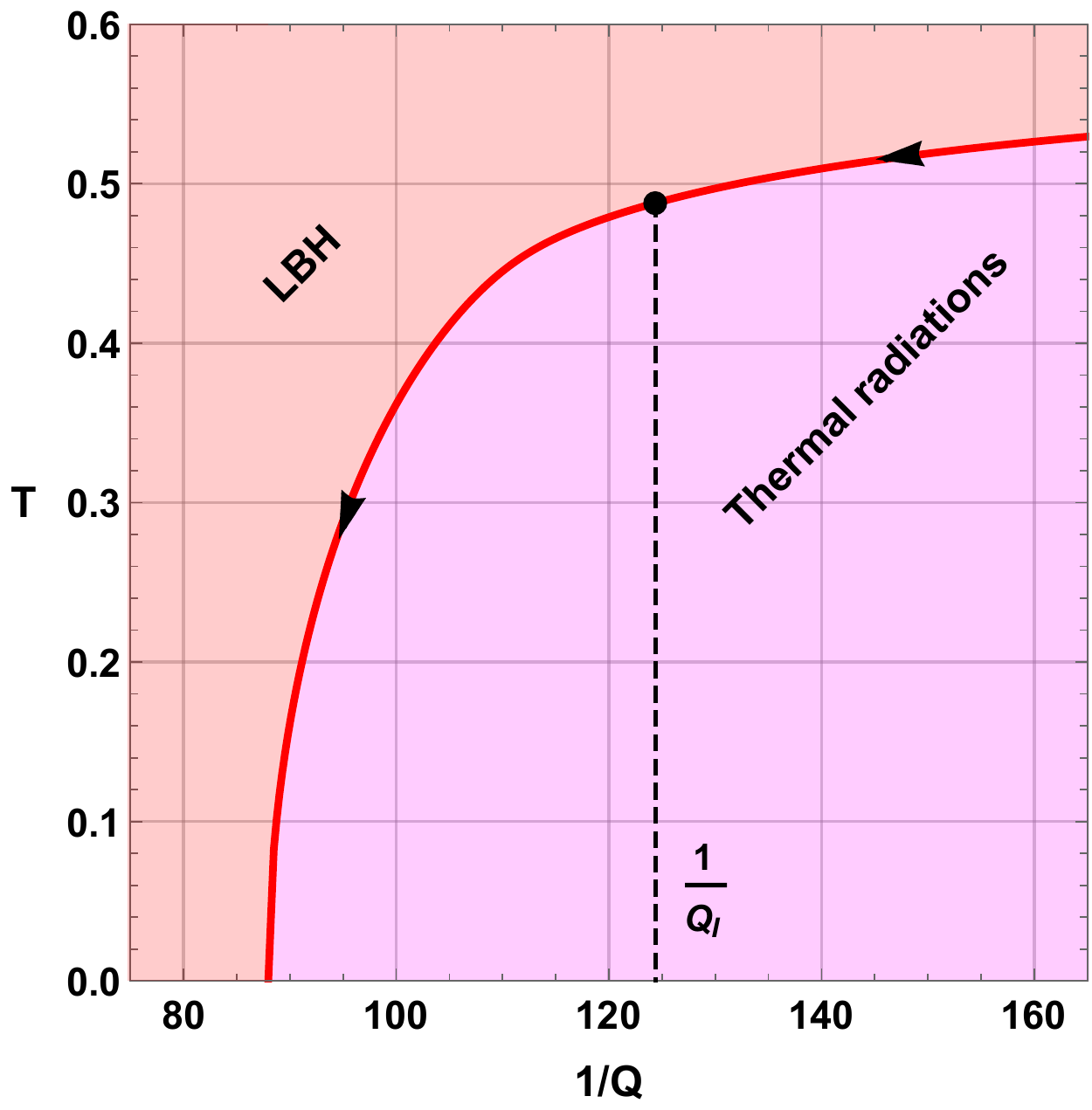}
		\caption{}
		\label{f32_1}
	\end{subfigure}
	\hspace{1pt}	
	\begin{subfigure}[h]{0.48\textwidth}
		\centering \includegraphics[scale=0.6]{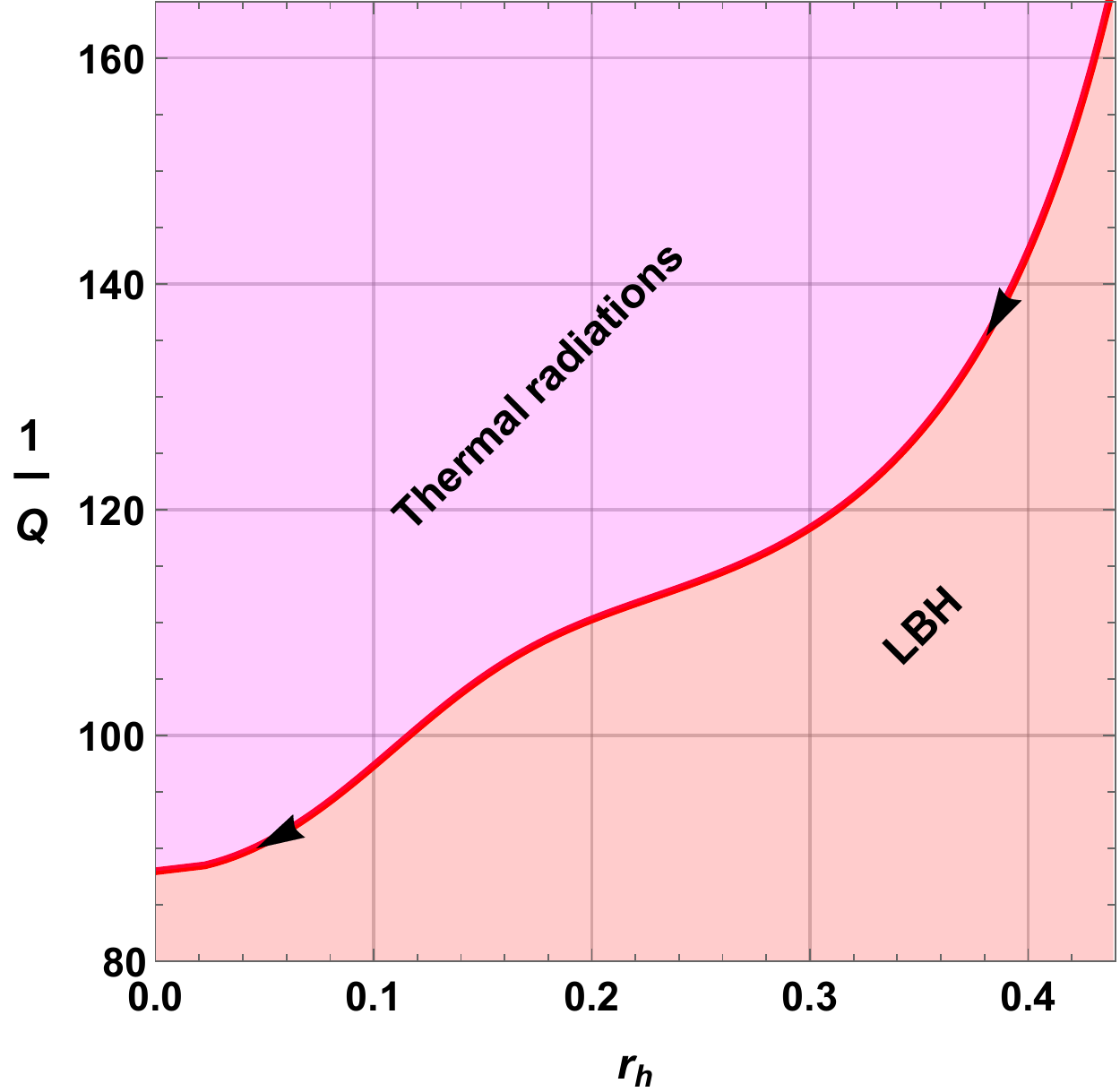}
		\caption{}
		\label{f32_2}	
	\end{subfigure}
	\caption{\footnotesize\it Stable phases annihilation points in (a) $(\frac{1}{Q}, T)$ and (b) $(r_h,\frac{1}{Q})$ planes for for the critical situation with $P = P_I = 0.403785 $ and $b= 3.5$.}
	\label{f32}
\end{figure}

Comparing Fig.\ref{f30} and Fig.\ref{f31}, we remark that the SSBH zone becomes smaller in the S-type black hole. At $P = P_I$, the SSBH phase will completely disappear. Indeed, we display in Fig.\ref{f32} the stable phases  annihilation points in  $(\frac{1}{Q}, T)$ and $(r_h,\frac{1}{Q})$ planes for $P = P_I$. 
There is only the LBH phase annihilation corresponding to the Hawking-Page transition and the system is perfectly equivalent AdS-Schwarzchild black hole. At the isolated critical point, $Q = Q_I$,  we do not see any critical behavior. Indeed, the temperature curve is a smooth function and does not show any irregularity which explains the topological charge of the isolated critical \eqref{20}.

\section{\textcolor{blue}{Topology in the dual conformal field theory (CFT)}}\label{adscft}

\begin{figure}[!ht]
 	\centering 
 	\begin{subfigure}[h]{0.48\textwidth}
 		\centering \includegraphics[scale=0.6]{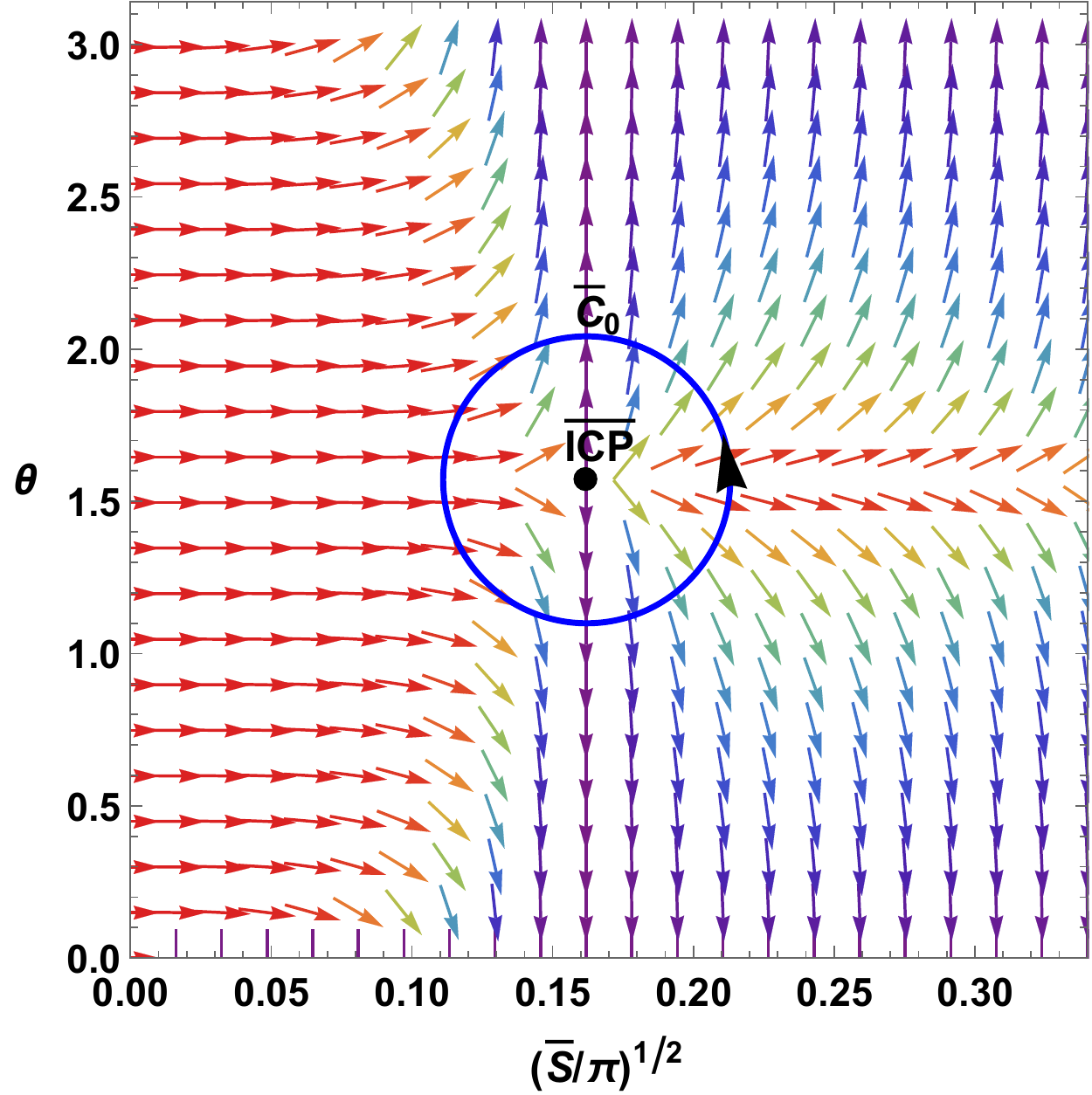}
 		\caption{$\bar{b} = \bar{b}_I$}
 		\label{f33_1}
 	\end{subfigure}
 	\hspace{1pt}	
 	\begin{subfigure}[h]{0.48\textwidth}
 		\centering \includegraphics[scale=0.6]{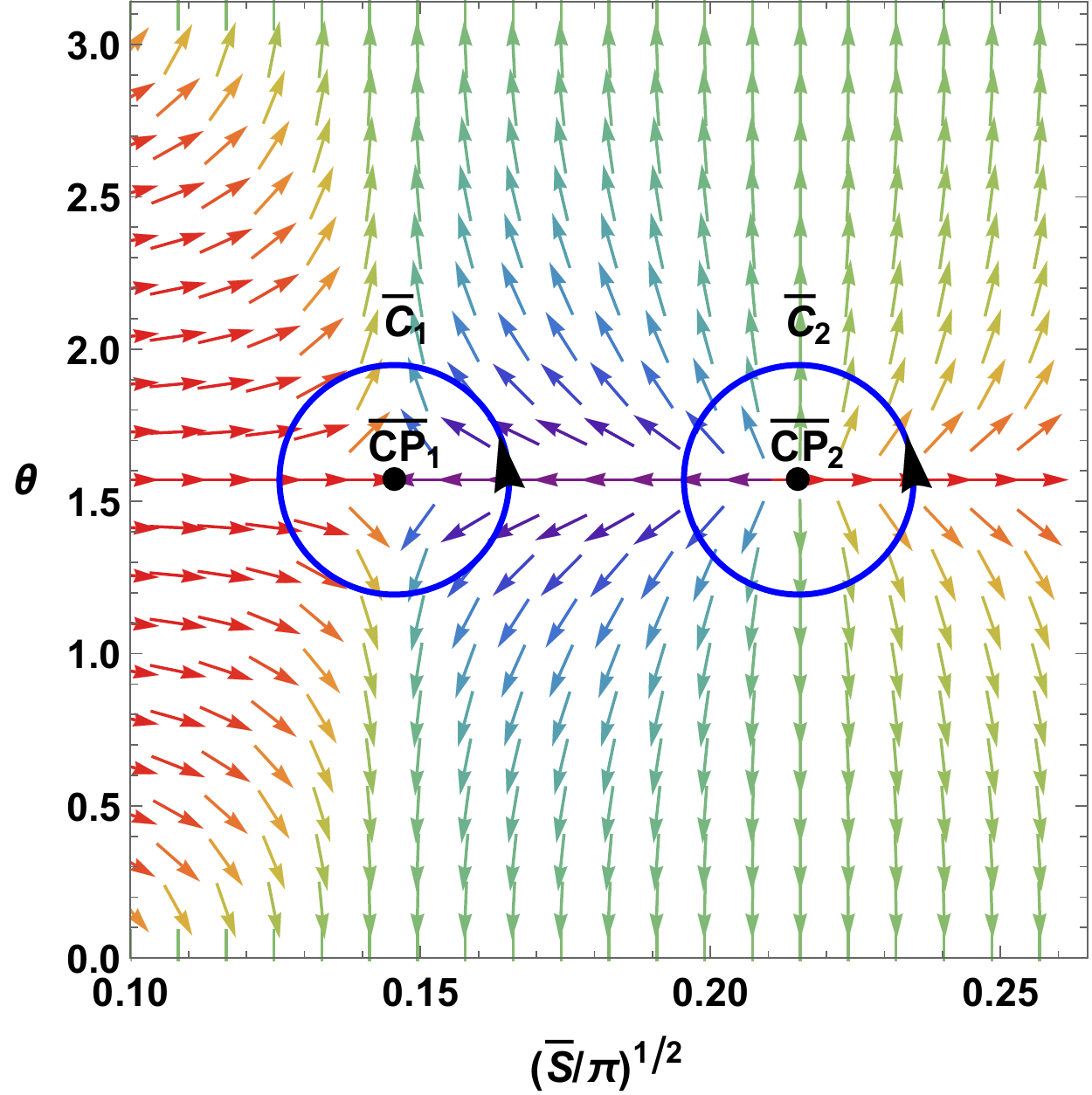}
 		\caption{$\bar{b}>\bar{b}_I$}
 		\label{f33_2}	
 	\end{subfigure}
 	\caption{\footnotesize\it Normalized vector field $n^i$ in
 		the $(\bar{S},\theta)$ plane for (a) $\bar{b}=\bar{b}_I$ with $\bar{Q} = 0.05$, and (b) $\bar{b} = 2.1 > \bar{b}_I$ with $\bar{Q} = 0.064$. }
 	\label{f33}
 \end{figure}

We advocate examining the holographic dual of Born-Infeld-AdS black holes thermodynamics within the framework of the AdS/CFT correspondence. Specifically, we propose investigating the topology within the dual conformal field theory (CFT) through the AdS/CFT correspondence. In alignment with the approach in \cite{Cong:2021jgb} and \cite{Bai:2022vmx}, we treat Newton’s constant $G$ as a parameter that is subjected to variation. Furthermore, to establish a mapping of the thermodynamic variables of AdS black holes to standard thermodynamic variables in the CFT, we redefine these variables in the bulk, accounting for the presence of Newton’s constant as \footnote{We have used the same convention as in \cite{Cong:2021jgb} and \cite{Bai:2022vmx}.}   
\begin{equation}\label{55}
	M' = \dfrac{4\pi}{G} M, \quad Q' = \dfrac{4\pi}{G} Q, \quad S' = \dfrac{4\pi}{G} S , \quad P' = \dfrac{P}{G} = \dfrac{3}{8 \pi G l^2}.
\end{equation}
When Newton’s constant $G$ is assumed to be fixed, the first law in the bulk reads
\begin{equation}\label{56}
	dM' = T' dS' + V'dP'+\phi' dQ' + \mathcal{B}'db,
\end{equation}
such that
\begin{equation}\label{57}
	T' = \dfrac{\partial M'}{\partial S'}, \quad V' = \dfrac{\partial M'}{\partial P'}, \quad \phi' = \dfrac{\partial M'}{\partial Q'}, \quad \mathcal{B}' = \dfrac{\partial M'}{\partial b}.
\end{equation}
But, the thermodynamics in the bulk admits a description of the dual CFT at finite temperature, which can be achieved by including the variation of Newton’s constant $G$, such that the first law becomes
 \begin{equation}\label{58}
 	dM' = T' dS' + V'dP'+\phi' dQ' + \mathcal{B}'db - (M' - T'S'-P'V'-\phi'Q')\dfrac{dG}{G}.
 \end{equation}
To construct a first law with a well-defined thermodynamic correspondence \footnote{The last term in Eq.\eqref{58} does not directly correspond to a thermodynamic interpretation \cite{Bai:2022vmx}.}, we rearrange Eq.\eqref{58} as
 \begin{equation}\label{59}
	dM' = T' dS' - \dfrac{M'}{2} \dfrac{dl^2}{l^2} +\dfrac{\phi'}{l} d(Q'l) + \dfrac{\mathcal{B}'}{l} d(bl) - (M' - T'S'-\phi'Q')\dfrac{d(l^2/G)}{(l^2/G)}.
\end{equation} 
Thus, the holographic dictionary can be identified  now as \cite{Cong:2021jgb,Bai:2022vmx} 
\begin{equation}\label{60}
	\begin{split}
	&E = M', \quad \bar{T} = T', \quad \bar{S} = S', \quad \bar{Q} = Q' l, \quad \bar{\phi} = \dfrac{\phi'}{l},\\
	&\quad \quad \quad  \bar{b} = b l, \quad \bar{\mathcal{B}} = \dfrac{\mathcal{B}'}{l}, \quad \mathcal{V}\sim l^2, \quad C \sim \dfrac{l^2}{G}, 
\end{split}
\end{equation}
where $\mathcal{V}$ and $C$ are the CFT thermodynamic volume and the central charge respectively. Inserting Eq.\eqref{60} into Eq.\eqref{59}, we obtain the thermodynamic first law in the dual CFT of Born-Infeld-AdS black holes
 \begin{equation}\label{61}
	dE = \bar{T} d\bar{S} - p d\mathcal{V} +\bar{\phi} d\bar{Q} + \bar{B} d\bar{b} + \mu dC,
\end{equation} 
with 
 \begin{equation}\label{62}
	p = \dfrac{M'}{2 \mathcal{V}}, \quad \mu =\dfrac{E-\bar{T}\bar{S}-\bar{Q}\bar{\phi}}{C}.
\end{equation}
Such that $p$ is the field theory pressure which satisfies the CFT equation of state
 \begin{equation}\label{63}
	E = 2 p\mathcal{V},
\end{equation}
and $\mu$ stands for the chemical potential associated with the central charge $C$. Therefore, the holographic Euler equation is obtained to be 
\begin{equation}\label{64}
	E = \bar{T}\bar{S} +\bar{Q}\bar{\phi}  + \mu C,
\end{equation}
which is consistent with the on-shell calculation of grand canonical free energy in the holographic field theory \cite{Bai:2022vmx,PhysRevD.77.124048,PhysRevD.74.104032}. 
More generally, when the boundary curvature radius $\mathcal{R}$ is different from the bulk curvature radius $l$, the CFT first law Eq.\eqref{61}, equation of state Eq.\eqref{63}  and Euler equation Eq.\eqref{64} hold true with the following holographic dictionary \cite{  2015JHEP...12..073K, PhysRevD.105.106014, Savonije:2001nd } 
\begin{equation}\label{65}
	\begin{split}
		&E = M'\dfrac{l}{\mathcal{R}}, \quad \bar{T} = T'\dfrac{l}{\mathcal{R}}, \quad \bar{S} = S', \quad \bar{Q} = Q' l, \quad \bar{\phi} = \dfrac{\phi'}{l} \dfrac{l}{\mathcal{R}},\\
		&\quad \quad \quad  \bar{b} = b l, \quad \bar{\mathcal{B}} = \dfrac{\mathcal{B}'}{l} \dfrac{l}{\mathcal{R}}, \quad \mathcal{V}\sim \mathcal{R}^2, \quad C \sim \dfrac{l^2}{G}. 
	\end{split}
\end{equation}
Here, the CFT volume $\mathcal{V}$ and central charge $C$ are now completely independent. The central charge $C$ represents the number of degrees of freedom in CFT and it corresponds to the number of particles in statistical physics\cite{Gao_2022}. Using Eqs.\eqref{65} and  \eqref{57}, the CFT temperature can be evaluated to be
\begin{equation}\label{66}
\bar{T}(\bar{S},\mathcal{V},C,\bar{Q},\bar{b})=\dfrac{1}{4\sqrt{\pi^3C\bar{S}\mathcal{V}}}\left[ \pi C + 3\bar{S}+2\bar{b}^2\bar{S}\left( 1-\sqrt{1+\dfrac{\pi^2\bar{Q}^2}{\bar{b}^2\bar{S}^2}}\right) \right].
\end{equation}
Let us first explore the topology of CFT thermodynamics through the first approach, which involves delineating the topology at critical points. At such a specific point, the temperature of a black hole follows the subsequent relation
\begin{equation}\label{67}
	\left. \dfrac{\partial \bar{T}}{\partial \bar{S}}\right)_{C,\mathcal{V},\bar{Q},\bar{b}}  = 0.
\end{equation}
Next, we introduce a novel thermodynamic function
\begin{equation}\label{68}
	\Phi = \dfrac{1}{\sin(\theta)} \bar{T}^*(\bar{S},\mathcal{V},\bar{Q},\bar{b}),
\end{equation}
in which $\bar{T}^*$ is the black hole temperature obtained via Eq.\eqref{67} upon eliminating $C$. Afterward, we construct a vector field $\phi = (\phi^{\bar{S}},\phi^{\theta})$, such that
\begin{equation}\label{69}
	\phi^{\bar{S}} = \left. \dfrac{\partial \Phi}{\partial \bar{S}}\right)_{\theta,\mathcal{V},\bar{Q},\bar{b}}, \quad \quad \quad \phi^{\theta} = \left. \dfrac{\partial \Phi}{\partial \theta}\right)_{\bar{S},\mathcal{V},\bar{Q},\bar{b}}, 
\end{equation}
 and the normlized vector $n$ still defined as $n= \frac{\phi}{\left| \left| \phi\right| \right| }$. 
 Considering the special case Eq.\eqref{19} in bulk, which corresponds to the existence of an isolated critical point, the condition reads in dual CFT as
 \begin{equation}\label{70}
 \bar{b}_I = \sqrt{\dfrac{3}{2}\left( 1+\sqrt{2}\right)}. 
 \end{equation} 
In this pivotal scenario, an isolated point is evident as illustrated in Fig.\ref{f1}. Consequently, for $\bar{b} < \bar{b}_I$, the absence of a critical point is noted. In contrast to the bulk scenario, the uniqueness of the isolated critical point in dual CFT is not guaranteed. For each electric charge $\bar{Q}$, there exists a corresponding central charge $C_I$ and entropy $\bar{S}_I$, giving rise to an isolated critical point. Importantly, the existence of this critical point remains independent of the CFT volume $\mathcal{V}$.  We depict in Fig.\ref{f33} the normalized vector field in the $(\bar{S}, \theta)$ plane for $\bar{b} = \bar{b}_I$ and $\bar{b} = 2.1> \bar{b}_I$ situations,
 we observe an isolated critical point, $\overline{ICP}$, and two critical points, $\overline{CP}_1$ and $\overline{CP}_2$ that merge as the parameter $\overline{b}$ increases.

\begin{figure}[!ht]
	\centering 
	\begin{subfigure}[h]{0.48\textwidth}
		\centering \includegraphics[scale=0.6]{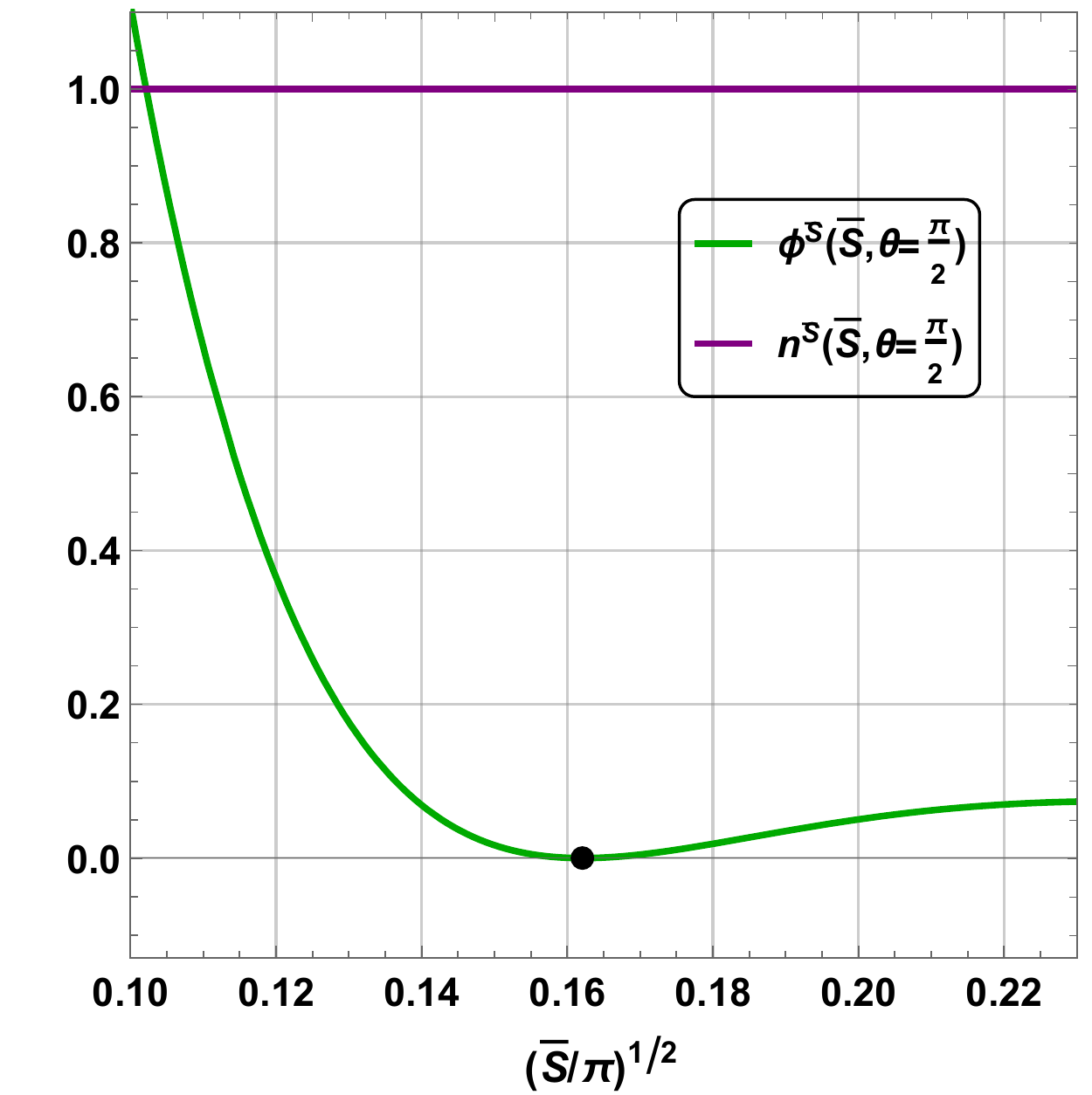}
		\caption{}
		\label{f34_1}
	\end{subfigure}
	\hspace{1pt}	
	\begin{subfigure}[h]{0.48\textwidth}
		\centering \includegraphics[scale=0.6]{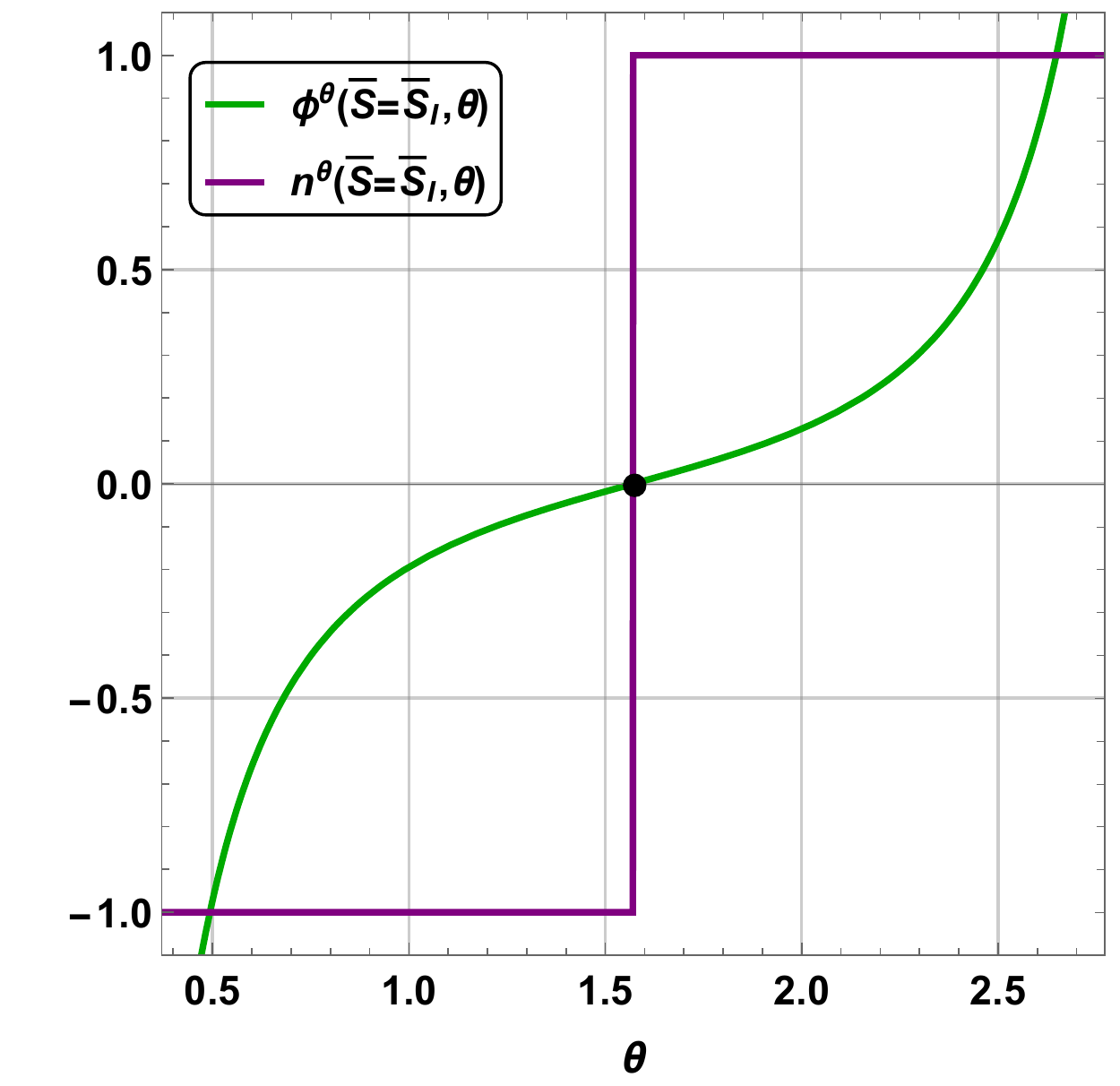}
		\caption{}
		\label{f34_2}	
	\end{subfigure}
	\caption{\footnotesize\it (a) Vector components $\phi^{\bar{S}}$ and $n^{\bar{S}}$ as a function of $\bar{S}$ for $\theta = \dfrac{\pi}{2}$. (b) Vector components $\phi^\theta$ and $n^\theta$ as a function of $\theta$ for $\bar{S} = \bar{S}_I = 0.0825442$. $\bar{Q} = 0.064$ and $\mathcal{V}=1$. }
	\label{f34}
\end{figure}

In order to examine the situation where $\bar{b} = \bar{b}_I$, indicating the presence of an isolated critical point, we depict the vector field component $\phi^{\bar{S}}$ and its normalized counterpart $n^{\bar{S}}$ in Fig.\ref{f34_1}. These visualizations are presented as functions of CFT entropy $\bar{S}$ with $\theta = \frac{\pi}{2}$ for the specific case of $\bar{b} = \bar{b}_I$. 


The evident observation is that the vector field component $\phi^{\bar{S}}$ precisely vanishes at $\bar{S}=\bar{S}_I$ (marked by a black dot), maintaining a consistent sign and establishing itself as a semi-stable point. Conversely, the normalized vector field component $n^{\bar{S}}$ does not vanish. This is attributed to the simultaneous approach of both $\phi^{\bar{S}}$ and $\left| \left| \phi\right| \right| $ towards zero while maintaining the constant and positive ratio $n^{\bar{S}} = \frac{\phi^{\bar{S}}}{\left| \left| \phi\right| \right| }$ ($n^{\bar{S}} (\bar{S},\frac{\pi}{2}) = +1$). Consequently, $\phi^{\bar{S}}$ vanishes at the isolated critical point.
 Additionally, from the panel Fig.\ref{f34_1}, it is evident that $\phi^\theta$ also vanishes at the isolated critical point $(\bar{S} = \bar{S}_I, \theta = \frac{\pi}{2})$, whereas $n^\theta$ experiences a discontinuity. To compute the topological charge $\mathcal{Q}(\bar{ICP})$, we consider the contour $C_0$ in Fig.\ref{f33_1} and employ Eq.\eqref{15}, yielding the following:
\begin{equation}\label{71}
	\mathcal{Q}(\overline{ICP}) = 0,
\end{equation} 
which is in agreement with the result found in the bulk.

\begin{figure}[!ht]
	\centering 
	\begin{subfigure}[h]{0.48\textwidth}
		\centering \includegraphics[scale=0.6]{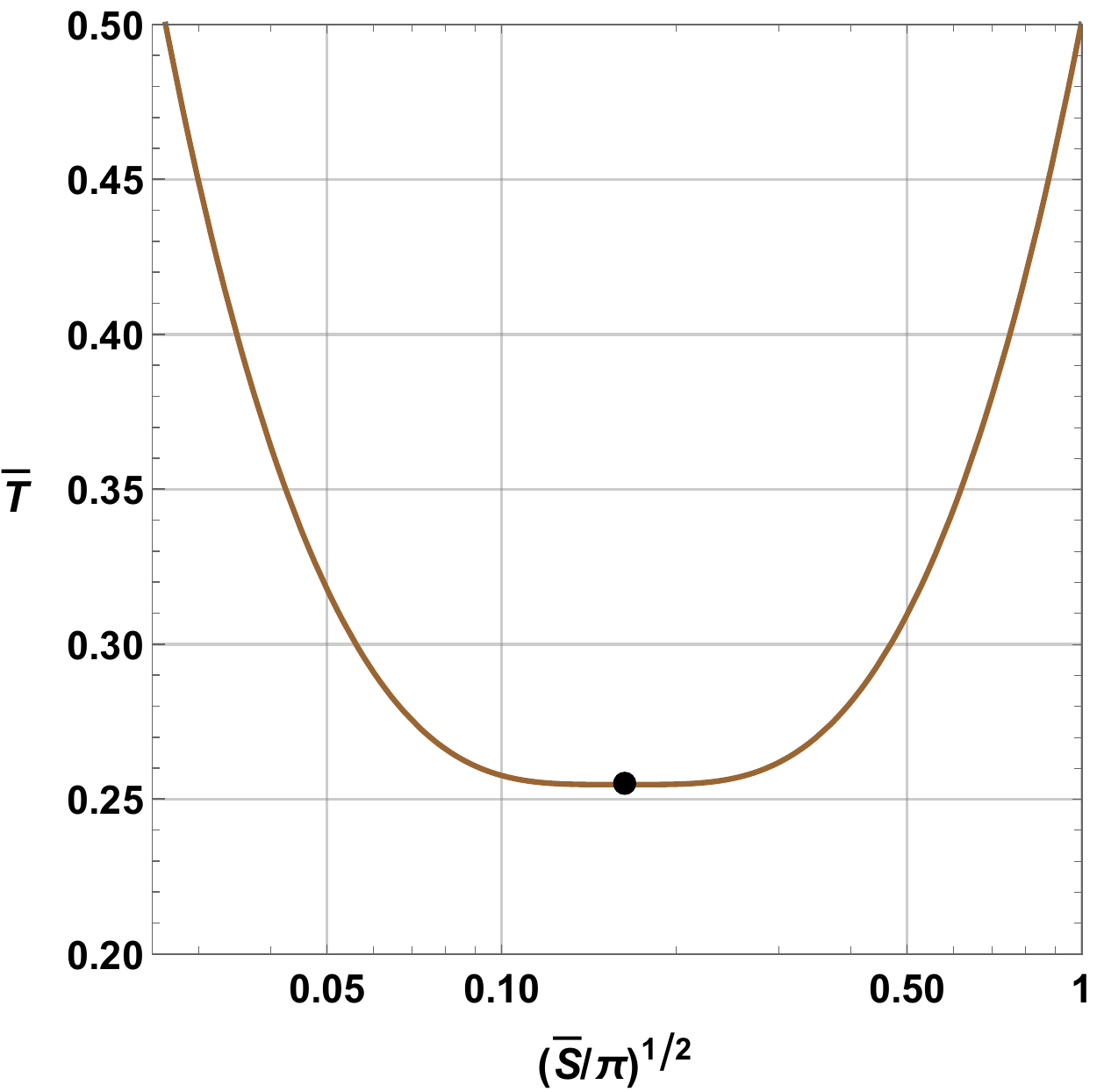}
		\caption{}
		\label{f35_1}
	\end{subfigure}
	\hspace{1pt}	
	\begin{subfigure}[h]{0.48\textwidth}
		\centering \includegraphics[scale=0.6]{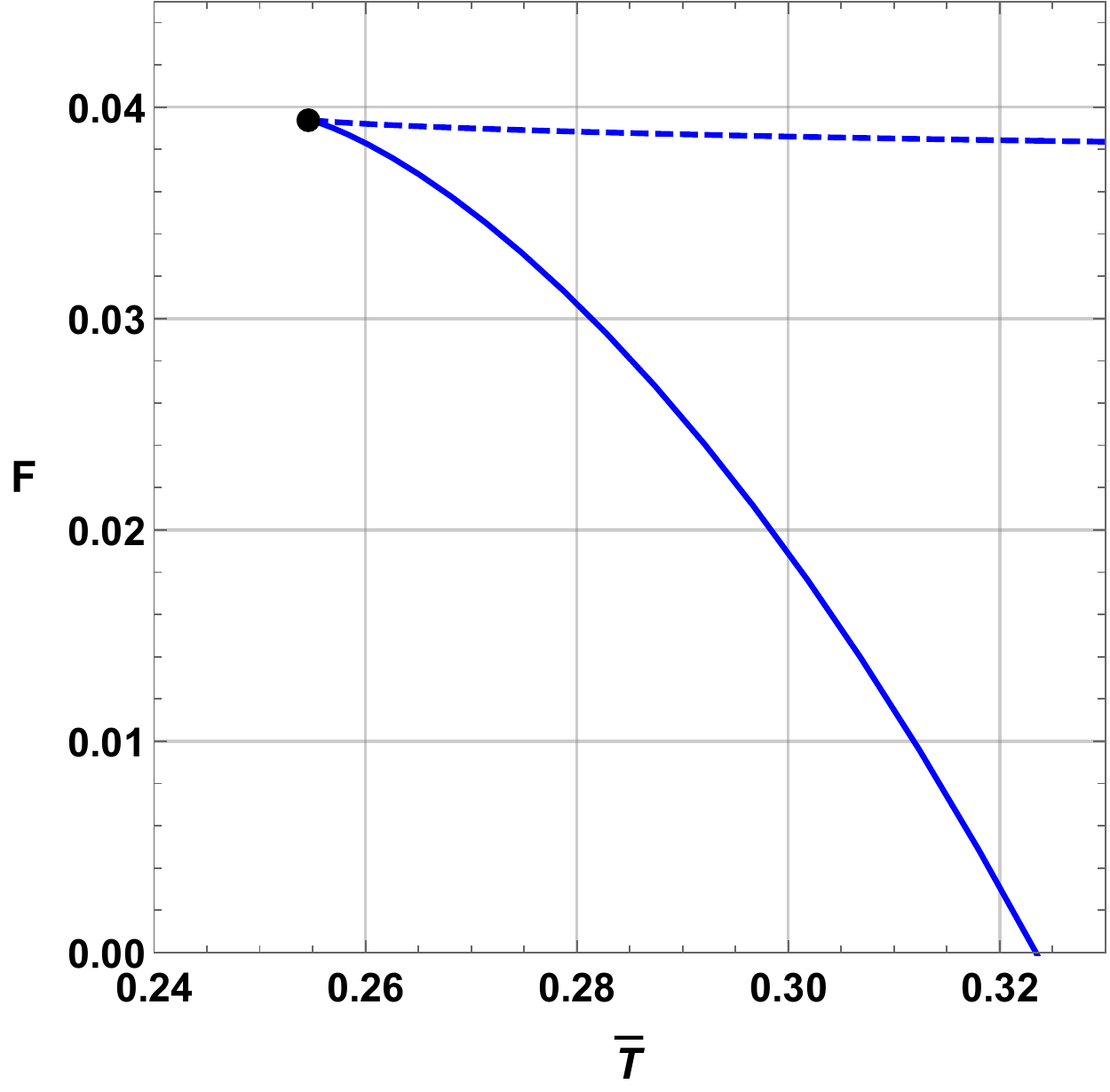}
		\caption{}
		\label{f35_2}	
	\end{subfigure}
	\caption{\footnotesize\it (a) CFT temperature as a function of entropy $\bar{S}$ and (b) and CFT free energy as a function of temperature with $\bar{b} = \bar{b}_I$, $\bar{Q} = 0.05$, $C = C_I = 0.269122$ and $\mathcal{V} = 1$. }
	\label{f35}
\end{figure}

The thermodynamics governing the isolated critical point mirrors that of its counterpart in the bulk. Specifically, in Fig.\ref{f35}, we depict the CFT temperature and free energy, where $F = E - \bar{T}\bar{S}$ for this scenario. Notably, the free energy exhibits a cusp, indicative of a transition between an unstable phase and a locally stable one. Additionally, the CFT temperature reveals an inflection point corresponding to its minimum, which aligns with the location of the isolated critical point.

\begin{figure}[!ht]
	\centering 
	\begin{subfigure}[h]{0.48\textwidth}
		\centering \includegraphics[scale=0.6]{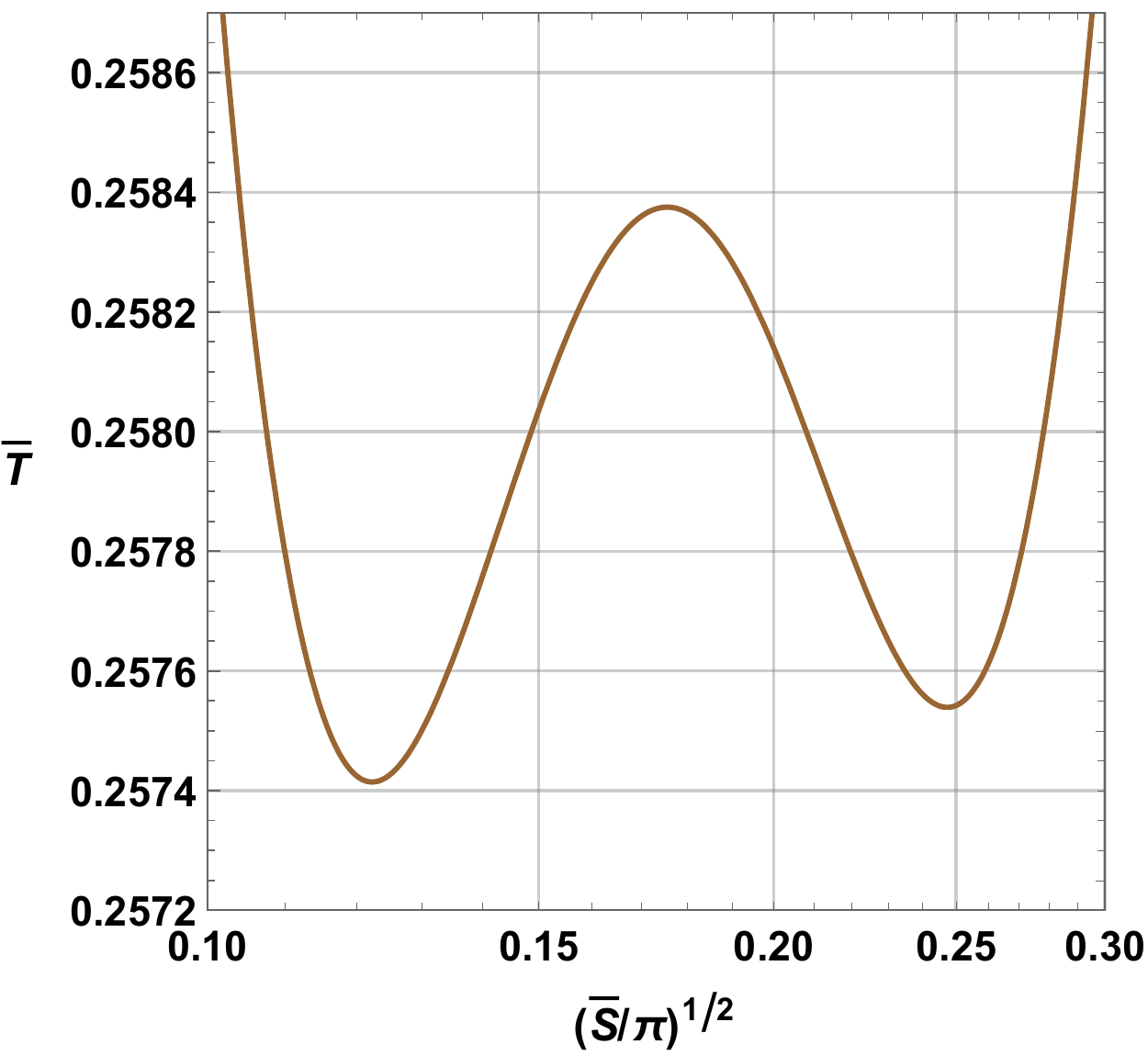}
		\caption{}
		\label{f36_1}
	\end{subfigure}
	\hspace{1pt}	
	\begin{subfigure}[h]{0.48\textwidth}
		\centering \includegraphics[scale=0.6]{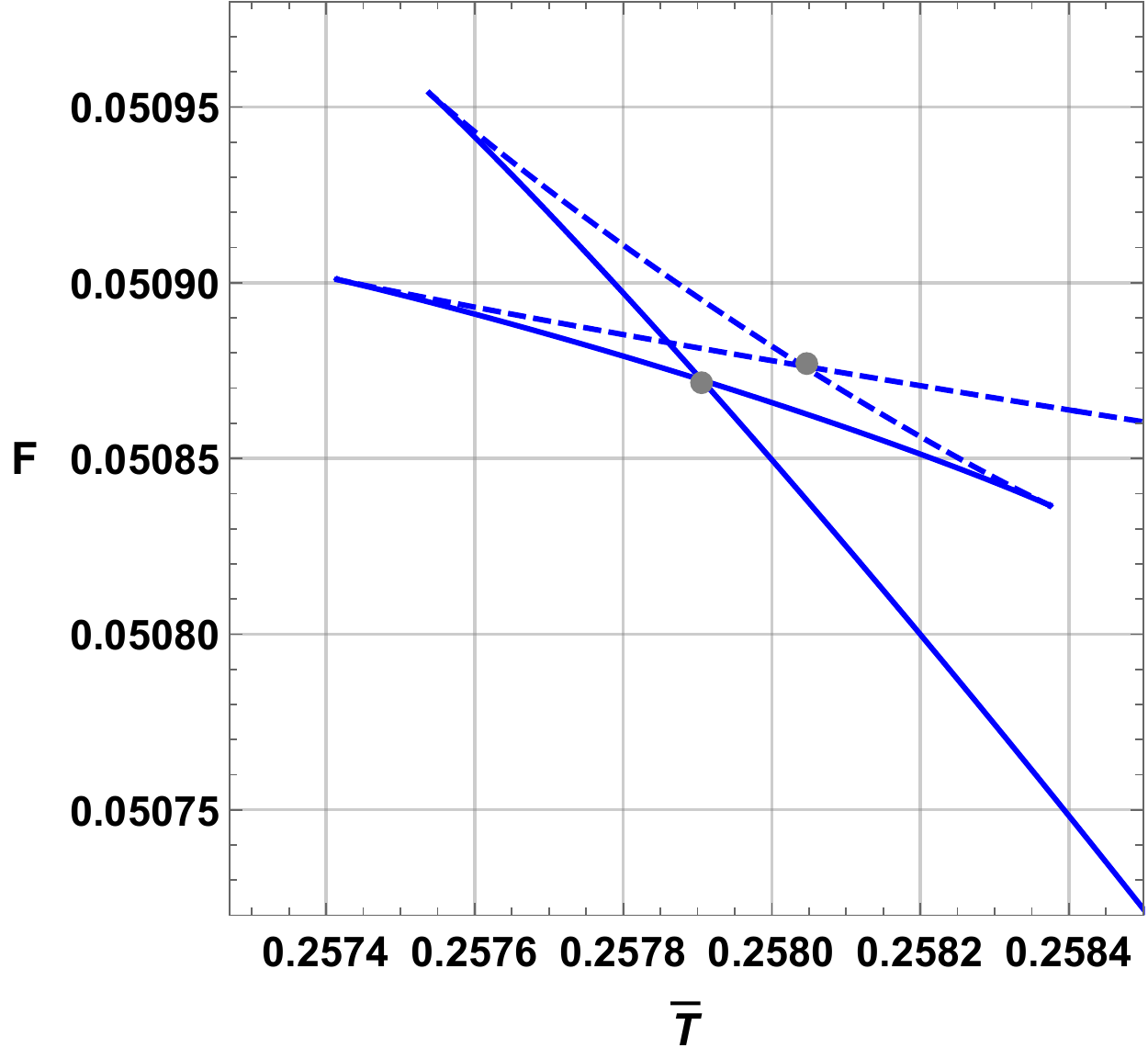}
		\caption{}
		\label{f36_2}	
	\end{subfigure}
	\caption{\footnotesize\it (a) CFT temperature as a function of entropy $\bar{S}$ and (b) and CFT free Gibbs energy as a function of temperature near the isolated critical point $\overline{ICP}$ with $\bar{b}= 2.1$, $Q = 0.064$, $C=0.36$ and $\mathcal{V} = 1$. }
	\label{f36}
\end{figure}

To elucidate the formation of the isolated critical point, we present in Fig.\ref{f36} the CFT temperature and free energy, utilizing parameters in close proximity to those characterizing the isolated critical point.
 Notably, two swallowtails emerge in the free energy-temperature diagram, signifying a phase transition between two unstable phases (represented by dashed curves) associated with a decreasing temperature concerning $\bar{S}$. The uniqueness of this phase transition is underscored by its occurrence between two unstable phases, as indicated by the upper gray dot in Fig.\ref{f36_2}. 
Furthermore, we observe a first-order phase transition between two stable phases, depicted by solid curves, corresponding to an increasing temperature with respect to $\bar{S}$. This conventional phase transition is marked by the lower gray dot in Fig.\ref{f36_2}. As the parameter $\bar{b}$ decreases to its critical value $\bar{S}_I$, the two critical points—conventional and unconventional—coincide, giving rise to the previously observed isolated critical point. Hence, the formation of the isolated critical point can be conceptualized as an annihilation process between a conventional critical point and an unconventional one.

Returning to Fig. \ref{f33}, we discern the emergence of two critical points when $\bar{b}>\bar{b}_I$. This appearance corresponds to the creation of a vortex/anti-vortex pair, as in the bulk. Utilizing \eqref{15} and considering the two contours $\bar{C}_1$ and $\bar{C}_2$, the topological charges associated with these two points are evaluated to be 
\begin{equation}\label{72}
	\mathcal{Q}(\overline{CP}_1) = -1, \quad \quad \mathcal{Q}(\overline{CP}_2) = +1.
\end{equation} 
 Apparently, the obtained result shows some contrast of our observations in the bulk. Specifically, the positive charge corresponds to the conventional critical point $\overline{CP}_2$, while the negative charge characterizes the unconventional critical point $\overline{CP}_1$. Furthermore, upon comparing Fig. \ref{f33} and Fig. \ref{f1}, it becomes apparent that the flow of the vector field is reversed. Whereas, analogous to the bulk, the overall topological charge of the entire system remains null
\begin{equation}\label{73}
	\mathcal{Q} = \mathcal{Q}(\overline{CP}_1) + \mathcal{Q}(\overline{CP}_2) = 0,
\end{equation} 
and which is the same as Eq.\eqref{71} when $\bar{b}=\bar{b}_I$. That is to say, the total topological charge is preserved and there is no topological transition. Hence, the two black holes ( $\bar{b}=\bar{b}_I$ and $\bar{b}>\bar{b}_I$) belong to the same topological class.

\begin{figure}[!ht]
	\centering 
	\begin{subfigure}[h]{0.48\textwidth}
		\centering \includegraphics[scale=0.6]{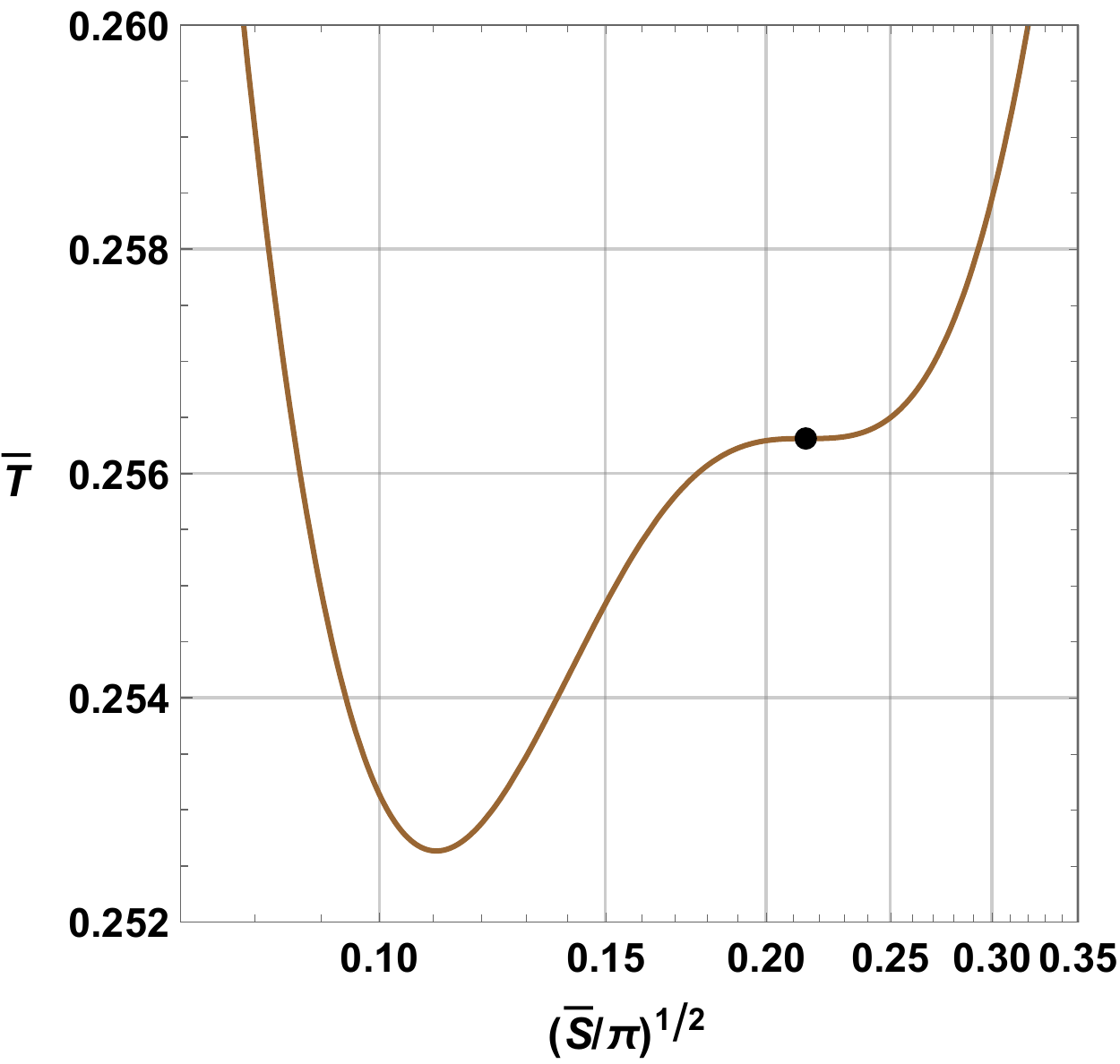}
		\caption{}
		\label{f37_1}
	\end{subfigure}
	\hspace{1pt}	
	\begin{subfigure}[h]{0.48\textwidth}
		\centering \includegraphics[scale=0.6]{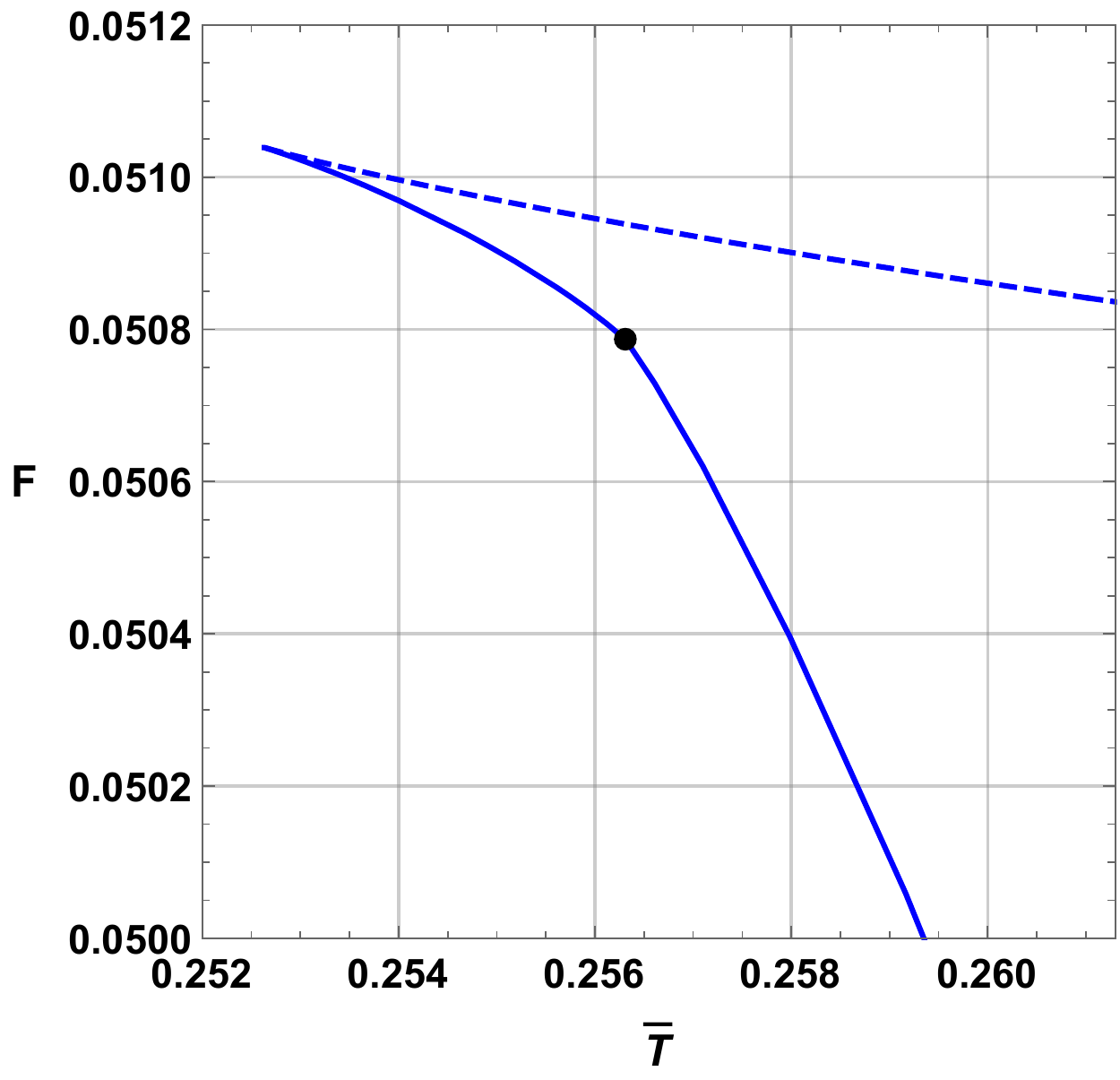}
		\caption{}
		\label{f37_2}	
	\end{subfigure}
	\caption{\footnotesize\it (a) CFT temperature as a function of entropy $\bar{S}$ and (b) and free energy as a function of temperature with $\bar{b}= 2.1$, $\bar{Q} = 0.064$, $C =C_{\overline{CP}_2} = 0.353953$ and $\mathcal{V} = 1$. }
	\label{f37}
\end{figure}

\begin{figure}[!ht]
	\centering 
	\begin{subfigure}[h]{0.48\textwidth}
		\centering \includegraphics[scale=0.6]{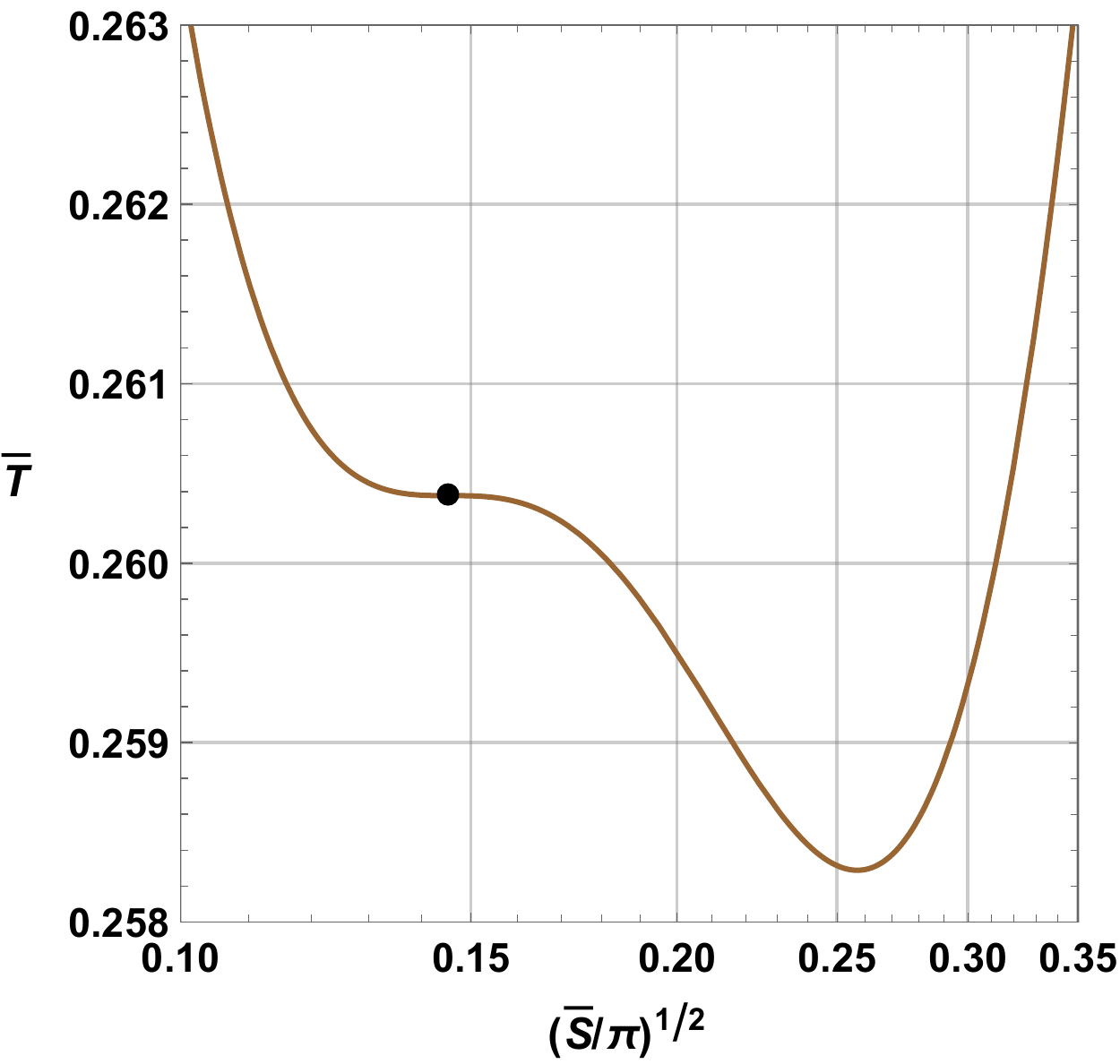}
		\caption{}
		\label{f38_1}
	\end{subfigure}
	\hspace{1pt}	
	\begin{subfigure}[h]{0.48\textwidth}
		\centering \includegraphics[scale=0.6]{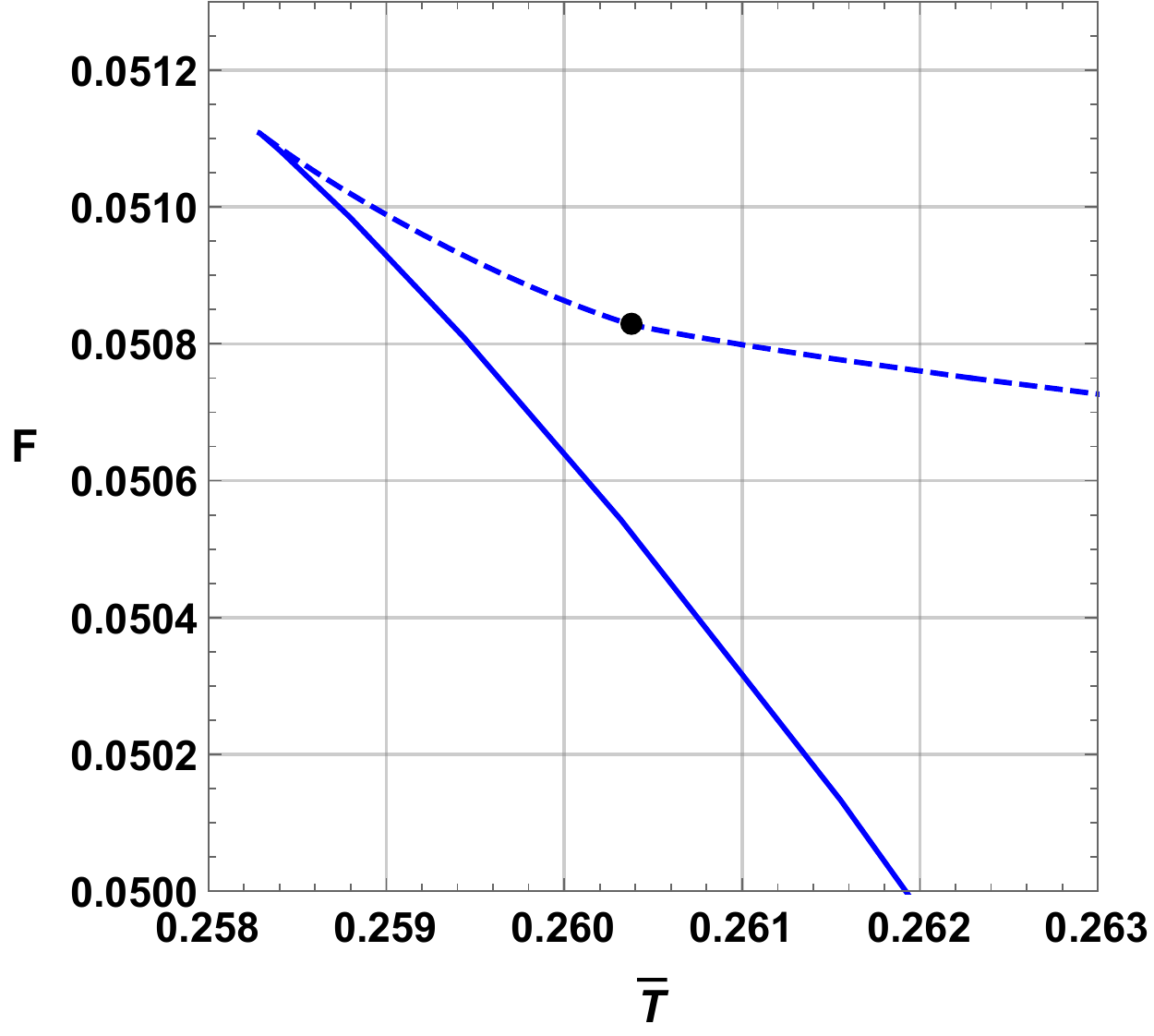}
		\caption{}
		\label{f38_2}	
	\end{subfigure}
	\caption{\footnotesize\it  (a) CFT temperature as a function of entropy $\bar{S}$ and (b) and free energy as a function of temperature with $\bar{b}= 2.1$, $\bar{Q} = 0.064$, $C =C_{\overline{CP}_1} = 0.364469$ and $\mathcal{V} = 1$. }
	\label{f38}
\end{figure}

Examining the thermodynamics associated with these two critical points, we depict the CFT temperature and free energy for the conventional critical point $\overline{CP}_2$ in Fig. \ref{f37}. Notably, we identify a secondary phase transition between two stable phases, marked by an inflection point in the right branch of the temperature curve, where it exhibits an increasing trend concerning entropy. The critical point $\overline{CP}_2$ is denoted by the black dot.
Regarding the unconventional critical point $\overline{CP}_1$, we illustrate the CFT temperature and free energy in Fig. \ref{f38} 
Notably, we identify an inflection point in the left branch of the temperature curve, indicating a decreasing trend concerning entropy. The critical point $\overline{CP}_1$ is marked by the black dot. This inflection point serves as a signature of the existence of a second-order phase transition. However, in the free energy-temperature diagram, we observe that this transition occurs between two unstable phases.

\begin{figure}[!ht]
	\centering 
 \includegraphics[scale=0.6]{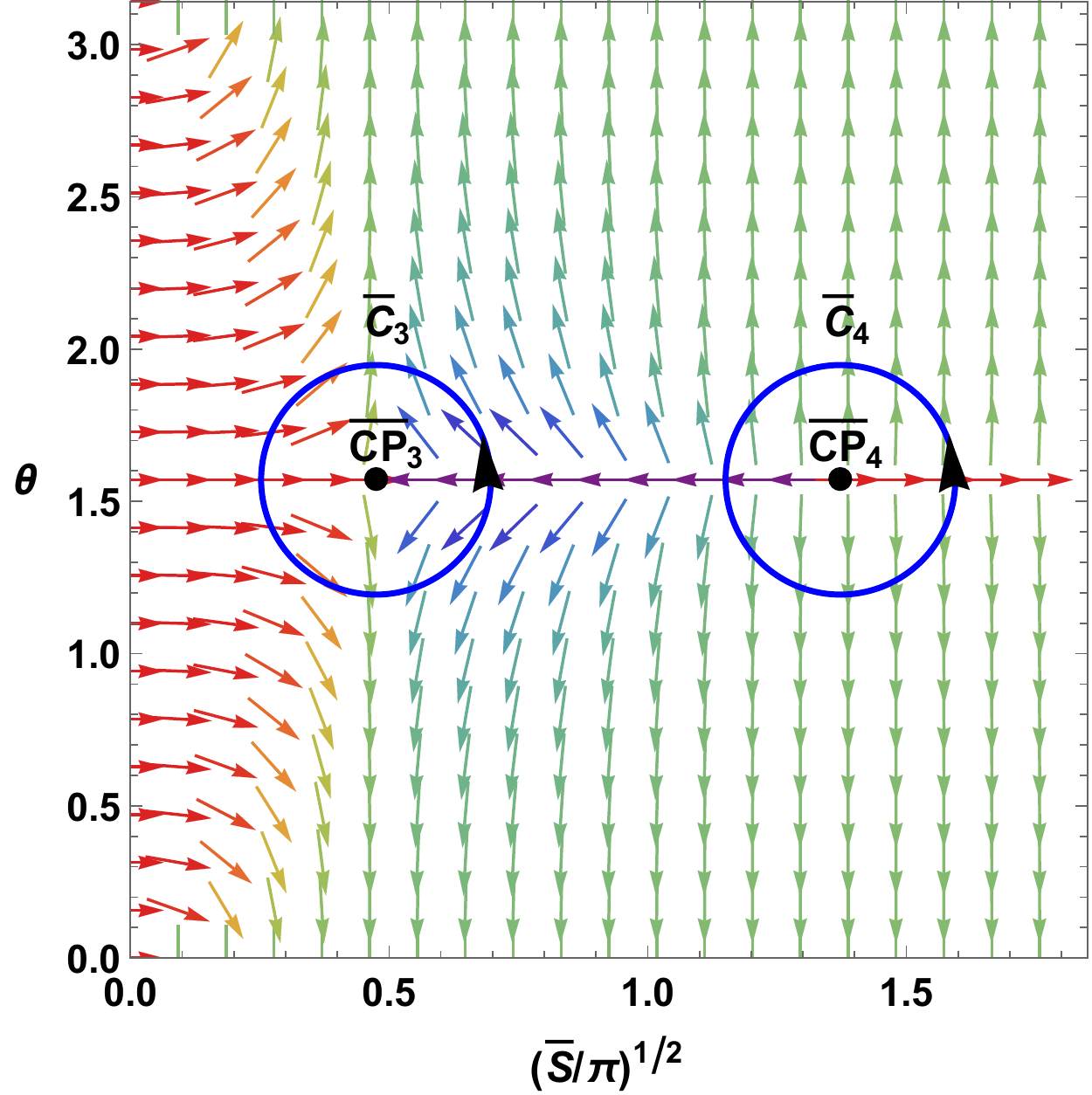}
	\caption{\footnotesize\it Normalized vector field $n^i$ in
		the $(\bar{S},\theta)$ plane for $\bar{b} = 4$ and $\bar{Q} = 2$ with $\mathcal{V}=1$. }
	\label{f39}
\end{figure}

Using, Eq.\eqref{23}, the topological transition in dual CFT would occurs when 
\begin{equation}\label{74}
	\left( \dfrac{\bar{Q}}{C} \right)_m = \dfrac{1}{2\bar{b}}.
\end{equation} 
In essence, the topology of the dual CFT is governed by $\frac{\bar{Q}}{C}$ for a fixed $\bar{b},$ whereas in the bulk, it is solely controlled by the electric charge. To observe a topological transition in the dual CFT, it is imperative to maintain $\frac{\bar{Q}}{C}$ at a constant value. Regrettably, such a transition cannot be discerned through this initial approach. Specifically, for a fixed electric charge, the central charge implicitly varies, causing the two critical points to belong to distinct topological classes.
 To illustrate this point, consider Fig. \ref{f39}, where we plot the normalized vector field in the $(\bar{S}, \theta)$ plane for $\bar{b} = 4$ and $\bar{Q} = 2$. Particularly, two critical points, $\overline{CP}_3$ and $\overline{CP}_4$, akin to those in Fig. \ref{f33_2}, emerge. These correspond to unconventional and conventional critical points, respectively. However, the global topological charge in this scenario is zero, indicating that there is no topological transition between this state and the one depicted in Fig. \ref{f33_2}. It is crucial to emphasize that despite the apparent similarity, these two situations are distinct. In the first situation shown in Fig. \ref{f33_2}, the two critical points belong to the same topological class, ensuring
 \begin{equation}\label{75}
	\left( \dfrac{\bar{Q}}{C} \right)_{\overline{CP}_1} < \dfrac{1}{2\bar{b}}, \quad \quad \left( \dfrac{\bar{Q}}{C} \right)_{\overline{CP}_2} < \dfrac{1}{2\bar{b}}.
\end{equation}
 That is to say, the system is topologically equivalent AdS-Schwarzchild one. Whereas, in the second situation depicted in Fig.\ref{f39} the two critical points do not belong to the same topological class such that
\begin{equation}\label{76}
	\left( \dfrac{\bar{Q}}{C} \right)_{\overline{CP}_3} < \dfrac{1}{2\bar{b}}, \quad \quad \left( \dfrac{\bar{Q}}{C} \right)_{\overline{CP}_4} > \dfrac{1}{2\bar{b}},
\end{equation}  
where $C_{\overline{CP}_3} = 0.820045 $ and $C_{\overline{CP}_4}=11.7734$.
 
\begin{figure}[!ht]
	\centering 
	\begin{subfigure}[h]{0.48\textwidth}
		\centering \includegraphics[scale=0.6]{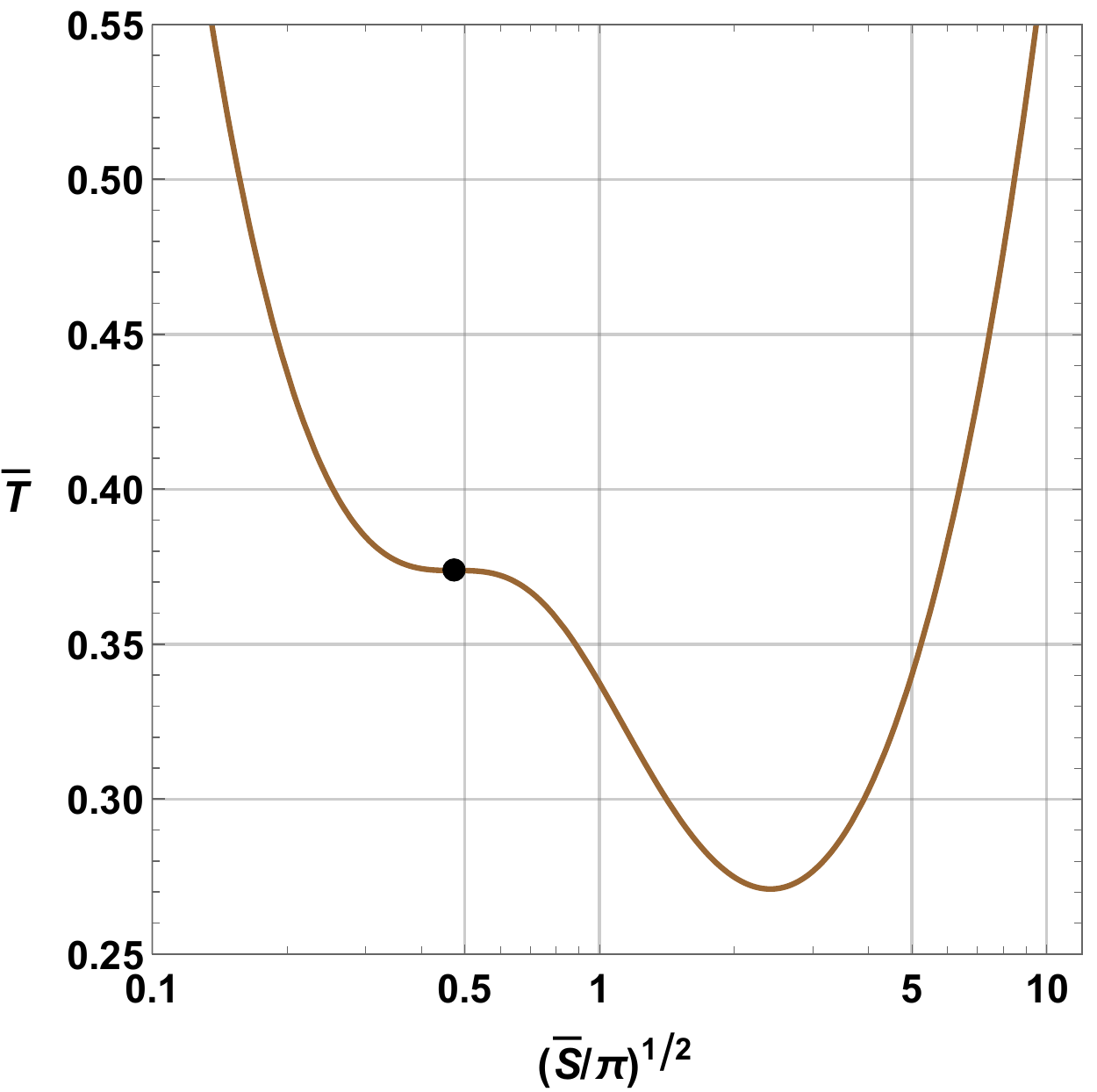}
		\caption{}
		\label{f40_1}
	\end{subfigure}
	\hspace{1pt}	
	\begin{subfigure}[h]{0.48\textwidth}
		\centering \includegraphics[scale=0.6]{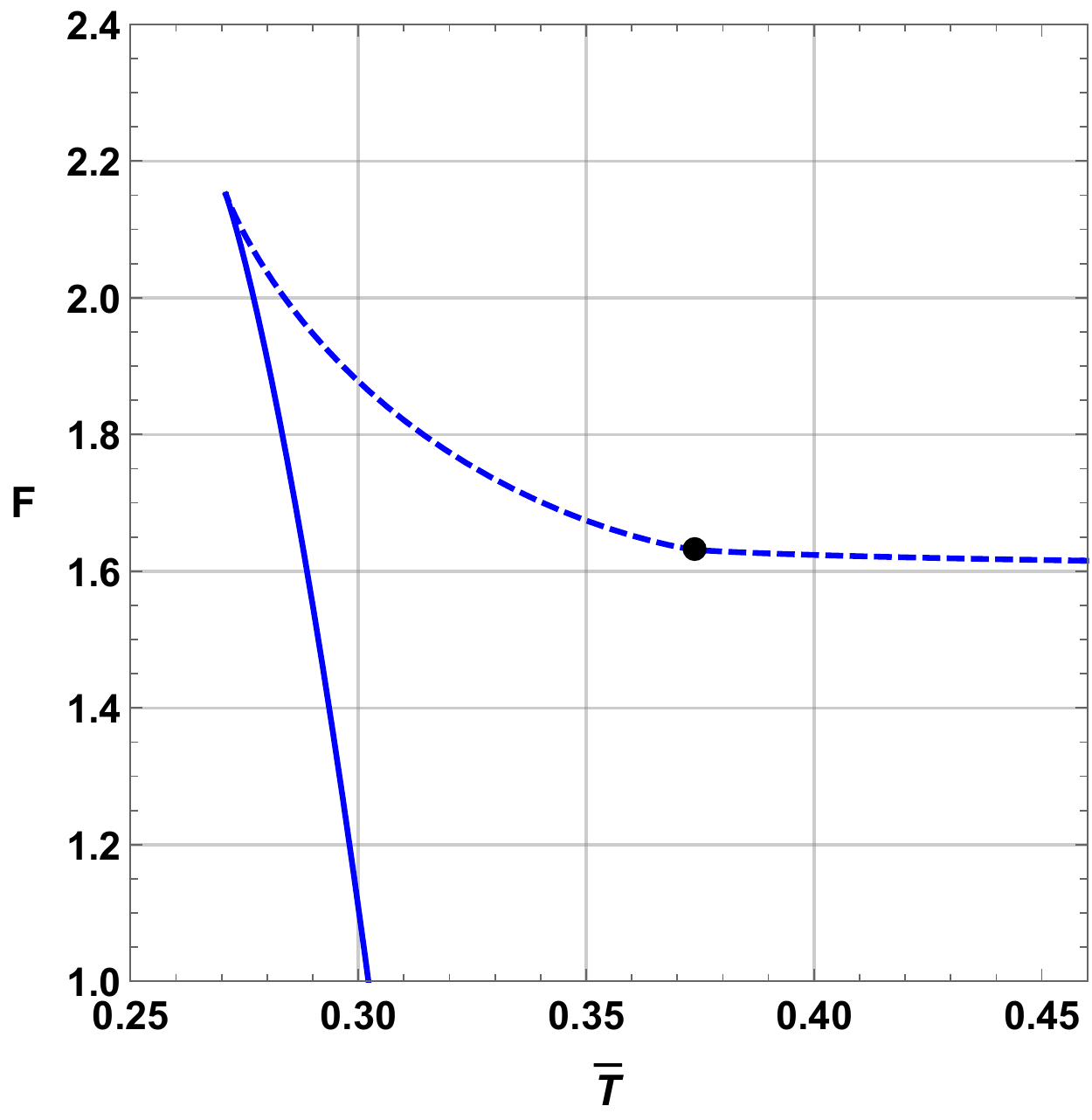}
		\caption{}
		\label{f40_2}	
	\end{subfigure}
	\caption{\footnotesize\it (a) CFT temperature as a function of entropy $\bar{S}$. (b) free energy as a function of temperature with $\bar{b}= 4$, $\bar{Q} = 2$, $C =C_{\overline{CP}_3} = 0.820045$ and $\mathcal{V} = 1$.}
	\label{f40}
\end{figure}

\begin{figure}[!ht]
	\centering 
	\begin{subfigure}[h]{0.48\textwidth}
		\centering \includegraphics[scale=0.6]{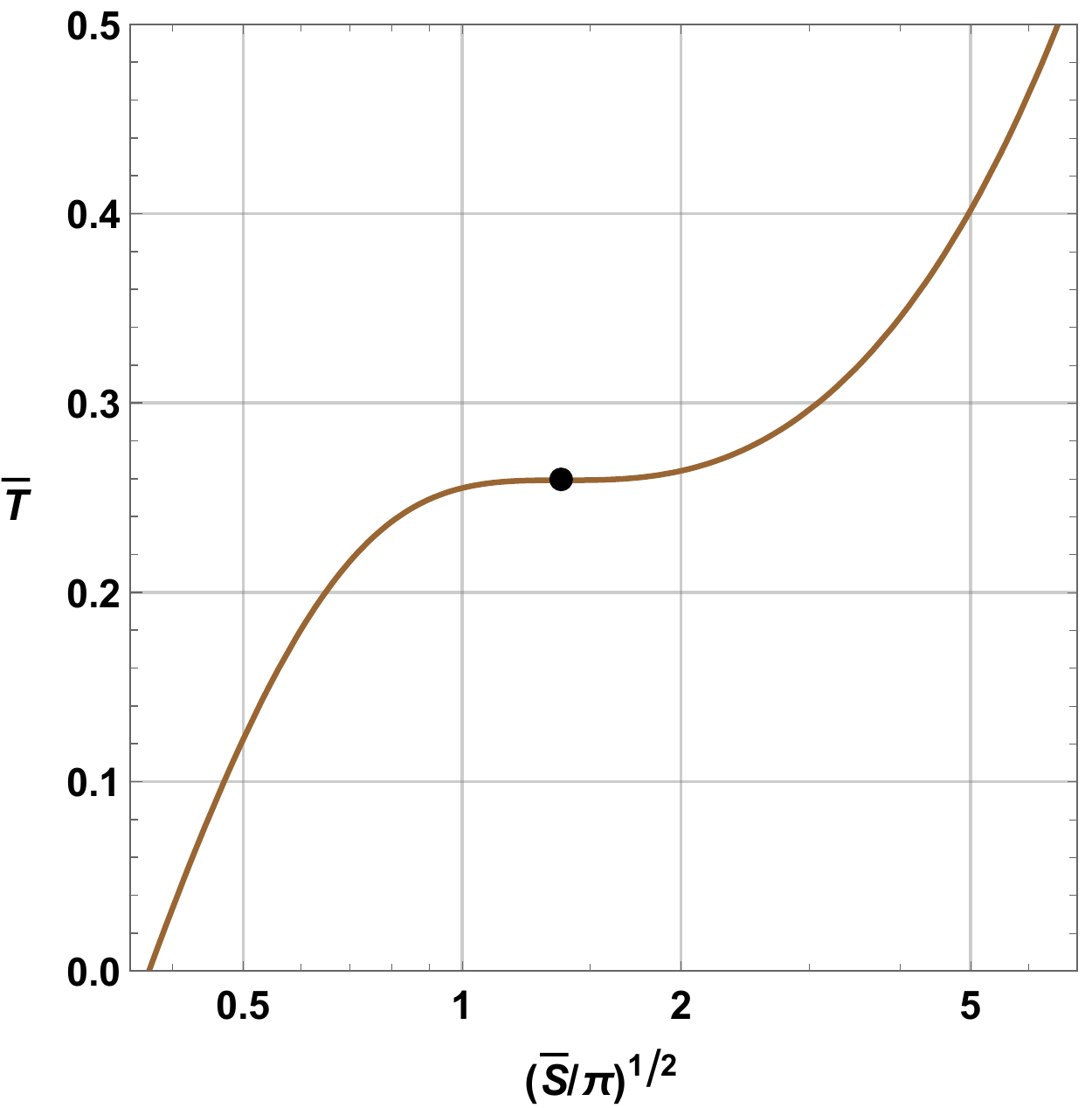}
		\caption{}
		\label{f41_1}
	\end{subfigure}
	\hspace{1pt}	
	\begin{subfigure}[h]{0.48\textwidth}
		\centering \includegraphics[scale=0.6]{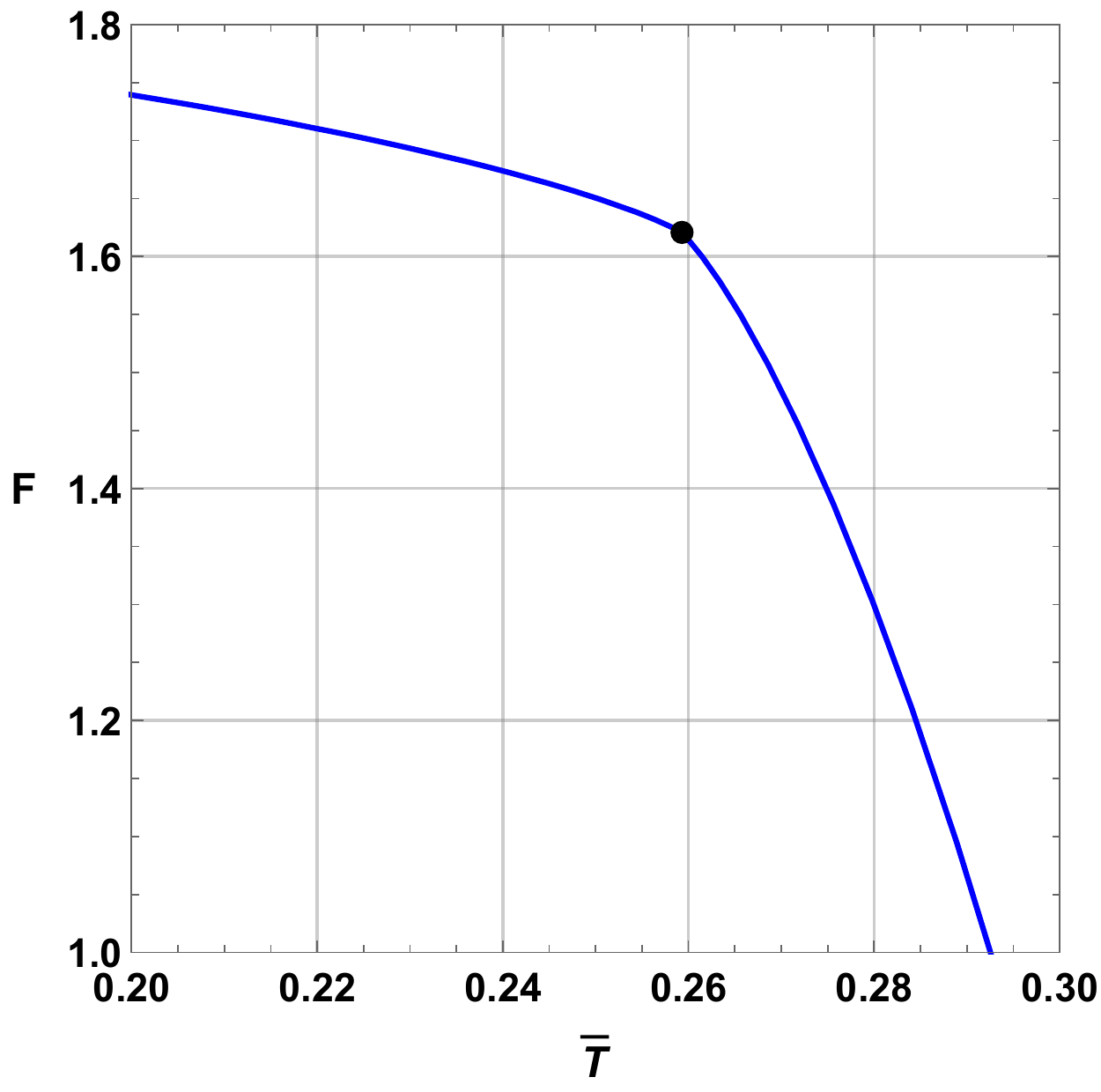}
		\caption{}
		\label{f41_2}	
	\end{subfigure}
	\caption{\footnotesize\it (a) CFT temperature as a function of entropy $\bar{S}$. (b) free energy as a function of temperature with $\bar{b}= 4$, $\bar{Q} = 2$, $C =C_{\overline{CP}_4} = 11.7734$ and $\mathcal{V} = 1$. }
	\label{f41}
\end{figure}

To demonstrate that the two critical points $\overline{CP}_3$ and $\overline{CP}_4$ do not fall within the same topological class, we present in Fig. \ref{f40} the CFT temperature and free energy corresponding to the unconventional critical point $\overline{CP}_3$.
It's observed that the situation depicted in Fig. \ref{f40} corresponds to a system topologically equivalent to AdS-Schwarzschild, where the temperature exhibits a minimum and cannot reach zero. Additionally, in Fig. \ref{f41}, we present the CFT temperature and free energy corresponding to the conventional critical point $\overline{CP}_4$.
Here, the system exhibits a different behavior compared to the first situation. It is topologically equivalent to AdS-Reissner-Nordström, featuring the usual second-order phase transition. The temperature is an increasing function with respect to entropy, reaching zero at a specific entropy value corresponding to an extremal state.
Therefore, it can be concluded that this approach is inadequate for studying the topology of the dual CFT, as the topological properties are contingent on $\frac{\bar{Q}}{C}$ rather than $\bar{Q}$ alone. An alternative approach is required to investigate the topology of the dual CFT while maintaining a fixed $\frac{\bar{Q}}{C}$. To address this, we must revisit the second approach, which centers on the free energy $F$ rather than temperature.

\begin{figure}[!ht]
	\centering 
	\begin{subfigure}[h]{0.48\textwidth}
		\centering \includegraphics[scale=0.6]{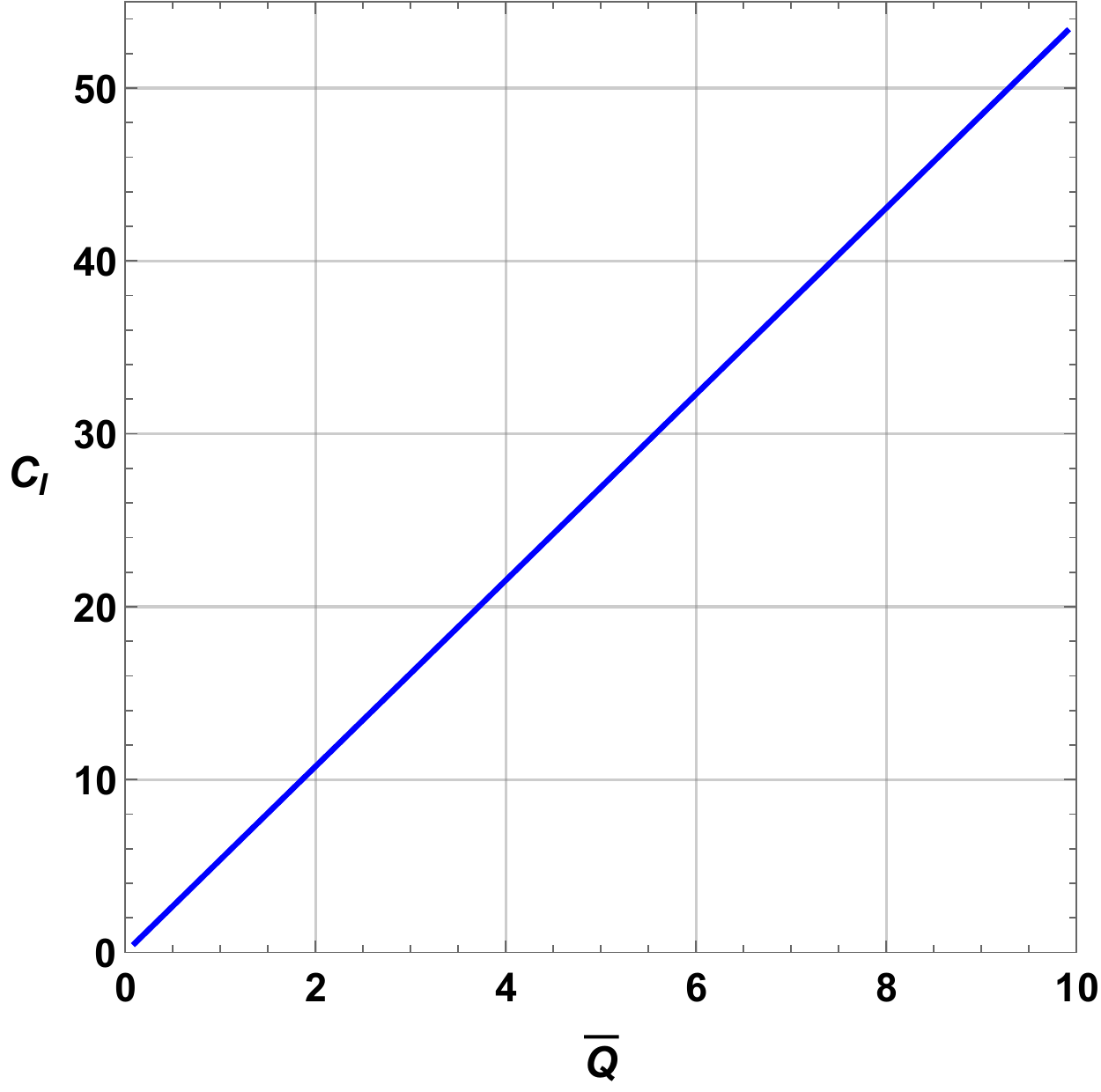}
		\caption{}
		\label{f42_1}
	\end{subfigure}
	\hspace{1pt}	
	\begin{subfigure}[h]{0.48\textwidth}
		\centering \includegraphics[scale=0.6]{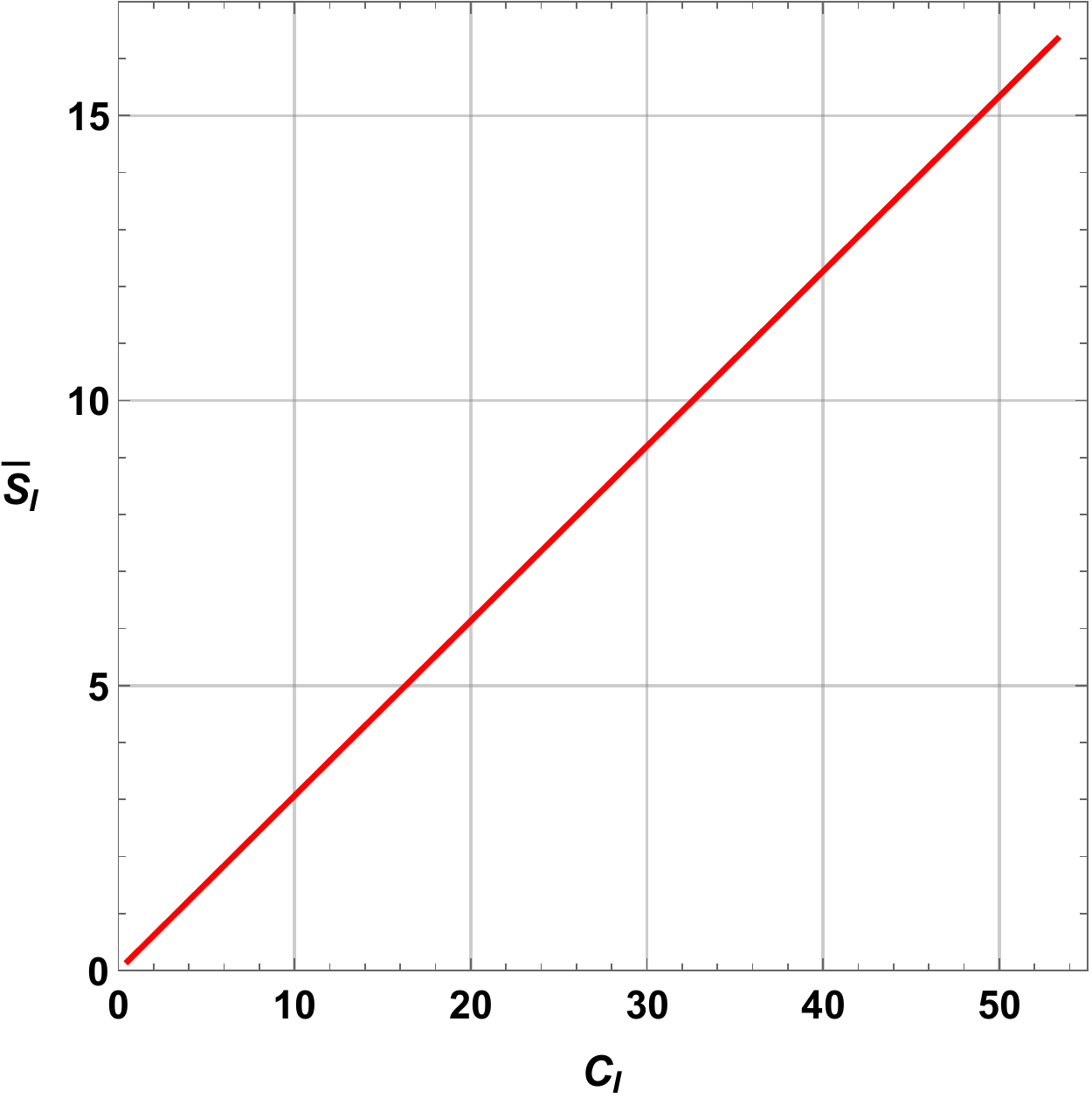}
		\caption{}
		\label{f42_2}	
	\end{subfigure}
	\caption{\footnotesize\it (a) Central charge as a function of electric charge and (b) entropy in terms of central charge at isolated critical situations for $\bar{b}= \bar{b}_I$. }
	\label{f42}
\end{figure}

Before delving further, let's revisit the issue of isolated critical points in the dual CFT. As previously mentioned, isolated critical points are not unique, and for any given electric charge, there can be an isolated critical point. To provide some insights into the characteristics of this isolated critical behavior, we depict in Fig. \ref{f42} the central charge $C_I$ as a function of electric charge and the corresponding entropy $\bar{S}_I$ in terms of central charge. 
Such a figure reveals that the central charge exhibits a linear dependence on electric charge in isolated critical situations. Consequently, the ratio $\left( \frac{\bar{Q}}{C} \right)_I = 0.18579$ remains constant and unique, serving as a distinctive characteristic of the isolated critical point. In addition, the entropy also displays a linear dependence on central charge in isolated critical situations. As a result, the ratio $\left( \frac{\bar{S}}{C} \right)_I = 0.306717$ remains constant and unique, providing another characteristic feature of the isolated critical point. 
Thus, to study the topology of dual CFT we should make a change of variables such that the isolated critical point is unique. We define two new variables as follows 
\begin{equation}\label{77}
	\bar{q} = \dfrac{\bar{Q}}{C} , \quad \quad 	\bar{s} = \dfrac{\bar{S}}{C}, \quad \quad 	\bar{F} = \dfrac{F}{C},
\end{equation}
which correspond to electric charge, entropy, and free energy per degree of liberty respectively. Then we expect that the topology of dual CFT depends only on $\bar{q}$ for a given $\bar{b}$. Using the second approach (off-shell free energy) that we have introduced previously in Sec.\ref{off-shell_FE}, we define now a new vector filed $\phi = (\phi^{\bar{s}},\phi^{\theta})$ by 
\begin{equation}\label{78}
	\phi = \left(- \dfrac{\partial \bar{F}}{\partial \bar{s}}, -\cot(\theta) \csc(\theta)\right).
\end{equation}
Then, we expect to observe a topological transition between AdS-Schwarzchild and AdS-Reissner-Nordström topological classes when
\begin{equation}\label{79}
	\bar{q}_m = \dfrac{1}{2 \bar{b}},
\end{equation}  
for a given $\bar{b}$. The condition  Eq.\eqref{29} to have a critical behavior in an S-type black hole becomes in dual CFT context 
\begin{equation}\label{80}
	\bar{b}_I<\bar{b} <\sqrt{\dfrac{3}{2\left( \sqrt{6\sqrt{3}-9}-1\right)}},
\end{equation} 
while the condition Eq.\eqref{30} to have a critical behavior in RN-type rereads as 
\begin{equation}\label{81}
	\bar{b} >\sqrt{\dfrac{3}{2\left( \sqrt{6\sqrt{3}-9}-1\right)}}.
\end{equation}

\begin{figure}[!ht]
	\centering 
	\begin{subfigure}[h]{0.48\textwidth}
		\centering \includegraphics[scale=0.6]{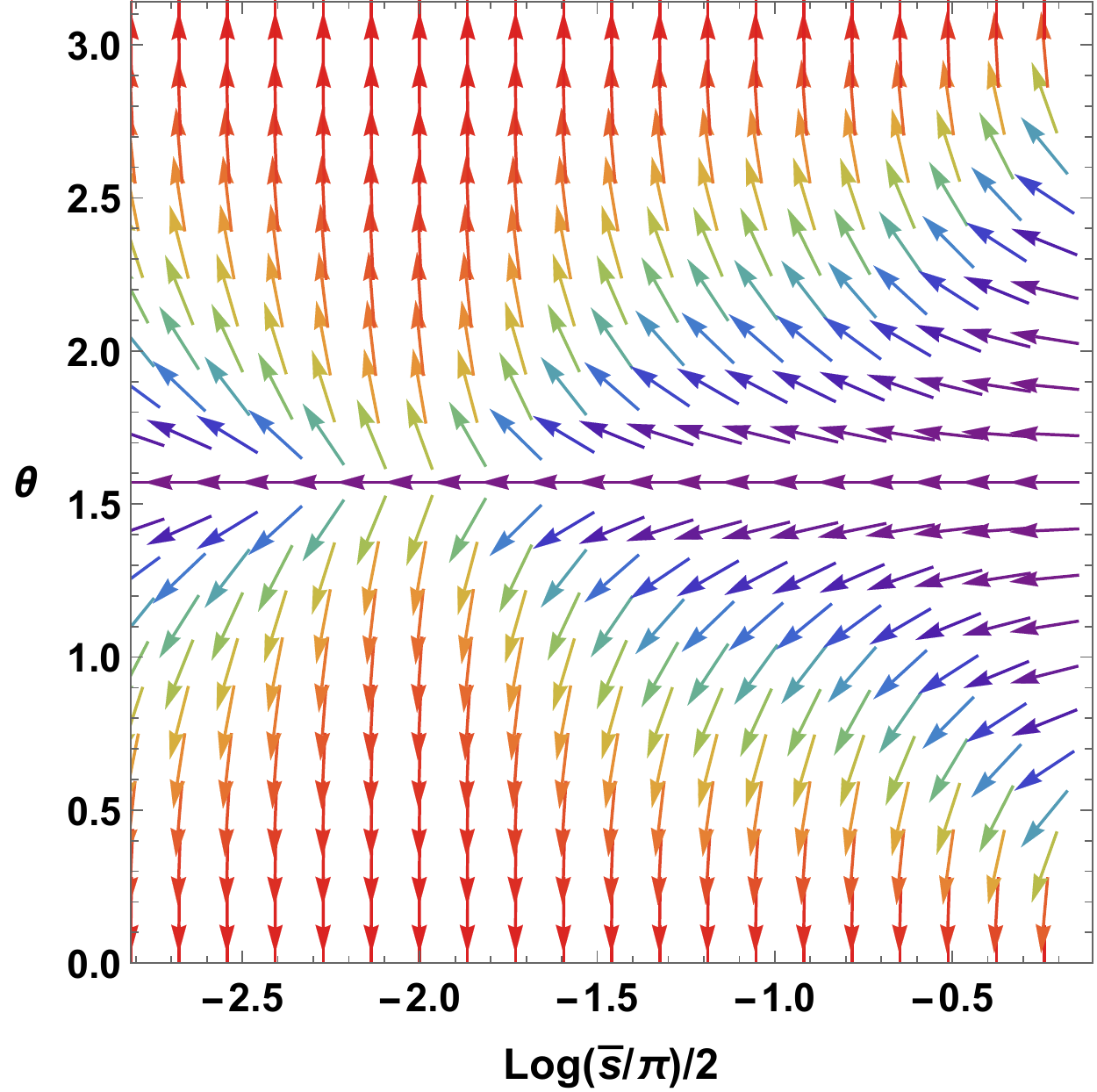}
		\caption{$\bar{T} =  0.245$}
		\label{f43_1}
	\end{subfigure}
	\hspace{1pt}	
	\begin{subfigure}[h]{0.48\textwidth}
		\centering \includegraphics[scale=0.6]{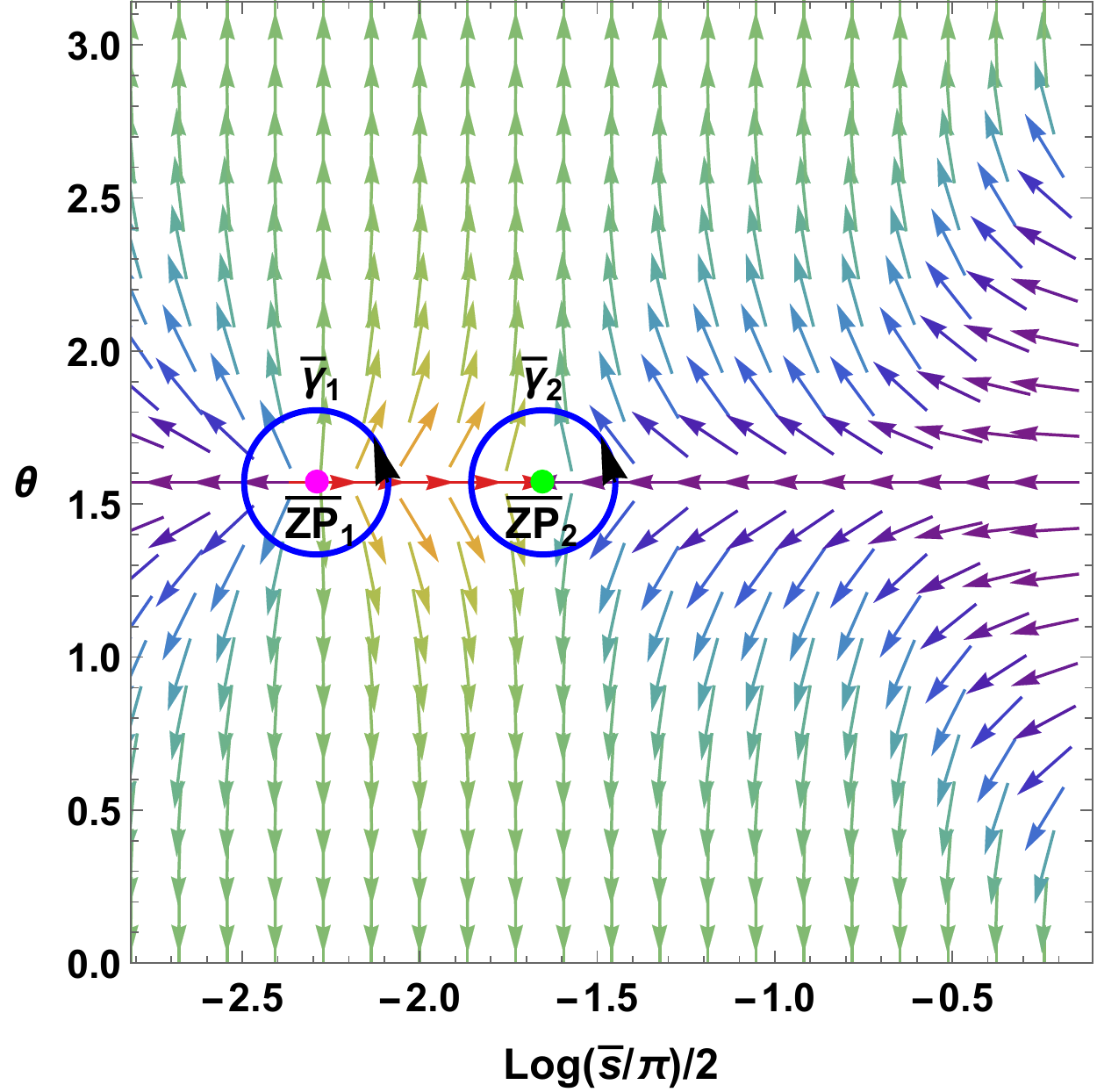}
		\caption{$\bar{T} =  0.255$}
		\label{f43_2}	
	\end{subfigure}
	\hspace{1pt}	
\begin{subfigure}[h]{0.48\textwidth}
	\centering \includegraphics[scale=0.6]{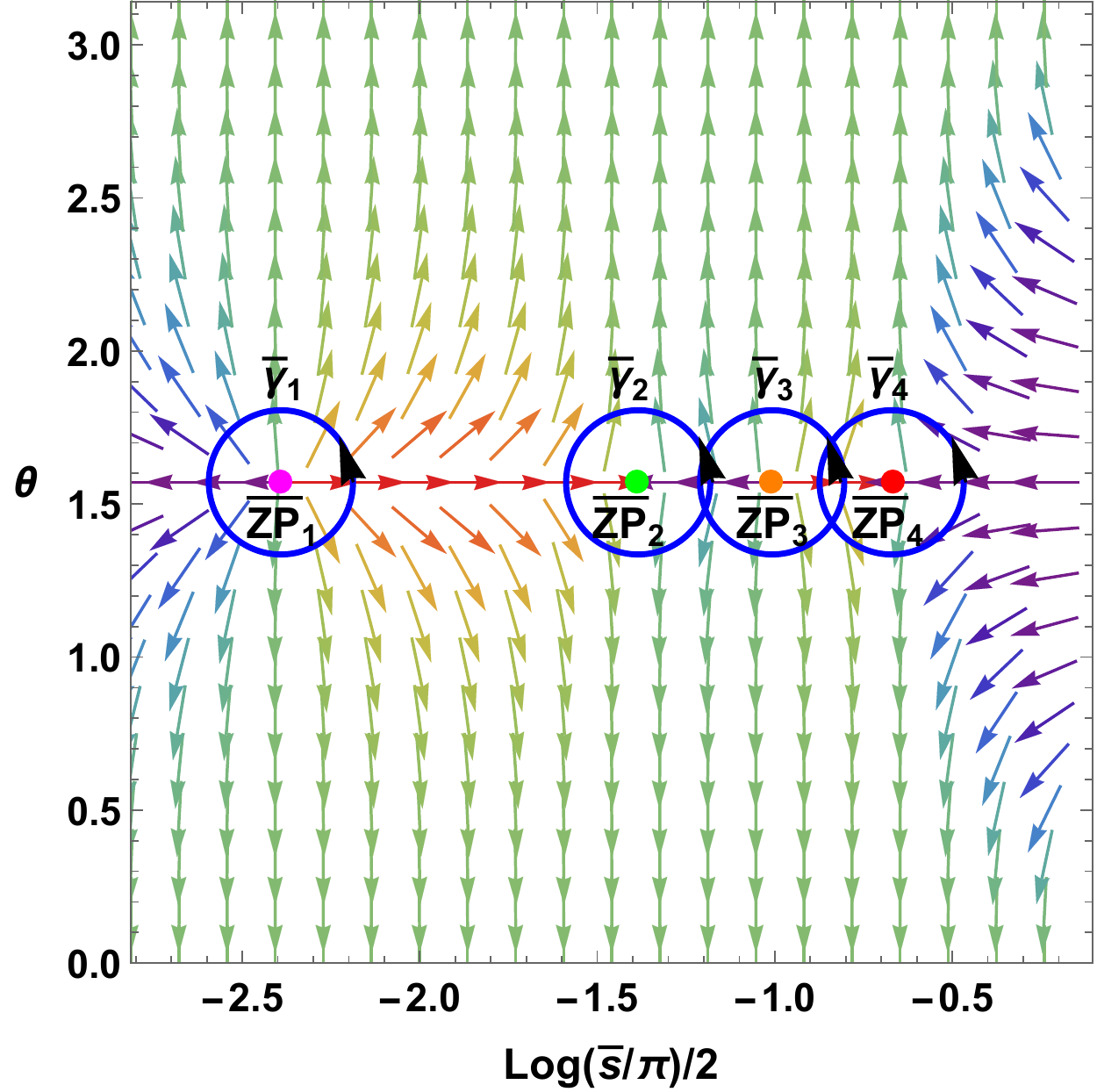}
	\caption{$\bar{T} =  0.2608$}
	\label{f43_3}	
\end{subfigure}
	\hspace{1pt}	
\begin{subfigure}[h]{0.48\textwidth}
	\centering \includegraphics[scale=0.6]{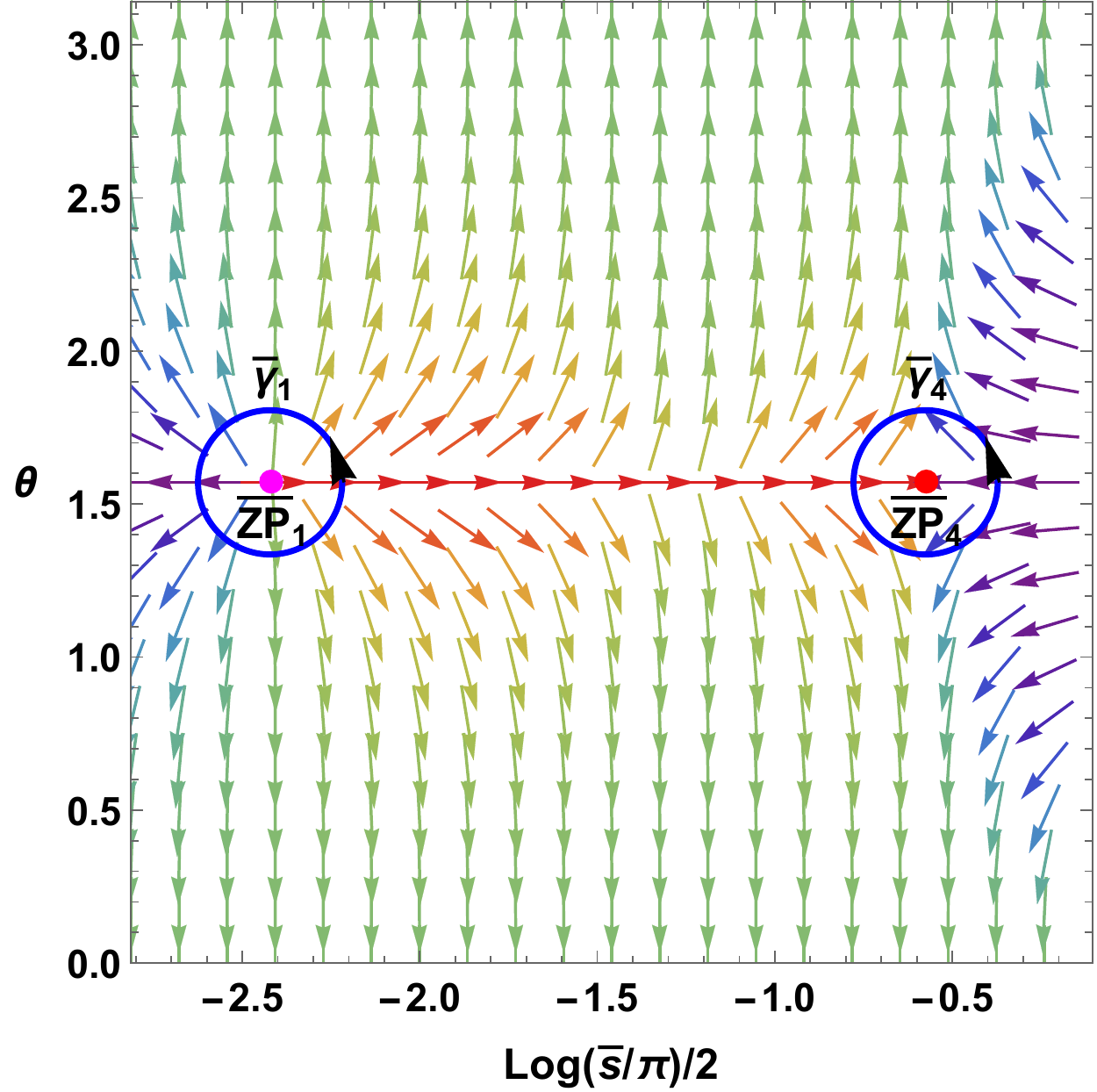}
	\caption{$\bar{T} =  0.263$}
	\label{f43_4}	
\end{subfigure}
	\caption{\footnotesize\it Normalized vector field $n^i$ in the $(\bar{s},\theta)$ plane for different CFT temperatures $\bar{T}$, with $\bar{b} = 2.4$, $\bar{q}= 0.17<\bar{q}_m$ and $\mathcal{V} =1$. }
	\label{f43}
\end{figure}

Within the S-type black hole consideration, we focus on a scenario with $b=2.4$, satisfying the condition Eq.\eqref{80}. In Fig.\ref{f43}, we present the normalized vector field $n^i$, defined as $\phi^i/\left| \left| \phi\right| \right|$, in the $(\bar{s},\theta)$ plane for $\bar{q}= 0.17<\bar{q}_m$.

Numerous remarks can be deduced 
\begin{itemize}
	\item For $\bar{T} = 0.245 $\big($\bar{T}<0.248774$\big) there is no zero point. Therefore, the total topological charge of the system is zero.
	\item For $\bar{T} = 0.255$ \big($0.248774<\bar{T}<0.259748$\big), we have vortex/anti-vortex creation and there are two zero points. Using Eq.\eqref{15},and the two contours $\bar{\gamma}_1$ and $\bar{\gamma}_2$, the topological charges of these two zero points are : 
	\begin{equation}\label{82}
		\mathcal{Q}(\overline{ZP}_1) = +1,  \quad \mathcal{Q}(\overline{ZP}_2) = -1.
	\end{equation}
The initial zero point $\overline{ZP}_1$ (magenta dot) with a positive topological charge $(+1)$ signifies an unstable phase, while the subsequent zero point $\overline{ZP}_2$ (green dot) with a negative topological charge $(-1)$ denotes a locally stable phase. Consequently, as in the bulk scenario, vortices correspond to stable phases, whereas anti-vortices correspond to unstable ones. The overall topological charge of the system is zero (\ref{eq:total_topological_charge}). 
	\begin{equation}\label{83}
	\mathcal{Q} = \mathcal{Q}(\overline{ZP}_1) + \mathcal{Q}(\overline{ZP}_2) = 0.
\end{equation}
	\item For $\bar{T} = 0.2608 $ \big($0.259748<\bar{T}<0.261922$) we are in the presence of another vortex/anti-vortex creation and there are four zero points. Using Eq.\eqref{15}, and the two contours $\bar{\gamma}_3$ and $\bar{\gamma}_4$, the topological charges of these two zero points are evaluated to 
\begin{equation}\label{84}
	\mathcal{Q}(\overline{ZP}_3) = +1,  \quad \mathcal{Q}(\overline{ZP}_4) = -1.
\end{equation}
The third zero point $\overline{ZP}_3$ (orange dot) with a positive topological charge $(+1)$ corresponds to an unstable phase, similar to the first zero point $\overline{ZP}_1$ (magenta dot). Additionally, the fourth zero point $\overline{ZP}_4$ (red dot) with a negative topological charge $(-1)$ corresponds to a locally stable phase as the second zero point $\overline{ZP}_2$ (green dot). In this configuration, the total topological charge of the system remains zero 
\begin{equation}\label{85}
	\mathcal{Q} = \mathcal{Q}(\overline{ZP}_1) + \mathcal{Q}(\overline{ZP}_2) +\mathcal{Q}(\overline{ZP}_3) + \mathcal{Q}(\overline{ZP}_4) = 0.
\end{equation}
	\item For $\bar{T} = 0.263 $\big($\bar{T}>0.259748$\big) we have a vortex/anti-vortex annihilation and there are only two zero points,$\overline{ZP}_1$ and $\overline{ZP}_4$, with opposite topological charges. The two zero points  $\overline{ZP}_2$ and $\overline{ZP}_3$ annihilate each other and disappear. The total topological charge of the system is still null  
\begin{equation}\label{86}
	\mathcal{Q} = \mathcal{Q}(\overline{ZP}_1) + \mathcal{Q}(\overline{ZP}_4) = 0.
\end{equation}
\end{itemize}
Thus, in the S-type case and for $\bar{q}<\bar{q}_m$ the total topological charge is always equal to zero independently of temperature and the system belongs to the topological class of AdS-Schwarzchild black hole.

\begin{figure}[!ht]
		\centering \includegraphics[scale=0.6]{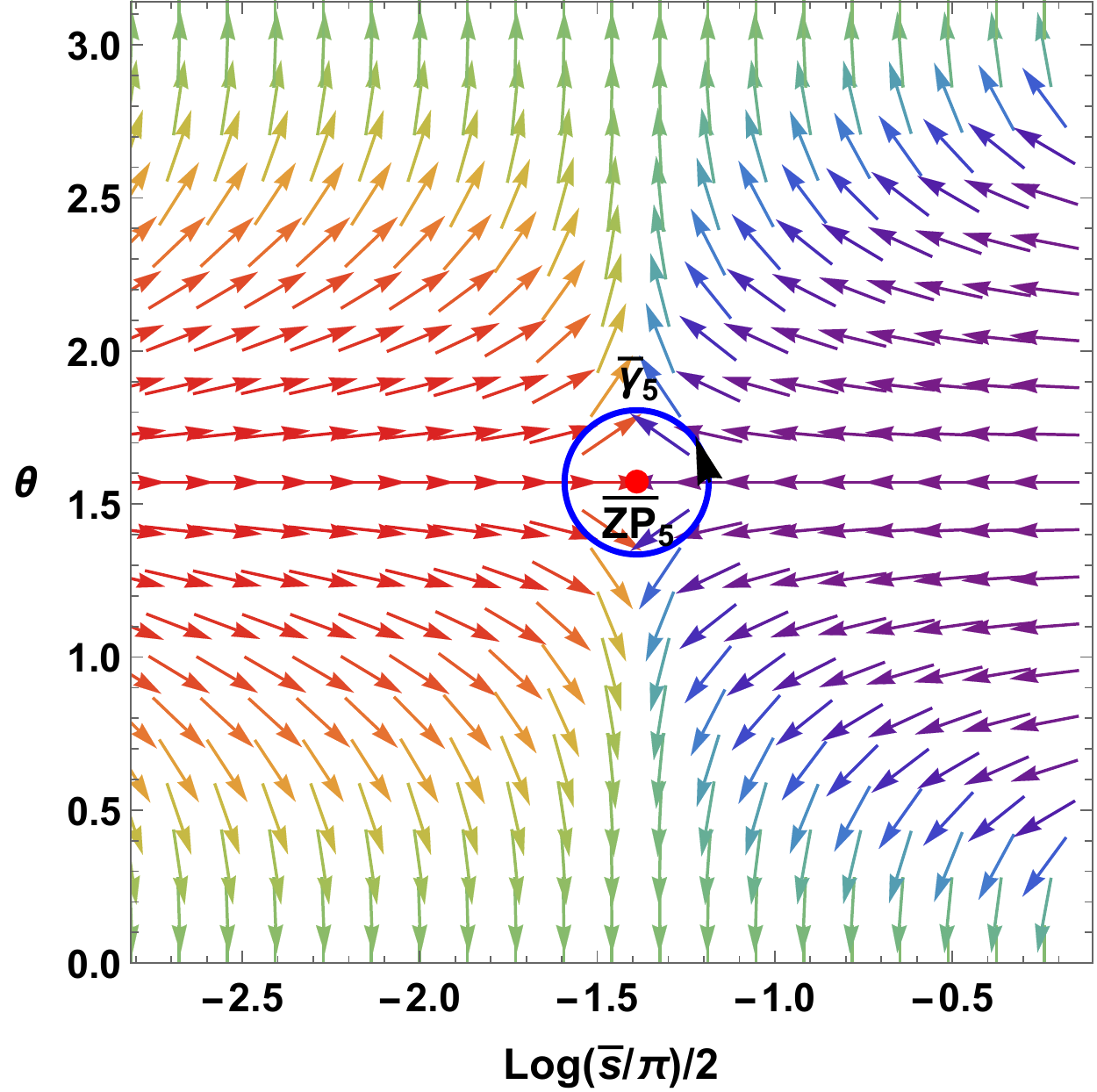}
	\caption{\footnotesize\it Normalized vector field $n^i$ in the $(\bar{s},\theta)$ plane for $\bar{T} = 0.2$, with $\bar{b} = 2.4$, $\bar{q}= 0.22>\bar{q}_m$ and $\mathcal{V} =1$. }
	\label{f44}
\end{figure}

Let's turn our attention to the case when $\bar{q}>\bar{q}_m$ in S-type black hole configuration, we illustrate in Fig.\ref{f44} the normalized vector field $n^i$ in the $(\bar{s},\theta)$ plane for $\bar{q}= 0.22>\bar{q}_m$ and $\bar{T} = 0.2$ \footnote{We can choose any temperature $\bar{T}>0$.}.
 We notice that just one zero point $\overline{ZP}_5)$ (vortex) which corresponds to a stable phase. With the help of Eq.\eqref{15} and the contour $\bar{\gamma}_5$, the topological charge of this zero point is found to be
	\begin{equation}\label{87}
	\mathcal{Q}(\overline{ZP}_5) = -1.
\end{equation}
Thus, the total topological charge of the system is equal to $-1$ :
	\begin{equation}\label{88}
	\mathcal{Q}=\mathcal{Q}(\overline{ZP}_5) = -1,
\end{equation} 
and the system belongs to the topological class of the AdS-Reissner-Nordström black hole. Hence, in the S-type case, there is a topological transition in dual CFT at $\bar{q} = \bar{q}_m$ and the system changes its topological class.

\begin{figure}[!ht]
	\centering 
	\begin{subfigure}[h]{0.48\textwidth}
		\centering \includegraphics[scale=0.6]{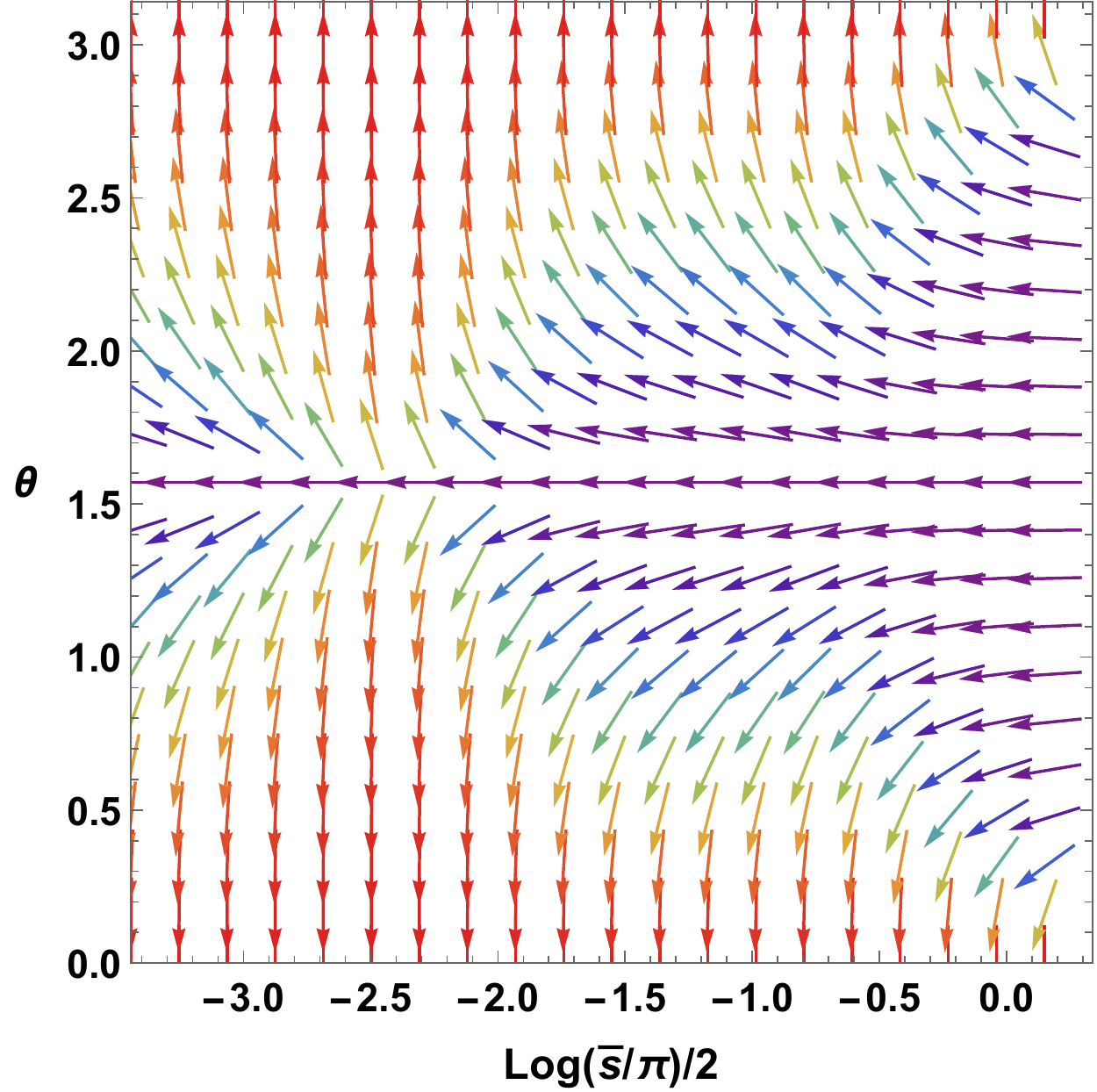}
		\caption{$\bar{T} =  0.25$}
		\label{f45_1}
	\end{subfigure}
	\hspace{1pt}	
	\begin{subfigure}[h]{0.48\textwidth}
		\centering \includegraphics[scale=0.6]{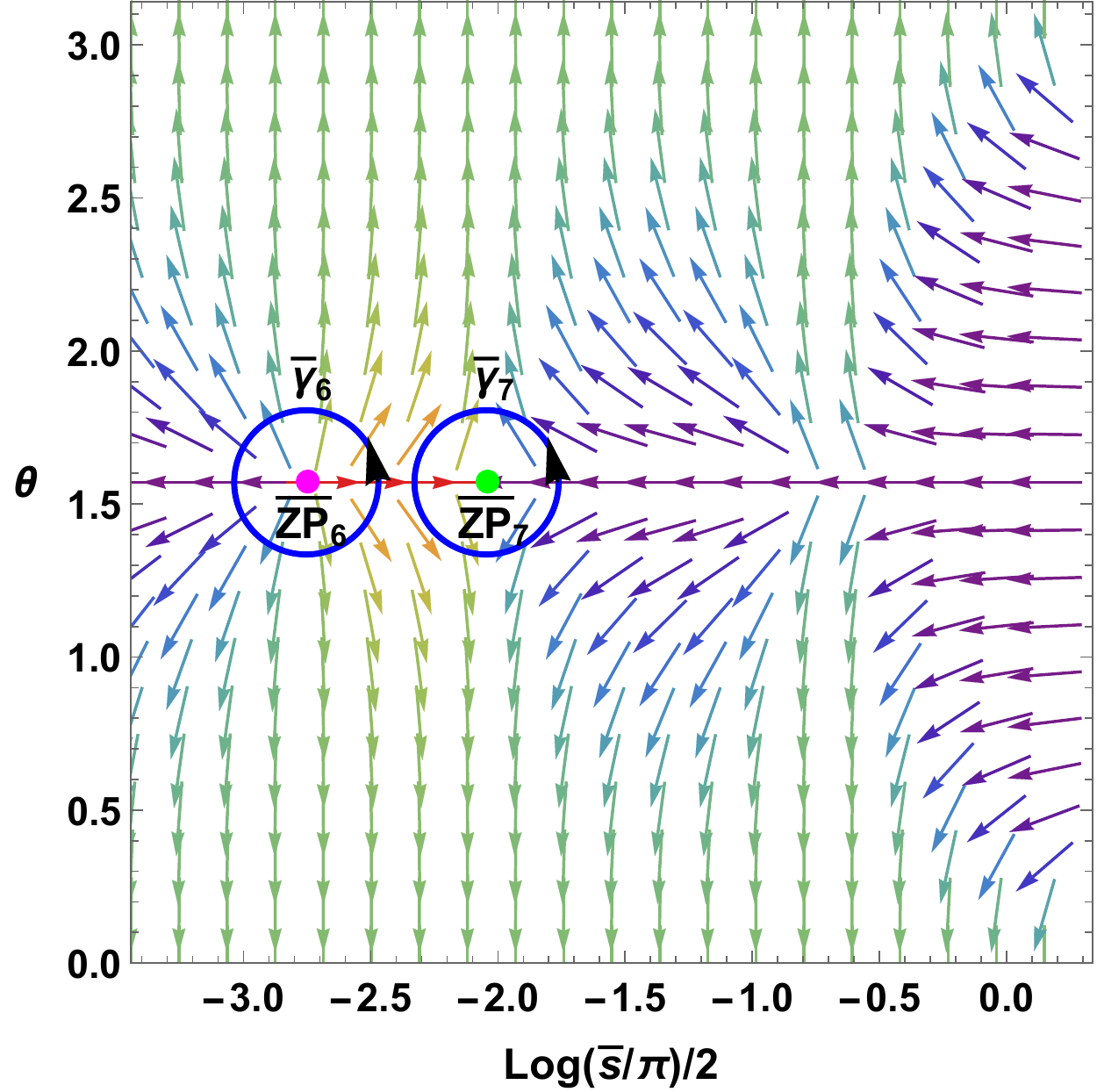}
		\caption{$\bar{T} =  0.265$}
		\label{f45_2}	
	\end{subfigure}
	\hspace{1pt}	
	\begin{subfigure}[h]{0.48\textwidth}
		\centering \includegraphics[scale=0.6]{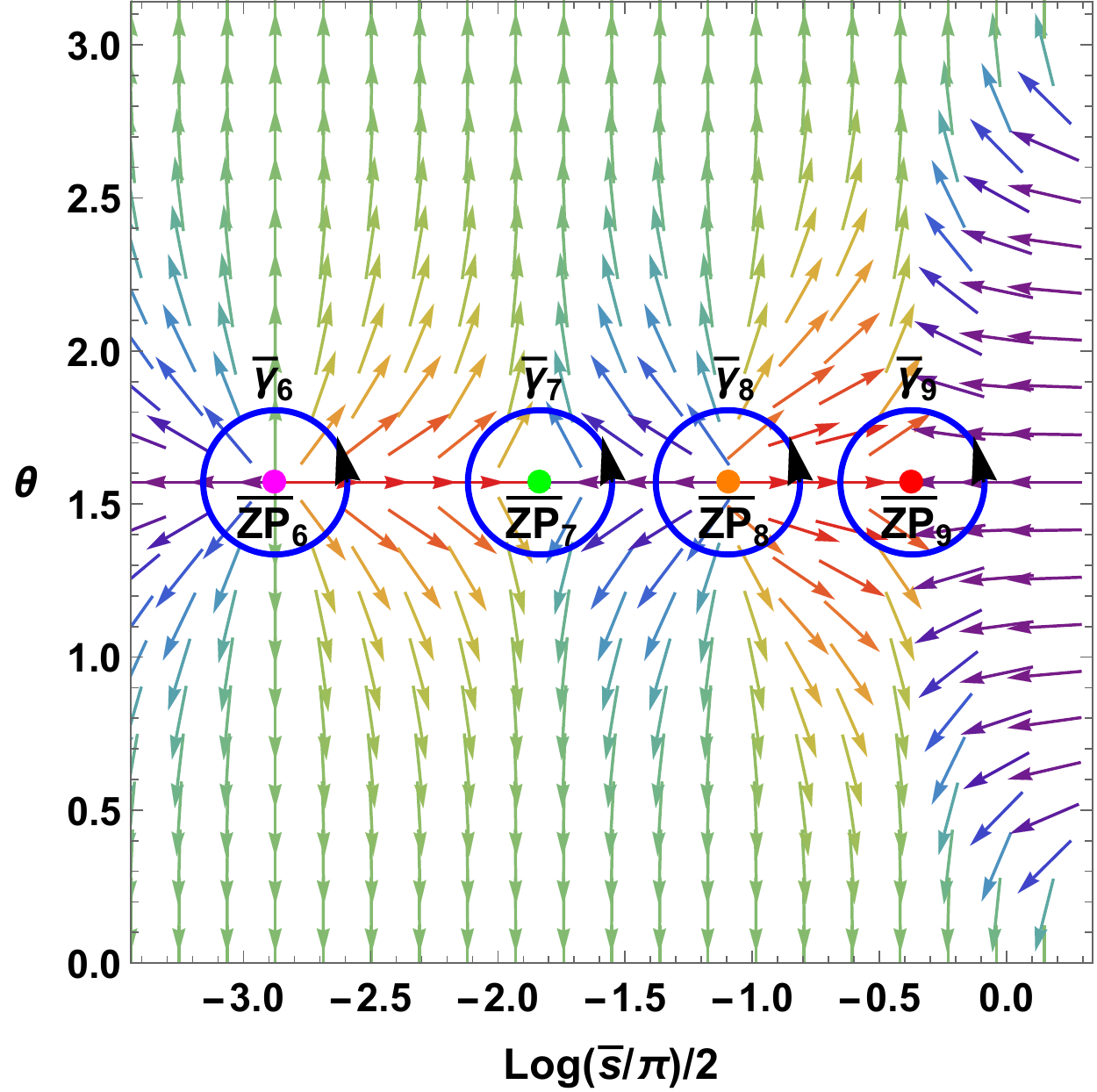}
		\caption{$\bar{T} =  0.275$}
		\label{f45_3}	
	\end{subfigure}
	\hspace{1pt}	
	\begin{subfigure}[h]{0.48\textwidth}
		\centering \includegraphics[scale=0.6]{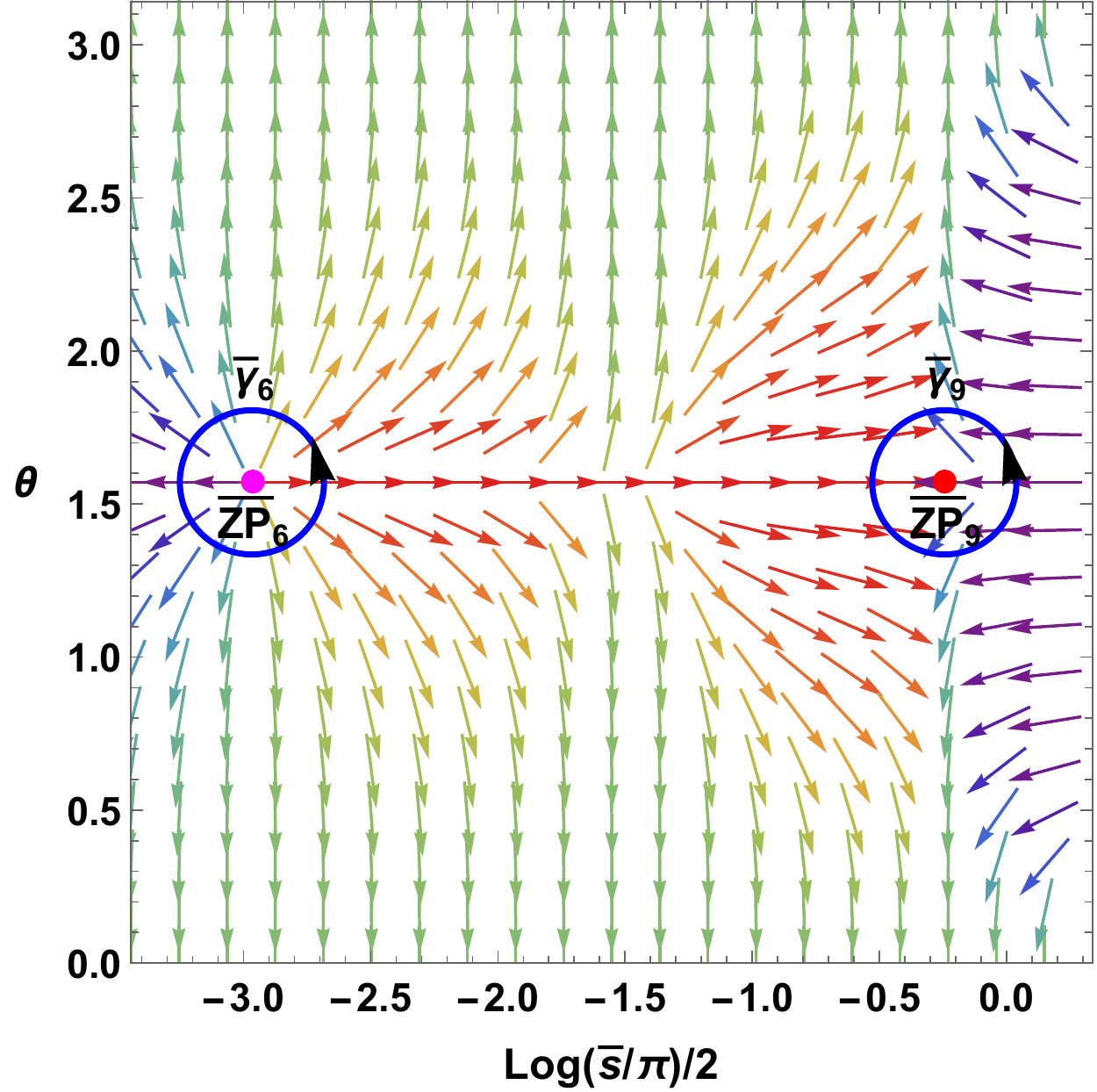}
		\caption{$\bar{T} =  0.285$}
		\label{f45_4}	
	\end{subfigure}
	\caption{\footnotesize\it Normalized vector field $n^i$ in the $(\bar{s},\theta)$ plane for different CFT temperatures $\bar{T}$, with $\bar{b} = 3$, $\bar{q}= 0.145<\bar{q}_m$ and $\mathcal{V} =1$. }
	\label{f45}
\end{figure}

Now, let us consider the scenario corresponding to an RN-type black hole with the chosen parameter $b=3$, respecting the condition given by Eq.\eqref{81}. In Fig.\ref{f45}, we present the normalized vector field $n^i$ in the $(\bar{s},\theta)$ plane for $\bar{q}= 0.145<\bar{q}_m$.
 Notably, the obtained figures exhibit equivalence to those in the S-type, as depicted in Fig.\ref{f43}. Consequently, the topology of the dual CFT in the RN-type case, for $\bar{q}<\bar{q}_m$, perfectly aligns with that in the S-type case. Thus, it is evident that for $\bar{q}<\bar{q}_m$, the total topological charge remains consistently equal to zero, irrespective of temperature and parameter $\bar{b}$.

\begin{figure}[!ht]
	\centering 
\begin{subfigure}[h]{0.48\textwidth}
	\centering \includegraphics[scale=0.6]{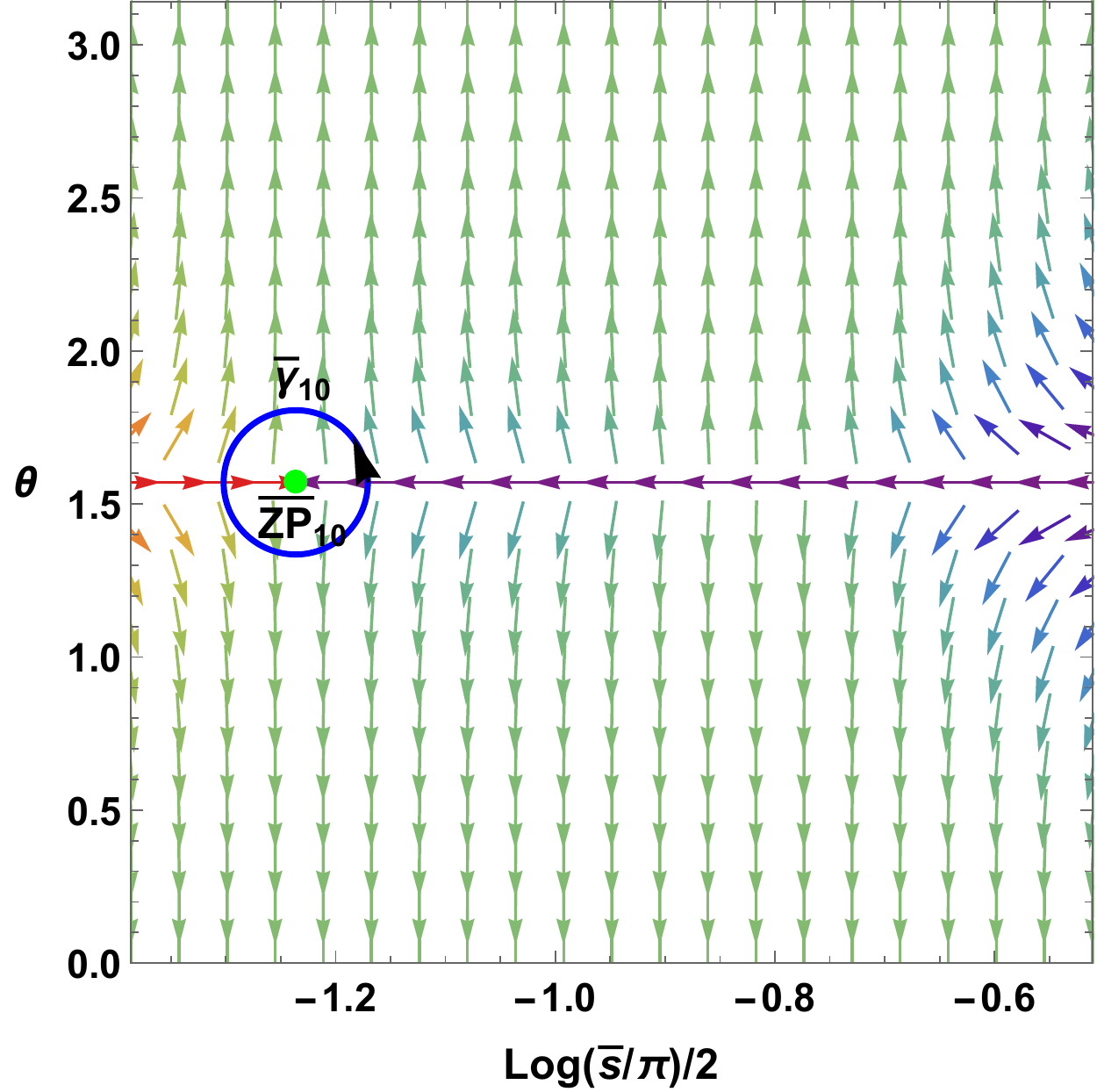}
	\caption{$\bar{T} =  0.26$}
	\label{f46_1}
\end{subfigure}
\hspace{1pt}	
\begin{subfigure}[h]{0.48\textwidth}
	\centering \includegraphics[scale=0.6]{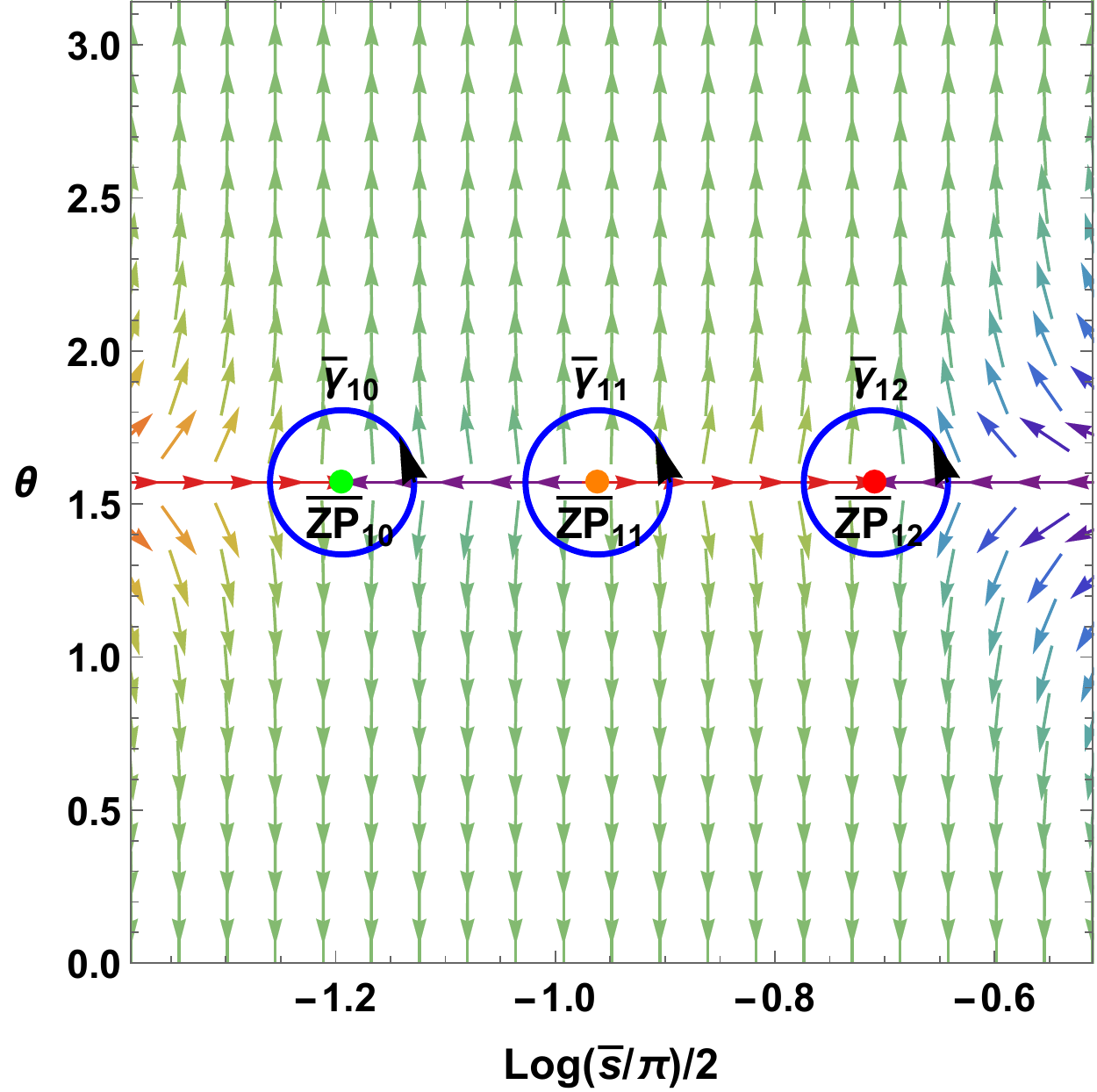}
	\caption{$\bar{T} =   0.2606$}
	\label{f46_2}	
\end{subfigure}
\hspace{1pt}	
\begin{subfigure}[h]{0.48\textwidth}
	\centering \includegraphics[scale=0.6]{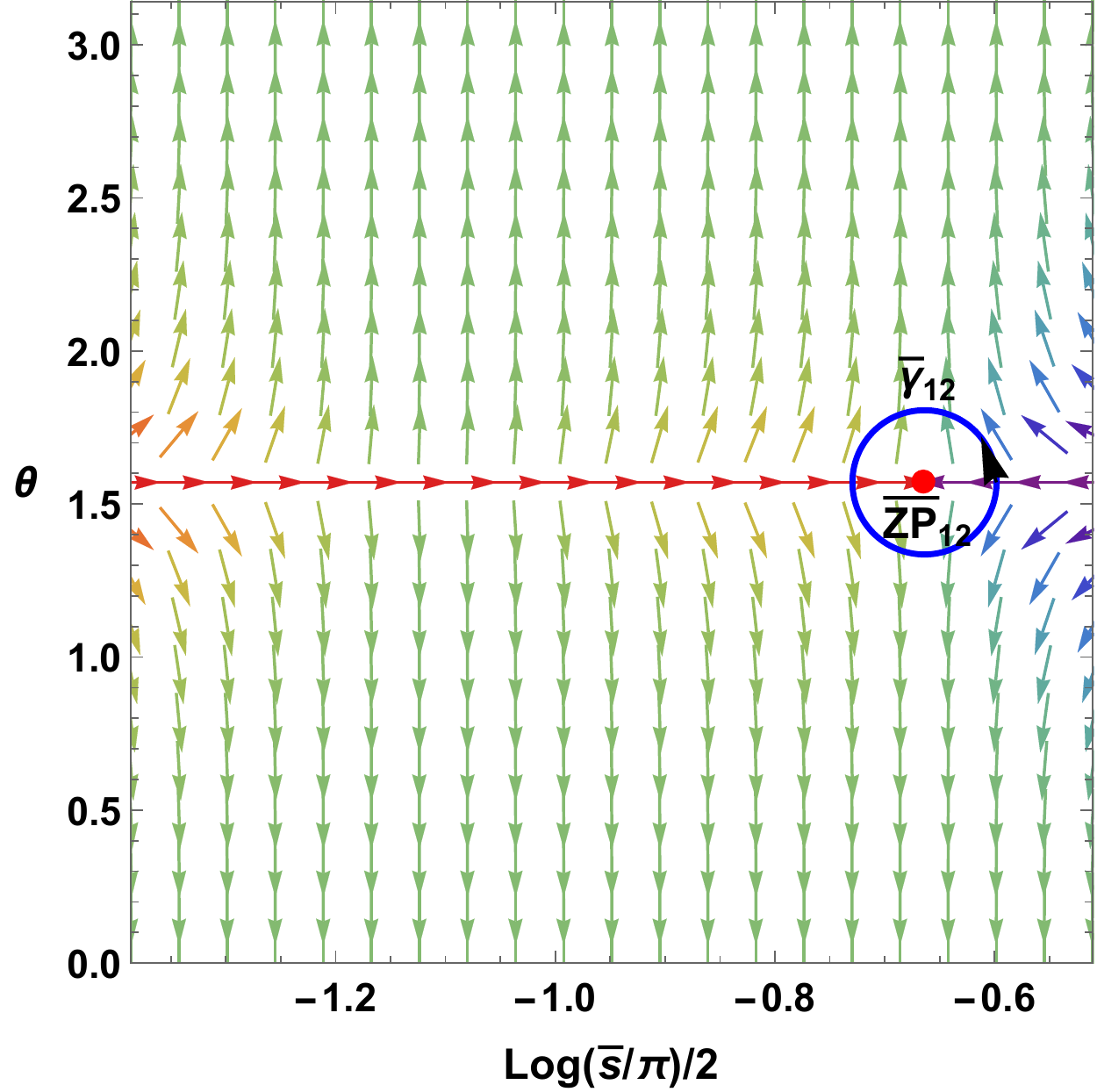}
	\caption{$\bar{T} =  0.2612$}
	\label{f46_3}	
\end{subfigure}
\caption{\footnotesize\it Normalized vector field $n^i$ in the $(\bar{s},\theta)$ plane for different CFT temperatures $\bar{T}$, with $\bar{b} = 3$, $\bar{q}= 0.168<\bar{q}_m$ and $\mathcal{V} =1$. }
\label{f46}
\end{figure}

Assuming the  $\bar{q}>\bar{q}_m$ scenario in RN-type, we depict in Fig.\ref{f46} the normalized vector field $n^i$ in the $(\bar{s},\theta)$ plane for $\bar{q}=  0.168>\bar{q}_m$ for different temperatures.

We observe that
\begin{itemize}
	\item $\bar{T} = 0.26 $\big($\bar{T}<0.260063$\big), we have just one zero point $\overline{ZP}_{10})$ (vortex) corresponding to a stable phase. Using Eq.\eqref{15} and the contour $\bar{\gamma}_{10}$, the topological charge of this zero point is
	\begin{equation}\label{89}
		\mathcal{Q}(\overline{ZP}_{10}) = -1.
	\end{equation}
 Leading to a total topological charge of the system equaling $-1$ :
\begin{equation}\label{90}
	\mathcal{Q}=\mathcal{Q}(\overline{ZP}_{10}) = -1,
\end{equation} 
\item For $\bar{T} = 0.2606 $\big($0.260063<\bar{T}<0.261096$\big) we observe a vortex/anti-vortex creation and there are three zero points. Using Eq.\eqref{15}, and the two contours $\bar{\gamma}_{11}$ and $\bar{\gamma}_{12}$, the topological charges of these last two zero points are : 
\begin{equation}\label{91}
	\mathcal{Q}(\overline{ZP}_{11}) = +1,  \quad \mathcal{Q}(\overline{ZP}_{12}) = -1.
\end{equation}
The third zero point $\overline{ZP}_{11}$ (orange dot)  with positive topological charge $(+1)$ corresponds to an unstable phase, and the fourth zero point $\overline{ZP}_{12}$ (red dot) is associated with negative topological charge $(-1)$ which corresponds to locally stable
phase as well as the first zero point $\overline{ZP}_{10}$ (green dot). Consequently, the total topological charge of the system in this situation is also equal to $-1$ : 
\begin{equation}\label{92}
	\mathcal{Q} = \mathcal{Q}(\overline{ZP}_{10}) + \mathcal{Q}(\overline{ZP}_{11}) +\mathcal{Q}(\overline{ZP}_{12})  = -1.
\end{equation}
\item For $\bar{T} = 0.2612$ \big($\bar{T}>0.261096$\big) we notice a vortex/anti-vortex annihilation and there is only one zero point,$\overline{ZP}_{12}$, with a negative topological charge. The two zero points  $\overline{ZP}_{10}$ and $\overline{ZP}_{11}$ annihilate each other and disappear. The total topological charge of the system is still equal to $-1$: 
\begin{equation}\label{93}
	\mathcal{Q} = \mathcal{Q}(\overline{ZP}_{12}) = -1.
\end{equation}
\end{itemize}
Therefore, for $\bar{q}>\bar{q}_m$, the total topological charge consistently equals $-1$, irrespective of temperature and the parameter $\bar{q}$ \big($\bar{b}>\bar{b}_I$\big). Consequently, a topological transition in the dual CFT occurs at $\bar{q} = \bar{q}_m$, for all $\bar{b}>\bar{b}_I$, leading to a change in the system's topological class.

Finally, in Table.\ref{Table4}, we summarize the topological characteristics of the dual CFT as a function of the electrical charge per degree of liberty $\bar{q}$,
\begin{table}[!ht]
	\begin{center}
		\begin{tabular}{|c||c|c|}
			\hline
			 Electric charge per degree of liberty $\bar{q}$   &   $\bar{q}<\bar{q}_m$  &    $\bar{q}>\bar{q}_m$ \\
	    	\hline
	    	 Topological charge $\mathcal{Q}$   &   $0$  &    $-1$\\
	    	 \hline
	    	 Topological class    &   AdS-Schwarzchild  &    AdS-Reissner-Nordström \\
	    	 \hline

		\end{tabular}
	\end{center}
	\caption{\footnotesize\it Topological characteristics of dual CFT for $\bar{b}>\bar{b}_I$.}
	\label{Table4}
\end{table}
 illustrating a topological transition between two distinct topological classes. This signifies that a Born-Infeld-AdS black hole and its dual CFT share the same topology. Therefore, when a topological transition takes place in the bulk, a corresponding transition occurs in the dual CFT as well.

\section{Conclusion}
\label{summary}
The thermodynamics structure at the critical points and its connection to the topology of the Born-Infeld-AdS black hole systems is worth studying one. In this paper, we adopted two different mechanisms to investigate the thermodynamics of the topological classes of the Born-Infeld-AdS black holes, namely, Duan's topological current $\phi$-mapping theory and the off-shell Gibbs free energy, respectively. In the first approach, we vividly described the thermodynamic phase transitions of the Schawrzschild-AdS type and the RN-AdS type black holes. Depending upon the values of the Born-Infeld parameter $b$, these black holes admitted one, two, or three critical points, which are connected to usual Van der Waals-type transitions between small and large black holes, and also an unconventional phase transition between the unstable small and the intermediate black holes or the intermediate and the large black holes. We identified the topological classes of such phase transitions that correspond to the generation or annihilation of the topological vortex/anti-vortex pairs. The unconventional phase transition was made disappeared when the Born-Infeld parameter reached $b=8\pi Q_m$, and the Born-Infeld AdS black hole behaved as the RN-AdS black holes. The topological charge for the critical point at $Q=Q_m$ is equal to $\mathcal{Q}=-1$, which subsequently confirmed that there exists a topological phase transition at this point. This topological phase transition is analogous to the BKT transition observed in the two-dimensional XY model and in such case, the parameter $Q_m$ played the role of the  BKT transition temperature $T_\text{BKT}$. 
The behaviour of the horizon temperature with respect to the variation of the horizon radius showed the RN-AdS type, which is a monotonically increasing function. Contrary to the $Q>Q_m$ case, we did not observe any unstable phase and thereby no cusp-like structure of the Gibbs free energy-temperature plot for $Q<Q_m$. The first derivative of the Gibbs free energy at $r_h\to 0$ is negative and therefore, we always have a Hawking-Page-like phase transition between the pure thermal radiations and the small/large black holes. Hence, Duan's formalism for the thermodynamic topology of the Born-Infeld AdS black hole indeed provided us with a beautiful insight into the topological classes during the phase transitions.\\

Next, we studied in a fully detailed way, the locally thermodynamic stable or unstable branches of the black hole solutions to probe the topological structures by constructing the off-shell free energy. We used the sense of the vector field definition as in Duan's case, but the field components have a direct correspondence with the Gibbs free energy. Similar to the previous case, we engineered a similar analysis of the S-type and the RN-type black holes. Keeping intact the condition on the pressure term (see Eq.~(\ref{29})), in terms of the Born-Infeld parameter, we first investigated the situation for the S-type black holes. In the plot of the vector field $n^i$ in the $(r_h,\theta)$ plane, we have four zeros corresponding to values of the charge parameter $Q_m=0.0093$ and the temperature $T=0.335$, and thereby correspondingly we have associated topological charges, yielding a total topological charge of value zero. These four zeros correspond to the USBH magenta line (first zero), the SSBH green line (second dot), the IBH orange line (third zero), and the LBH red line (fourth zero), respectively. A summary of the topological classes of the S-type Born-Infeld-AdS is given in Tab.~\ref{Table2}, for different sets of values of the electric charge parameter and the temperature. It is to be emphasized that for a specific choice of the parameter set $(Q,T)$, we had conventional and unconventional phase transitions. A similar analysis of the RN-type Born-Infeld-AdS black holes are summarized in Tab.~\ref{Table3}. Unlike the S-type black holes, for RN-type black holes we observed a forbidden region where the conditions for the extremality is highlighted in Figs.~\ref{f22} and \ref{f23}. \\

To add a rich flavour to the vortex/anti-vortex pairs creation and annihilation processes, we probed the thermodynamics of the relevant quantities. For our analysis, we plotted the vortex/anti-vortex pairs creation and annihilation points in the $(\frac{1}{Q},T)$ plane. We made the analogy that the curve in the $(\frac{1}{Q},T)$ plane has a similar structure to the familiar $(G,T)$ diagram that occurred during the first order phase transition characterizing a swallowtail behavior. This way, we interpreted the inverse of the electric charge as the temperature of the annihilation/creation points and the temperature as the free energy. Such an analogy led us to categorize three thermodynamic processes, namely, LBH and SSBH annihilation phases which
are both locally stable (cf. blue lines in Figs.~\ref{f24} and \ref{f25}), and SSBH creation phase which is
an unstable phase (cf. purple lines Figs.~\ref{f24} and \ref{f25}). A detailed study of the different thermodynamic phases of the vortex/anti-vortex pairs annihilation and creation processes have been discussed from both the local and global thermodynamic point of view.
\textcolor{blue}{At the end, we have reexamined the 
 the phase structure from the conformal field theory perception, underscoring the fact that the Born-Infeld-AdS black hole and its dual CFT share an identical topology. This underscores the crucial observation that any topological transition occurring in the bulk corresponds to a parallel transition in the dual CFT.}
This work comes up with certain open questions. One of them may concern other methods being used to probe the topology of the black hole phase transitions. It would be of interest to investigate such topological tools by applying the Riemann surface and the number of foliations exploited in many places. We hope to come back to such an issue in future works.

\section*{Acknowledgment}
 This work was supported in part by the National Natural Science Foundation of China (Grants  No. 12347177, Grants  No. 12247101 and No. 12247178), the 111 Project (Grant No. B20063), and Lanzhou City’s scientific research funding subsidy to Lanzhou University.

\bibliographystyle{unsrt}
\bibliography{BITopology.bib}

\end{document}